\documentclass[article,a4paper]{elsarticle}

\usepackage{lineno,hyperref}
\journal{International Journal of Impact Engineering}

\usepackage[top=2cm, bottom=2cm, left=2cm, right=2cm]{geometry}

\usepackage{subfigure}
\usepackage{xcolor}
\usepackage{float}
\usepackage{sectsty}

\graphicspath{{Figures/}}

\begin{document}
\allsectionsfont{\sffamily}

\begin{frontmatter}

\title{\huge{\textbf{\textsf{Experimental Investigation of the Fluid-Structure Interaction during the Water Impact of Thin Aluminium Plates at High Horizontal Speed}}}}
%\tnotetext[mytitlenote]{Fully documented templates are available in the elsarticle package on 
%\href{http://www.ctan.org/tex-archive/macros/latex/contrib/elsarticle}{CTAN}.}

\author{Emanuele Spinosa\corref{cor1}}
\cortext[mycorrespondingauthor]{Corresponding author}
\address{CNR-INM (Institute of Marine Engineering) Via di Vallerano 139, 00128 Roma (RM), Italy}
\ead{emanuele.spinosa@inm.cnr.it}   
%\telephone{+39 06 502991}
%\fax{+39 06 5070619}

\author{Alessandro Iafrati \fnref{myfootnote}}
\address{CNR-INM (Institute of Marine Engineering) Via di Vallerano 139, 00128 Roma (RM), Italy}
\ead{alessandro.iafrati@cnr.it}  


\begin{abstract}
The water impact of an inclined flat plate and at high horizontal velocity is experimentally investigated with focus on the fluid-structure interaction aspects. Several test conditions have been examined by varying the vertical to horizontal velocity ratio, the pitch angle and the plate thickness. Measurements are performed in terms of strains, loads and local pressure. The study highlights the significant changes in the strains and, more in general, in the structural behaviour when varying the plate stiffness and the test conditions. For some of the test presented, permanent deformation are also found. The strong fluid-structure interaction is analysed by comparing the simultaneous measurements of strains and pressures, and it is shown that the deformation of the plate leads to a reduction of the pressure peak and to a corresponding pressure rise behind it. The variation in the shape of the spray root caused by the structural deformation are discussed based on both pressure measurements and underwater images. Despite the reduction of the pressure peak intensity, it is shown that the structural deformation leads to an increase in the total loading up to 50\% for the test conditions examined in this study. It is also observed that in presence of large structural deformations the hydrodynamic loads do not obey to the scaling that works in the case of the thick plates, and some practical conclusions about the scaling of tests in presence of a strong fluid-structure interaction are provided.
\\
\\
\emph{$\copyright$ 2021. This manuscript version is made available under the CC-BY-NC-ND 4.0 license}

\url{http://creativecommons.org/licenses/by-nc-nd/4.0/}

\end{abstract}

\begin{keyword}
Fluid-Structure Interaction, Water impact, Aircraft ditching, Hydroelasticity
\end{keyword}

% \setcounter{tocdepth}{4}
% \tableofcontents

\end{frontmatter}

%\linenumbers

\section{Introduction}

In this paper the fluid-structure interaction effects taking place during
the water entry of inclined flat thin Aluminium plates at high horizontal velocity are
investigated experimentally. The basic motivation for the study stems from
the aircraft ditching problem, i.e. the emergency landing on water of
aircraft in controlled conditions. Although being, fortunately, a rare
event, this problem still imposes aircraft manufacturers to carefully consider its
occurrence during the design stage, as the aircraft has to be certified at
ditching according to the regulation of the airworthiness authorities. Full
scale tests on aircraft are of course impractical, being too expensive, therefore scaled model tests are generally considered 
\cite{Climent2006,Zhang2012}. However, as discussed in \cite{Iafratiaiaa2019}, scaled model experiments are not fully 
representative. Generally, the structure is not scaled,
 thus the main struts and the skin are much stiffer than in reality. The poor reliability of the scaled model tests is also 
associated with the low test speed, which prevents the occurrence of some important hydrodynamic phenomena \citep{iafrati2019cavitation,Climent2006}.

Detailed reviews of the studies on water entry problems with application to aerospace structures can be found in 
\cite{seddon2006review} and in \cite{hughes2013aerospace}. In \cite{hughes2008application} and \cite{ren2016verification} some problems related to helicopter ditching are examined. The use of computational approaches for the design and certification of aircraft at ditching is of course very attractive, and 
significant steps forward have been taken recently \cite{bisagni2018modelling,siemann2018advances,siemann2017coupled,anghileri2011rigid,anghileri2014survey}. Nonetheless, a further 
enhancement of the capabilities of those tools to reproduce both the hydrodynamics and the strong fluid-structure interaction during 
the ditching phase is advisable. For such a development, experimental data truly representative of what happens in a real scenario 
are needed. 

Water impact problems are far more common in the naval field, as the ship often moves in a wavy sea (hull slamming), see \cite{faltinsen2000hydroelastic,abrate2011hull,facci2015assessment,facci2016three,campbell2012simulating}. The impact on the water surface is always accompanied by an intense pressure peak moving along the hull. Nevertheless, in \cite{faltinsen1997wave} it is shown that the pressure peak on its own, especially when it is large, is quite irrelevant for the determination of the structural response, because it is strongly sensitive to the external conditions and, moreover, is too concentrated in space and time to affect significantly the structural behaviour. In this respect, it is generally observed that the hydro-elastic effects are of great importance, and therefore it is essential to gather information on the local deformations of the body, using for example strain measurements, and to examine their coupling with the overall velocity and the pressure fields, as well as with the shape of the free surface during the water entry. For some specific cases, it is also essential to account for the occurrence of ventilation and cavitation \cite{faltinsen2000hydroelastic,faltinsen2005hydrodynamics,iafrati2019cavitation}.

The relevance of fluid-structure interaction phenomena during the water
entry of a plate or wedge, with focus on slamming, is assessed in several
works. 
In \cite{loarn2010simplified} a simplified model to investigate the
effect of curvature, characteristic of the bow part of racing yachts, is
developed. The hydroelastic interaction during panel-water impact is also 
studied in \cite{stenius2011hydroelastic}. By using numerical
simulations, kinematic effects related to the structural deformation of the
fluid-solid boundary, and inertia effects, related to the rate at which the structure responds to the external loading, are 
distinguished. These effects are highlighted by comparing the results with numerical simulations in which the structural problem and the hydrodynamic one are not coupled. In 
\cite{stenius2013experimental} a series of experiments aimed at validating the numerical simulations performed in 
\cite{stenius2011hydroelastic} is presented. 
It is shown that the largest hydroelastic effect appears to be a time-lag
of the peak found in the measured strains with respect to the one derived from 
the simulations. This time-lag is assumed to be inertia related. Kinematic 
hydroelastic effects, instead, are mainly concentrated in the upper part
of the panel, where a local decrease in the deadrise angle is observed.
In \cite{panciroli2012hydroelasticity, panciroli2013dynamic} a comparison
between experiments on highly deformable wedges entering water and
numerical simulations using SPH (Smooth Particle Hydrodynamics) is
presented. It is observed that the intensity of the hydroelastic effects is
governed by the ratio of the wetting time to the natural period of the
structure, as already pointed out in \cite{faltinsen1997wave,faltinsen2000hydroelastic}.
It is also found that more than one shape mode can be excited. However, at least for the conditions
examined, the structural deformation does not appear to influence the
hydrodynamic pressure distribution, unless separation occurs as a
consequence of cavitation and ventilation, which are phenomena that cannot
be captured using their numerical model.
Khabakhpasheva and Korobkin \cite{khabakhpasheva2013elastic} developed a semi-analytical method to
solve the two-dimensional problem of the water impact of an elastic wedge. Wagner theory is used for
the determination of the hydrodynamic loads and the structural response is derived using modal analysis.
Shams and Porfiri \cite{shams2015treatment} 
developed a novel method for the simulation of the hydroelastic impact of 
flexible wedges in free fall with arbitrary boundary conditions.
The model is able to capture the footprints of the hydroelastic
coupling, including the oscillations of the impact forces due to the appearance of the
instantaneous suction on the wedge and the pressure fluctuations in the flow field.
A rather detailed investigation of the slamming response of stiffened Aluminium plates and of the 
effect of small imperfections generated by the welding process is provided in \cite{khedmati2014numerical}, 
where some design recommendations are given as well.

All the works cited above refer to the pure vertical water entry case.
However, the presence of a horizontal velocity component may affect both
the intensity of the pressure and the wetting time, and its
effect may be very significant, particularly if the horizontal component is very high, 
which is the case of aircraft ditching. A theoretical model for the fluid-structure 
interaction during the water impact of a two-dimensional elastic plate at high horizontal velocity is proposed in \cite{reinhard2013water}. 
The plate is modelled as an Euler beam with free-free boundary conditions. 
It is found that the plate deformation may lead to a significant increase in the hydrodynamic 
loading, and thus the hydrodynamic and structural problems have to be fully coupled 
in order to achieve a physically accurate result. 
Wang et al. \cite{wang2016hydroelastic} performed a numerical and theoretical study of the 
water impact of a horizontal plate with both horizontal and vertical speed. The plate is modelled as a series of 
longitudinal strips, each of which has the behaviour of an Euler beam, and the study is mainly focused on the longitudinal compressive force. It was found that the structural response of the plate and
the compressive force strongly depend on the boundary conditions and on the horizontal speed.
Experimentally, the water impact with horizontal velocity of a flat plate is investigated in \cite{smiley1951experimental}, but no 
information about the structural deformation is available.
However, the pressure data measured by the different probes seems rather
coherent, and therefore it is expected that the plate used for the tests is thick enough to remain in the elastic regime.
Water impact tests at high horizontal velocity of elastic plates are presented in \cite{iafrati2015high} and 
\cite{iafrati2016experimental}. The latter work is focused on hydrodynamic aspects, whereas in the former one some aspects concerning the 
plate deformation are also addressed. Results of tests performed with a horizontal velocity in the range 30 to 45~m/s, vertical 
velocity of 1.5~m/s and pitch angles from 
4 to 10 degrees are provided. In order to assess the test-to-test
repeatability, a plate made of Aluminium alloy AL2024 T3, 15~mm thick is 
employed, thus ensuring that deformations are always within the elastic range and 
are small enough not to affect the hydrodynamics. Beside thick plates, much thinner Aluminium 
plates, 3~mm and 0.8~mm thick, are also tested. A preliminary analysis of the strains
in some test conditions is presented in \cite{iafrati2015nav}, where significant fluid-structure interaction aspects are highlighted.

In this paper, the structural deformations of the plates and the subsequent changes in the hydrodynamics are analysed based on the 
data of the tests performed at different impact velocities, pitch angles and for different plate stiffness's. Results are mainly 
presented in terms of strains. However, for a few cases the simultaneous measurements of both strains and pressures are provided in 
order to highlight the changes in the hydrodynamics induced by the structural deformation. The effects of the structural deformation on the hydrodynamics are clearly highlighted by the underwater images, whereas more relevant consequences can be observed based on the loads measured in different conditions.

\section{Experimental Setup and Test Conditions}
\label{setup}

Experiments were performed on the High Speed Ditching Facility (HSDF), available at the CNR-INM (Institute of Marine Engineering). The facility consists of a guide, which is suspended on a water basin through five bridges. The guide can be rotated in order to achieve different horizontal to vertical velocity ratios. 
Detailed information on the facility can be found in Iafrati et al.\cite{iafrati2015high}. The flat plate (the test model) was installed onto a box structure, which was mounted underneath a trolley running along the guide.

\begin{figure}[htbp]
\centering
\includegraphics[width=0.75\textwidth]{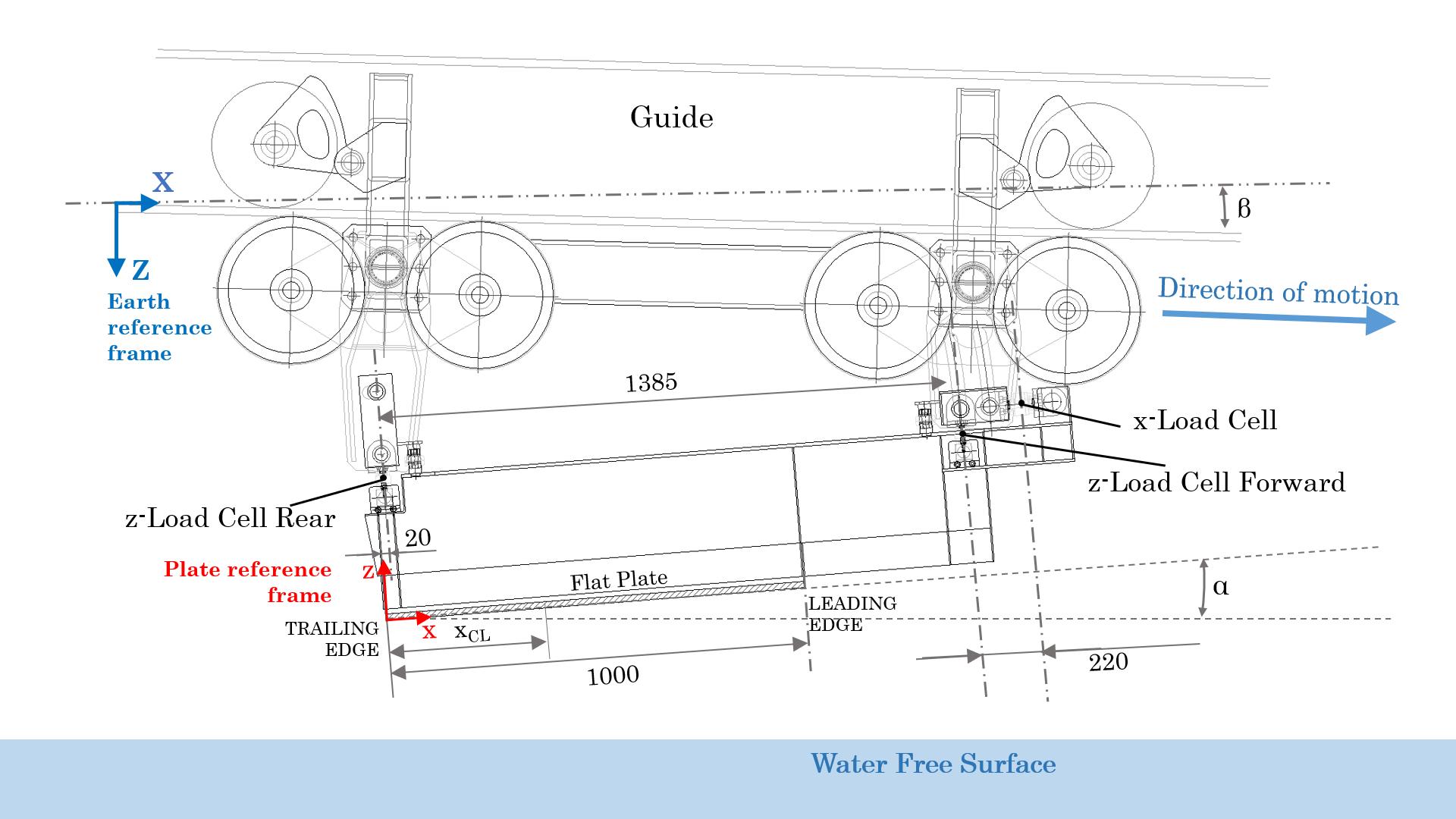}
\caption{Technical drawing of the trolley installation on the guide (dimensions are in~mm).}
\label{fig:trolley_drawing}
\end{figure}

It is useful to distinguish the two reference frames used in the following and shown in
Figure \ref{fig:trolley_drawing}: an earth reference frame, with coordinate system $(X,Y,Z)$, 
and a reference frame with coordinates $(x,y,z)$ fixed to the plate and moving with the trolley. 
The horizontal and vertical velocity of the trolley, denoted as $U$ and $V$ respectively, are referred to the 
earth-reference frame.

A technical drawing of the trolley is shown in Figure \ref{fig:trolley_drawing}. 
$\alpha$ is the pitch angle, $\beta$ is the angle of inclination of the guide, which 
corresponds to $\arctan(V/U)$.
The trolley was accelerated to the test speed by a system of elastic ropes,
and it was left free to run along the guide a short time before the impact. 
Tests were performed on Aluminium plates of different thickness. Firstly,
a plate 15~mm thick was tested: in this case the structural deformations
remained within the elastic range, thus allowing to explore a wide range of conditions, 
assuring a negligible effect of the structural deformation on the hydrodynamics. 
The results of these tests are detailed in Iafrati et al.
\cite{iafrati2015high} and in Iafrati \cite{iafrati2016experimental}. Such
results are referred to as thick plate in the following. 
In addition, thin panels, 3~mm and 0.8~mm thick, were considered, for which  
larger and even permanent deformations were expected. 

\begin{figure}[htbp]
\centering
\includegraphics[width=0.70\textwidth]{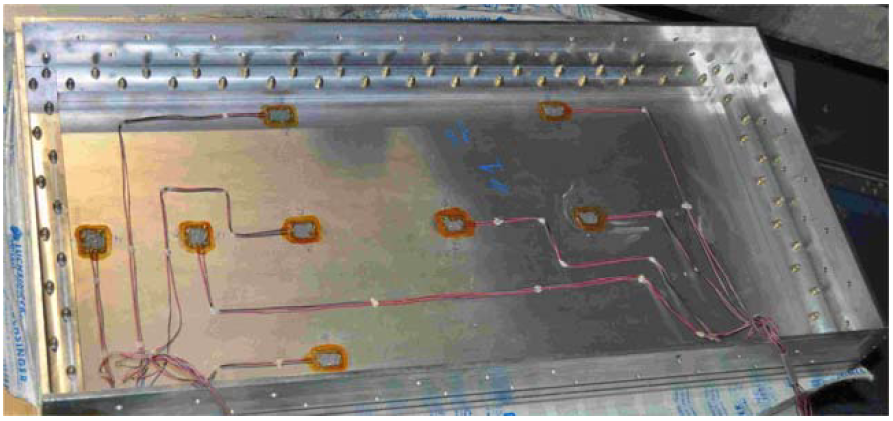}
\caption{Flat plate instrumented with the strain gauges. The
trailing edge of the plate is on the left.}
\label{fig:instrumentedplate_picture}
\end{figure}

All plates were made of the Aluminium alloy AL2024-T3, characterized by a 
yielding stress of approximately 340 MPa, which corresponds to a
yielding strain of about 4600 $\mu$m/m. The plates were 500~mm
wide and 1000~mm long and were connected to a thicker Aluminium frame by
two rows of bolts, as shown in Figure \ref{fig:instrumentedplate_picture},
mimicking a quasi-clamped boundary condition all around. The width of the
frames was 75~mm and 50~mm for the thick and thin plates respectively. The different width
 was motivated by the different scopes of the tests. For the thick plates the wider frame 
was chosen to maximize the stiffness. In the thin plates the frame border was smaller in order 
to maximize the deformation, still keeping the two-row bolt connection and mimicking the 
clamped boundary condition as much as possible. Bolts diameters of 5/16" and 1/4" were used 
for the thick and thin plates, respectively. In
order to prevent the pull-out failure at the bolts heads, for the 0.8~mm
plates a 3~mm thick reinforcement, 54~mm wide all around, was bonded with a 
structural adhesive to the tested plates before connecting 
them to the frame. 

\begin{figure}[htbp]
\centering
\includegraphics[width=1.00\textwidth]{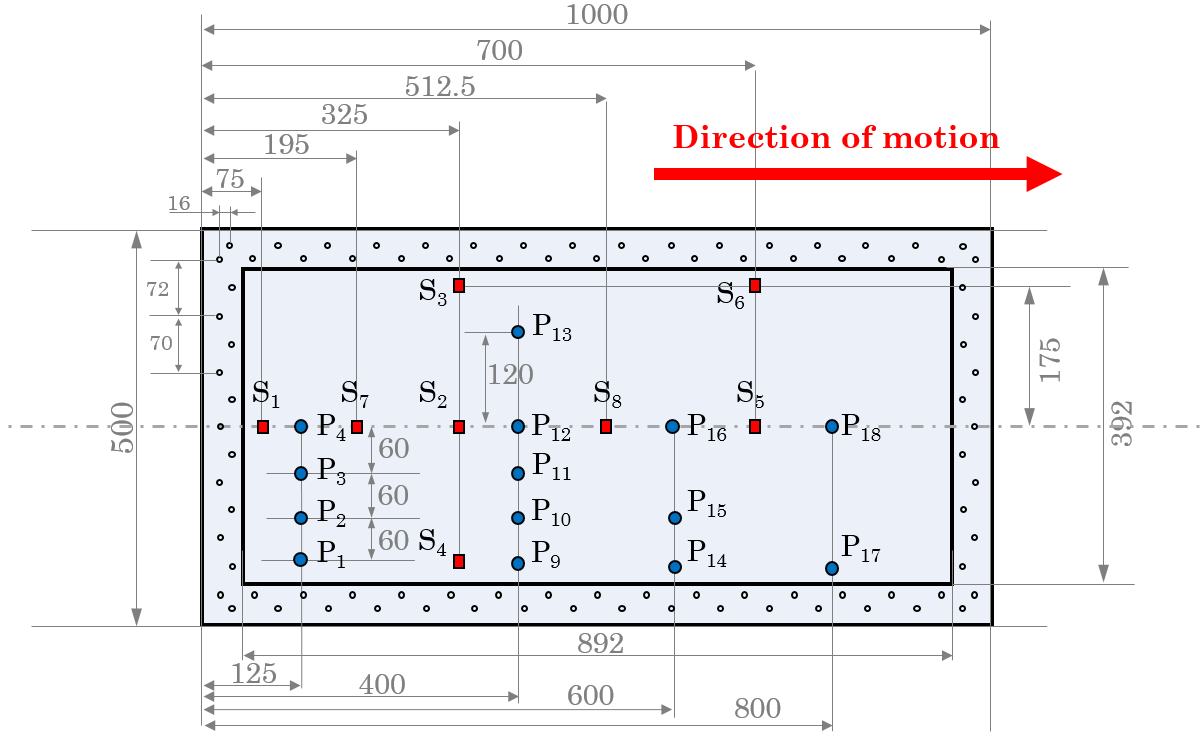}
\caption{Top view of the plate with the position of the strain gauges and of the pressure probes. The
trailing edge of the plate is on the left (dimensions are in~mm).}
\label{fig:Sensor_Positions}
\end{figure}

The plates were instrumented by biaxial strain gauges, type Vishay C2A-12-125LT-350 for the thick plates and Vishay EP-08-125TQ-350 
on the thin plates, the former allowing a maximum deformation of 3\%, the latter arriving up to 15\%. 
Six and eight strain gauges were used for the thick and for the thin
plates respectively. In the following S$_x$ and S$_y$ denote the strain
measurement in the $x$ and $y$ directions, respectively. 
In all tests with the thick plates and in some of the tests with the thin plates, 
pressures were measured using Kulite XTL-123B probes, 300 psi full-scale
range. The position of the strain gauges and of the pressure probes is 
shown in Figure \ref{fig:Sensor_Positions}.
A picture of a 0.8 mm plate instrumented with the strain gauges only is shown in Figure \ref{fig:instrumentedplate_picture}. 
The total loads acting on the plates were measured by six load cells: four 
Kistler piezo-electric load cells, type 9343A, full-scale range of 70 kN, were 
used to measure the load component acting normal to the plate 
(\emph{z}-direction), whereas two Kistler 9363A, full-scale range of 120 kN,
were used to measure the tangential component 
(\emph{x}-direction). The forces are referred to the reference
frame fixed to the plate.
All signals were acquired by four Sirius and one Dewe43 Dewesoft DAQ modules, 
connected to a ruggedized computer. Pressure signals were sampled at 
200 kHz, whereas all other channels were sampled at 20 kHz. 
The DAQ modules include an anti-aliasing filter and an A/D converter. 
After the data have been stored, no further filtering
was performed.
All measurement are zeroed by taking the average over a short time interval 
before the impact, and therefore a null value of the pressure 
corresponds to the atmospheric value. 

Tests were recorded by two high-speed cameras, 1024x1024 pixel resolution.
One camera, operated at 5000 frame per second (fps), was located at 
the side and was used to measure the speed at the impact, whereas the other camera,
operated at 2000 fps, was located underwater to visualize the impact phase 
and the flow features beneath the plate. 

Tests were performed by varying the horizontal velocity, the vertical to 
horizontal velocity ratio $V/U$ and the pitch angle. Here attention is mainly 
focused on the tests performed on the 3~mm and 0.8~mm Aluminium plates, but a 
comparison with the corresponding tests on the 15~mm Aluminium
plates \citep{iafrati2016experimental} is also presented in order to highlight 
the role played by the structural deformation on the loading and on the hydrodynamics.
Each test condition is identified by a code
\textbf{TT\_PP\_VV\_l}. \textbf{TT} is a one or two-digit number that indicates the plate thickness (15, 3 and 08
stand for 15~mm, 3~mm and 0.8~mm respectively). \textbf{PP} is a two-digit number that indicates the pitch angle at impact (04 for 4$^{\circ}$, 06 is for 6$^{\circ}$ and 10 is for 10$^{\circ}$). \textbf{VV} is a two-digit number that indicates the nominal test velocity (30 for 30~m/s, 40 for 40~m/s and 45 for 45~m/s). The final letter \textbf{l} denotes the repeat in alphabetical order (a,b, ... for the first, second repeat and so on). The complete list of the tests performed on the thick (15~mm) and thin plates (3~mm and 0.8~mm thick) is given in Table 
\ref{tab:AlTestConditions}.
\begin{table}
\centering
\begin{tabular}{cccccc} \hline
\textbf{Thick Plates}  \\ \hline \hline
ID code & Thickness [mm] & U [m/s] & V [m/s] & Pitch [$^{\circ}$] & Pressure probes \\ \hline
15\_04\_45\_a & 15     & 44.89   & 1.5 & 4   & yes  \\ \hline % 1113\_03
15\_04\_45\_b & 15     & 45.27   & 1.5 & 4   & yes  \\ \hline % 1113\_03 
15\_04\_45\_c & 15     & 45.27   & 1.5 & 4   & yes  \\ \hline % 1113\_04
15\_10\_30\_a  & 15     & 29.88   & 1.5 & 10   & yes  \\ \hline % 1131\_02
15\_10\_30\_b  & 15     & 30.92   & 1.5 & 10   & yes  \\ \hline % 1131\_03
15\_10\_30\_c  & 15     & 30.74   & 1.5 & 10  & yes  \\ \hline % 1131\_04
15\_10\_45\_a  & 15     & 45.85   & 1.5 & 10  & yes  \\ \hline % 1133\_01
15\_10\_45\_b  & 15     & 45.85   & 1.5 & 10  & yes  \\ \hline % 1133\_02
15\_10\_45\_c  & 15     & 45.85   & 1.5 & 10  & yes  \\ \hline % 1133\_03
15\_06\_40\_a  & 15     & 40.43   & 1.5 & 6  & yes  \\ \hline % 1122\_01 
15\_06\_40\_b  & 15     & 40.10   & 1.5 & 6  & yes  \\ \hline % 1122\_03 
15\_06\_40\_c  & 15     & 39.78   & 1.5 & 6  & yes  \\ \hline % 1122\_03
   \\ \hline
\textbf{Thin Plates}  \\ \hline \hline
ID code & Thickness [mm] & U [m/s] & V [m/s] & Pitch [$^{\circ}$] & Pressure probes \\ \hline
3\_06\_40\_a   & 3     & 40.08   & 1.5 & 6   & no  \\ \hline % 2122\_03
08\_06\_40\_a   & 0.8   & 40.08   & 1.5 & 6   & no  \\ \hline % 3122\_01
08\_06\_40\_b  & 0.8   & 40.08   & 1.5 & 6   & no  \\ \hline % 3122\_02
3\_04\_45\_a  & 3     & 46.04   & 1.5 & 4   & no  \\ \hline % 2113\_01
08\_04\_45\_a   & 0.8   & 45.65   & 1.5 & 4   & no  \\ \hline % 3113\_01 
3\_04\_45\_b   & 3     & 46.04   & 1.5 & 4   & yes \\ \hline % 2113\_02 
08\_10\_45\_a   & 0.8   & 45.85   & 1.5 & 10  & no  \\ \hline % 3133\_01
3\_10\_30\_a   & 3     & 30.92   & 1.5 & 10  & yes \\ \hline % 2131\_01 
08\_10\_30\_a  & 0.8   & 31.10   & 1.5 & 10  & no  \\ \hline % 3131\_01
3\_10\_30\_b  & 3     & 30.92   & 1.5 & 10  & yes \\ \hline % 2131\_02
08\_10\_30\_b  & 0.8   & 31.28   & 1.5 & 10  & yes \\ \hline % 3131\_02                 
\end{tabular}
\caption{Test conditions for the thick (15~mm) and thin (3~mm and 0.8~mm) plates.}
\label{tab:AlTestConditions}
\end{table}
The use of elastic ropes to accelerate the trolley does not allow a precise control of the 
nominal test velocity, which can be matched with an accuracy
of $\pm$~0.50~m/s. 
% REPEATABILITY -- reference to appendix
A detailed analysis of the test repeatability, based on the 
results obtained for the thick plates, is provided in \cite{iafrati2015high}. 
Therein, the data of ten repeats of the tests performed at two different pitch 
angles were used to retrieve the ensemble average and the standard deviation in 
terms of pressures, strains and loads. 
A similar analysis was performed on the thin plates, even though,
due to the occurrence of permanent deformations, the number of repeats was 
limited to two. More details on test repeatability are found in \ref{repeatability_section}

% TIME ORIGIN and IMPACT PHASE
In order to compare the results of
the different tests or repeats, an origin of the time axis has to be chosen. 
For the thick plates the origin of the time was set at the passage
of the pressure peak at the pressure probe $P_4$, i.e. the first probe located
along the midline. For the thin plates, since pressures are only available for a limited number of tests, 
the peak of the strain $S_{1x}$ is used instead.
Of course, the first contact of the plate with the water surface happens
somewhat before the chosen time origin, as it can be argued by the growth 
of the normal load measured at the rear. 
It is worth remarking that when comparing the results of thick and thin plates,
a small misalignment in the time histories may occur due to the different 
positions of the strain gauge S$_1$.
For the purpose of the study, the most interesting part of the test is the 
so-called 'impact phase', which lasts from the first contact of the plate 
with water up to the time at which the spray root reaches the leading edge, 
and the front face of the acquisition box impacts with water, thus leading 
to a sharp deceleration. 
In the following, the beginning of the impact phase is assumed to coincide 
with the time origin defined above, whereas the end is identified from the 
test data as the time $t_E$ at which a sharp rise in the tangential loads 
$F_x$ occurs. 

\section{Experimental results and discussion}

\subsection{Strain measurements}
\label{strains}

The present study is mainly aimed at investigating how the plate
deformation changes when varying the plate stiffness and the test
conditions. It is worth noticing that in some conditions permanent 
deformation occurs. As anticipated, for the specific material adopted in the tests, 
the yielding limit is estimated about 4600 $\mu m/m$.
It is expected that fluid-structure interaction phenomena can have a significant
effect on the dynamics of the impact for the thin plates (3~mm and~ 0.8 mm). This is due to both the low thickness and to the low value, less than or equal to 1, of the ratio of the wetting time to the first natural frequency of the plate, which is estimated in \ref{nat_frequencies}.

\begin{figure}[htbp]
	\centering
		\includegraphics[width=0.7\textwidth]{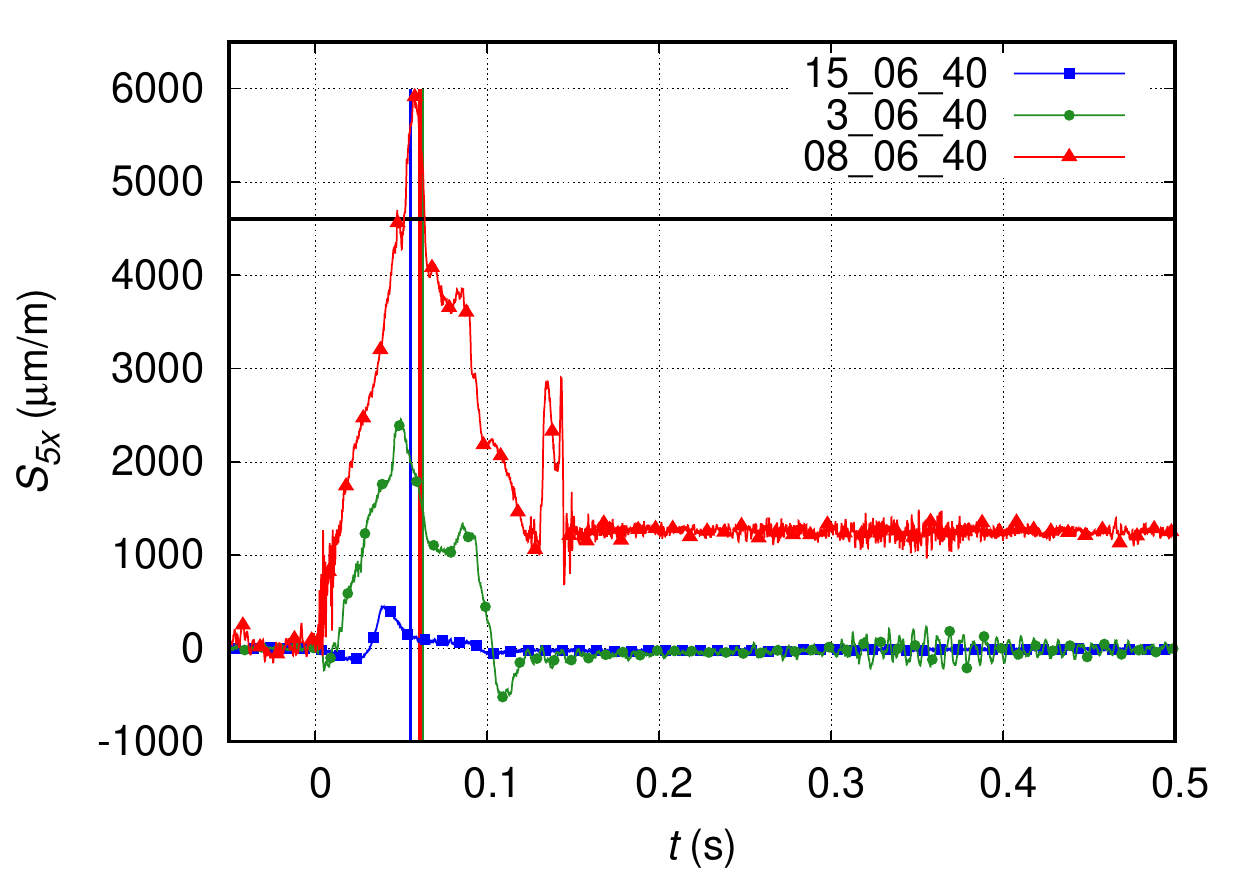}
	\caption{Time histories of S$_{5x}$ for the cases XX\_06\_40, i.e. at U=40~m/s, pitch angle~6$^{\circ}$ and different thickness (i.e. X=15,3,08). The end of the impact phases for each test is indicated by vertical lines of the same colour as that of the corresponding plot line. The black horizontal line indicates the value of the yielding strain.}
	\label{fig:S05X_X122_long_comparison}
\end{figure}

As already mentioned, the attention is focused here on the impact phase only. 
However, the plate is still highly loaded at the end of the impact phase,
which is the reason why the strains do not return to zero even for the thick plates.
This is clearly seen in Figure 
\ref{fig:S05X_X122_long_comparison}, where the time histories of S$_{5x}$ 
for the cases XX\_06\_40, i.e. U=40~m/s, $\alpha=6^{\circ}$ and different 
thickness's (XX =15, 3 or 08), are drawn for a time interval much longer than the impact 
phase. The data indicate that in all cases the strain is non-zero at $t=t_E$, 
whereas it returns to zero for $t>>t_E$ for the 15 mm and 3 mm plates only. For the 0.8 mm plate, 
the yielding limit is exceeded and the strain remains at a finite non-zero value at the end of the tests.
It is worth noticing that out-of-plane deformations are also observed for the 3 mm plate, in some cases at least. 
However, they were much smaller than those observed for the 0.8 mm plate, 
and they were mainly localized close to the frame, which explains why no residual strains appear in the middle of the plate.

\begin{figure}[htbp]
\centering
\subfigure[Strain Gauge Positions]{\includegraphics[width=0.75\textwidth]{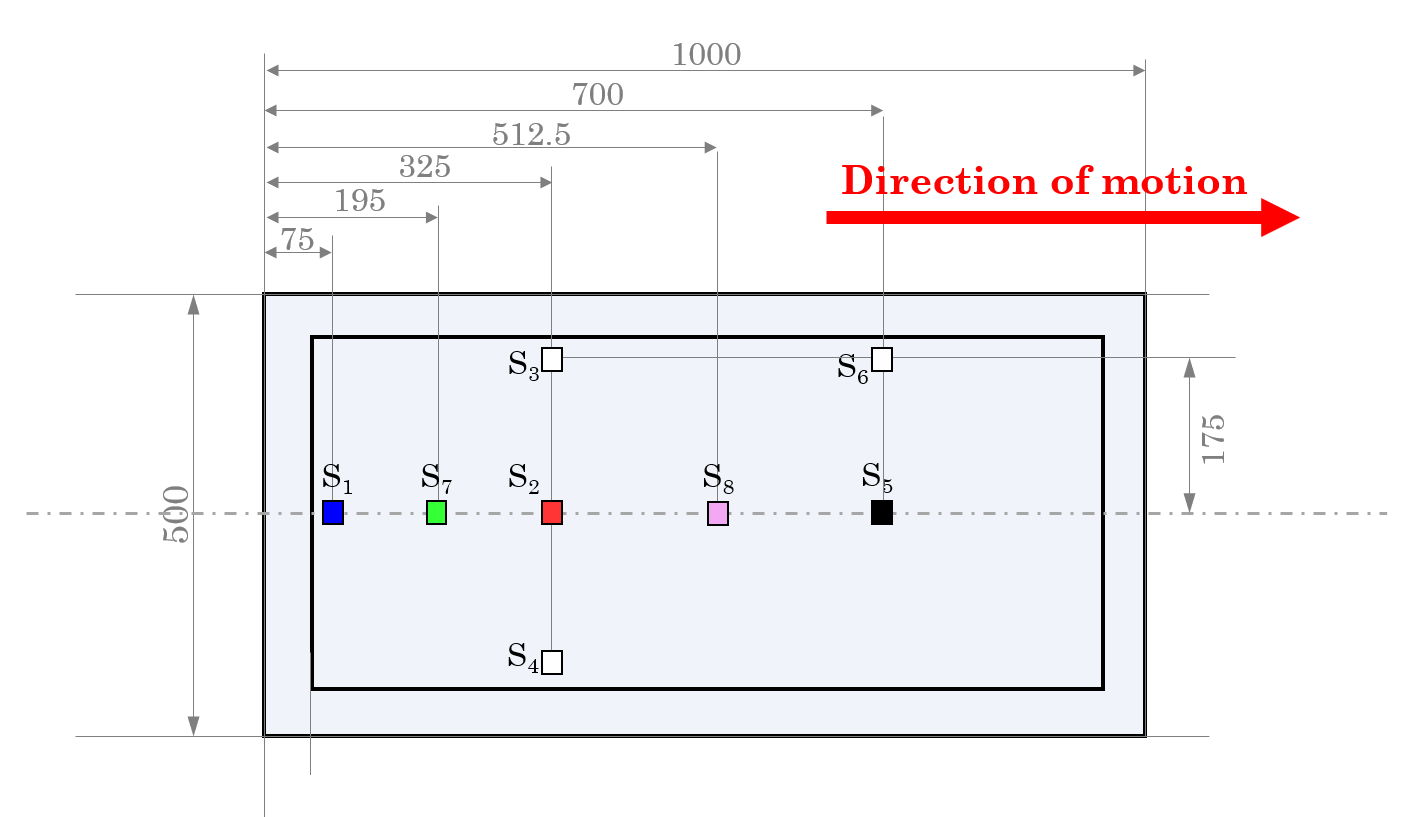}} 
\\
\subfigure[15\_10\_30]{\includegraphics[width=0.31\textwidth]{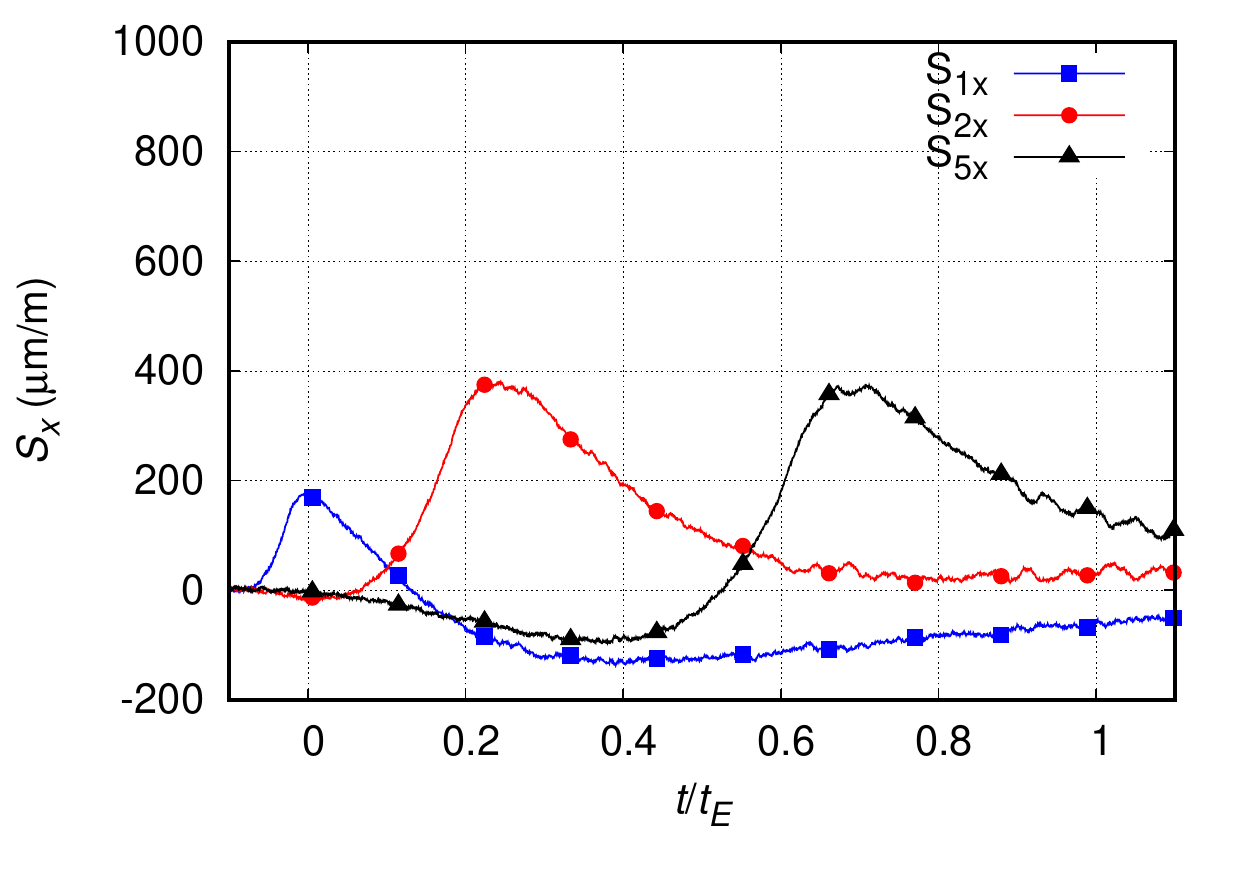}} \quad
\subfigure[15\_06\_40]{\includegraphics[width=0.31\textwidth]{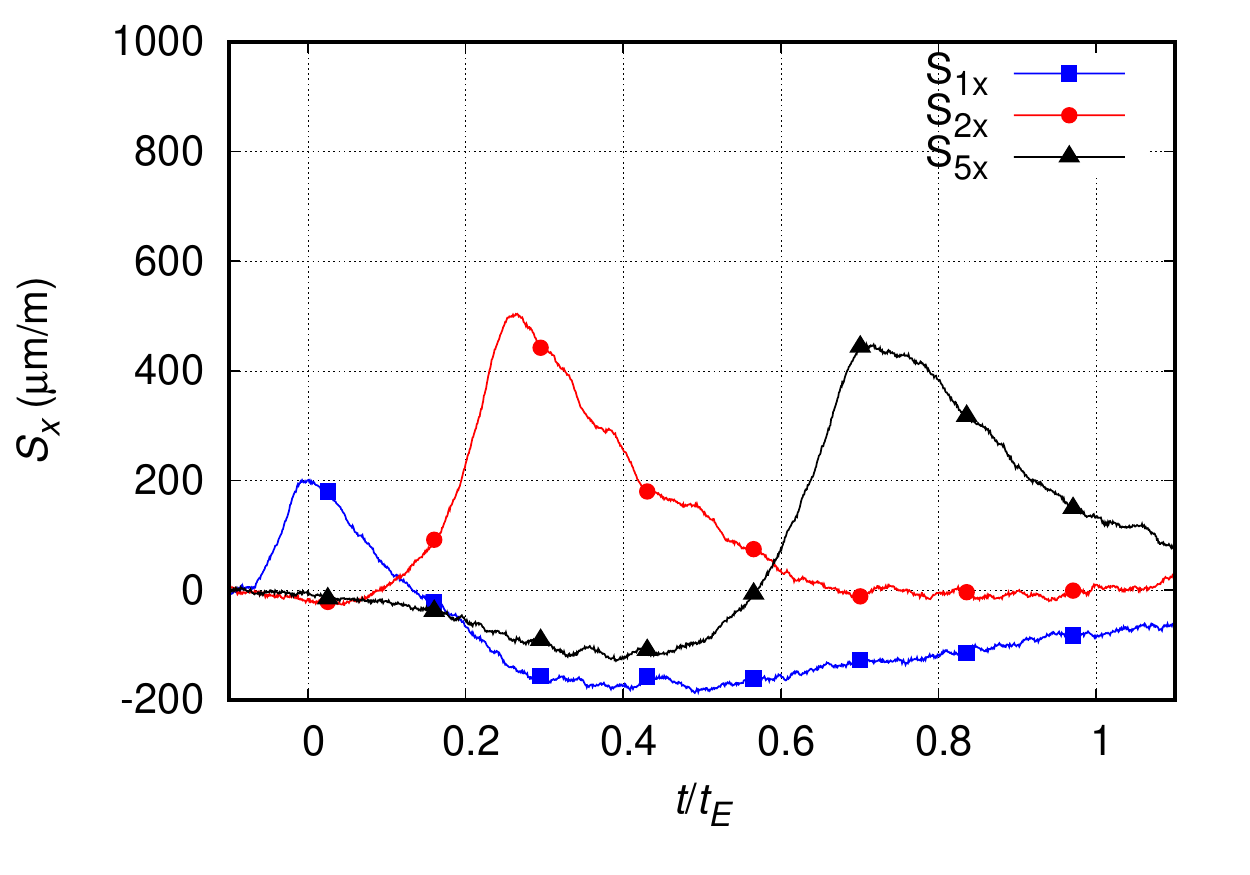}} \quad
\subfigure[15\_04\_45]{\includegraphics[width=0.31\textwidth]{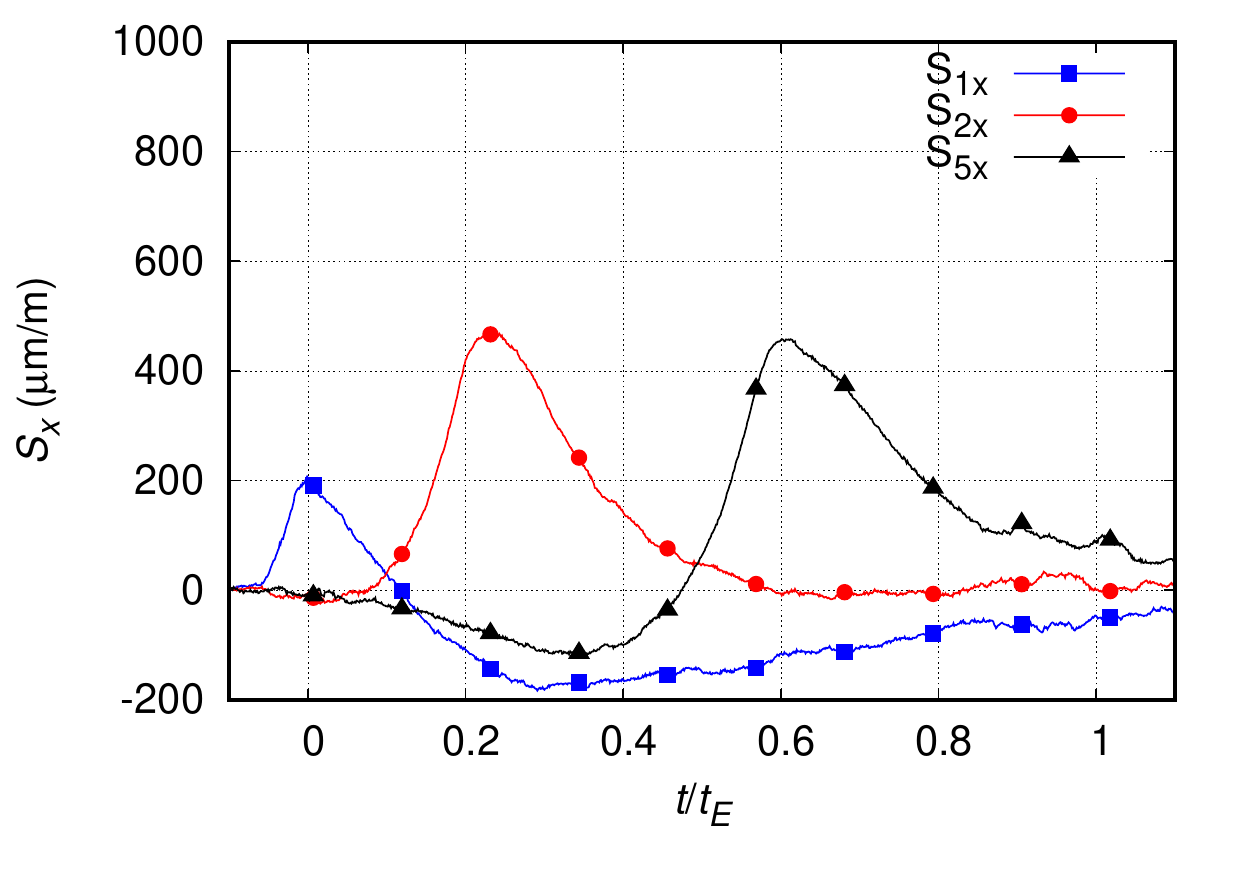}} \\
\subfigure[3\_10\_30]{\includegraphics[width=0.31\textwidth]{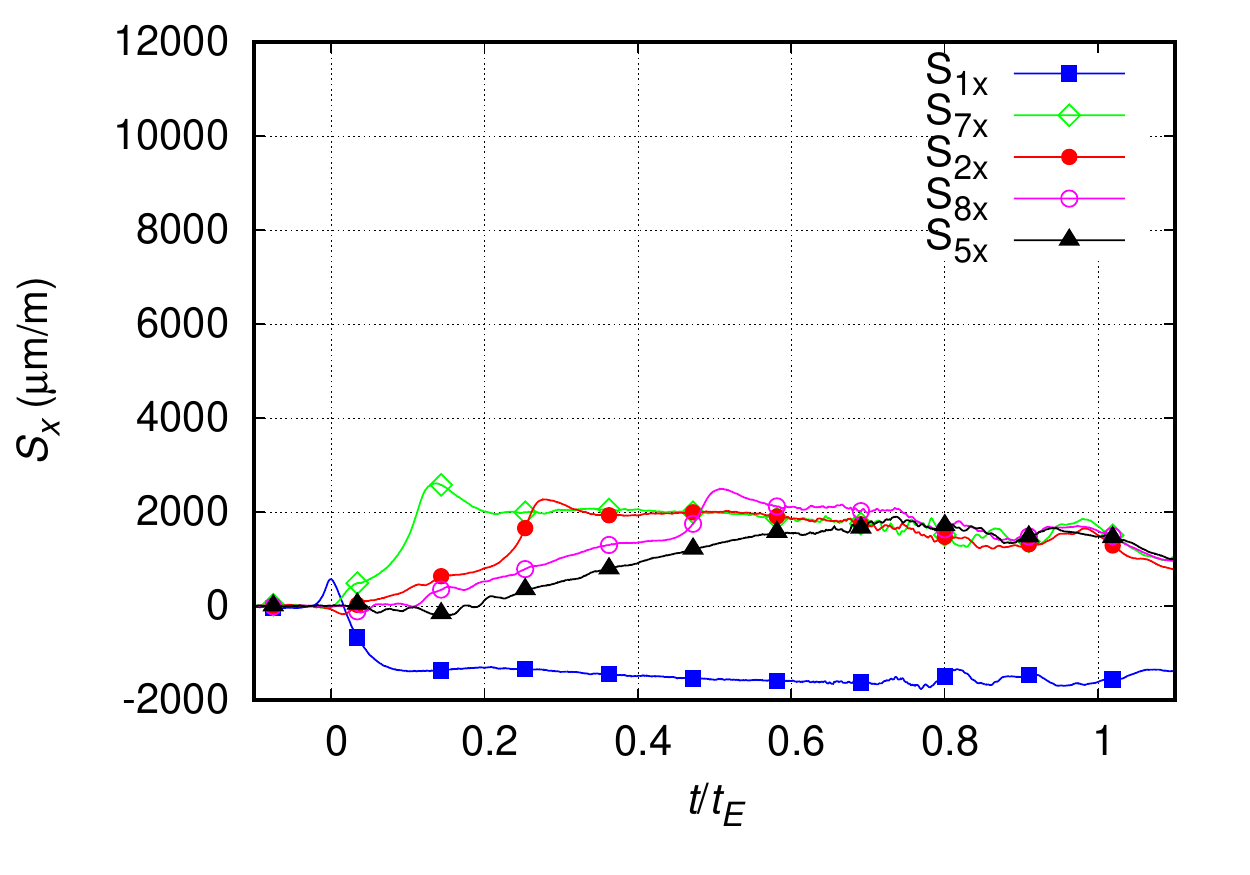}} \quad
\subfigure[3\_06\_40]{\includegraphics[width=0.31\textwidth]{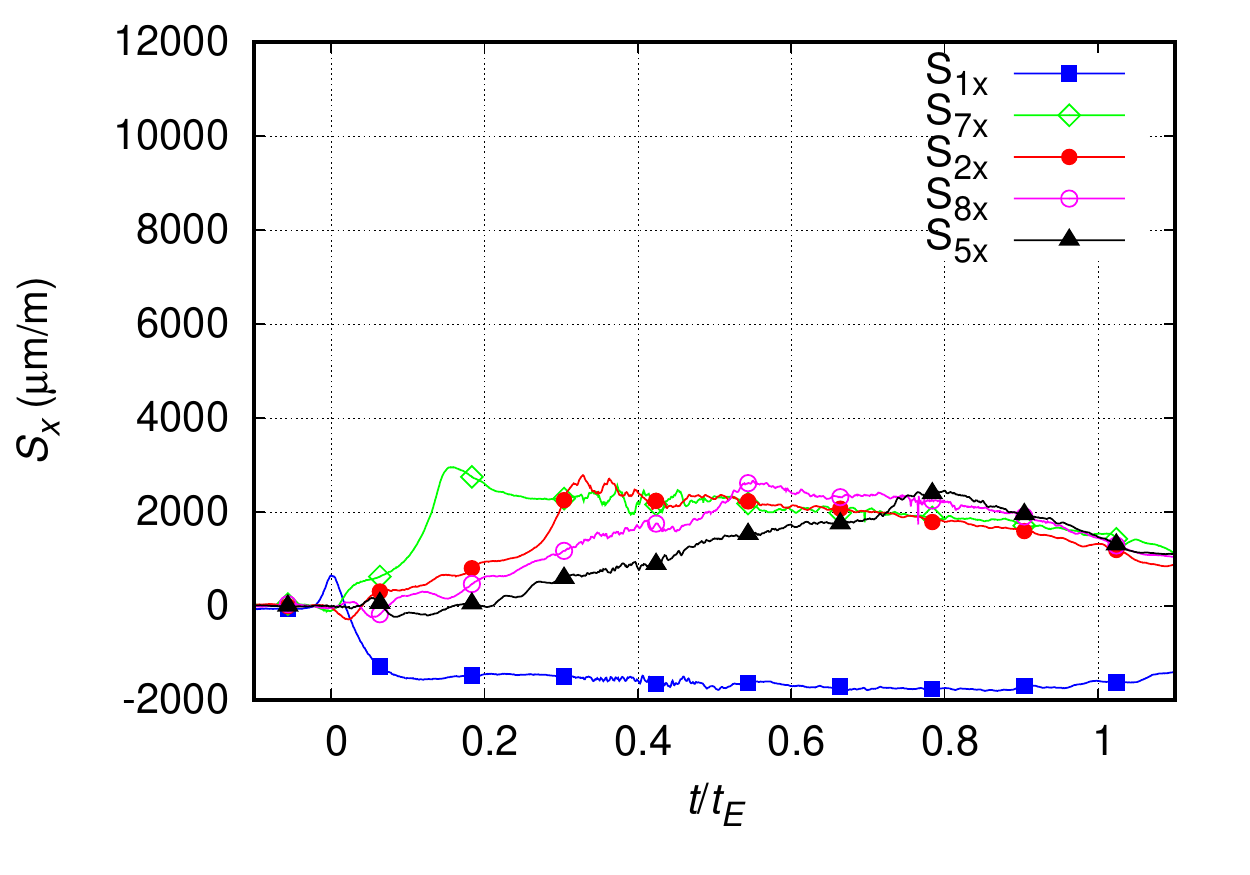}} \quad
\subfigure[3\_04\_45]{\includegraphics[width=0.31\textwidth]{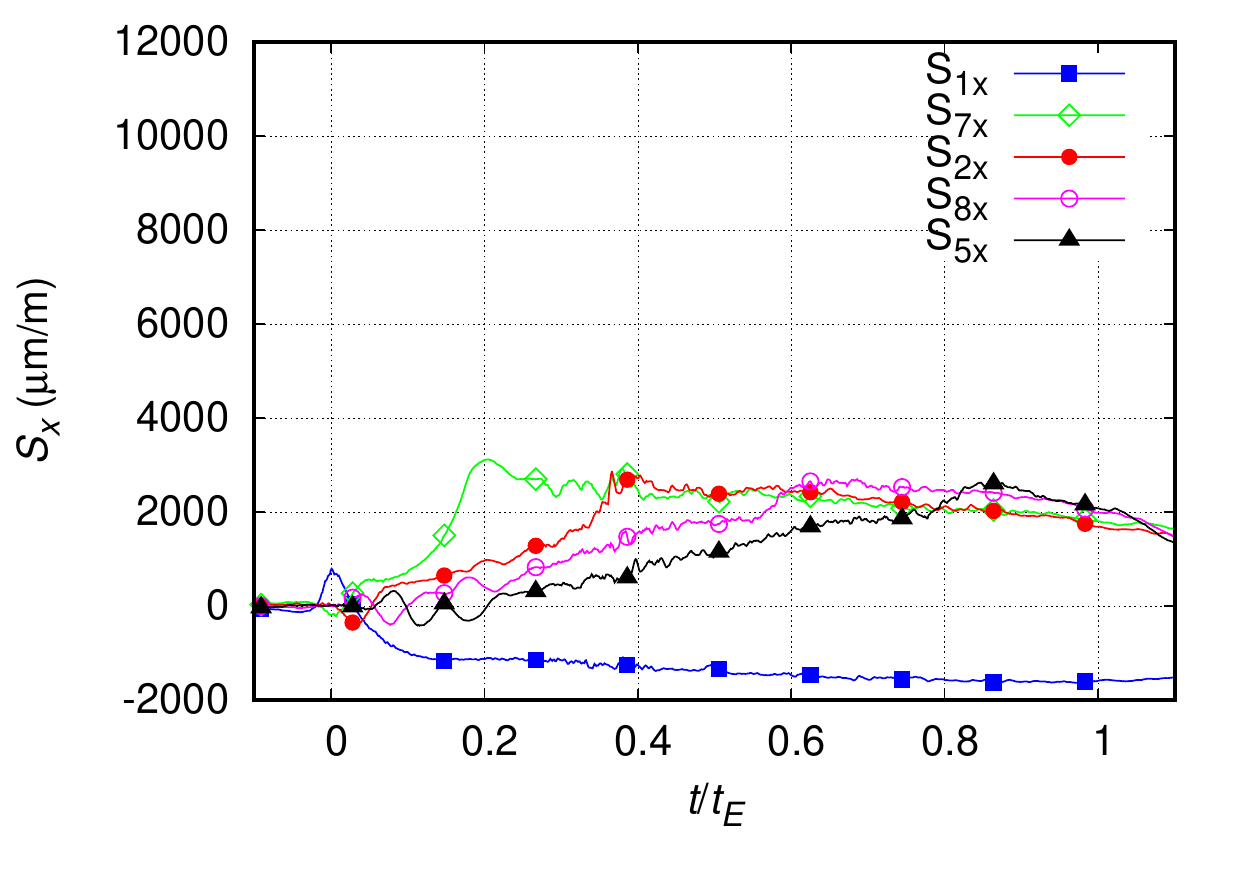}} \\
\subfigure[08\_10\_30]{\includegraphics[width=0.31\textwidth]{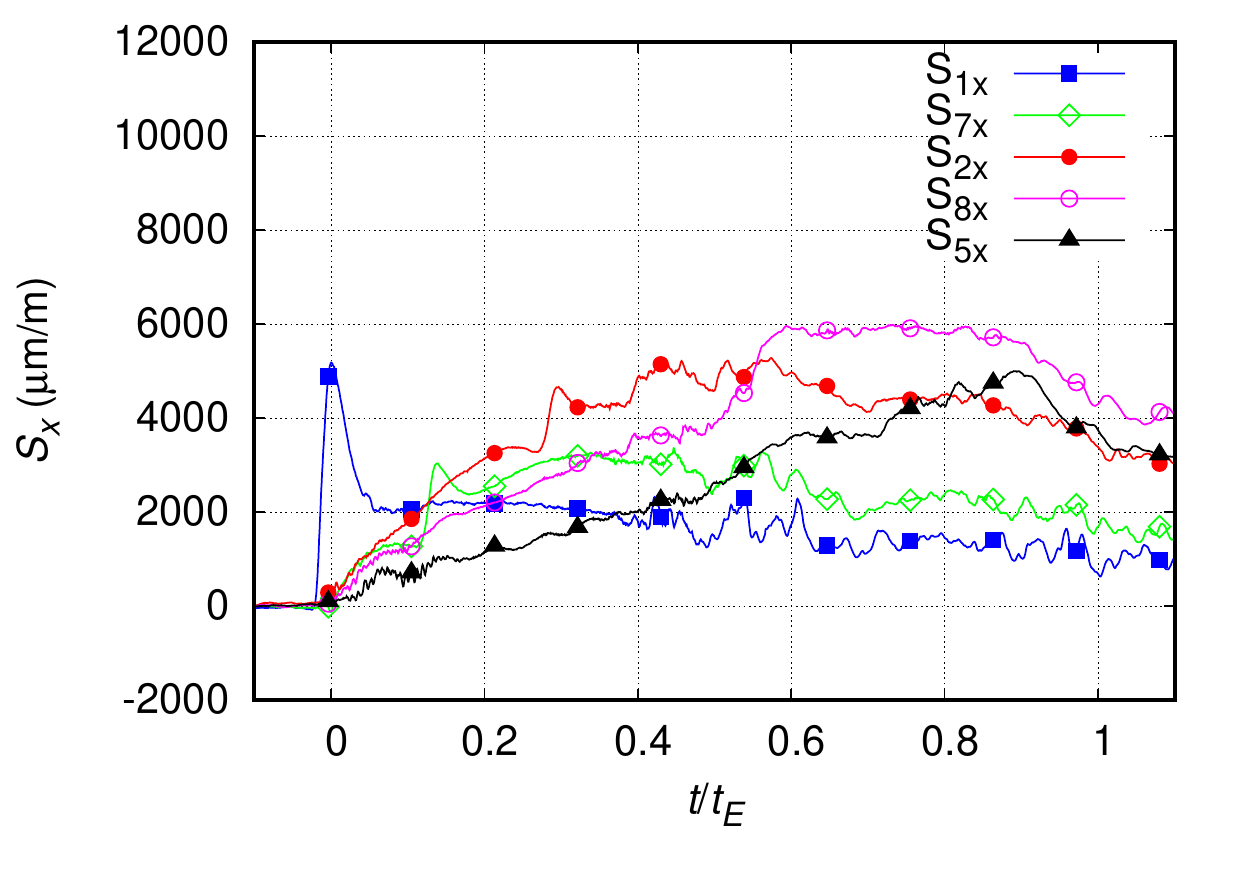}} \quad
\subfigure[08\_06\_40]{\includegraphics[width=0.31\textwidth]{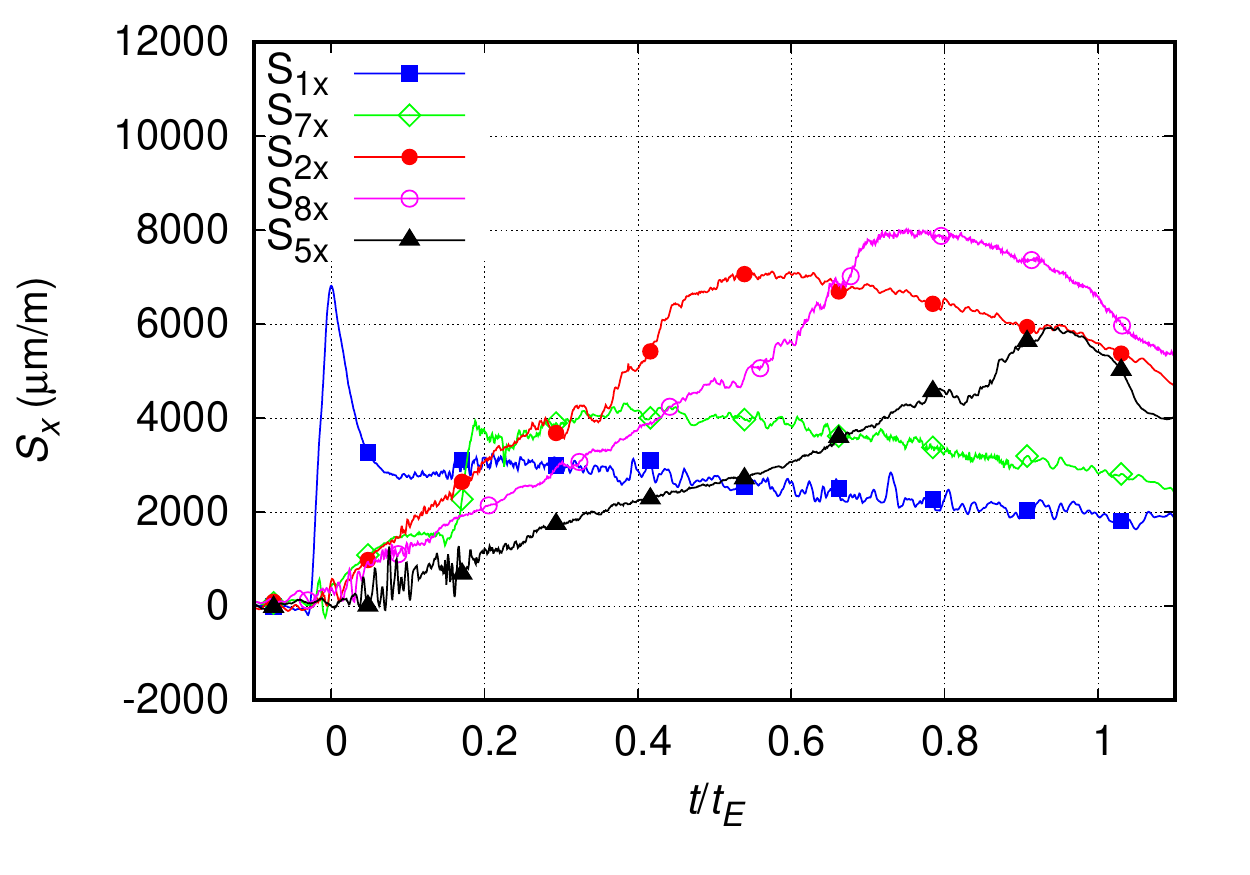}} \quad
\subfigure[08\_04\_45]{\includegraphics[width=0.31\textwidth]{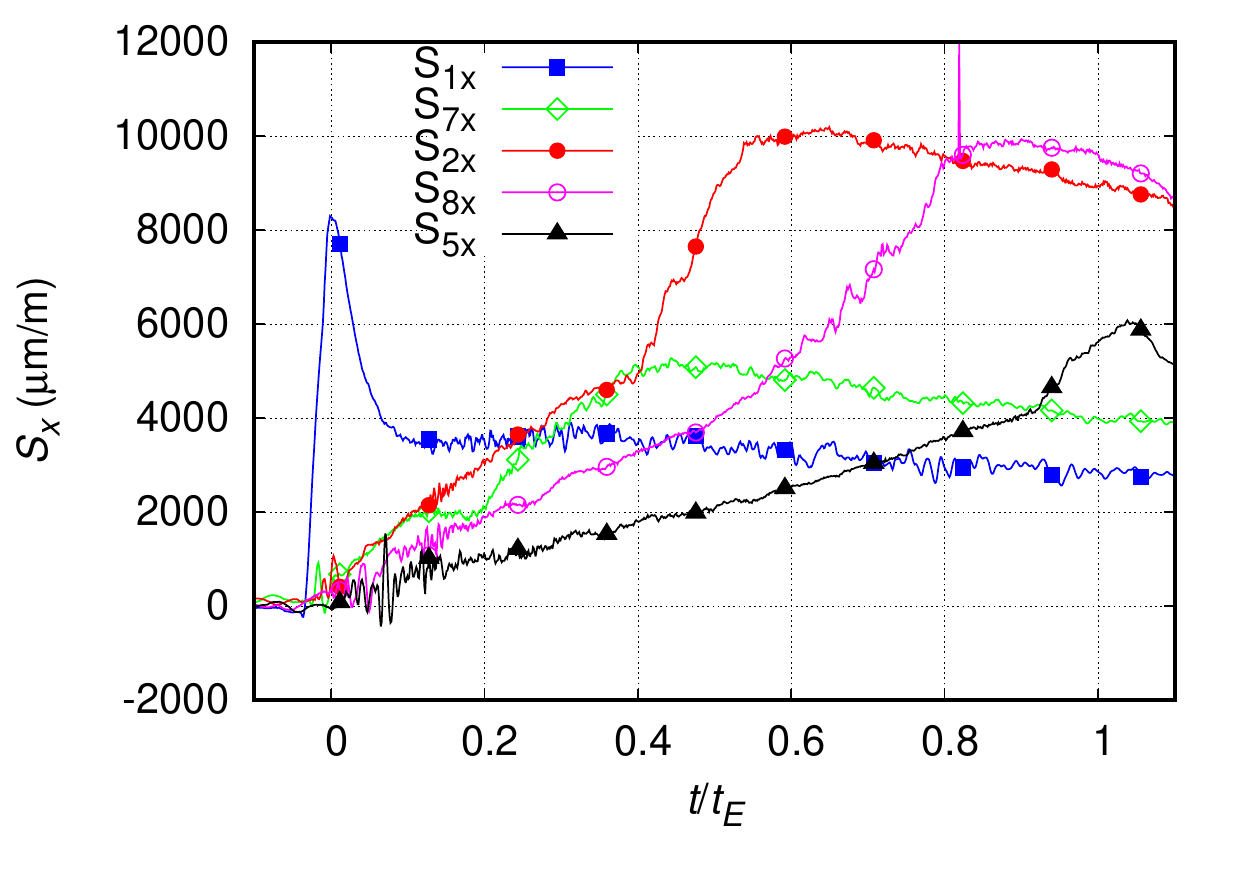}} 
\caption{Comparison of the longitudinal strains measured along the midline for the test conditions XX\_10\_30, XX\_06\_40 and
XX\_04\_45.}
\label{fig:x_strains_3Cases}
\end{figure}

In Figure \ref{fig:x_strains_3Cases} the longitudinal strain measured 
by the gauges located along the midline are shown both for the thick
and thin plates, for three test conditions XX\_10\_30, XX\_06\_40 and
XX\_04\_45 (i.e. for XX=15, 3, 08). In order to 
help the reader, a drawing of the gauge installation is provided in Fig.
\ref{fig:x_strains_3Cases}(a) with 
the position of the gauges indicated with the same colour used in the plots.
For the thin plates, i.e. the 3~mm and 0.8~mm thickness, in order to 
facilitate the
comparison of the data among the different test conditions, the same vertical 
scale is adopted. The same vertical range, albeit much smaller than that
used for the thin plates, is also
adopted for the comparison with the thick plate data.
As anticipated in Section \ref{setup}, the origin of the time axis is
taken at the first peak of strain gauge S$_{1x}$. 
Strains are drawn versus $t/t_E$, $t_E$ denoting the end of the impact phase 
for each condition, so that all data are meaningful in the same range 
$t/t_E \in [0,1]$.
From Figure \ref{fig:x_strains_3Cases} it is seen that for the 
thick plate (15~mm) the strains exhibit a quite similar behaviour in the
different cases. 
For the gauges located in the most internal part the plate, i.e. S$_2$ and S$_5$, the $x$-strains 
start from zero, diminish in the early stage, then rise up to a maximum and 
gently decay to zero afterwards. For the gauge S$_1$, located very close to the
frame and therefore strongly affected by the boundary conditions, 
the $x$-strain rises rapidly, then decays gently, reaches a minimum and rises 
again to zero.
It is worth noticing the asymmetry of the strain time histories about the peak.
As shown later on in Section \ref{pressurecameras}, the peak in the strains
occurs at about the time when the pressure peak passes beneath the gauge.
The pressure is zero ahead of the spray root, rises quite sharply up to the
peak and decays more gently afterwards \cite{iafrati2016experimental}.
The lower decay rate, which results in a higher loading of the plate,  
is the reason for the higher strain levels measured after the passage of the
pressure peak.
The behaviour is quite different for the thin plates. The strain values
are, of course, much higher and, furthermore, at least for the gauges
far from the boundaries, the strain level remains constant or diminishes
rather slowly after the passage of the pressure peak. 
Partly, this is due to the fact that the deformation is in the plastic
regime, particularly for the 0.8~mm plates, when the yielding limit is exceeded. 
However, the flatter strains are also a consequence of
the large out-of-plane deformation, which disrupts the spray and causes a
spreading of the pressure distribution. 
Especially for the 0.8~mm plates some hydroelastic effects occur, as it
can be seen for $t/t_E \ge 0.4$.
The longitudinal strain measured by gauge S$_1$ is much different from that
measured by the other probes along the midline. For the 3~mm plates  
it shows a peak and decays to negative values afterwards,
whereas for the 0.8~mm plates the strain exceeds the yielding limit and 
remains positive for the remaining part of the test.
The strain signals exhibit an oscillatory behaviour, which
is sometimes more evident after the first peak. This can be a consequence of the fluid-structure interaction.

\begin{figure}[htbp]
\centering
\subfigure[08\_10\_XX]{\includegraphics[width=0.48\textwidth]{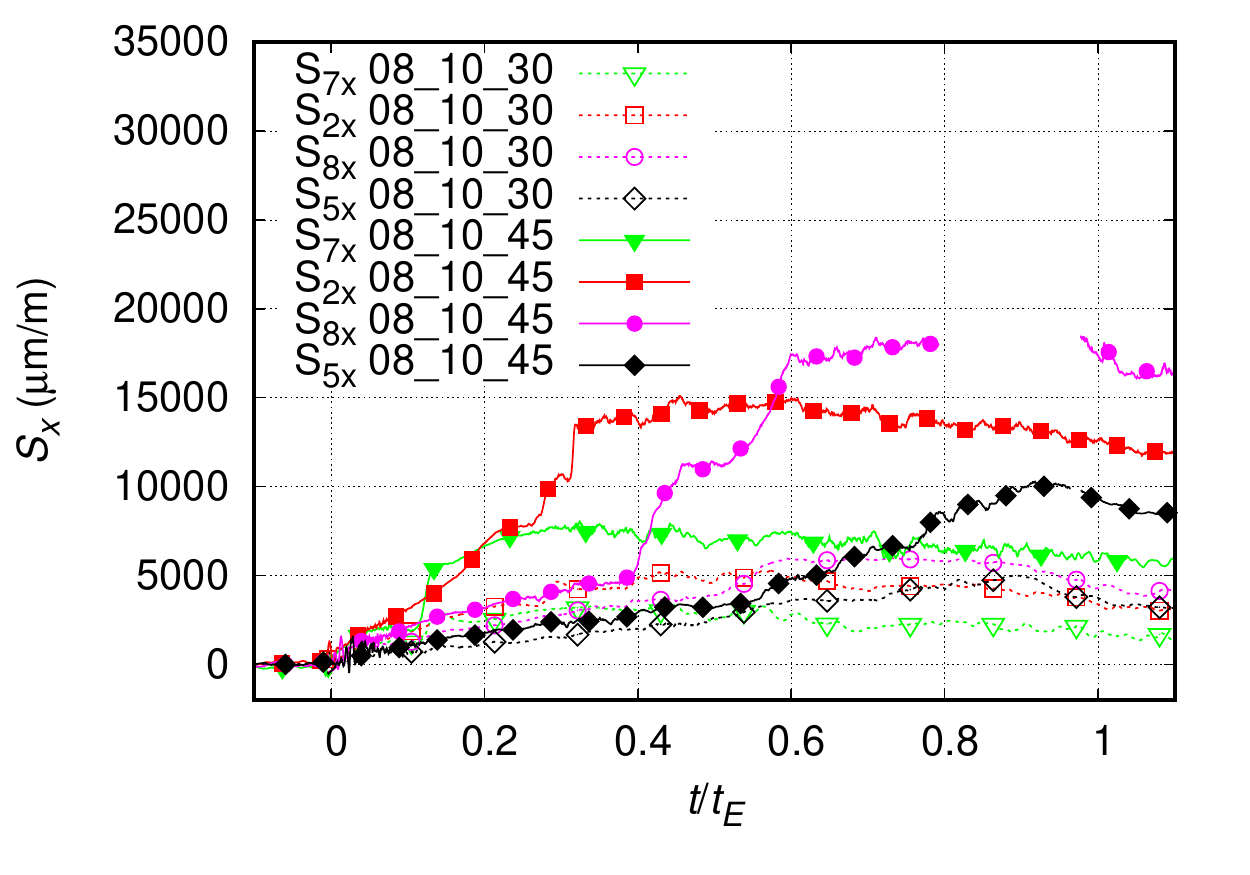}} \quad
\subfigure[08\_XX\_45]{\includegraphics[width=0.48\textwidth]{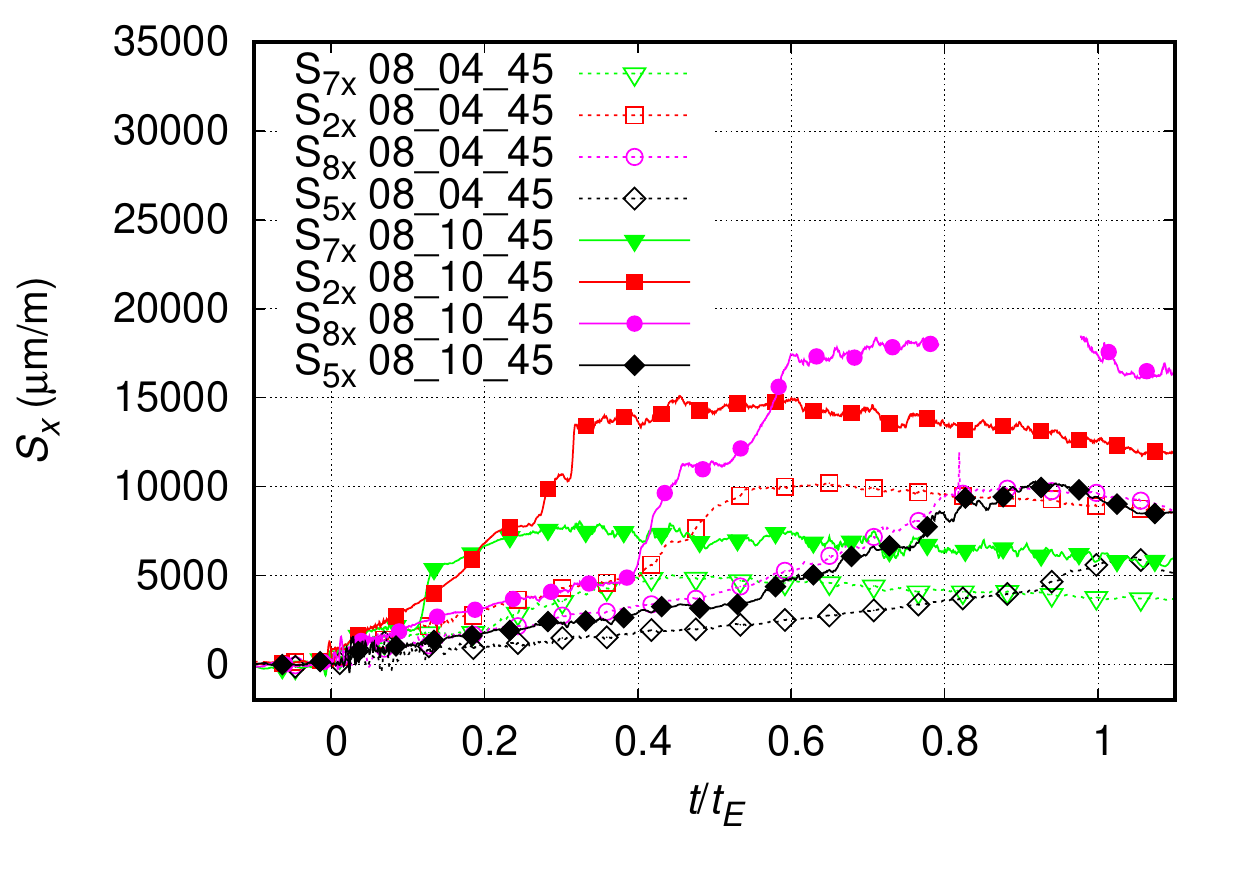}} 
\caption{Comparison of the longitudinal strains measured along the midline between the test conditions 08\_10\_30 (dashed lines) and 08\_10\_45 (solid lines) (a) and between the test conditions 08\_04\_45 (dashed lines) and 08\_10\_45 (solid lines) (b).}
\label{fig:31XX_strains}
\end{figure}

\begin{figure}[htbp]       
\centering
\subfigure[08\_10\_XX]{\includegraphics[width=0.65\textwidth]{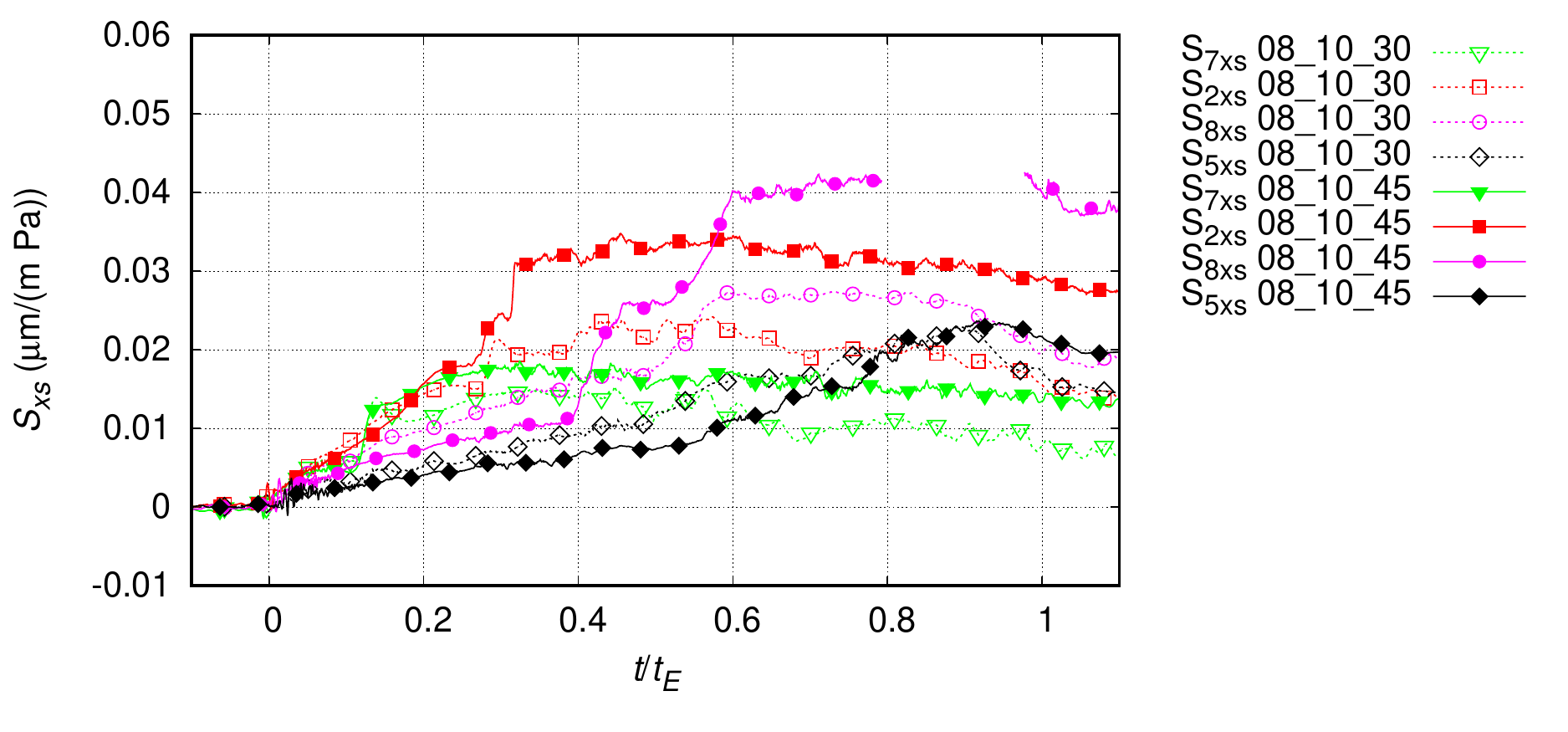}} \\
\subfigure[08\_XX\_45]{\includegraphics[width=0.65\textwidth]{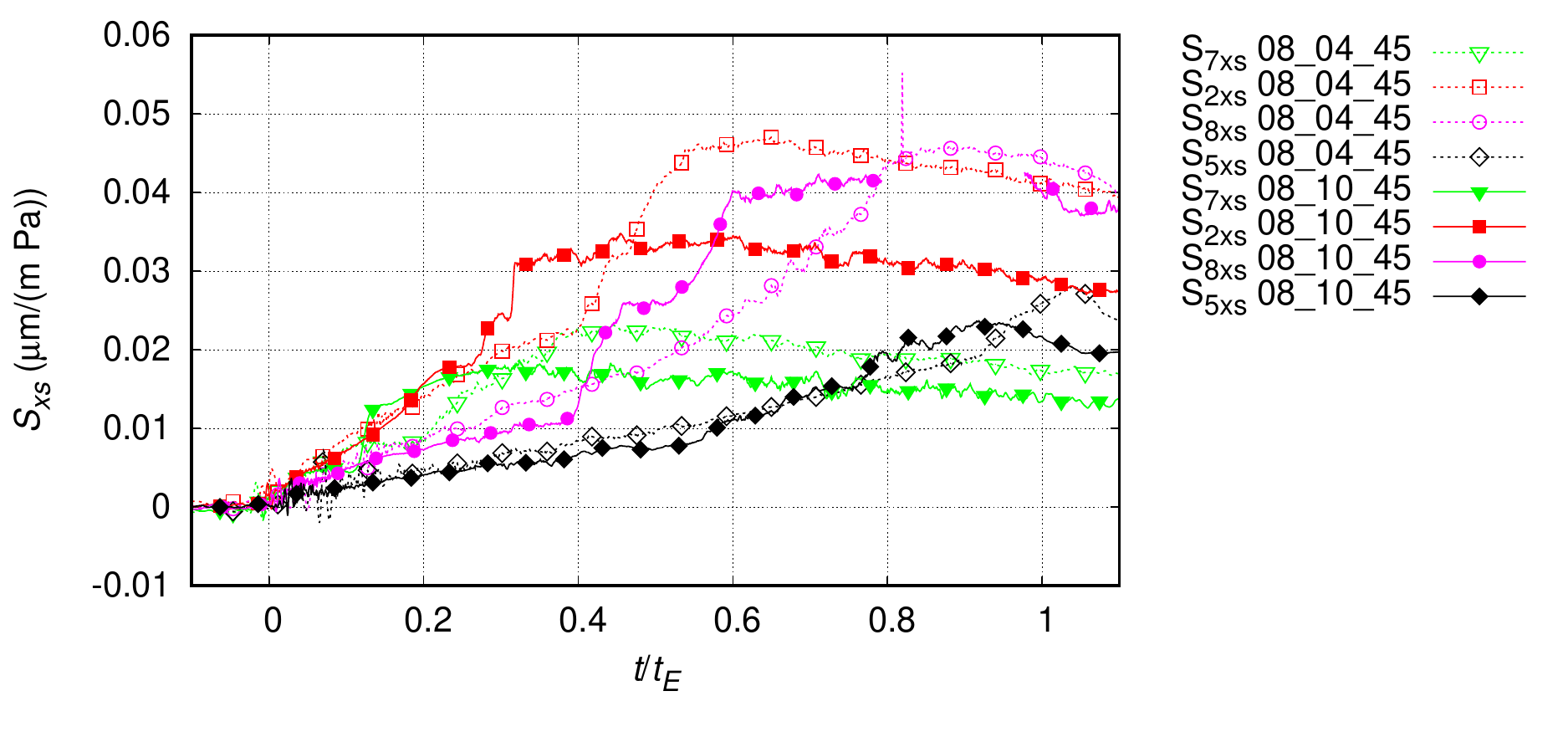}} 
\caption{Comparison of (a) the scaled longitudinal strains along the midline between the test conditions 08\_10\_30 (dashed lines) and 08\_10\_45 (solid lines) and (b) between the test conditions 08\_04\_45 (dahsed lines) and 08\_10\_45 (solid lines).}
\label{fig:31XX_strains_scaled}
\end{figure}

The test conditions of the data provided in Figure \ref{fig:x_strains_3Cases} do not allow to distinguish the effects of the pitch or of the horizontal velocity. In fact, considering the limited number of tests available for the deformable specimens, test conditions were chosen in order to be as representative as possible of a realistic scenario. Hence, tests at lower horizontal speed were conducted at a higher pitch angle, which is how the aircraft operates in order to guarantee the same lift.
Nonetheless, from Figure \ref{fig:x_strains_3Cases} it can be seen that the pitch angle and test speed do not affect much the strains for the 15~mm and 3~mm plates, whereas they have a quite significant effect on the thinnest plates, particularly for the gauges located in the middle of the plate. At least for the thinnest plates, the effects of the pitch angle and test speed can be distinguished by comparing the data of the test conditions 08\_10\_30 and 08\_04\_45 with 08\_10\_45, as shown in Figure \ref{fig:31XX_strains}.
In particular, the comparison in Figure \ref{fig:31XX_strains}(a) shows that the longitudinal strains, especially for the gauges S$_2$ and S$_8$\footnote{The data recorded by the gauge S$_8$ for $t \in (0.8 \quad 0.975)$ present a spike, therefore they were not plotted}, increase with the horizontal speed and with the pitch angle. The larger strain is of course associated with a higher loading. As discussed in \cite{iafrati2016experimental}, the loads during the water impact of the thick plates scale with the factor $\rho (U^2+V^2)\sin(\alpha+\beta)$, so a load increase with $U$ and $\alpha$ could reasonably be expected also for the thin plates. It is interesting to scale the longitudinal strains with the same quantity, as shown in Figure \ref{fig:31XX_strains_scaled}, where the scaled strains are denoted by S$_{xs}$. In particular, Figure \ref{fig:31XX_strains_scaled}(a) shows that especially S$_{2xs}$ and S$_{8xs}$ at 10$^{\circ}$ and U=45 m/s are higher than those at the same pitch angle and at U=30 m/s. Such a behaviour is expected because, as discussed in Section \ref{hyrodynamicloads} and shown in Figure \ref{fig:ZLoadsScaling}(c), the scaled normal load is higher at 10$^{\circ}$ and U=45 m/s.
The interpretation of the effect of the pitch angle is, instead, less straightforward. As observed in Figure \ref{fig:31XX_strains_scaled}(b), the scaled strains S$_{5xs}$ are nicely overlapped. As for the other strain time histories, there is an initial phase in which all the scaled strains are overlapped. Successively, there is an intermediate phase, in which the scaled strains at 10$^{\circ}$ are higher than those at 4$^{\circ}$. In the final part of the impact phase, the strains at 4$^{\circ}$ overcome those at 10$^{\circ}$. This is especially evident for S$_{2xs}$. This behaviour is in agreement with the scaled loads, shown in Figure \ref{fig:ZLoadsScaling}(c), where it is observed that the scaled load at 4$^{\circ}$ and U=45 m/s results higher than that at 10$^{\circ}$ and U=45 m/s. It is worth noticing that, however, the gap between the scaled strains, especially in the last part of the impact phase, could be increased by the fact that at 10$^{\circ}$ the velocity decrease within the impact phase (about 2 m/s) is slightly higher than at 4$^{\circ}$ (about 0.5 m/s), as detailed in \cite{iafrati2015high}. Even though such a velocity variation cannot be easily considered in the scaling, it is estimated that it may be responsible for an increase of about 10\% of the scaled strains measured at 10$^{\circ}$, which reduces the gap.

\begin{figure}[htbp]
\centering
\subfigure[Strain Gauge Positions]{\includegraphics[width=0.75\textwidth]{Figures/SG_schematic_midline.png}} 
\\
\subfigure[15\_10\_30]{\includegraphics[width=0.31\textwidth]{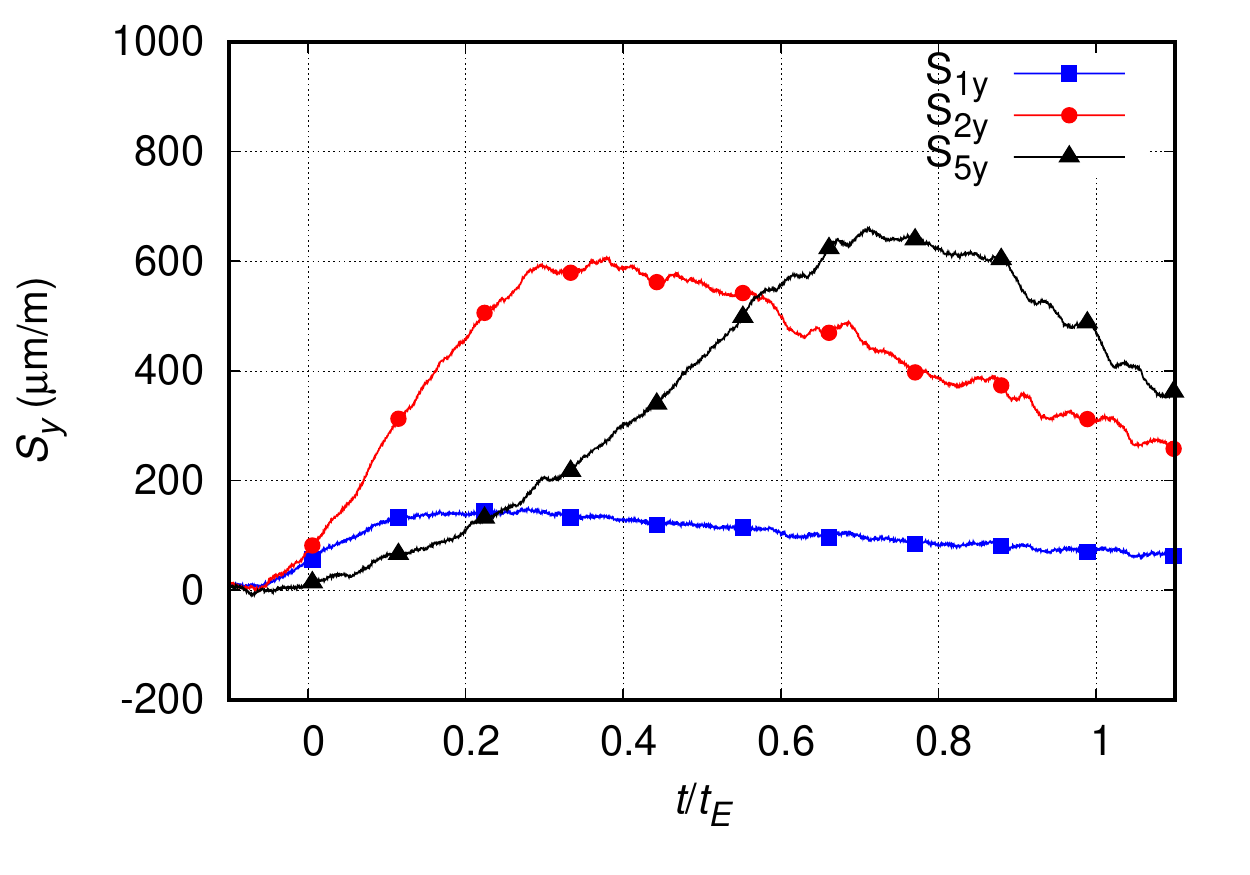}} \quad
\subfigure[15\_06\_40]{\includegraphics[width=0.31\textwidth]{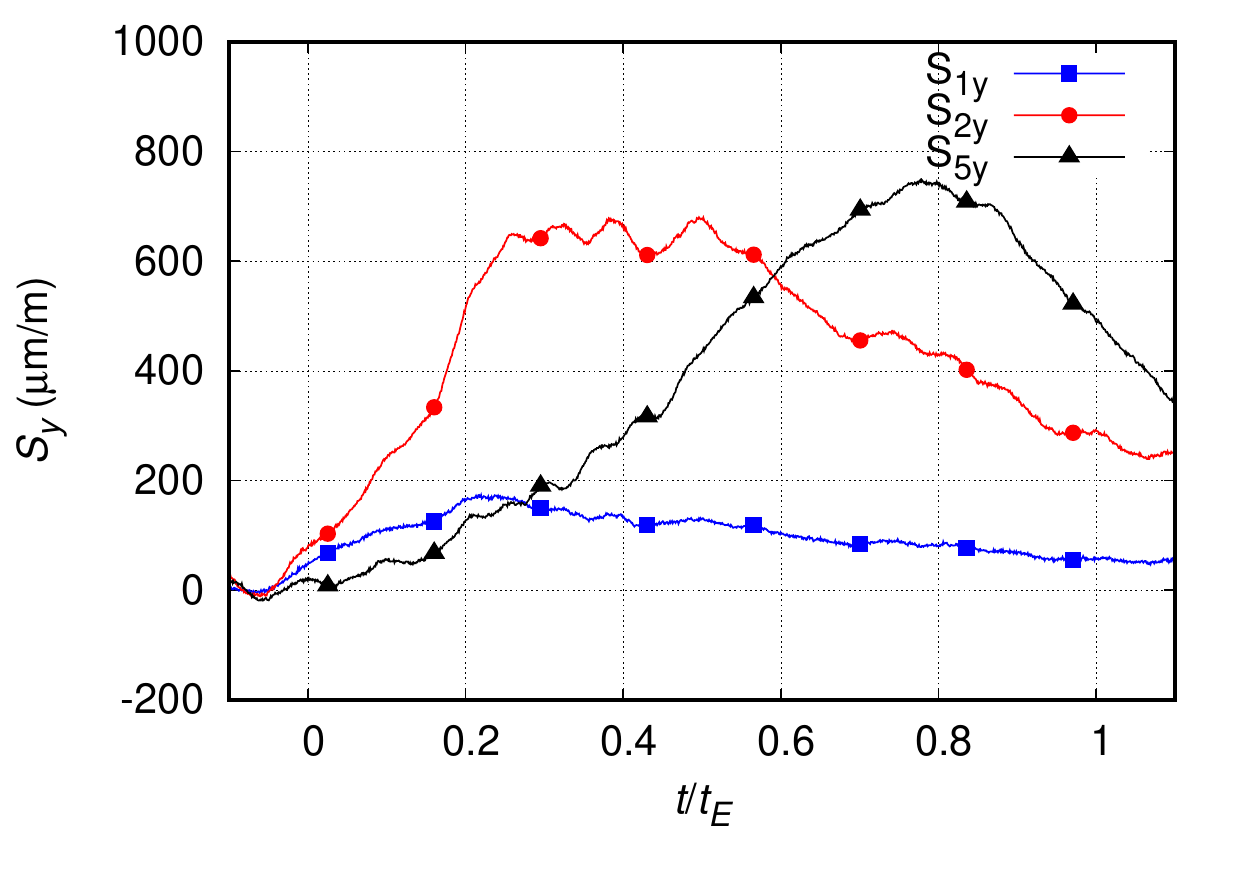}} \quad
\subfigure[15\_04\_45]{\includegraphics[width=0.31\textwidth]{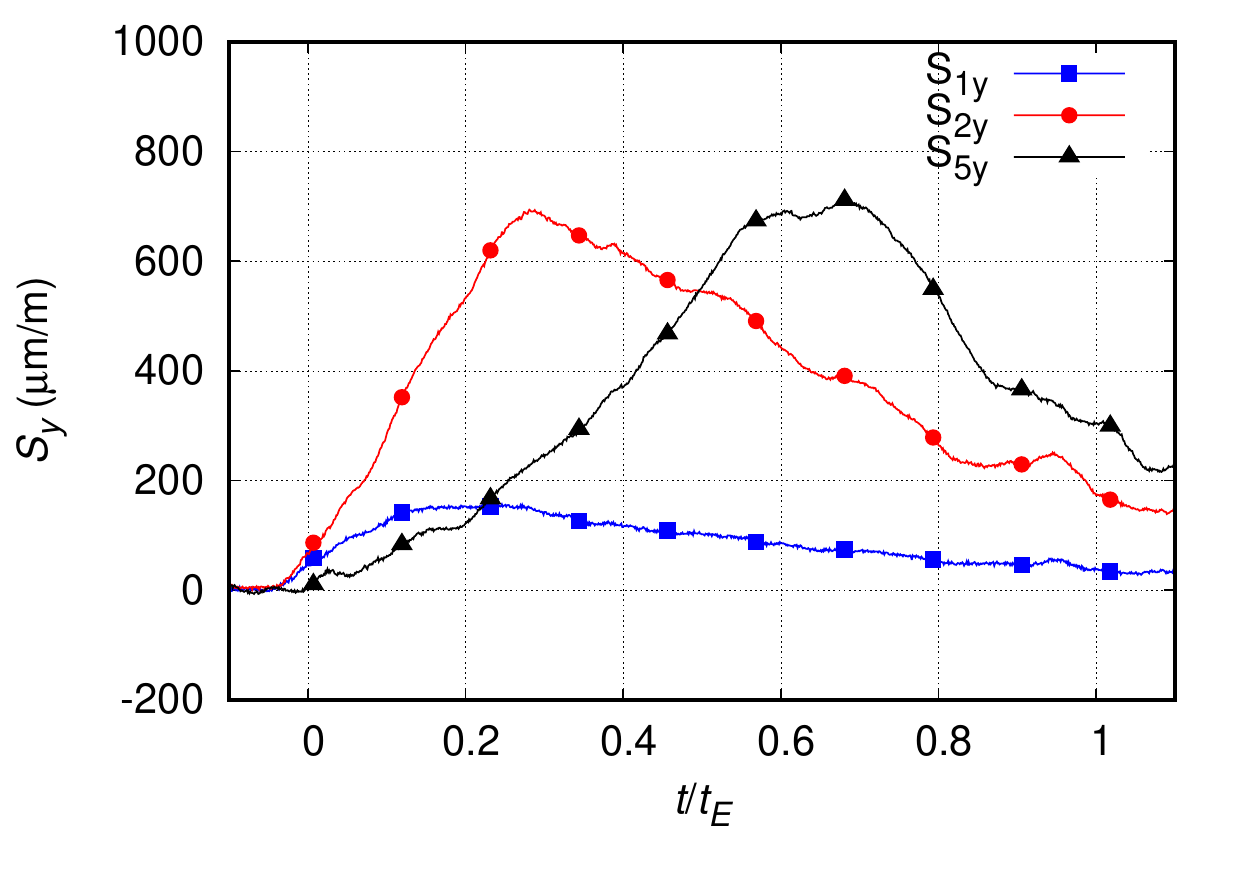}} \\
\subfigure[3\_10\_30]{\includegraphics[width=0.31\textwidth]{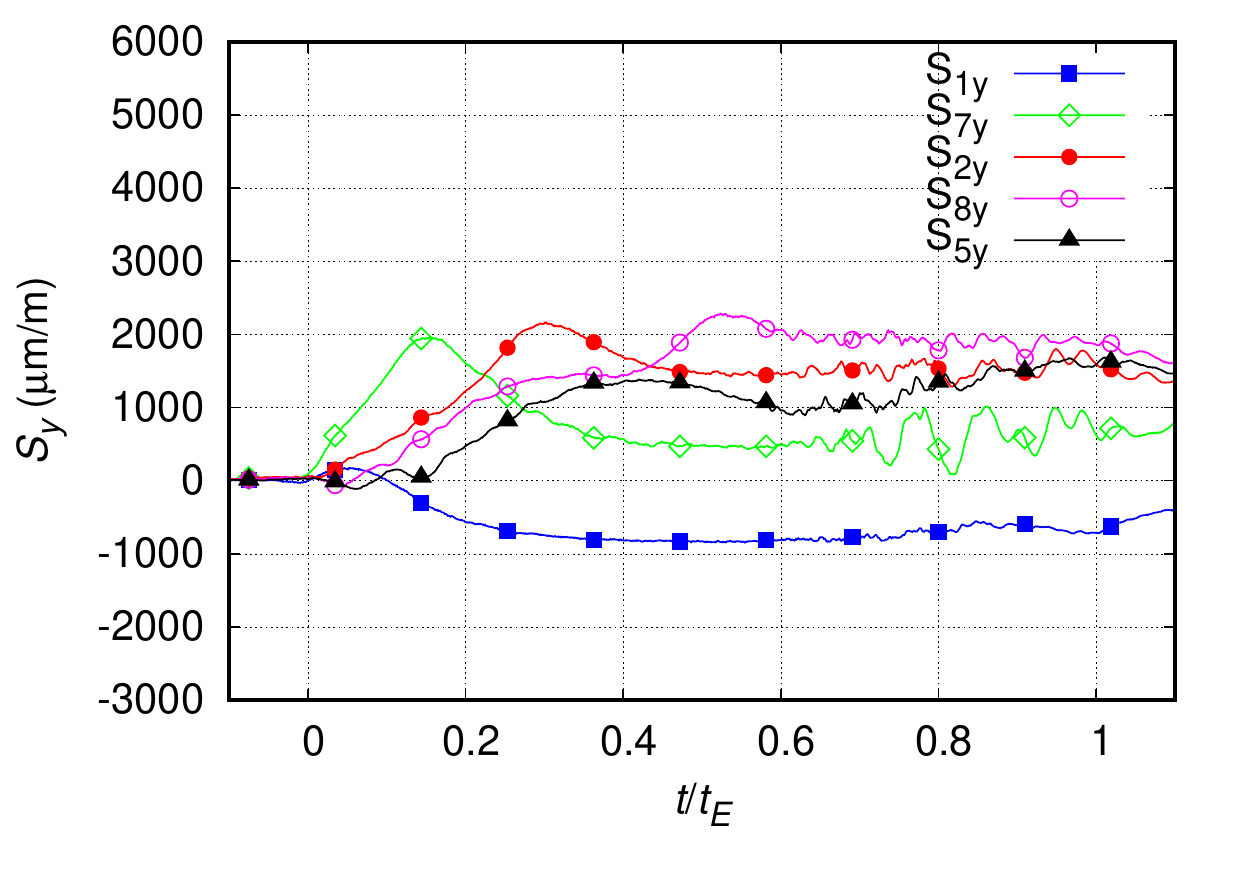}} \quad
\subfigure[3\_06\_40]{\includegraphics[width=0.31\textwidth]{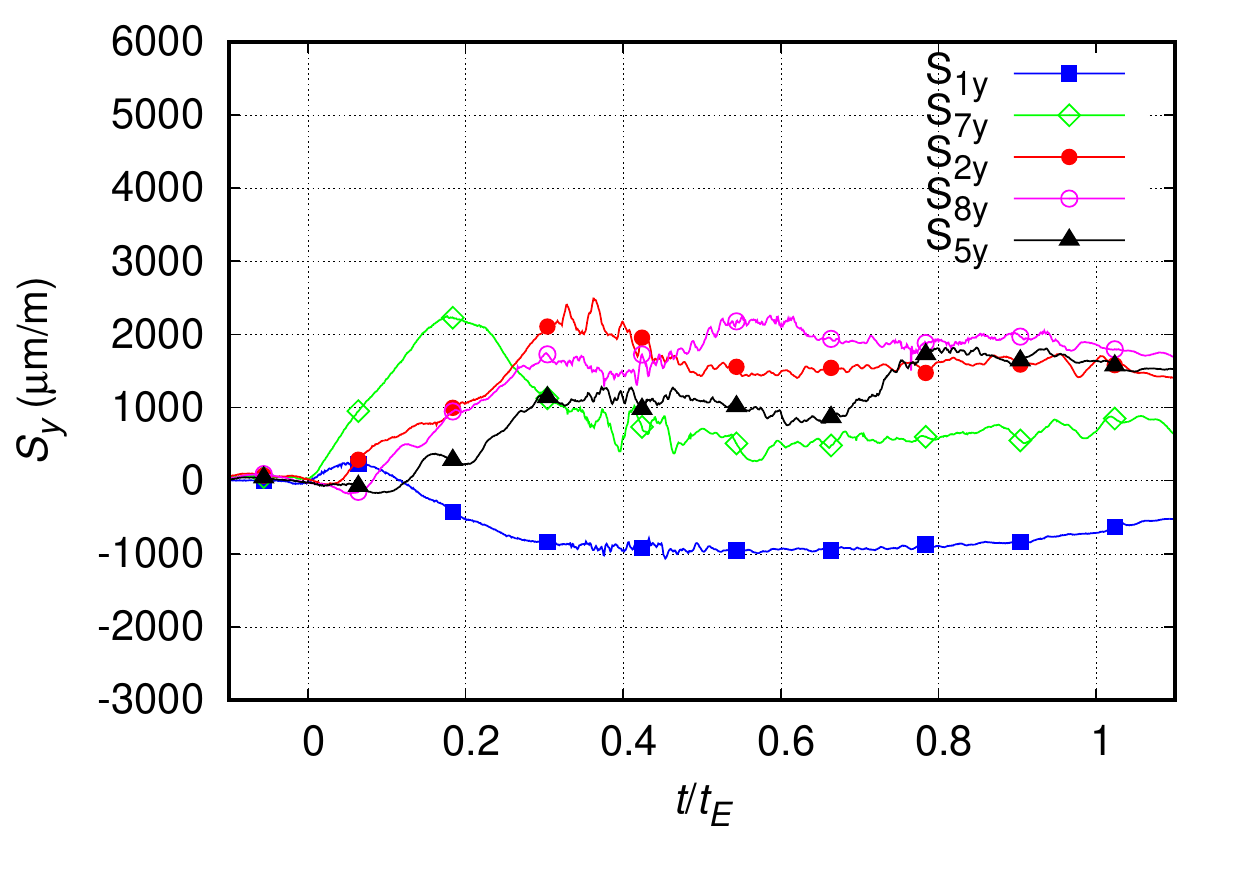}} \quad
\subfigure[3\_04\_45]{\includegraphics[width=0.31\textwidth]{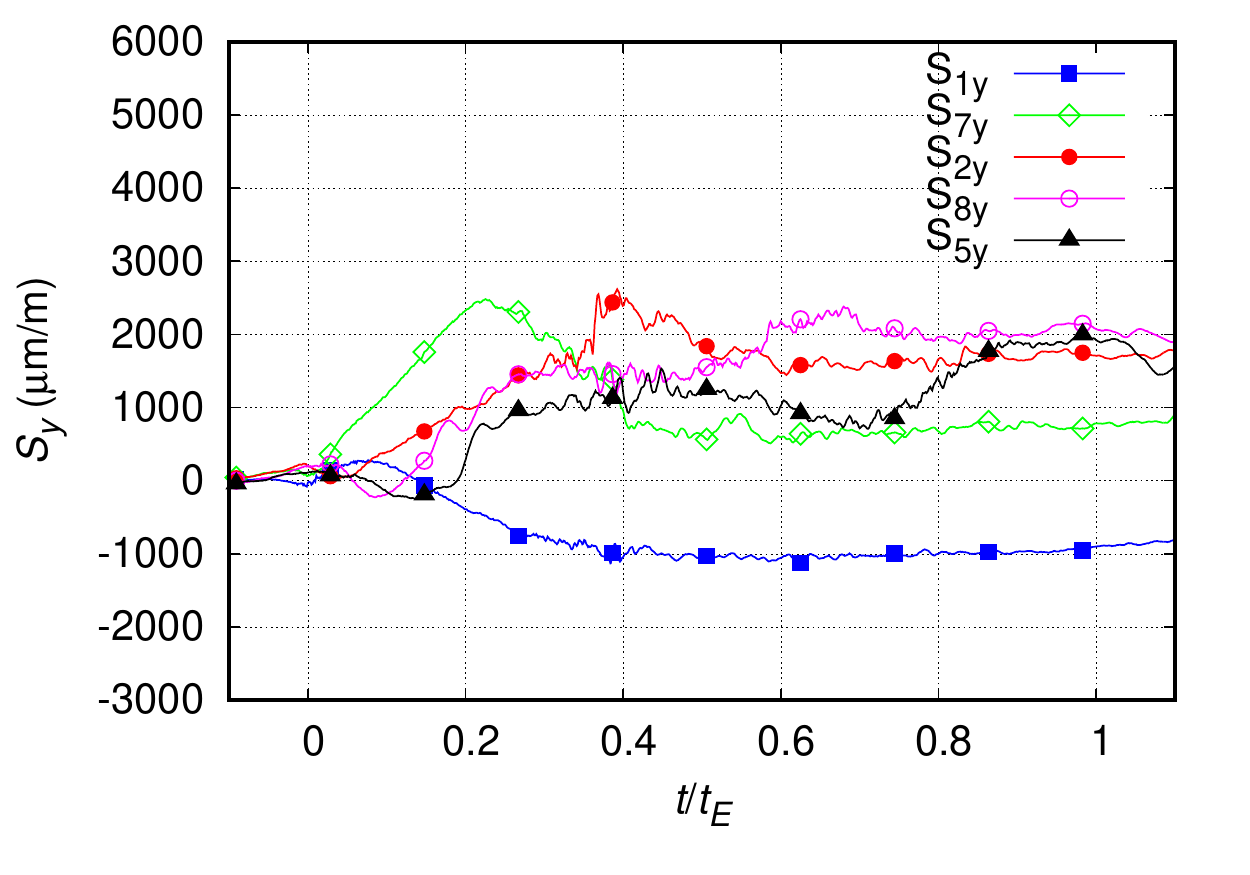}} \\
\subfigure[08\_10\_30]{\includegraphics[width=0.31\textwidth]{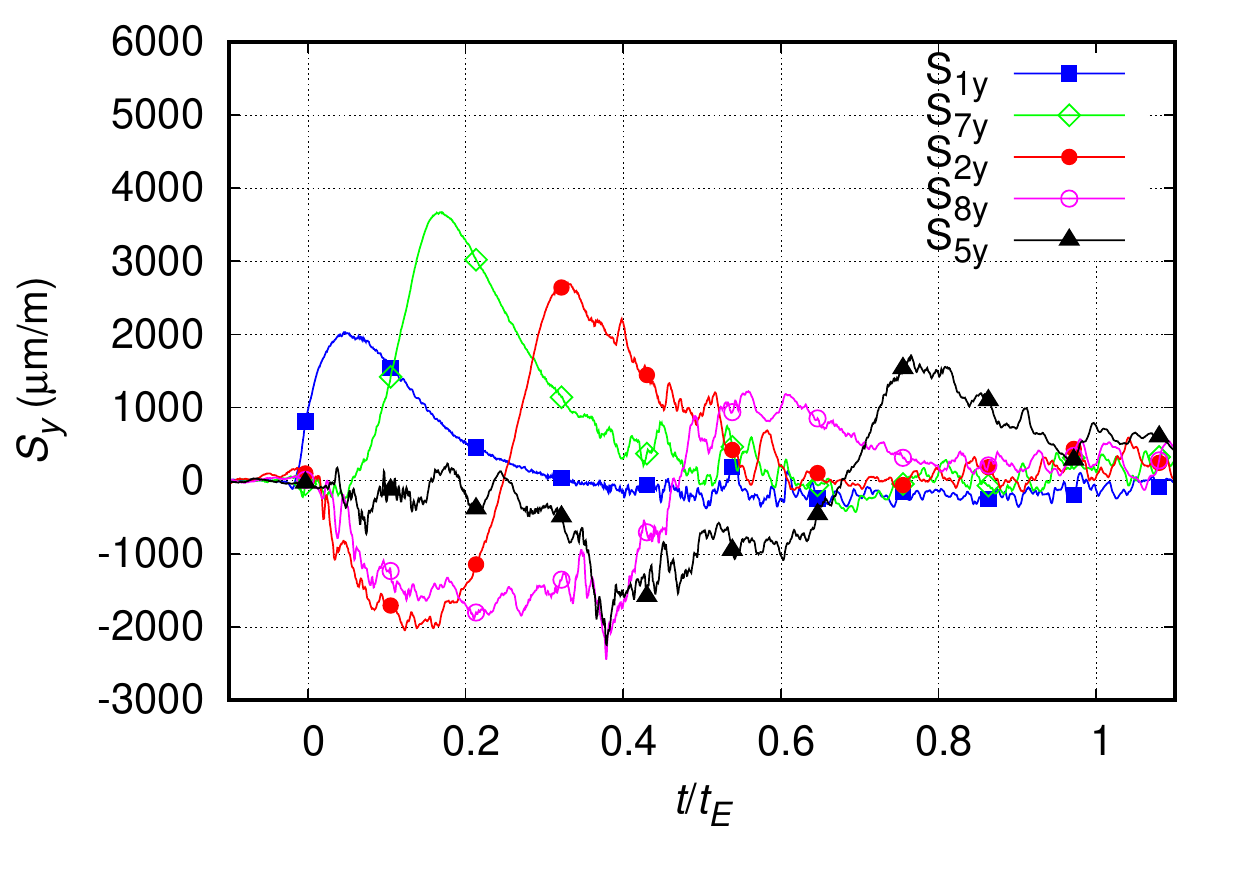}} \quad
\subfigure[08\_06\_40]{\includegraphics[width=0.31\textwidth]{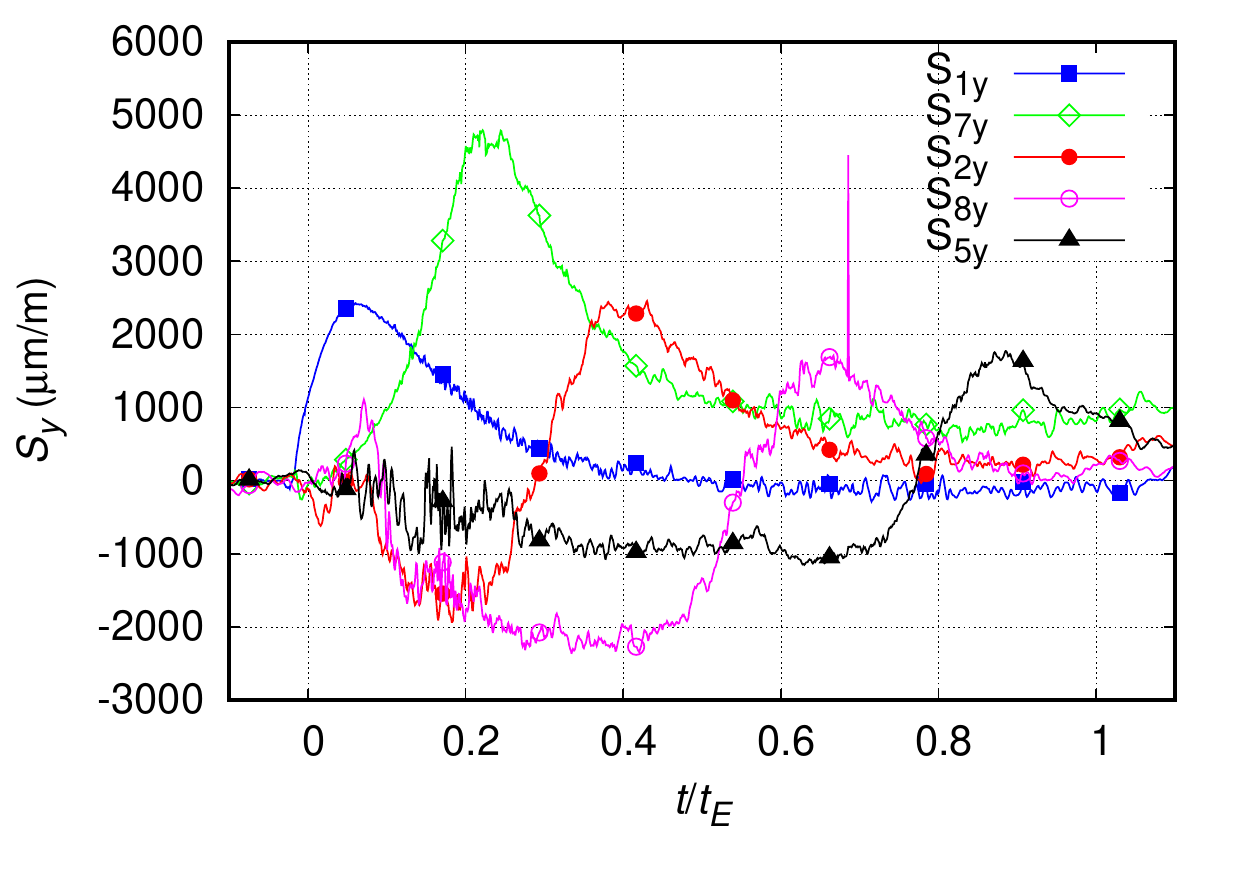}} \quad
\subfigure[08\_04\_45]{\includegraphics[width=0.31\textwidth]{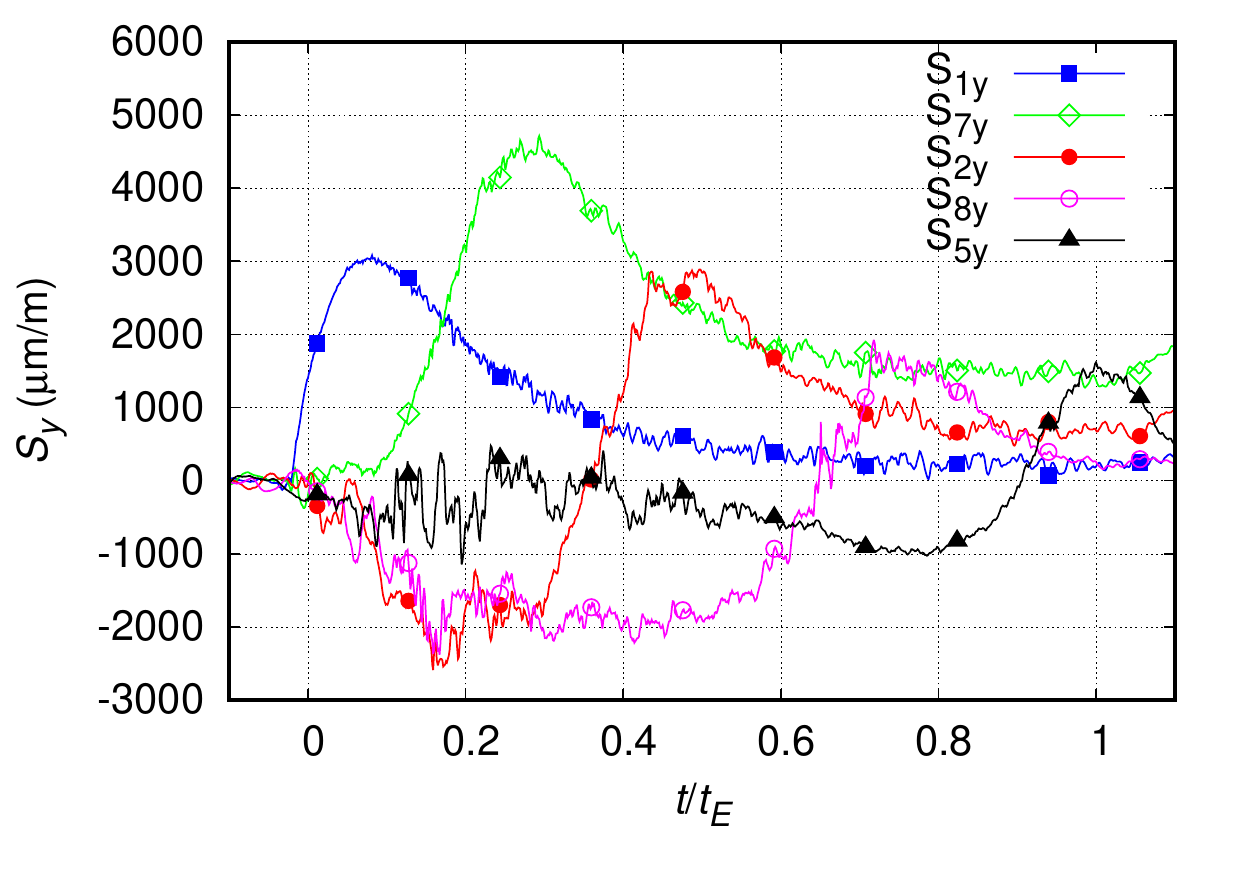}}  
\caption{Comparison of the transverse strains measured along the midline for the test conditions XX\_10\_30, XX\_06\_40 and
XX\_04\_45.}
\label{fig:y_strains_3Cases}
\end{figure}

In Figure \ref{fig:y_strains_3Cases} the $y$-strains measured by the gauges 
located along the midline are shown both for the thick and thin plates
for three test conditions i.e XX\_10\_30, XX\_06\_40 and
XX\_04\_45.
For the thick plate, as observed already for the strains in the longitudinal 
direction, the $y$-strains display a maximum followed by a gentle decay.
However, differently from the longitudinal strain, the strains in the
transverse direction remain always positive, both before and after the
passage of the peak, indicating that the out-of-plane deformation is 
always upwards. Moreover, the strains are more symmetric about the peak,
which is a consequence of the more uniform pressure loading in the $y$-direction compared to 
the $x$-direction. By comparing the above data with the corresponding strains measured in the
longitudinal direction, shown in Figure \ref{fig:x_strains_3Cases}, it is
observed that the peak in the $y$-direction is delayed with respect to the one
measured in $x$-direction, which is of course a consequence of the pressure
distribution. The strain level increases as the plate thickness is reduced. For the 3~mm
case, the gauges located far from the boundaries display a gentle rise
followed by a peak. After the peak, the strains undergo a small reduction
and remain almost constant until the end of the impact phase.
The gauge S$_1$ exhibits a small rise and then diminishes, reaching an
almost constant negative level.
Much more irregular are the time histories of the strains recorded for 
the 0.8~mm plates.
Gauges S$_{1y}$ and S$_{7y}$ exhibit an initial rise, a peak and a
decay to zero. The other gauges show negative strains, quite high
in amplitude in the early phase, denoting a downward curvature.
Next, the strains rise up to a positive peak and decay to
zero. For the 15~mm and 3~mm the peak of the strains have about the same
amplitude for the different gauges, whereas for the 0.8~mm plates the strain 
level decreases moving towards the leading edge. 
The fact that strains tend to return zero at the end for the impact phase means
that the deformation is basically elastic. The decreasing value of the strain peaks is
associated with the reduction of the pressure peaks, 
which is observed while going towards the leading edge, as discussed in more 
detail in Section \ref{pressurecameras}.
At a given thickness, the effects of test speed and pitch angle on the
transverse strain, although present, it is not as relevant as for the the longitudinal strains, 
apart from the higher peak of S$_7$ for the 0.8 mm thickness.

\begin{figure}[htbp]
\centering
\subfigure[Strain Gauge Positions]{\includegraphics[width=0.75\textwidth]{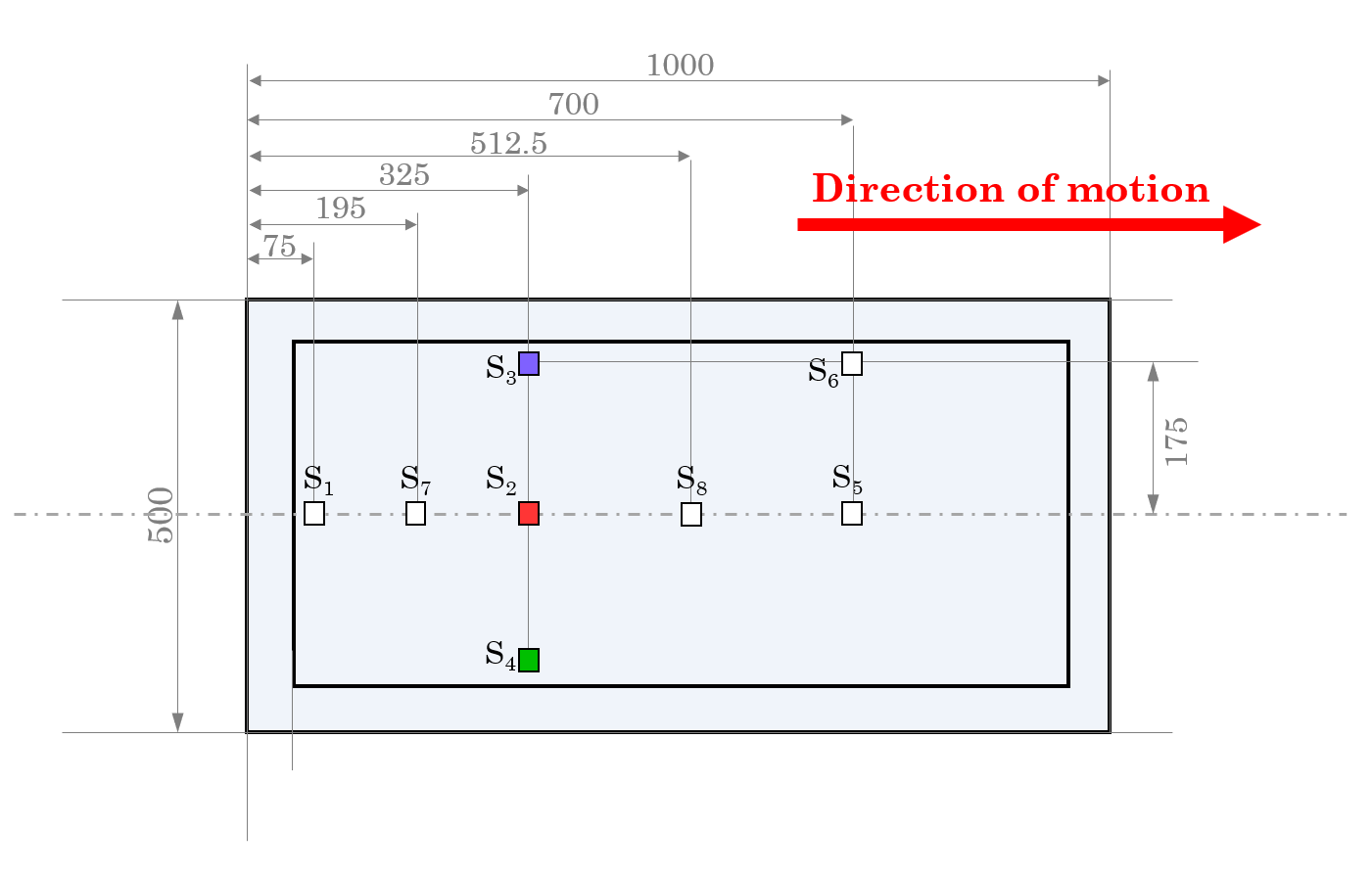}} 
\\
\subfigure[15\_10\_30]{\includegraphics[width=0.31\textwidth]{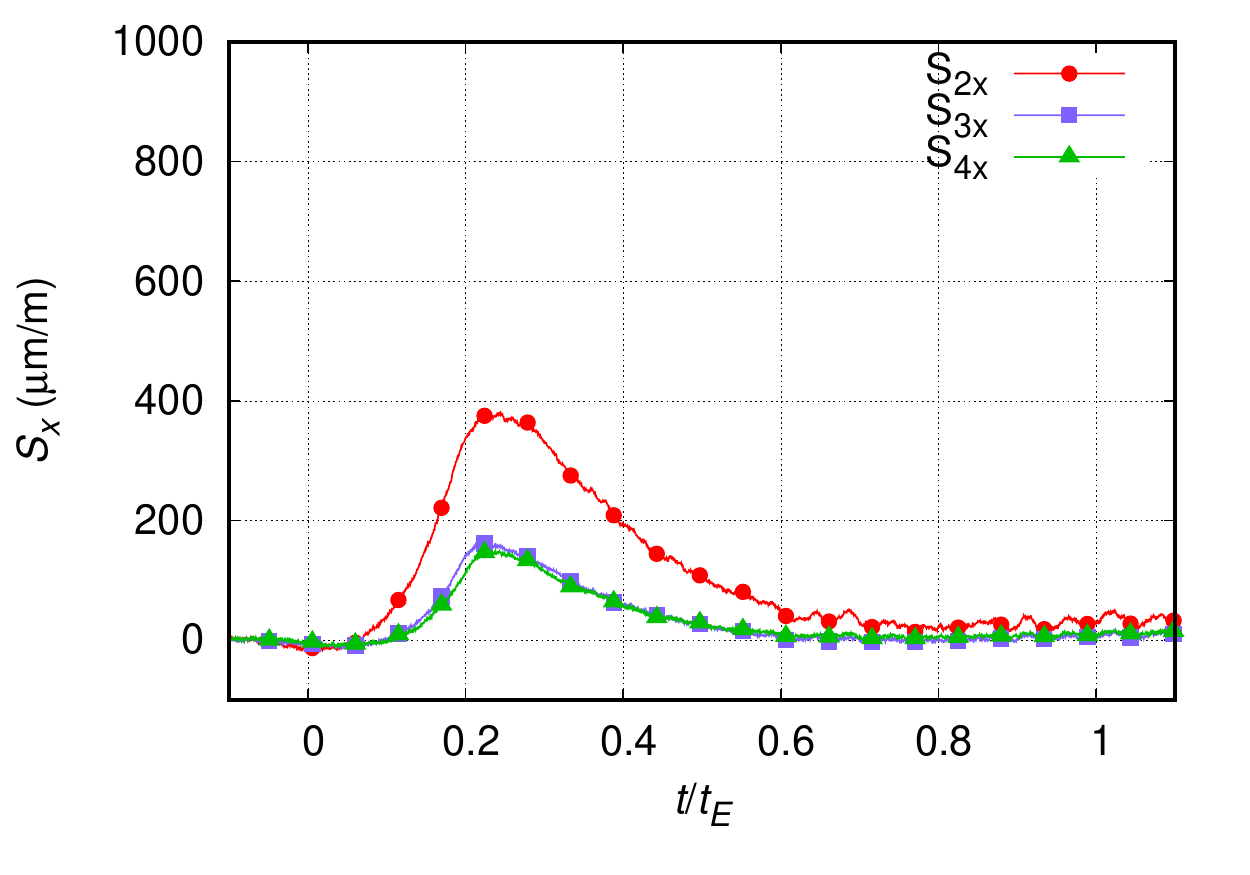}} \quad
\subfigure[15\_06\_40]{\includegraphics[width=0.31\textwidth]{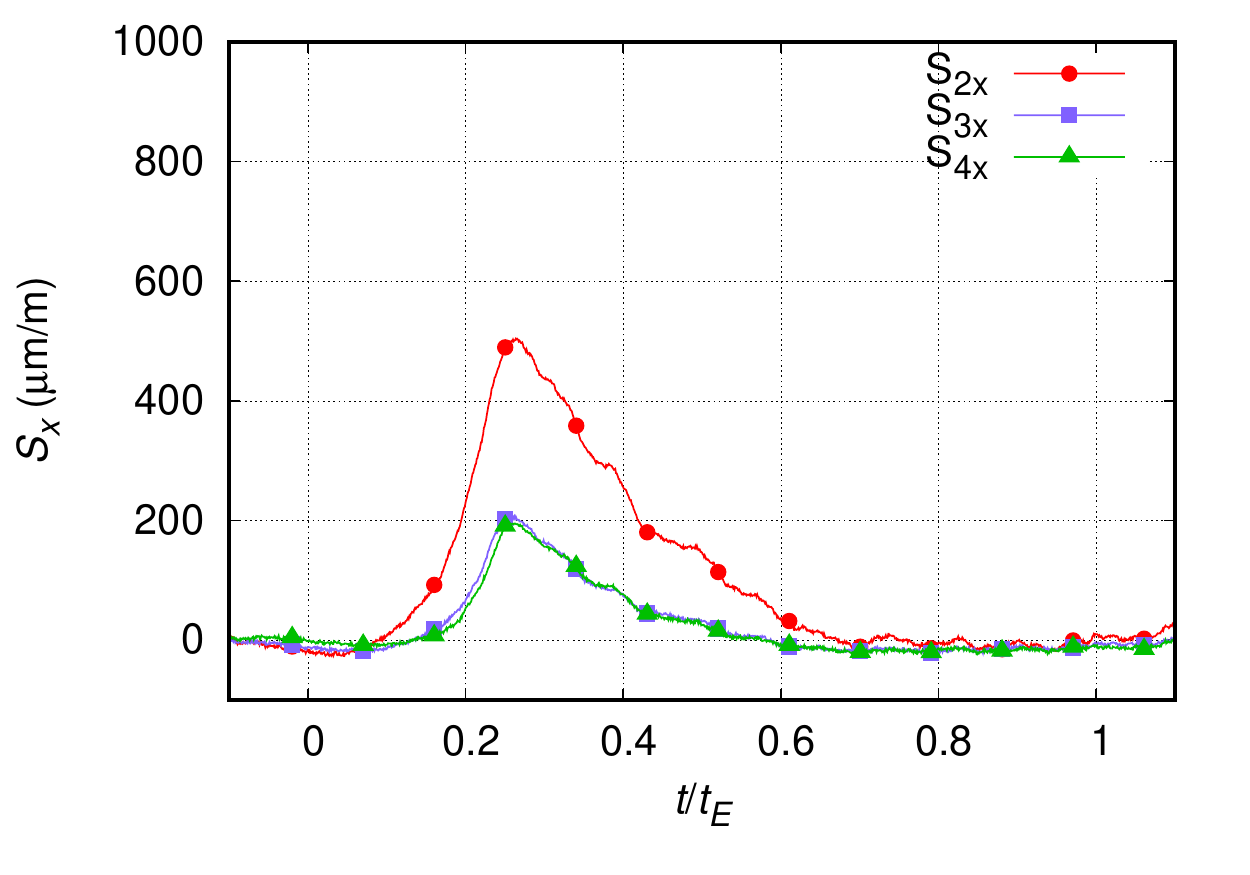}} \quad
\subfigure[15\_04\_45]{\includegraphics[width=0.31\textwidth]{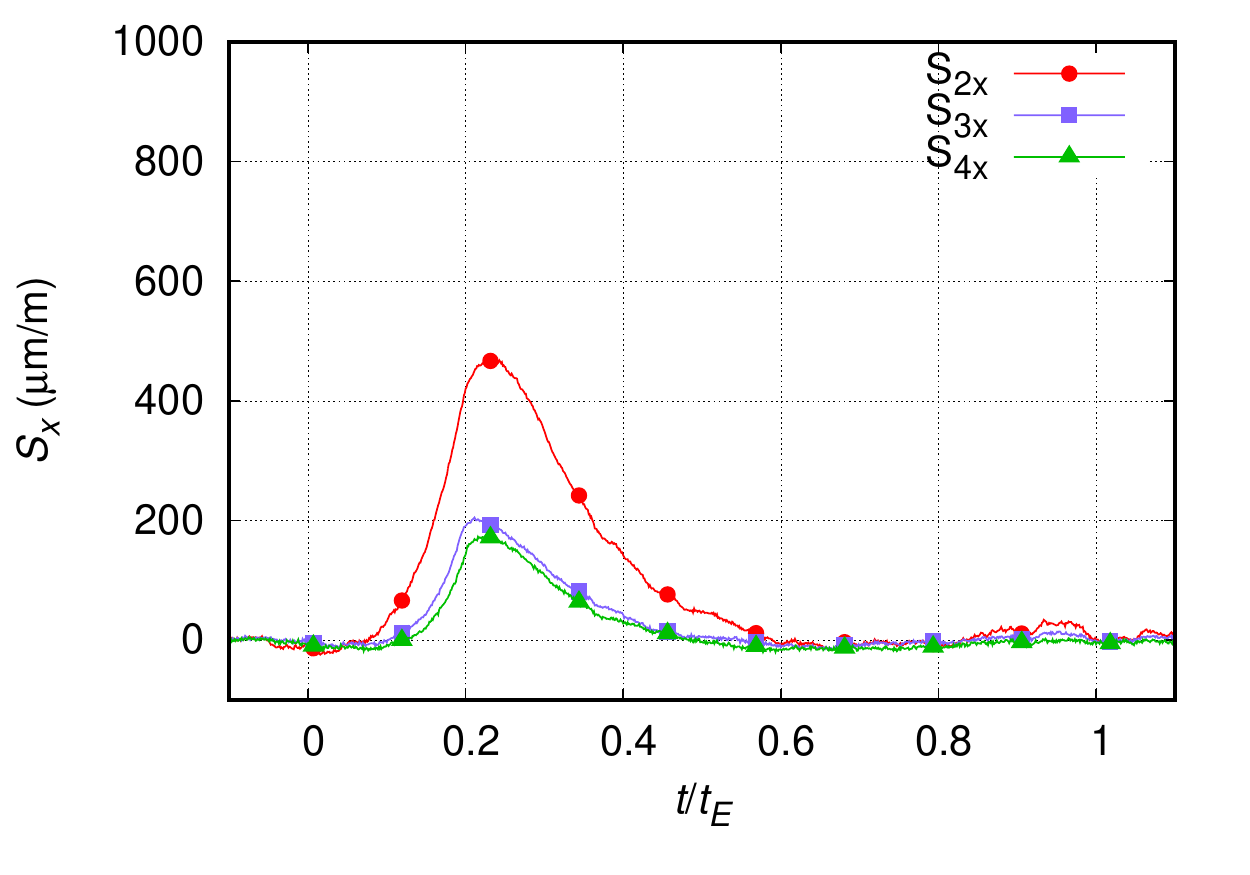}} \\
\subfigure[3\_10\_30]{\includegraphics[width=0.31\textwidth]{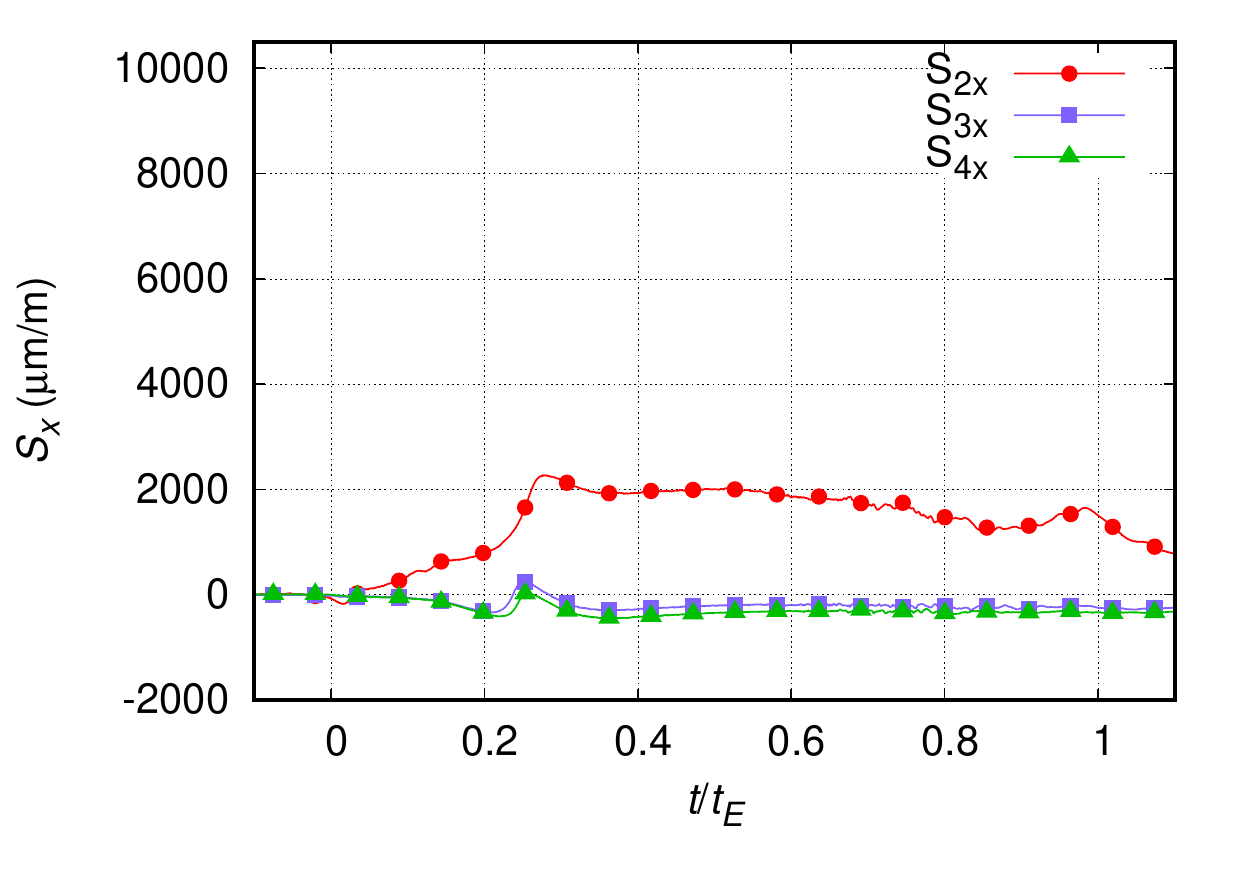}} \quad
\subfigure[3\_06\_40]{\includegraphics[width=0.31\textwidth]{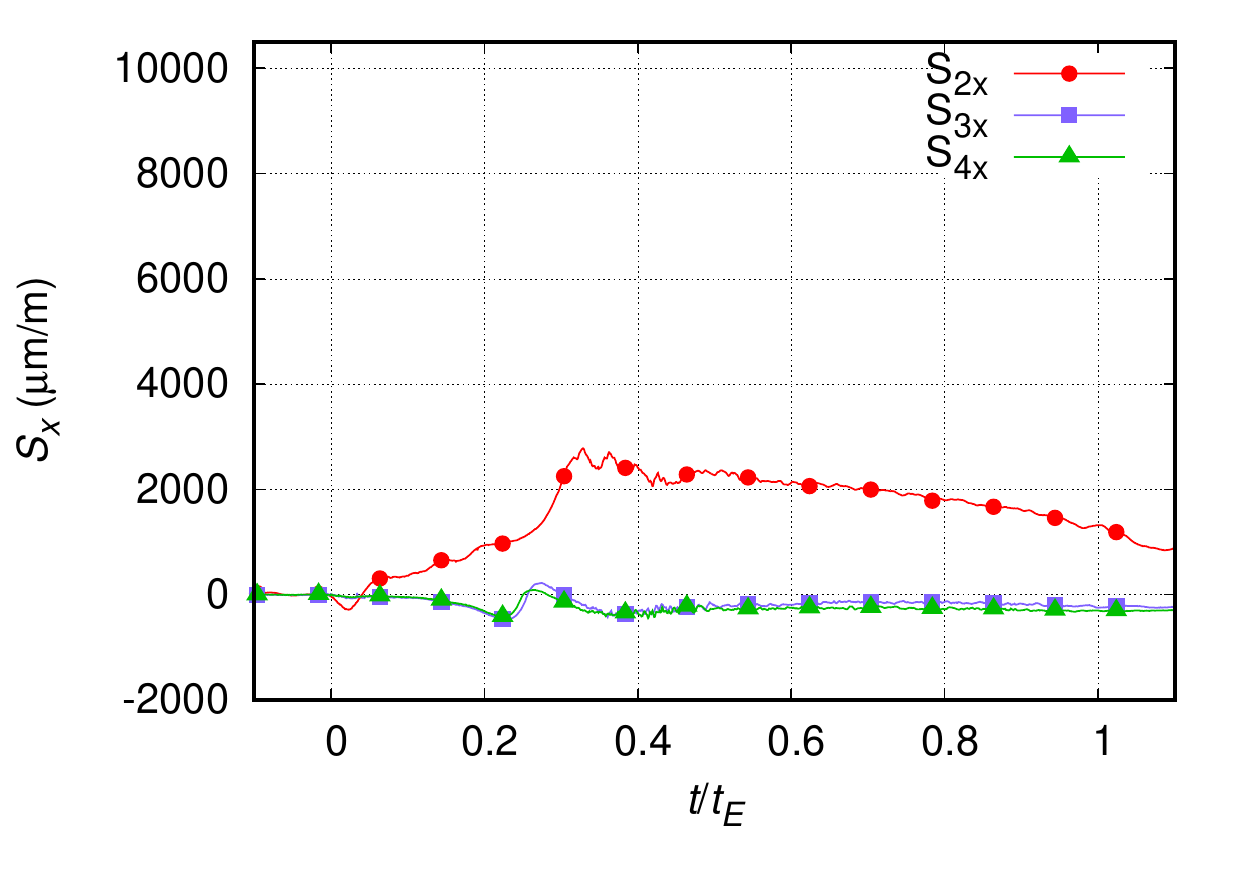}} \quad
\subfigure[3\_04\_45]{\includegraphics[width=0.31\textwidth]{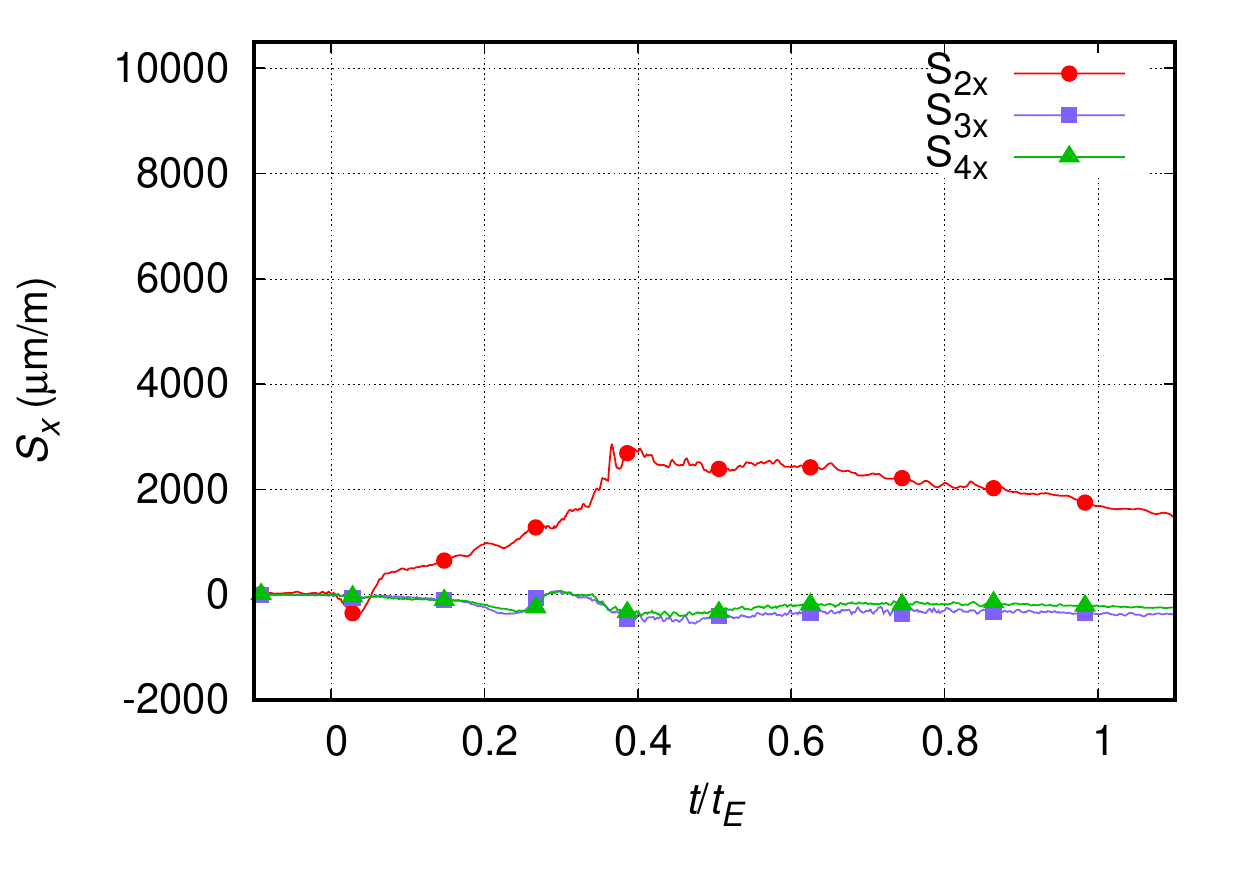}} \\
\subfigure[08\_10\_30]{\includegraphics[width=0.31\textwidth]{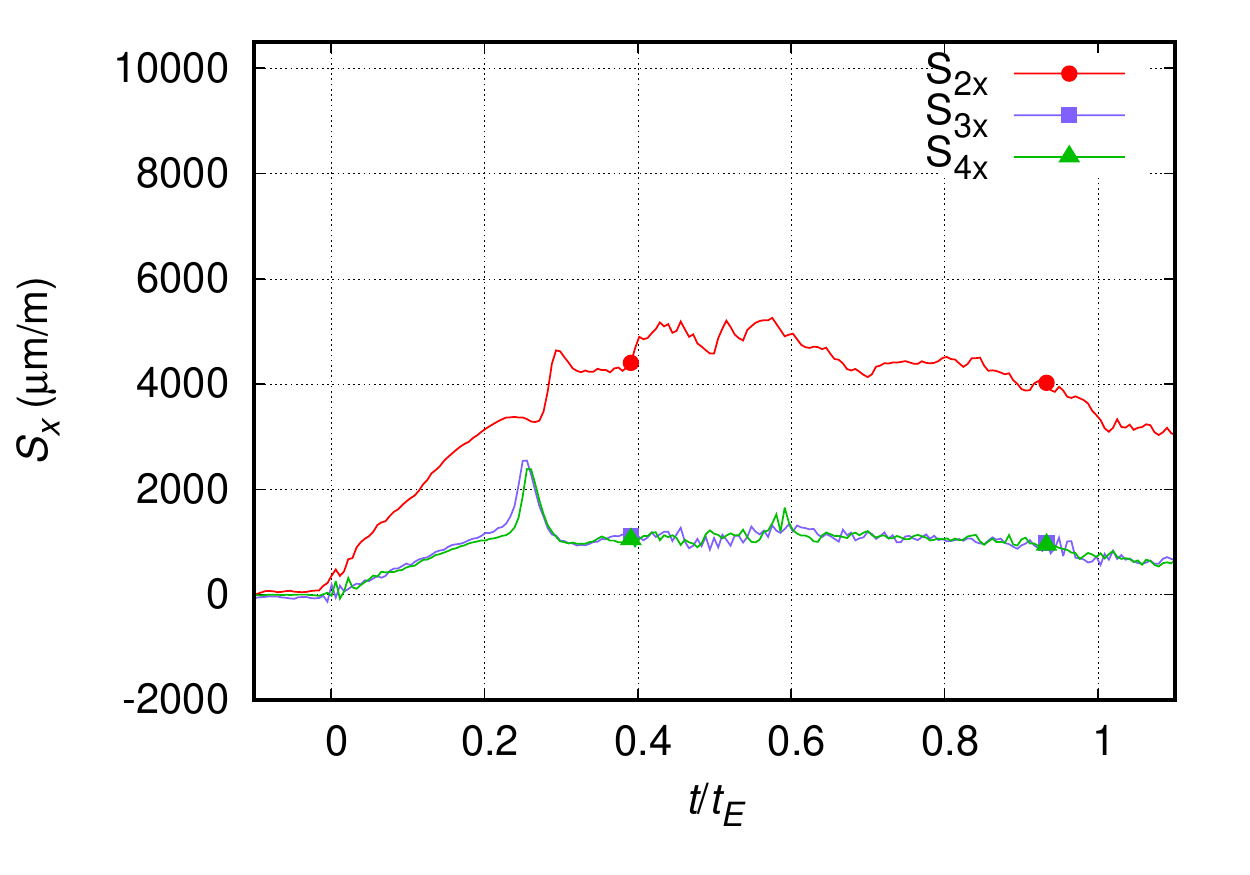}} \quad
\subfigure[08\_06\_40]{\includegraphics[width=0.31\textwidth]{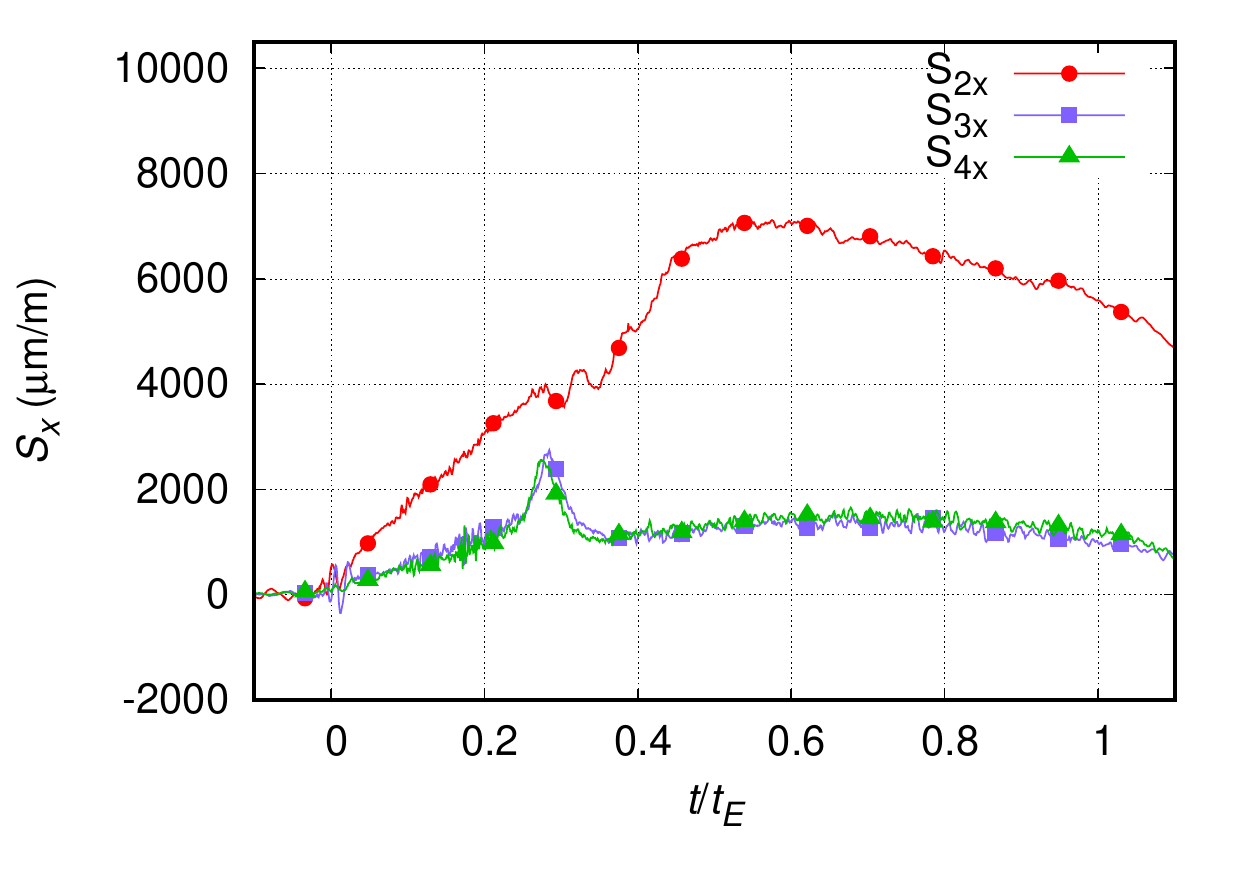}} \quad
\subfigure[08\_04\_45]{\includegraphics[width=0.31\textwidth]{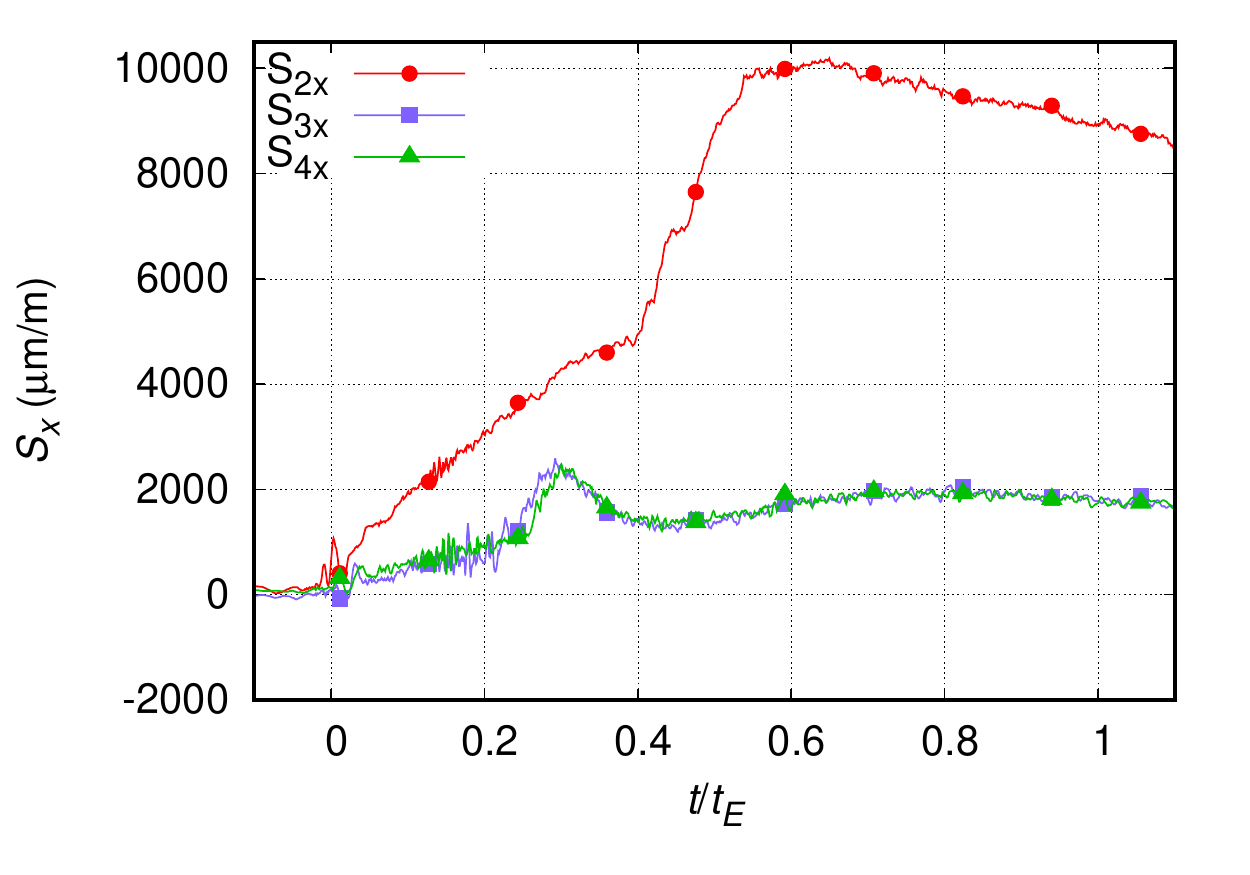}} 
\caption{Comparison of the longitudinal strains measured at x=0.325~m for the test conditions 
XX\_10\_30, XX\_06\_40 and XX\_04\_45.}
\label{fig:x_strains_3Cases_spanwise}
\end{figure}

\begin{figure}[htbp]
\centering
\subfigure[Strain Gauge Positions]{\includegraphics[width=0.75\textwidth]{Figures/SG_schematic_spanwise.png}} 
\\
\subfigure[15\_10\_30]{\includegraphics[width=0.31\textwidth]{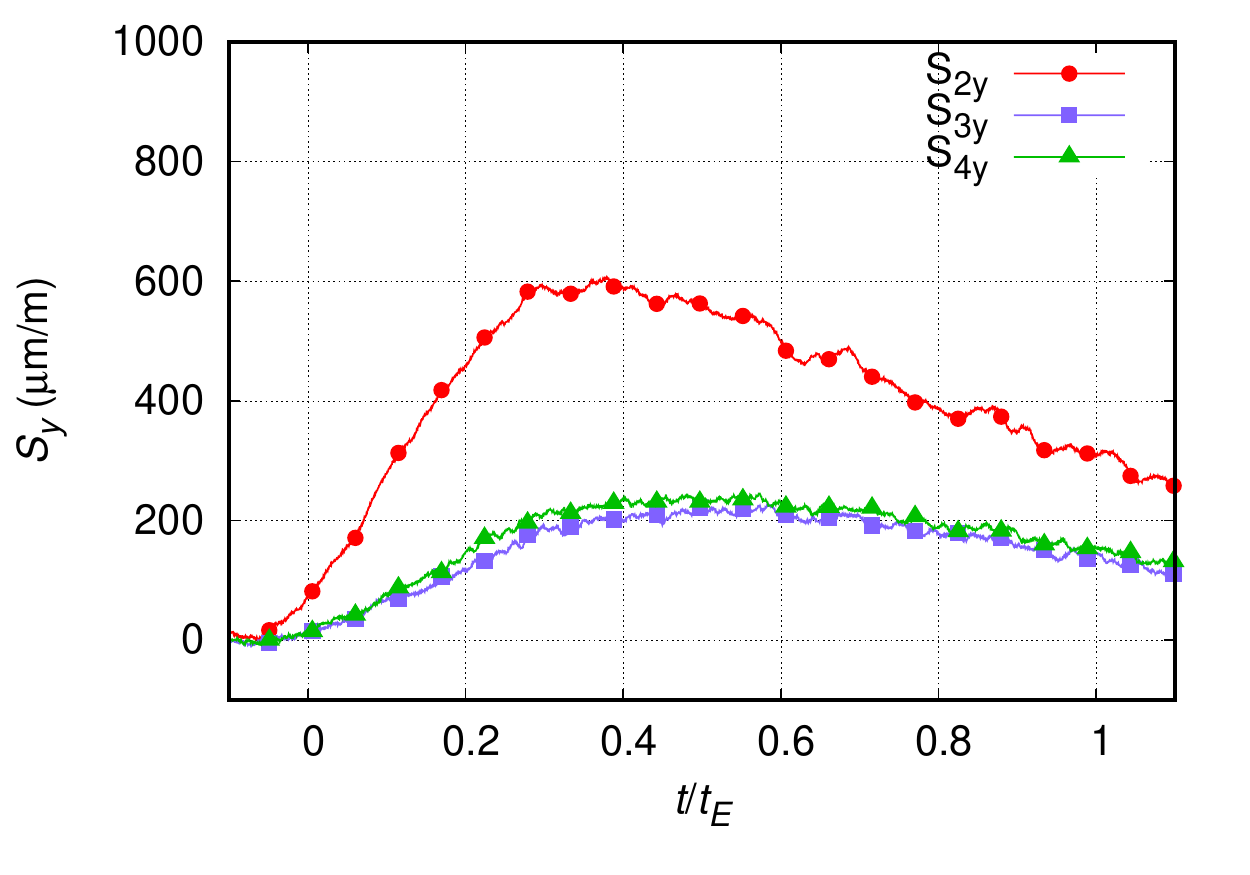}} \quad
\subfigure[15\_06\_40]{\includegraphics[width=0.31\textwidth]{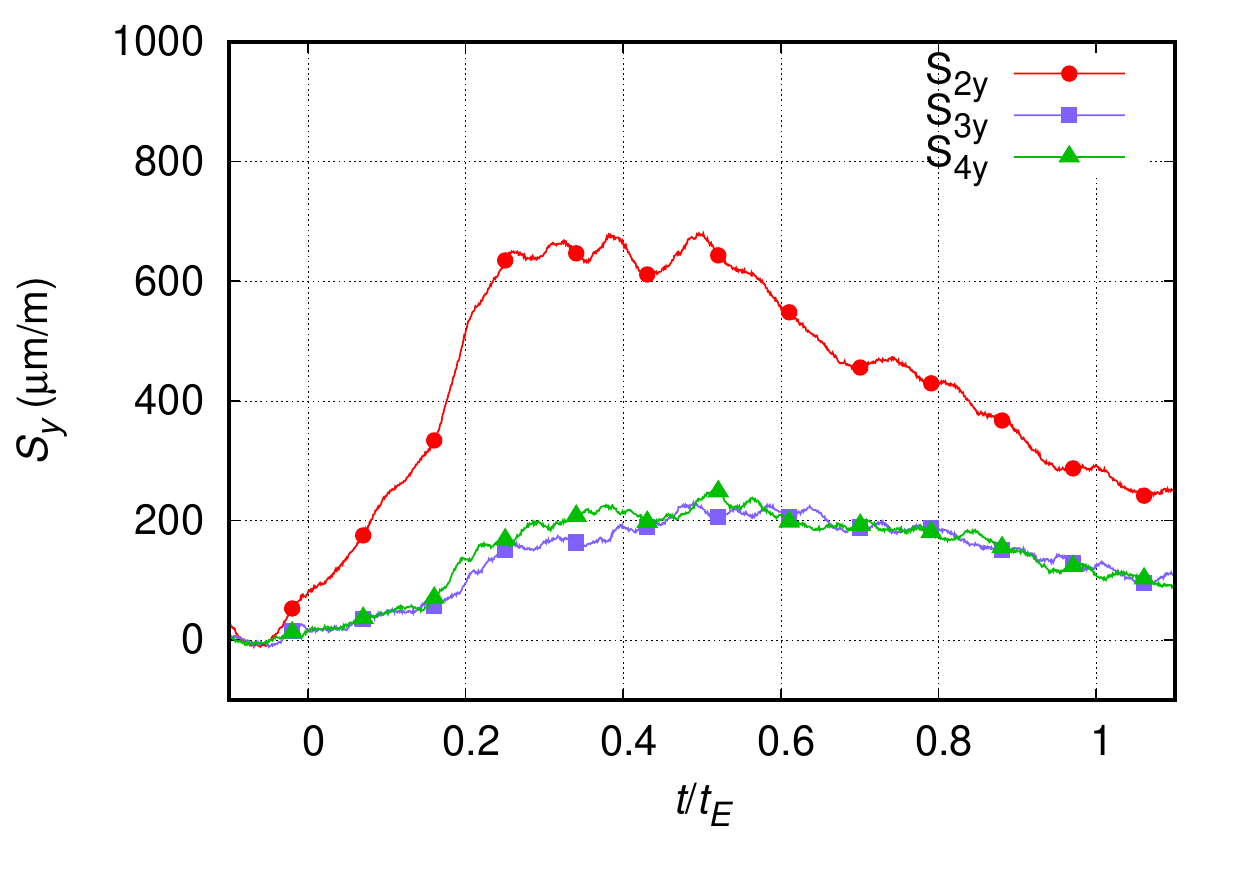}} \quad
\subfigure[15\_04\_45]{\includegraphics[width=0.31\textwidth]{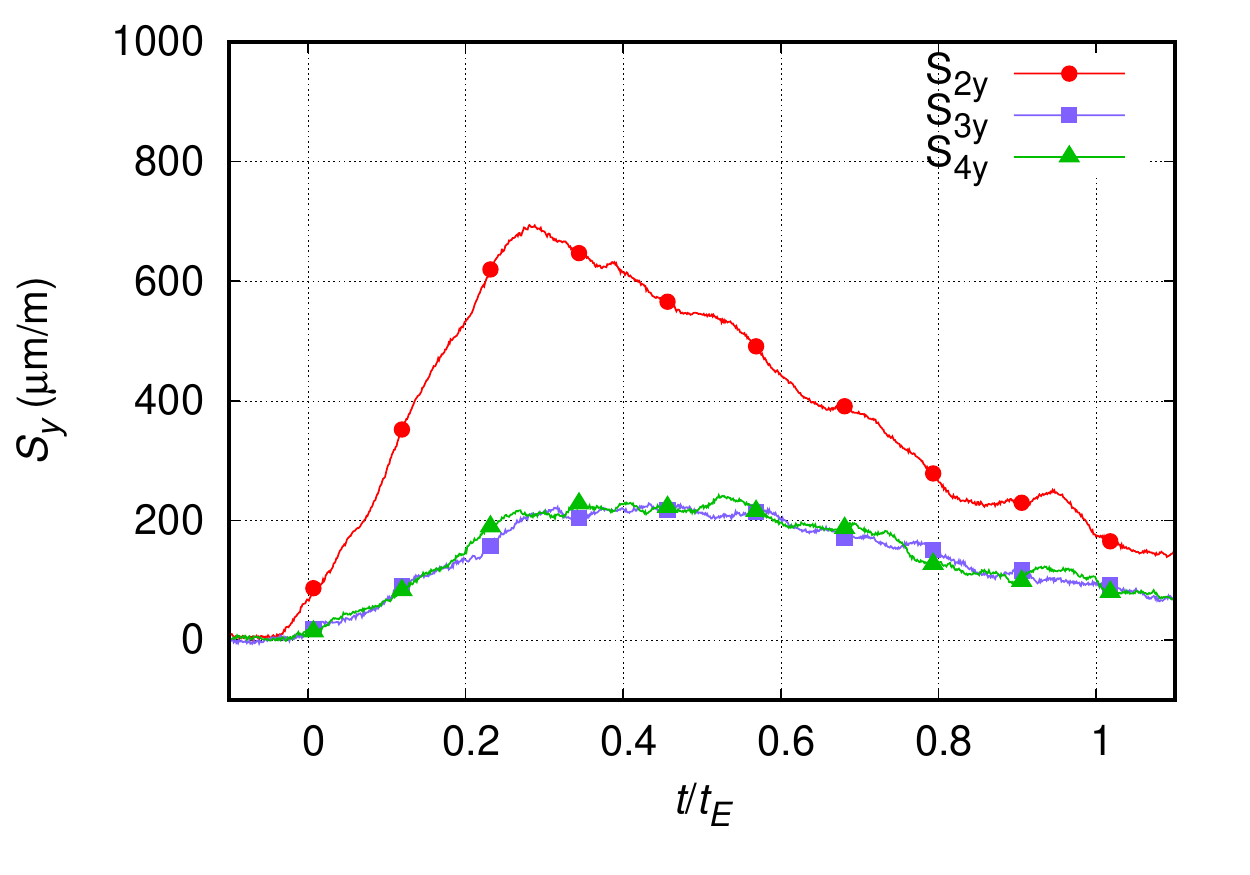}} \\
\subfigure[3\_10\_30]{\includegraphics[width=0.31\textwidth]{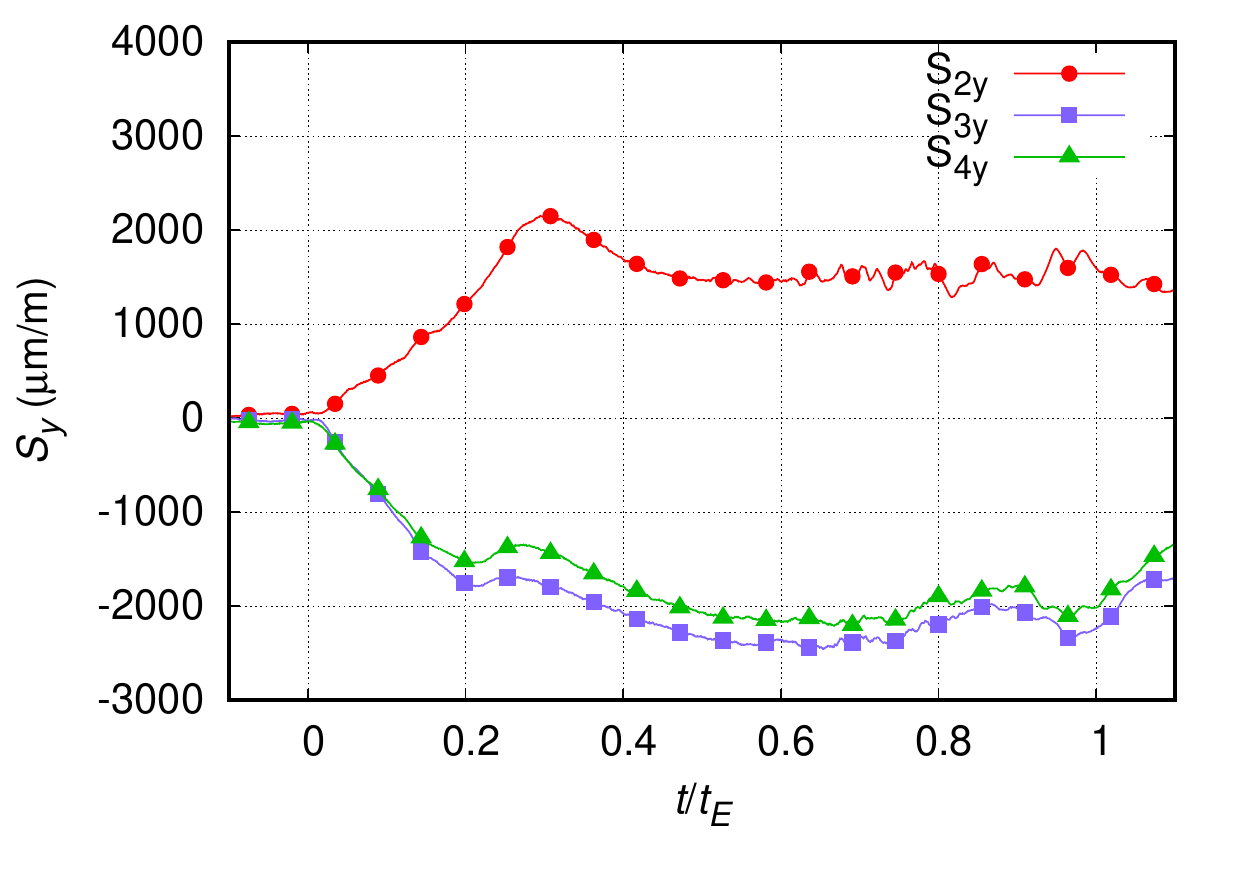}} \quad
\subfigure[3\_06\_40]{\includegraphics[width=0.31\textwidth]{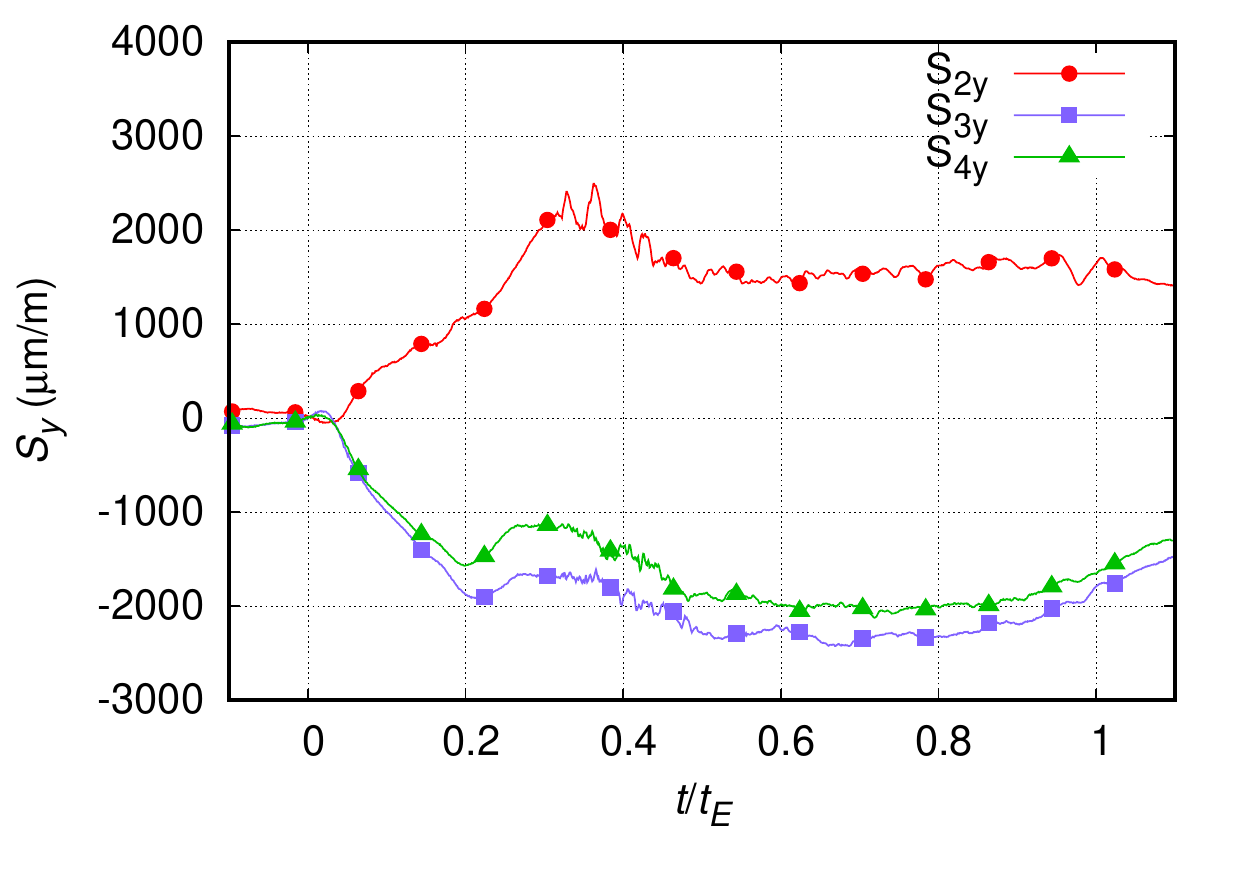}} \quad
\subfigure[3\_04\_45]{\includegraphics[width=0.31\textwidth]{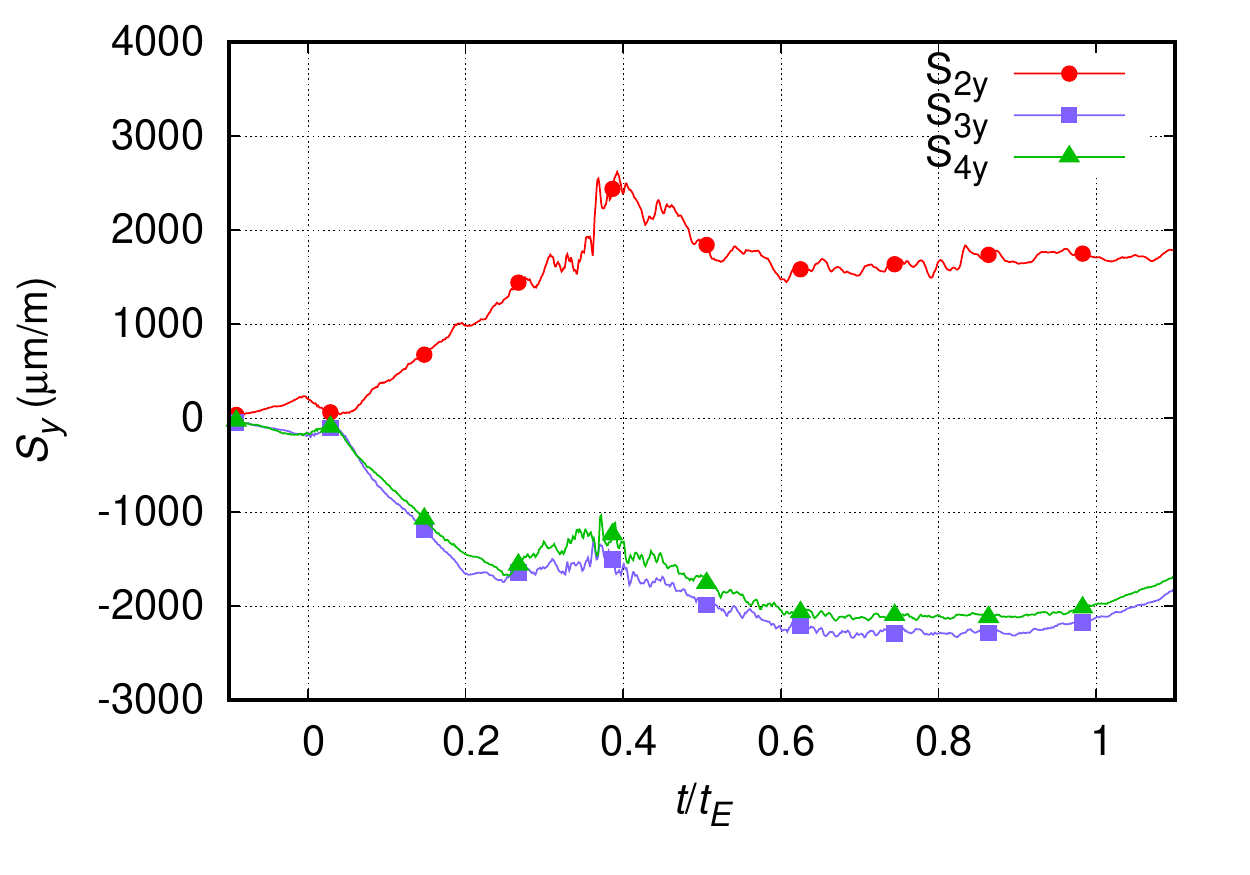}} \\
\subfigure[08\_10\_30]{\includegraphics[width=0.31\textwidth]{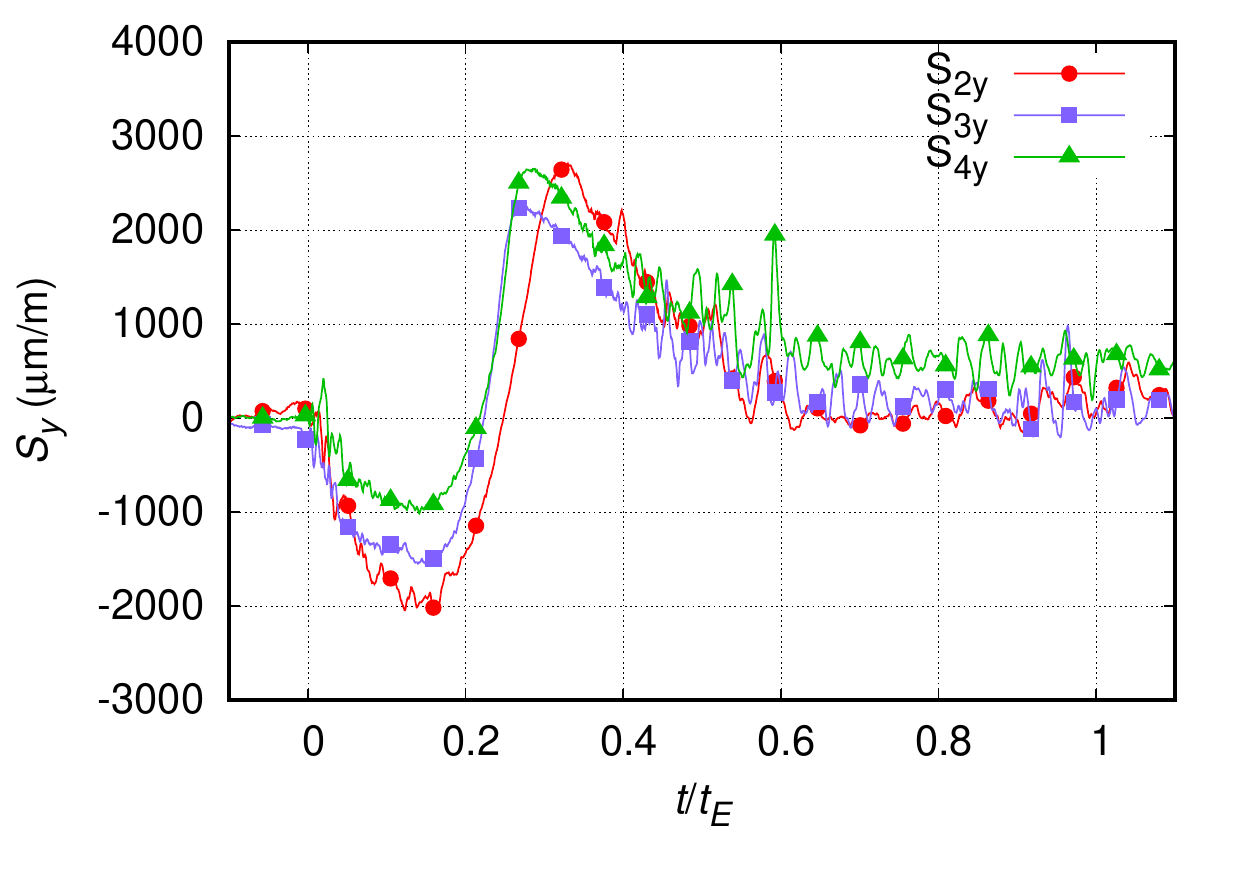}} \quad
\subfigure[08\_06\_40]{\includegraphics[width=0.31\textwidth]{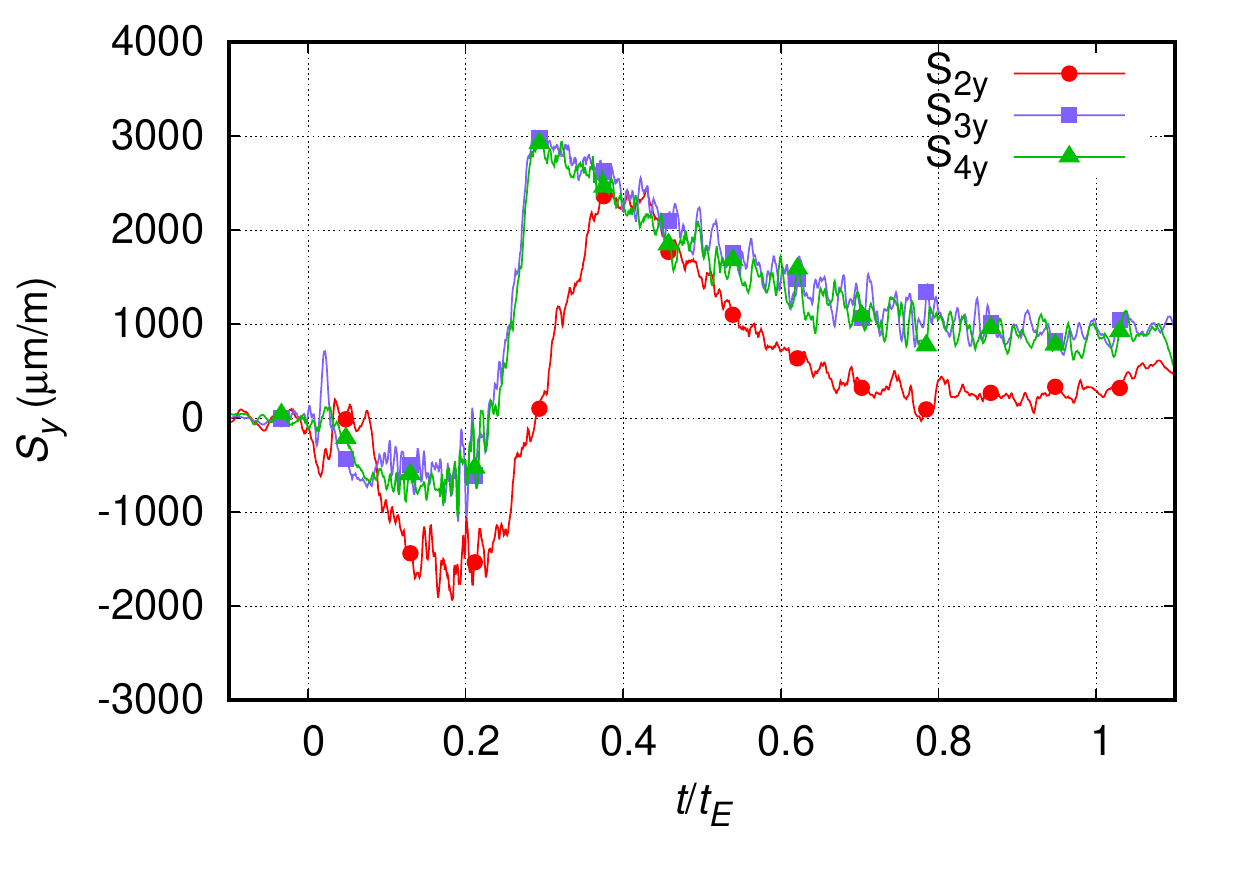}} \quad
\subfigure[08\_04\_45]{\includegraphics[width=0.31\textwidth]{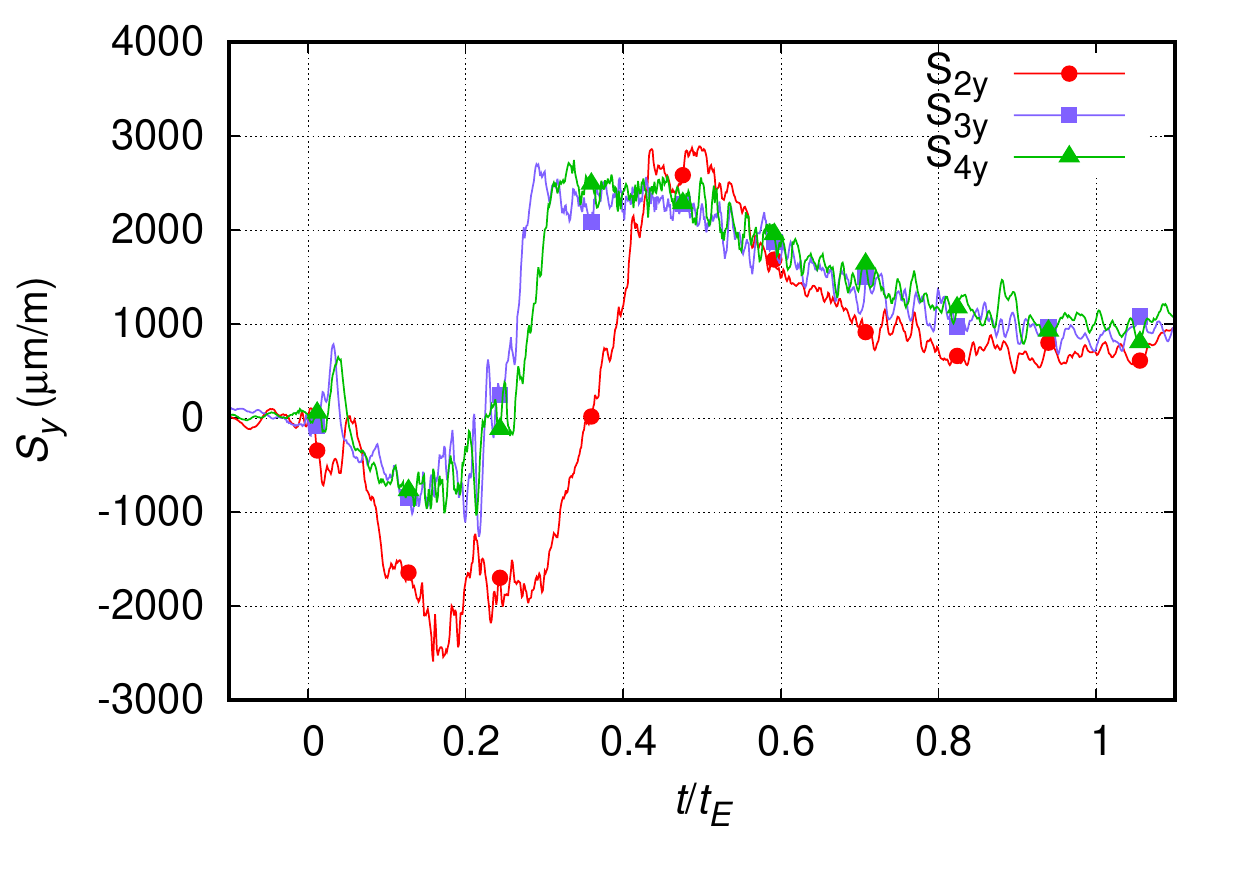}} 
\caption{Comparison of the transverse strains measured at x=0.325~m for the test conditions XX\_10\_30, XX\_06\_40 and
XX\_04\_45.}
\label{fig:y_strains_3Cases_spanwise}
\end{figure}

In order to investigate the behaviour of the strains at a fixed distance from the
 trailing edge, the data of the gauges S$_2$, S$_3$ and S$_4$ are provided in Figure
\ref{fig:x_strains_3Cases_spanwise} and 
\ref{fig:y_strains_3Cases_spanwise} for the $x$ and $y$ components,
respectively. As above, the data for the different plate thickness are
shown for the three different test conditions.
In the $x$-direction, due to the boundary conditions, the strains measured 
in the middle are higher than those at the sides. For the 15~mm
plates the strain in the middle is about twice as that at the sides.
Something similar also happens for the 0.8~mm plates, at least until the 
peak is reached at the side. Next, the strain at the side diminishes
whereas the strain in the middle continues to grow up for half the duration 
of the impact phase. This is more evident if the test speed increases and
if the pitch angle reduces.
A quite different behaviour characterizes the strains for the 3~mm plates.
In this case the strain in the middle exhibits a positive growth before
the passage of the pressure peak and remains almost constant, or diminishes
very slowly, after the pressure peak.
On the contrary, the $x$-strains at the sides show a small peak associated 
with the passage of the pressure peak, but their values are always rather
small and are negative for most of the time.
It is worth noticing that the peak of the strains for the thin plates are
slightly delayed compared to the 15~mm plate. This is presumably due to the
larger out-of-plane deformation of the plates, which induces a change in the
pressure distribution, as described in Section \ref{pressurecameras}. In all cases, a quite satisfactory overlapping of the measurements at the
two sides can be observed, which confirms the symmetry of the loading and, thus,
of the test conditions.
The data shown in Figure \ref{fig:x_strains_3Cases_spanwise} also confirm that the effect of pitch angle and of the  horizontal speed on the longitudinal strains is much more relevant at the middle (S$_{2x}$) than at the sides (S$_{3x}$ and S$_{4x}$).

Strains in the $y$-direction are shown in Figure 
\ref{fig:y_strains_3Cases_spanwise} for the three test conditions 
XX\_10\_30, XX\_06\_40 and XX\_04\_45.
As already found for the gauges located along the midline,
also at these locations the strains in $y$-direction are smoother that
those in the $x$-direction.
For the 15~mm plates, the peak of the strains in the middle of the plate is
about three times higher than that at the side and there are no substantial changes
with the test condition.
In the 3~mm plates there is a quite a significant difference between the
strains in the middle, which are positive, and those at the sides, which
are negative, thus denoting a difference in the curvature and the presence
of an inflection point. The passage of the pressure peak is not so evident
in the curves.
Also in this case there are no significant changes in the trends and in the
peak values when varying the test condition. The  most remarkable
difference concerns a delay in the occurrence of the peak at S$_2$ when 
increasing the speed and reducing the pitch angle. 
A completely different behaviour characterizes the $y$-strains at the side
for the 0.8~mm plates. In all three cases the peak values at the sides are
comparable to those in the middle. The peak in the middle is delayed
with respect to the sides, which is an effect of the forward bending of the
the spray root, as it is further discussed in Section \ref{pressurecameras}.
The fact that the strains are rather uniform in the transverse direction is
because, due to the small thickness, the effect of the boundary condition
is confined to a small neighbourhood of the frame.
The data for the 0.8~mm plates also show that the strains, which are negative
before the passage of the pressure peak, turn positive afterwards.
In a spatial reference frame, this indicates that the strains are positive
behind the pressure peak and are negative ahead of the spray root.
Hence, for such a small plate thickness, the pressure peak generates an
inflection of the plate, which propagates with it.

\subsection{Residual deformations}
\label{permdef}
The thin plates display residual deformations at the end of the tests,
thus indicating that the strains go beyond the yielding limit in some areas. In the
previous section it is shown that this is revealed by some
of the gauges in the 0.8~mm plates at least. However, residual
deformations are also observed for the 3~mm plates, even though the 
measured strains are always below the yielding limit. 
Very likely, the maximum strains occur next to the 
inner boundary of the frame and are not recorded by the installed gauges, because for the thin plates they 
are located at 25~mm from the boundary itself, see Figure \ref{fig:Sensor_Positions}.

The residual deformation is provided in terms of the out-of-plane
displacement $w$ measured on a matrix of points upon the plate, before
dismounting the plate from the Aluminium frame. 
It is worth remarking that the plate is loaded also after the impact
phase. Although the highest normal loads are generated during that stage, 
it cannot be excluded that, for the 0.8~mm plates in particular, the 
residual deformation is affected by the loading and by the fluid-structure 
interaction taking place in the final stage of the test. 

\begin{figure}[htbp]
\centering
\subfigure[3\_10\_30]{\includegraphics[width=0.48\textwidth]{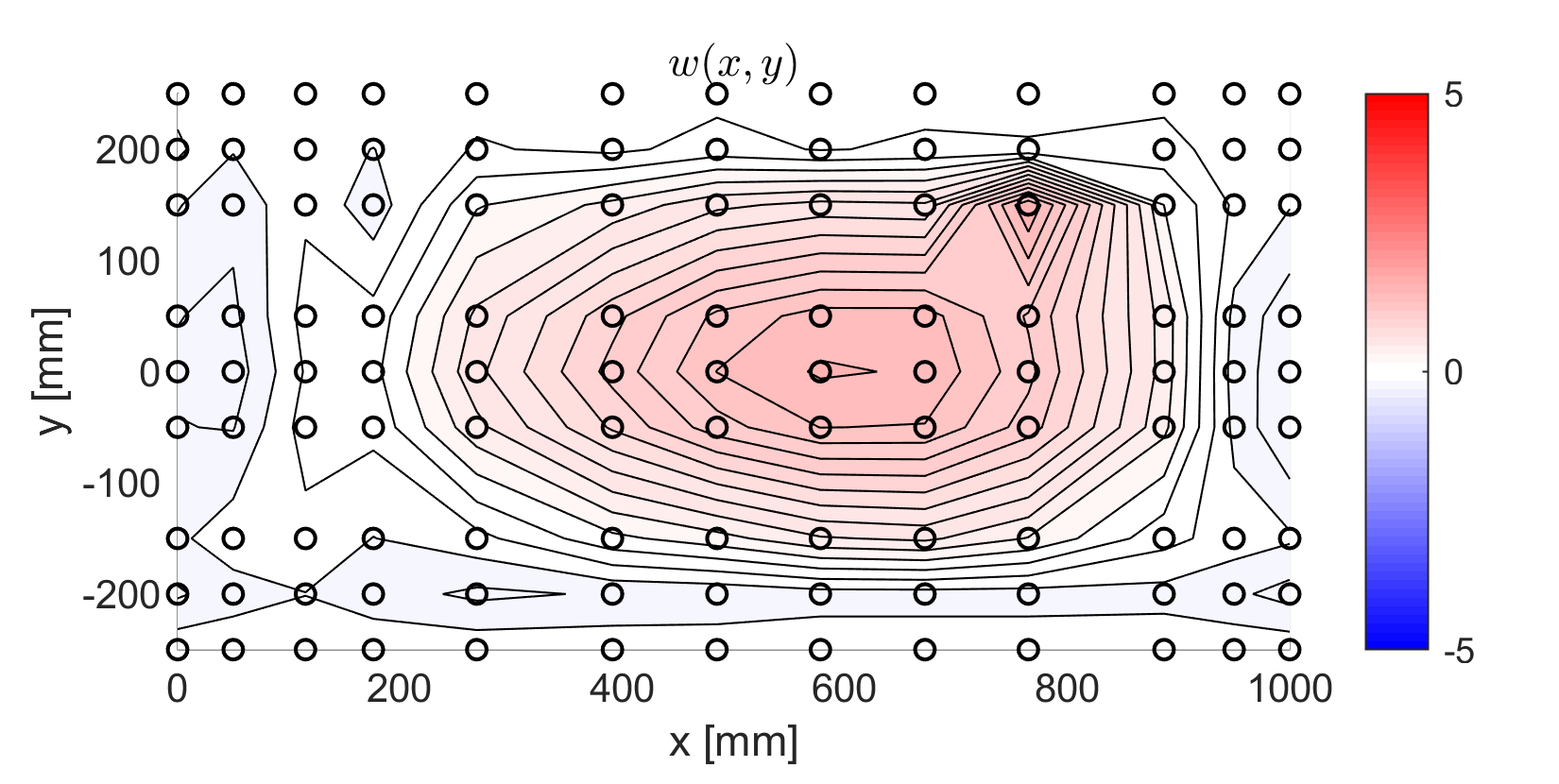}} \quad
\subfigure[08\_10\_30]{\includegraphics[width=0.48\textwidth]{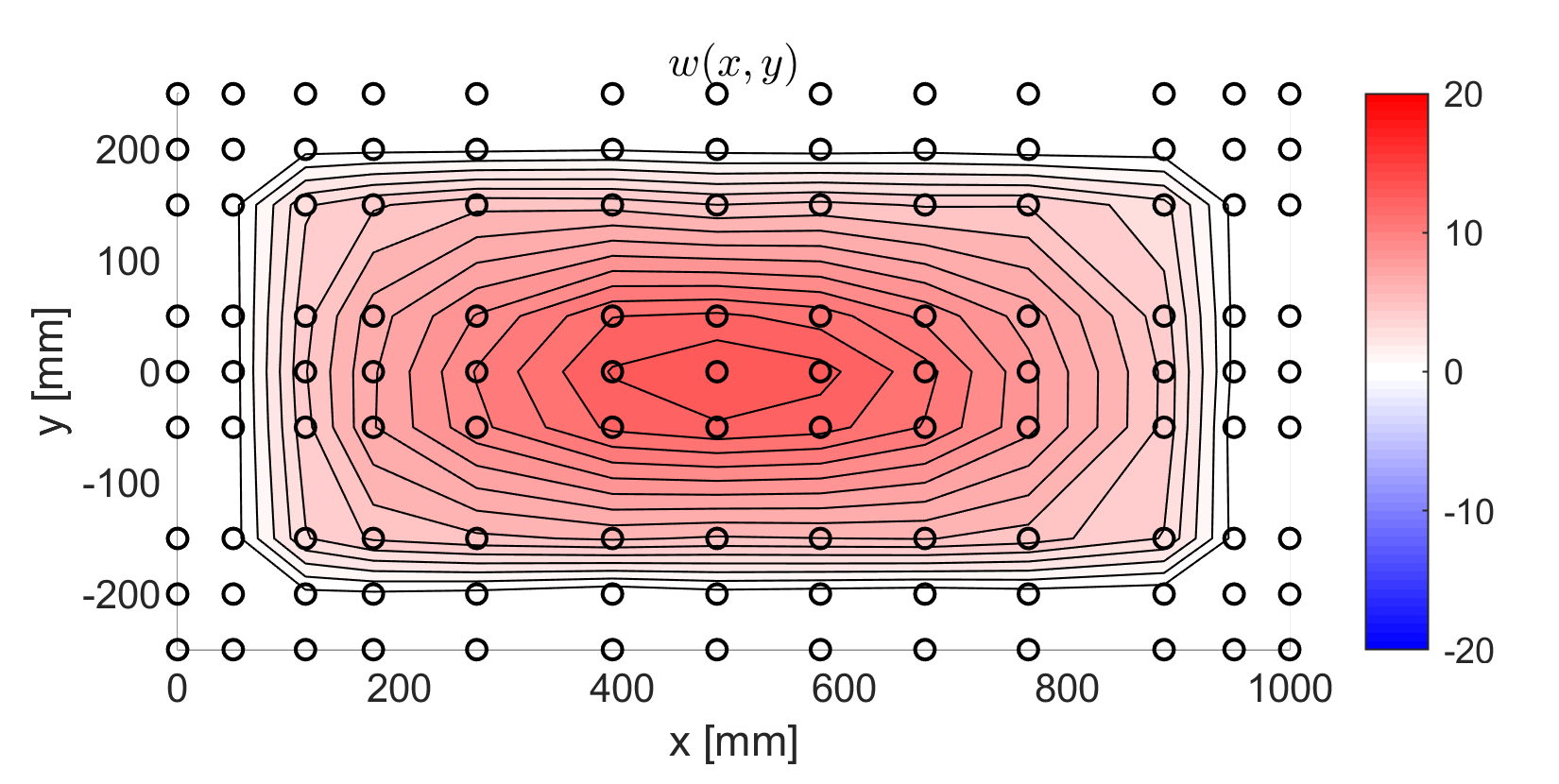}} \\
\subfigure[3\_06\_40]{\includegraphics[width=0.48\textwidth]{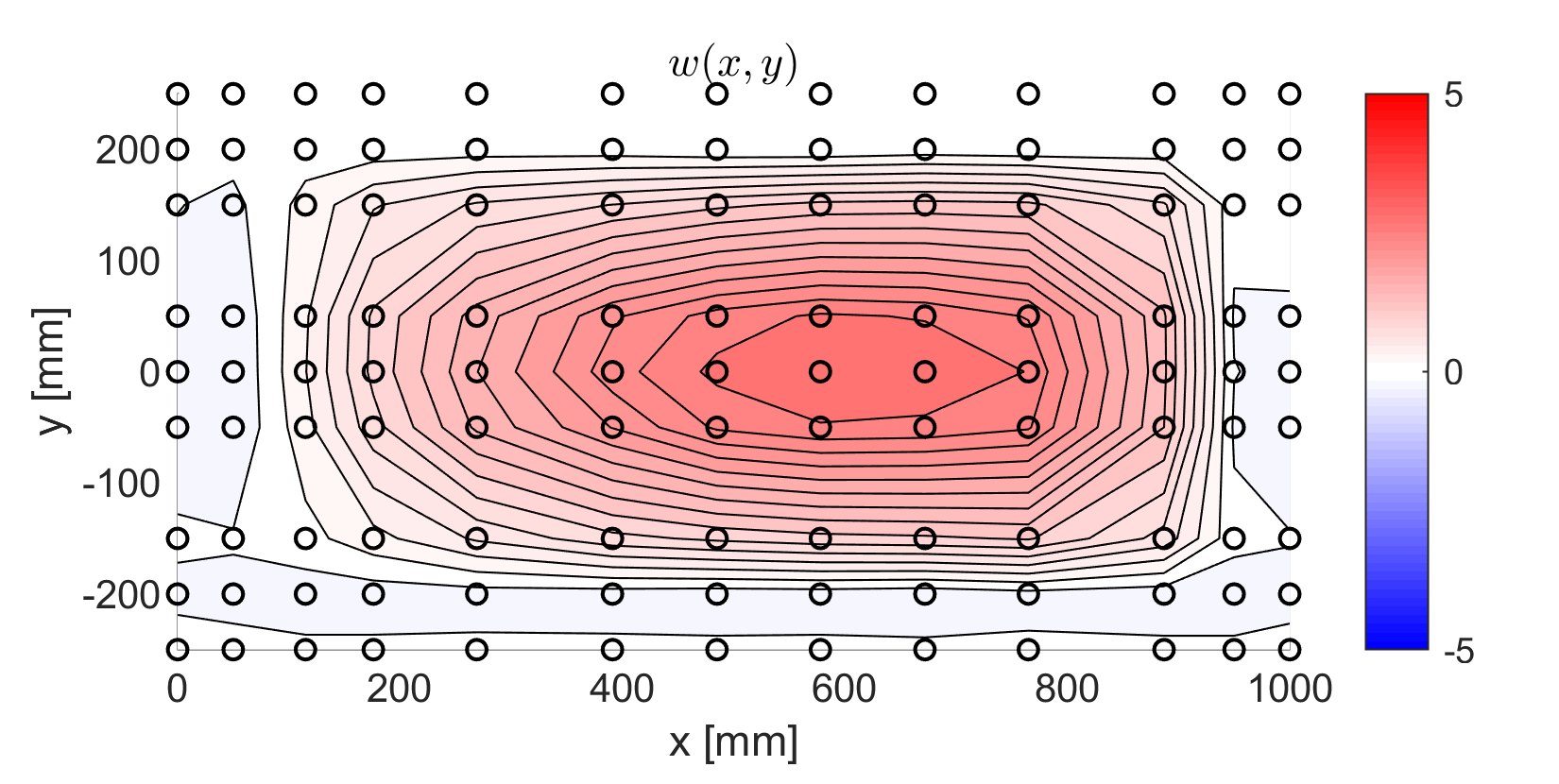}} \quad 
\subfigure[08\_06\_40]{\includegraphics[width=0.48\textwidth]{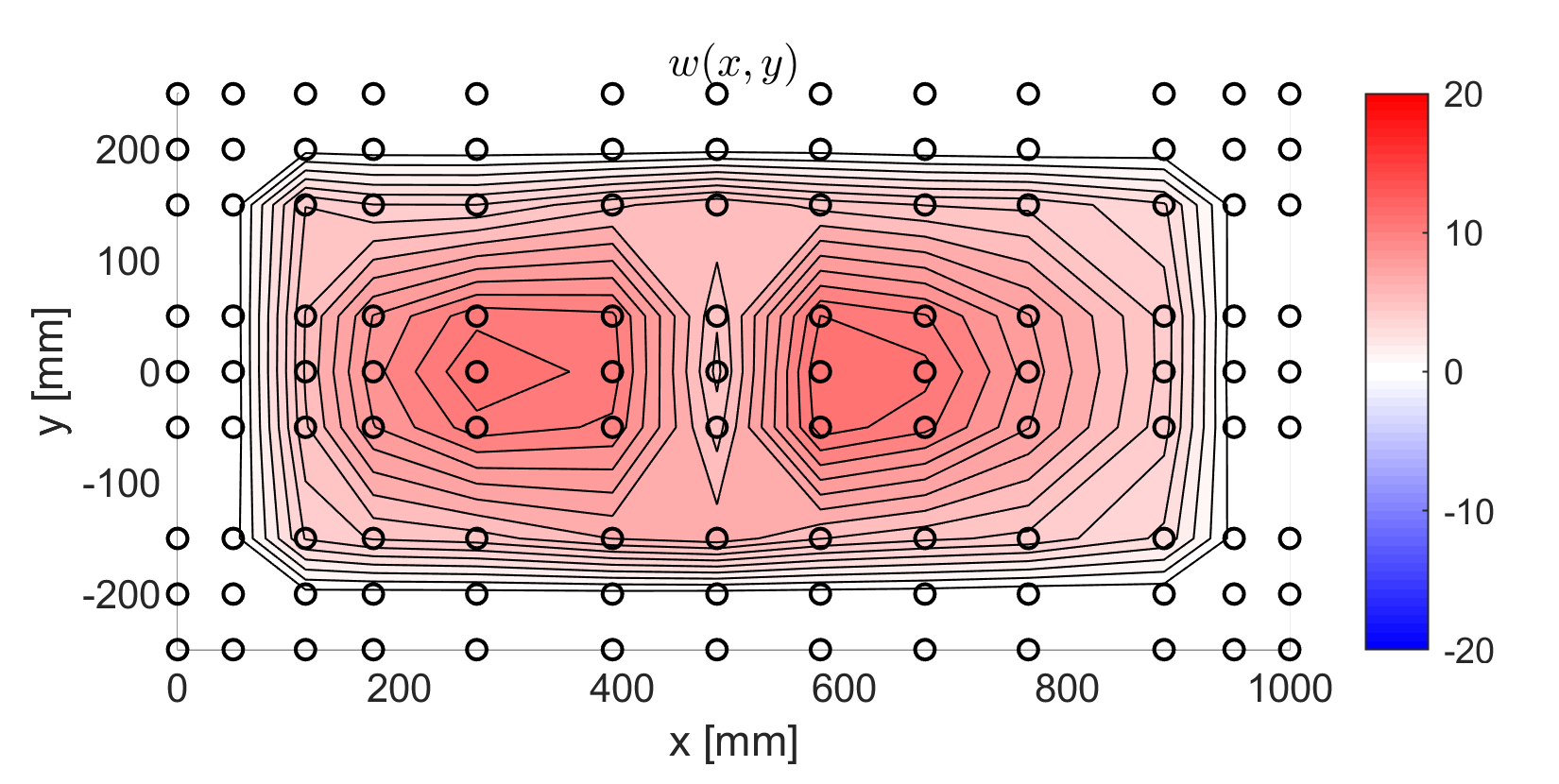}} \\
\subfigure[3\_04\_45]{\includegraphics[width=0.48\textwidth]{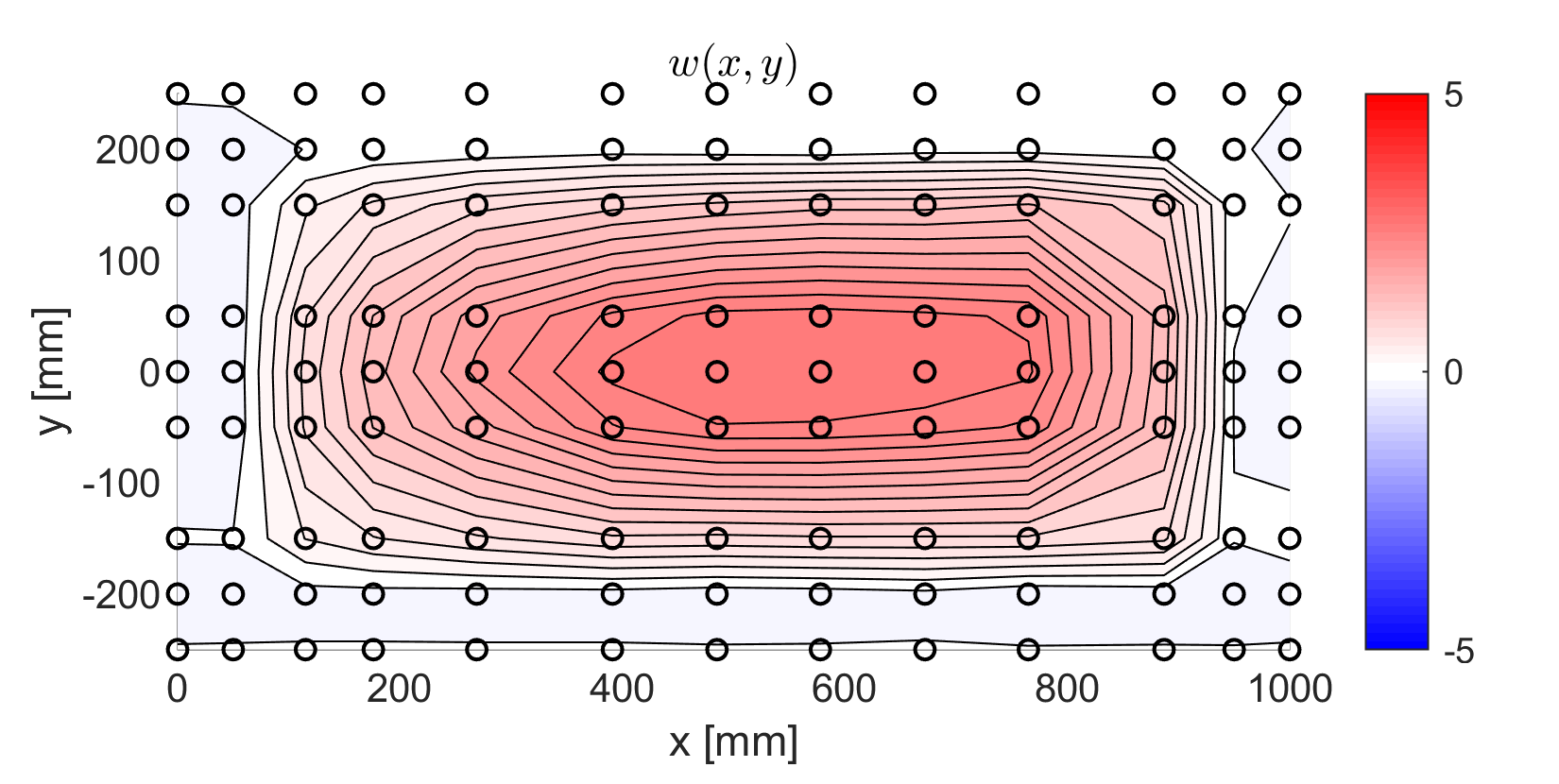}} \quad
\subfigure[08\_04\_45]{\includegraphics[width=0.48\textwidth]{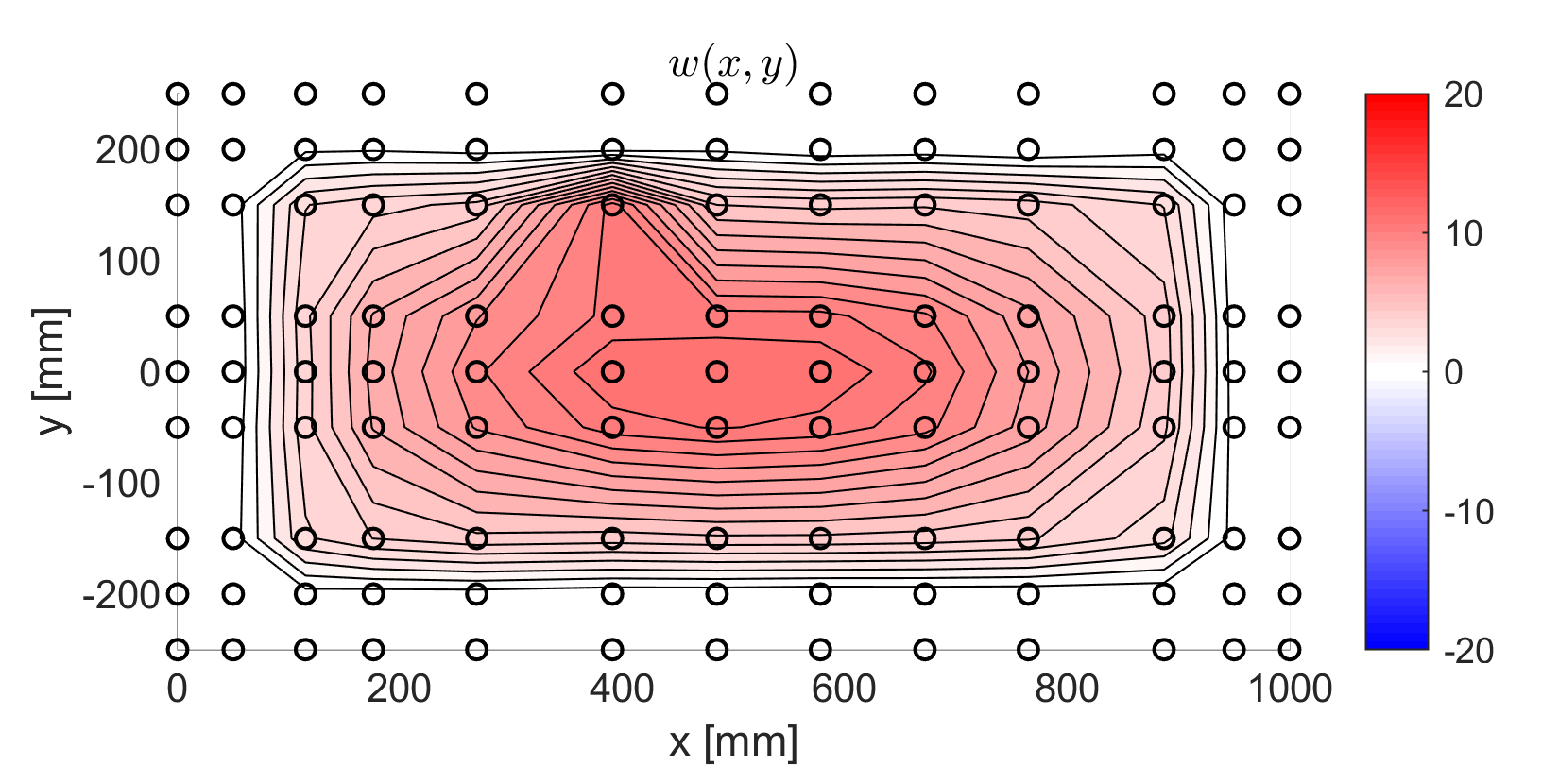}} \\
\raggedleft
\subfigure[08\_10\_45]{\includegraphics[width=0.48\textwidth]{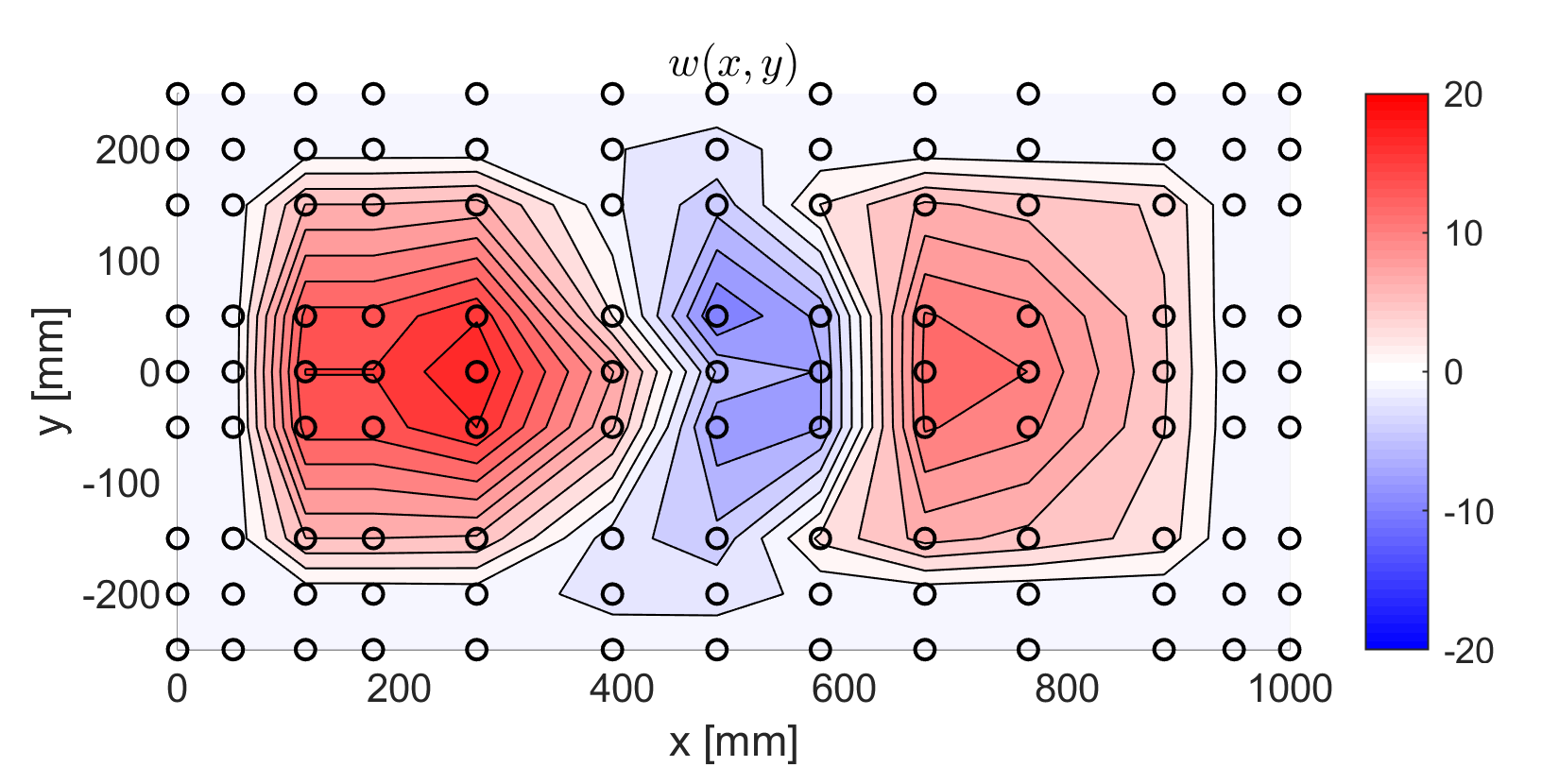}}
\caption{Residual out-of-plane displacements (mm) measured on the flat plates at the end of the tests. The trailing edge, corresponding to $x=0$, is on the left. The black round circles indicate the positions of the measurement points.}
\label{fig:PermDef}
\end{figure}

The colour contours of the out-of-plane displacement are shown in Figure
\ref{fig:PermDef}, where the position of the measurement points are indicated with black round circles. Positive values mean upward displacement. 
The data for the 3~mm and 0.8~mm plates are shown on the left and right
columns, respectively. Different scales are adopted for the two cases. 
In all cases, the displacement is zero in the bolted region, which is 50~mm wide all around.
The data show that for the same plate thickness the residual deformations increase if the 
plate loading increases, hence with the horizontal speed and the pitch angle, and if the plate thickness is lower, see also Section \ref{strains} and Section \ref{hyrodynamicloads}.

\begin{figure}[htbp]
 \centering
 \subfigure[08\_06\_40\_a]{\includegraphics[width=0.47\textwidth]{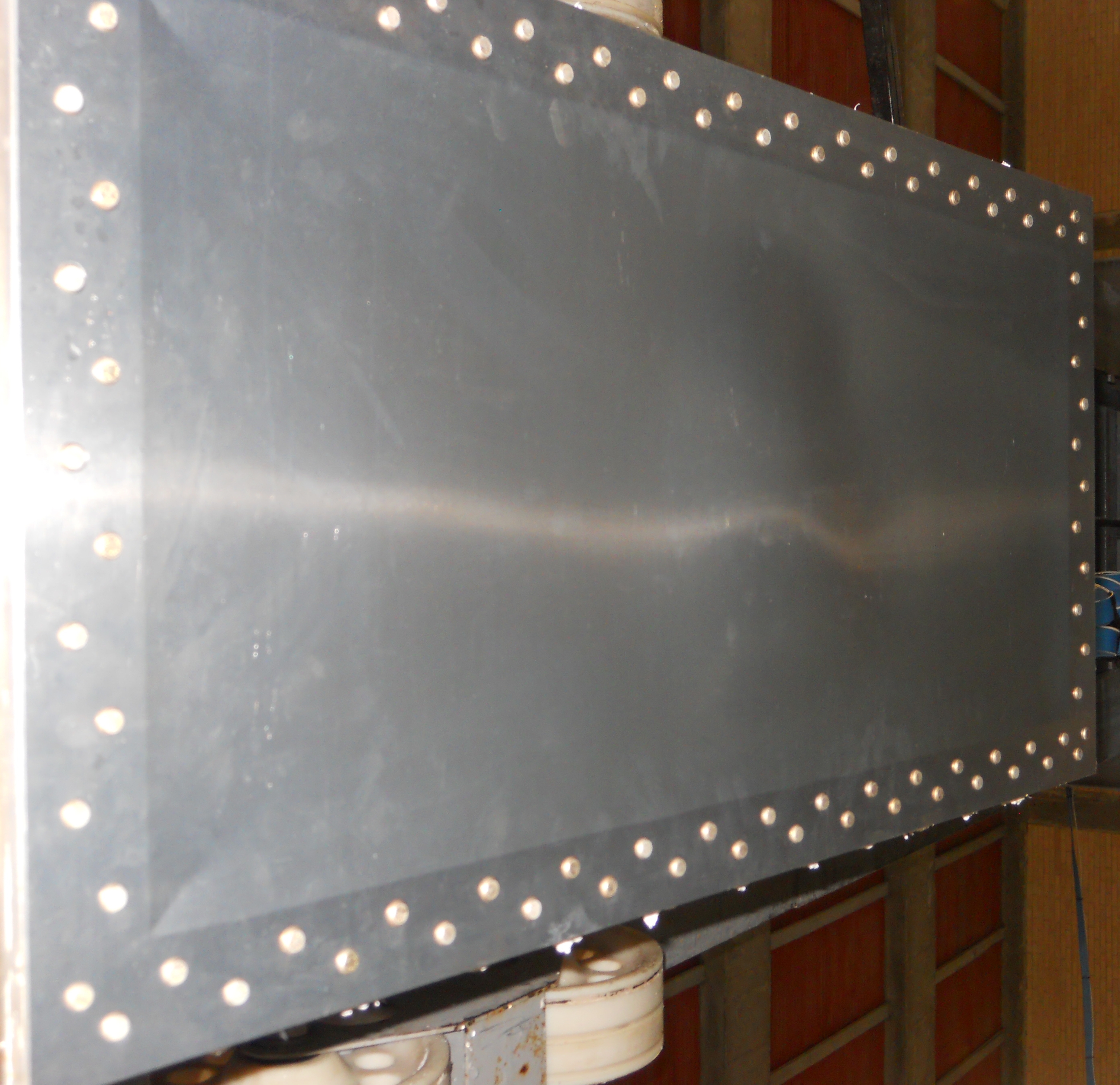}}
 \subfigure[08\_10\_45\_a]{\includegraphics[width=0.47\textwidth]{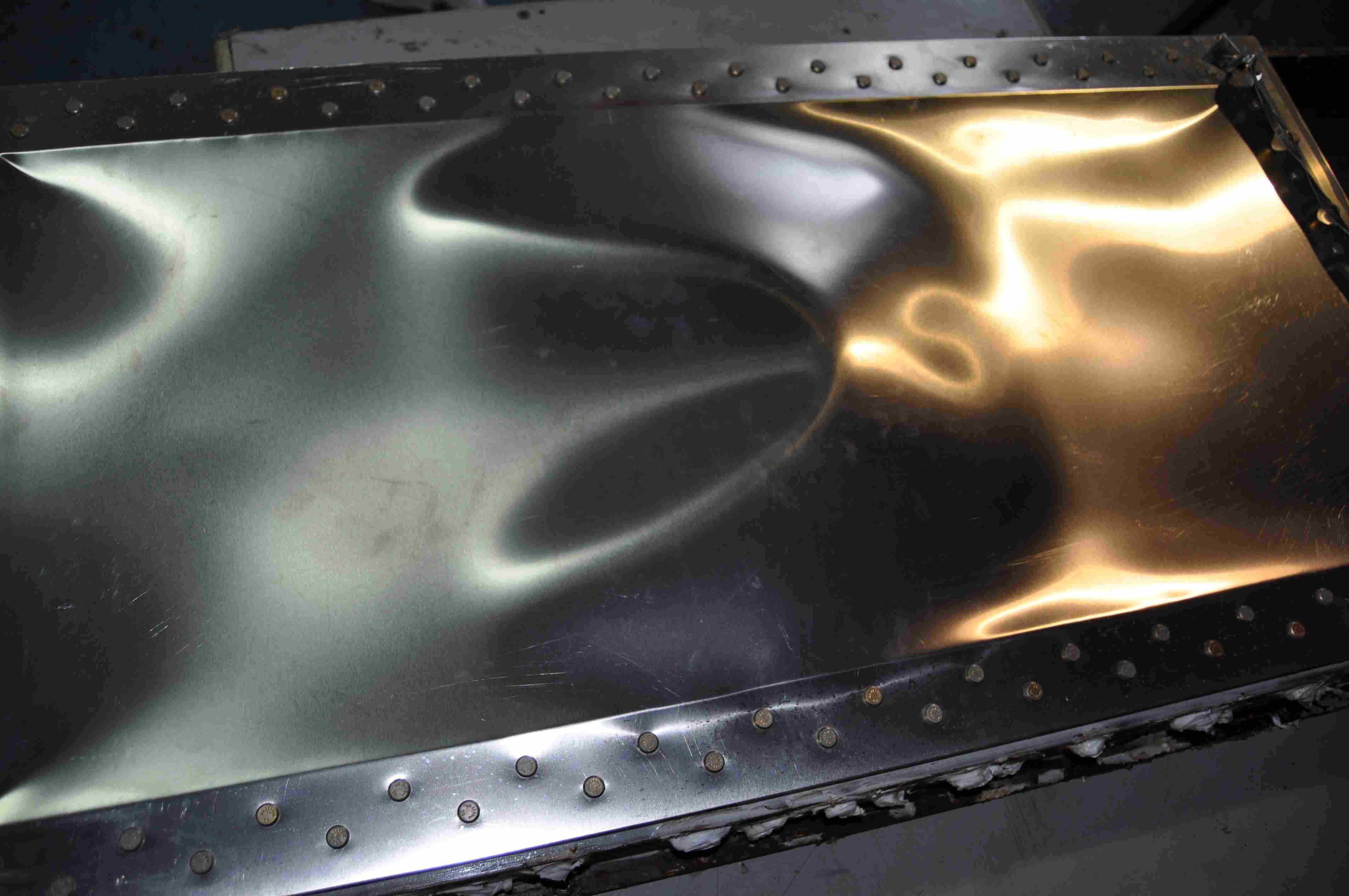}}
 \caption{Permanent bumps found on the 0.8~mm plates at the end of two tests. The trailing edge is on the left.}
 \label{fig:deformed}
\end{figure}

In some cases the thin plates exhibits a residual bump in the middle, and this is
particularly evident in the test 08\_10\_30\_a, i.e. 45~m/s and $10^\circ$ and to a smaller extent in the test 08\_06\_40\_a.
A picture of the plates after these tests is shown in Figure \ref{fig:deformed}.
The formation of the bump can be either the result of buckling
or a consequence of the suction forces induced by the vertical
oscillations of the last portion of the guide at the end of the test.
Partly, the bump is also a consequence of a small shift of the plate beneath 
the bolts, which is not restored at the end of the test. 

\begin{figure}[htbp]        
\centering
\subfigure[3\_10\_30 $\partial^2 w / \partial x^2$]{\includegraphics[width=0.48\textwidth]{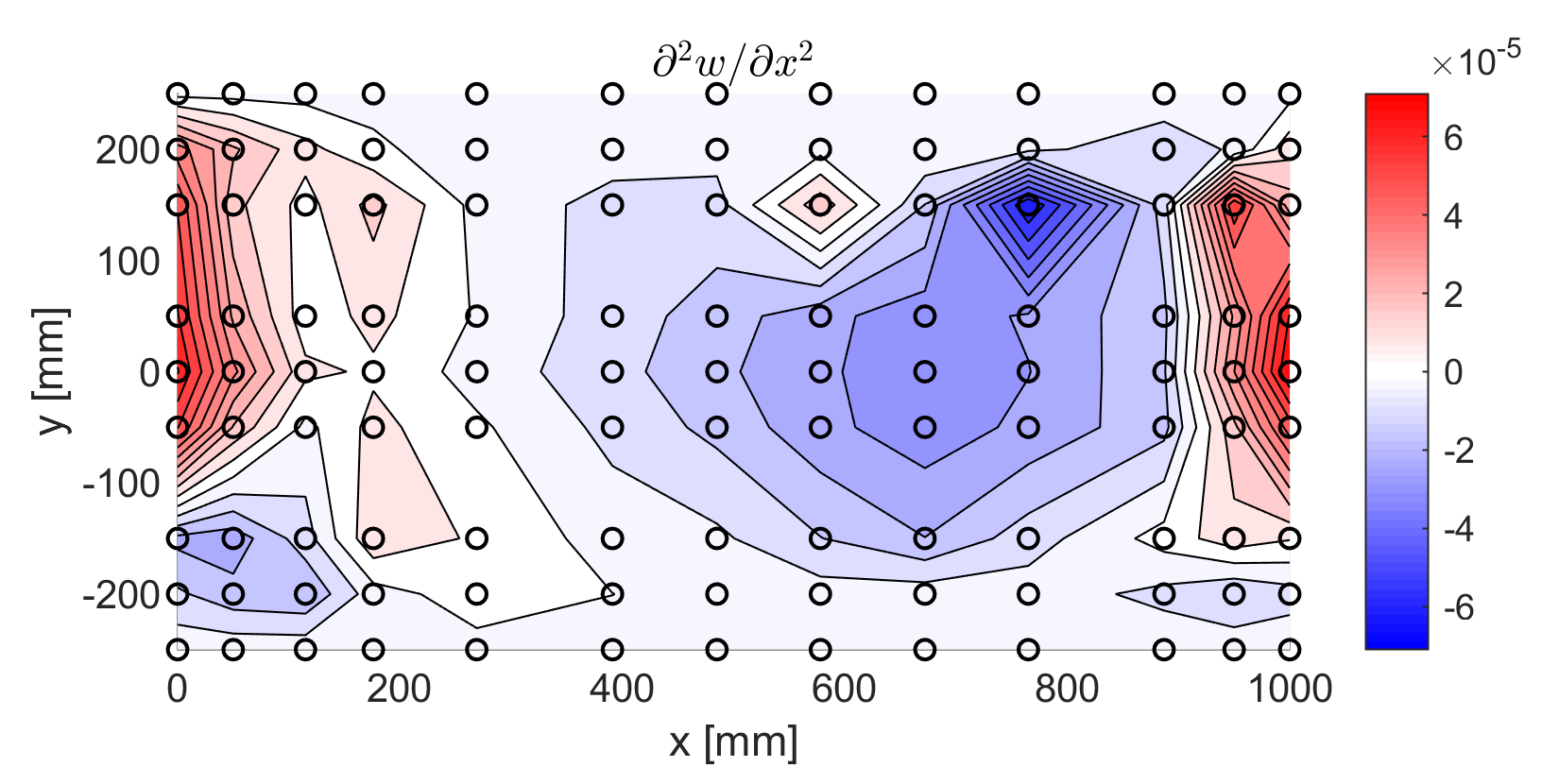}} \quad
\subfigure[3\_10\_30 $\partial^2 w / \partial y^2$]{\includegraphics[width=0.48\textwidth]{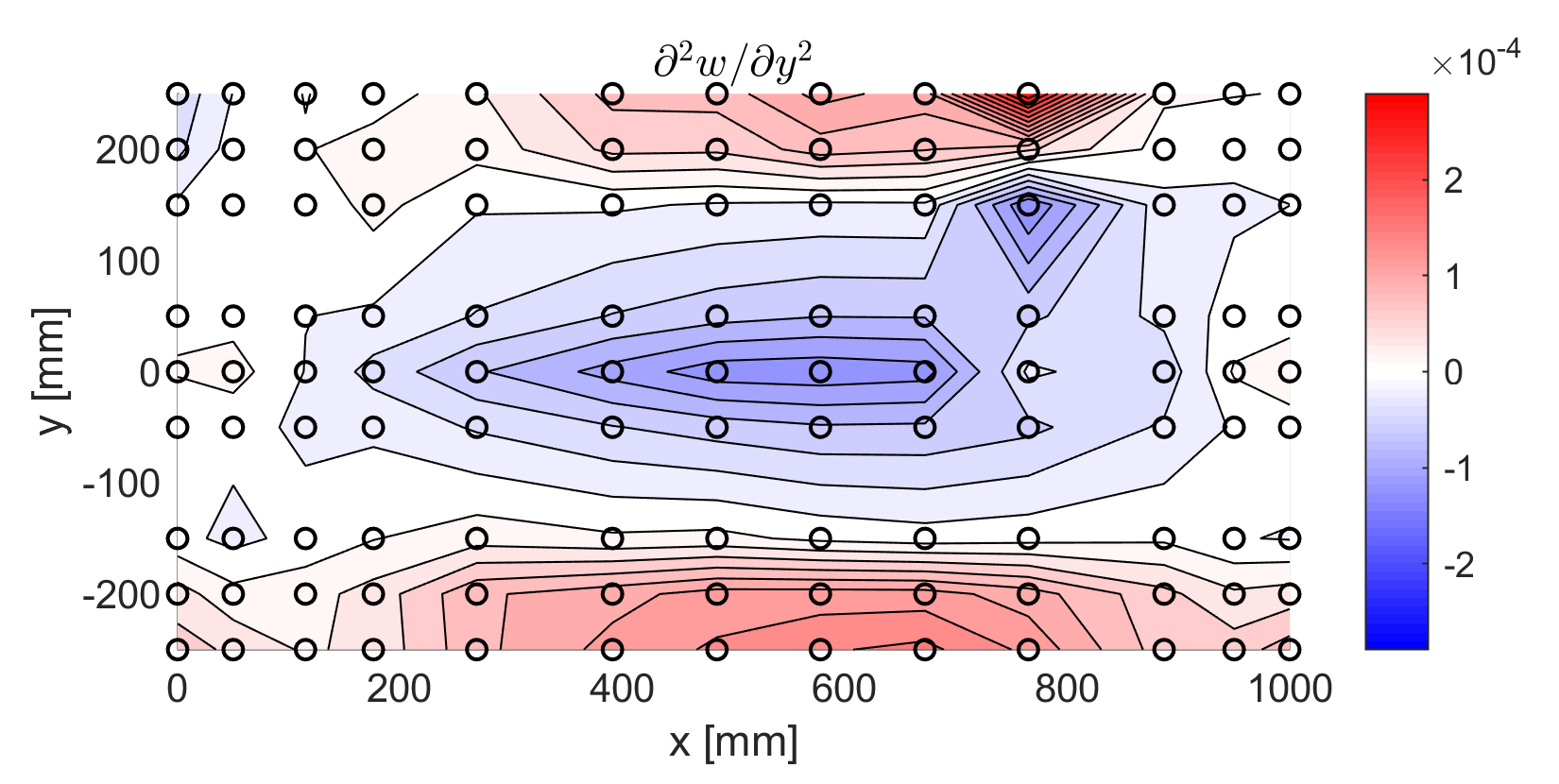}}  \\
\subfigure[08\_04\_45 $\partial^2 w / \partial x^2$]{\includegraphics[width=0.48\textwidth]{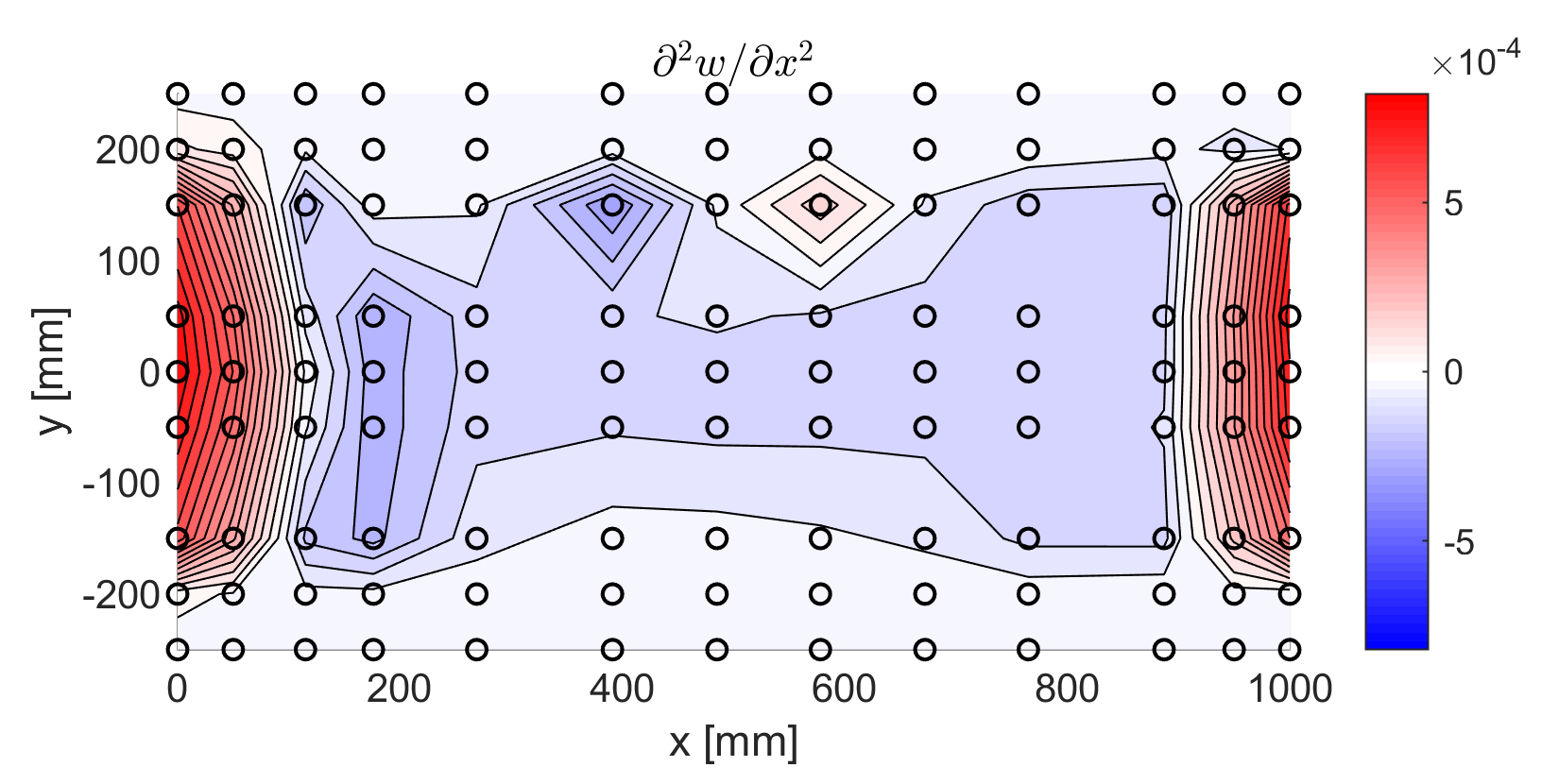}} \quad
\subfigure[08\_04\_45 $\partial^2 w / \partial y^2$]{\includegraphics[width=0.48\textwidth]{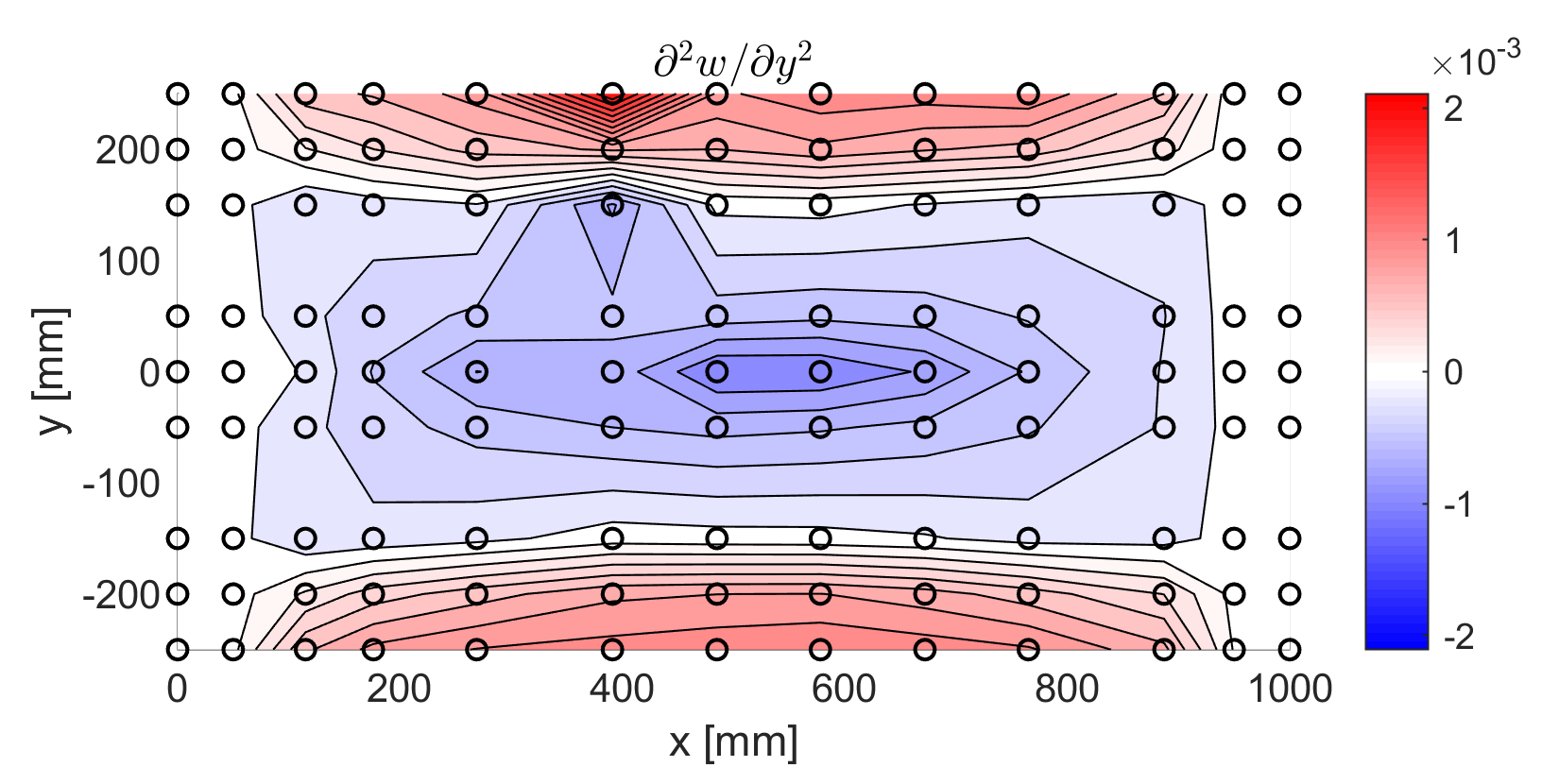}}  \\
\subfigure[08\_10\_45 $\partial^2 w / \partial x^2$]{\includegraphics[width=0.48\textwidth]{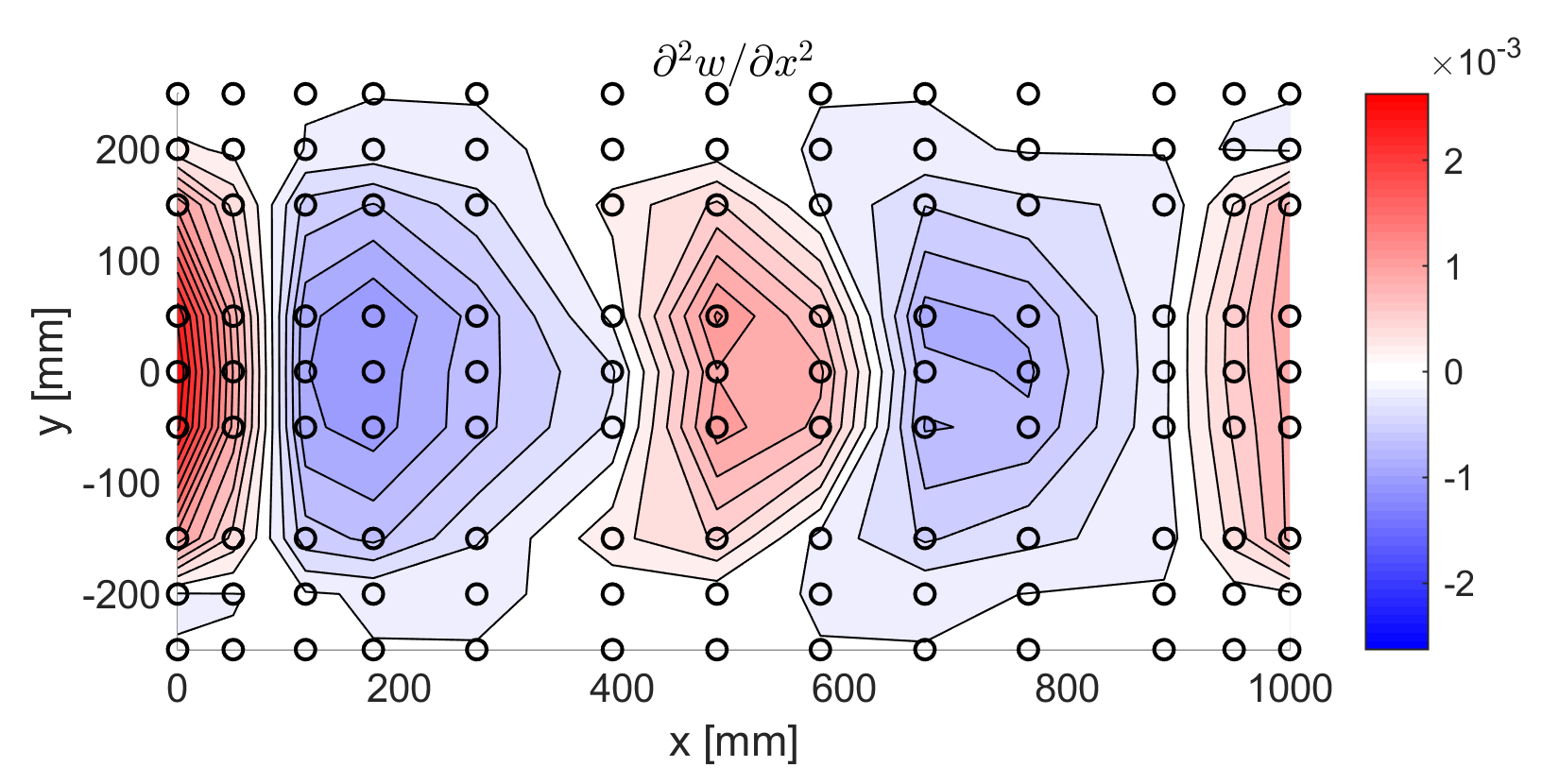}} \quad
\subfigure[08\_10\_45 $\partial^2 w / \partial y^2$]{\includegraphics[width=0.48\textwidth]{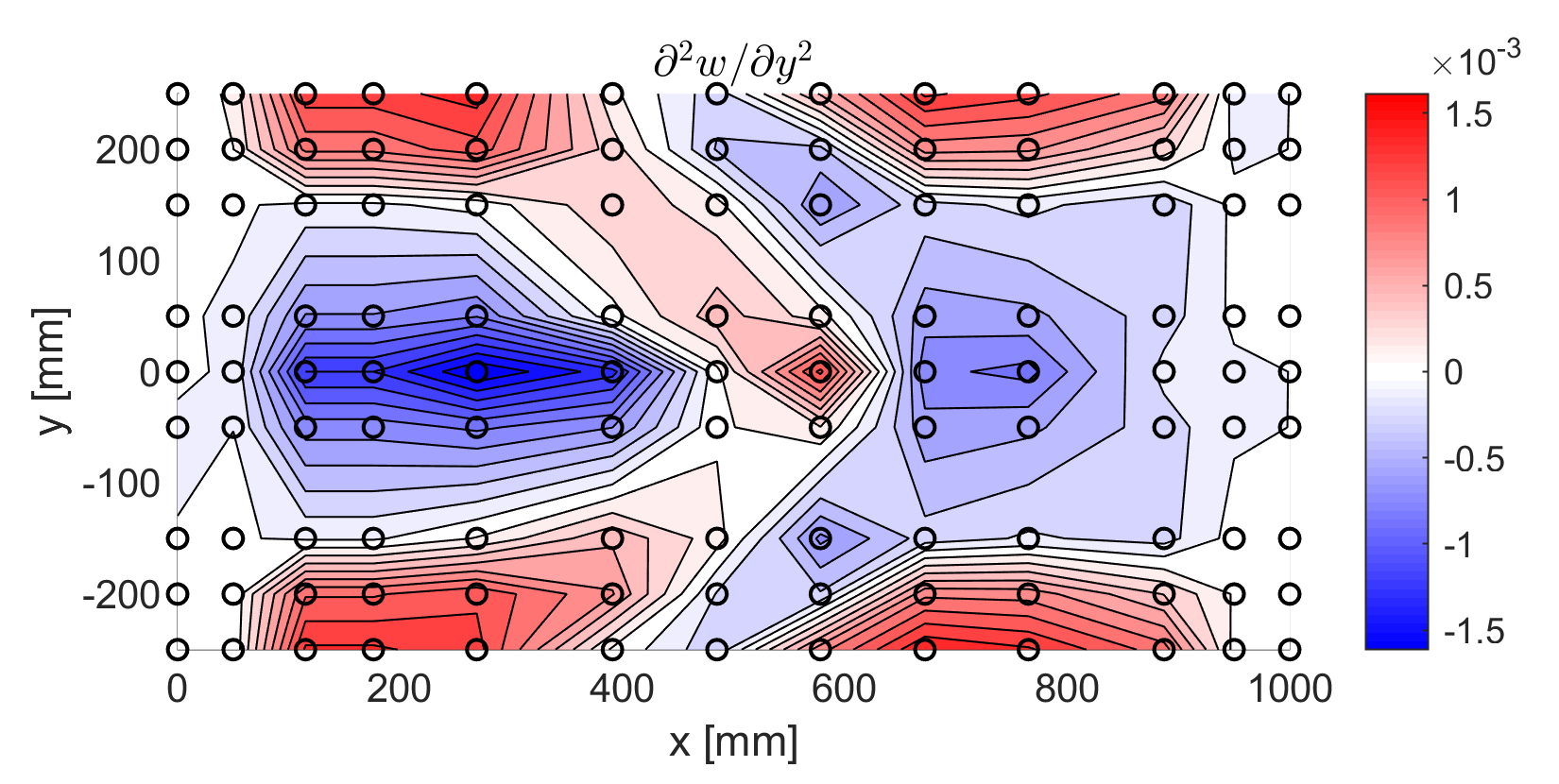}}  \\
\raggedleft
\caption{Second derivatives of the residual out-of-displacement fields in the $x$ direction $\frac{\partial^2 w}{\partial x^2}$ (1/mm) on the left column and in the $y$ direction $\frac{\partial^2 w}{\partial y^2}$ (1/mm) on the right column for the test conditions 3\_10\_30, 08\_04\_45 and 08\_10\_45. The trailing edge, corresponding to $x=0$, is on the left. The black round circles indicate the positions of the measurement points.}
\label{fig:SecondDer}
\end{figure}

The relationship between the strains and the residual deformations can be analysed in more detail using the
second derivatives of the out-of-plane displacement in $x$ and $y$ direction, which are directly related to S$_x$ and S$_y$
respectively. The signs of the curvature are based on the coordinate system fixed to the trolley shown in Figure \ref{fig:trolley_drawing}. Using this sign convention, a negative (upward) curvature is associated with a positive value of the measured strain, whereas a positive (downward) curvature is associated with a negative value of the measured strain. The colour contours of the second derivatives for the test conditions 3\_10\_30, 08\_04\_45 and 08\_10\_45 are shown in Figure \ref{fig:SecondDer}. The second derivatives are computed from the original measured displacement grid by a double application of a central finite difference scheme, apart from the point at the sides, where a single-sided scheme is used. Of course, due the poor resolution of the data, the second derivative cannot be very accurate.
Looking at the second derivatives in the $x$ direction for the test condition 08\_10\_30, there is a region with downward (positive) 
curvature at the trailing and leading edges. This is consistent with the negative value of the time history of $S_{1x}$, shown in Figure \ref{fig:x_strains_3Cases}(a). 
The central area exhibits an upward curvature (negative), which is consistent with the positive values recorded by other strain gauges. 
The transition from an area of positive to an area of negative second derivative indicates an inflection point in the permanent deformation. In the case 08\_04\_45 a similar behaviour of the second derivative is observed. 
As for the second derivative in $y$ direction, in both cases 3\_10\_30 and 08\_04\_45 the appearance of an area of downward (positive)
curvature is quite evident at the top and at the bottom sides. This is in agreement with the strain time histories shown in Figure \ref{fig:y_strains_3Cases_spanwise}(e) for the case 3\_10\_30, where S$_{3y}$ and S$_{4y}$ present significant negative values over the whole impact phase. This could lead to a permanent deformation in this sense, even though the recorded strains are not observed to overcome the yielding limit. For the case 08\_04\_45, shown in Figure \ref{fig:y_strains_3Cases_spanwise}(j), however, the time histories of S$_{3y}$ and S$_{4y}$ are always positive, which would imply an upward curvature during the impact phase. The discrepancy could be due to the stress and the loads experienced by the plate after the impact phase, as mentioned above. Finally, the case 08\_04\_45 shows that the formation of the bump creates significant changes in the plate curvature, as explained above.

\subsection{Pressure Measurements}
\label{pressurecameras}

As discussed in \cite{iafrati2016experimental} as soon as the plate
touches the water, a thin spray forms and propagates along the plate.
A sharp pressure peak is formed just behind the spray root, with the
pressure peak being proportional to the square of the spray root
propagation velocity \cite{iafrati2016experimental,cointe1987hydrodynamic,
howison1991incompressible}.
The aim of this section is to show how and to which extent the structural
deformation changes the pressure distribution and the loading. It is worth
noticing that pressure probes were used in few selected cases only (see
Table \ref{tab:AlTestConditions}). This is because before performing the tests, 
the possibility that the plate could not resist the impact could not be excluded,
and there was also the risk that the holes needed for the installation 
of the pressure probes could have triggered the rupture.
As anticipated, the results presented in this work indicate that the presence of the
sockets and of the probes do not introduce any significant change in the 
structural response. 

\subsubsection{Pressure changes in the longitudinal direction}
\label{pressurechangeslongitudinal}
\begin{figure}[htbp]
\centering
\subfigure[Pressure probe positions (top view)]{\includegraphics[width=0.58\textwidth]{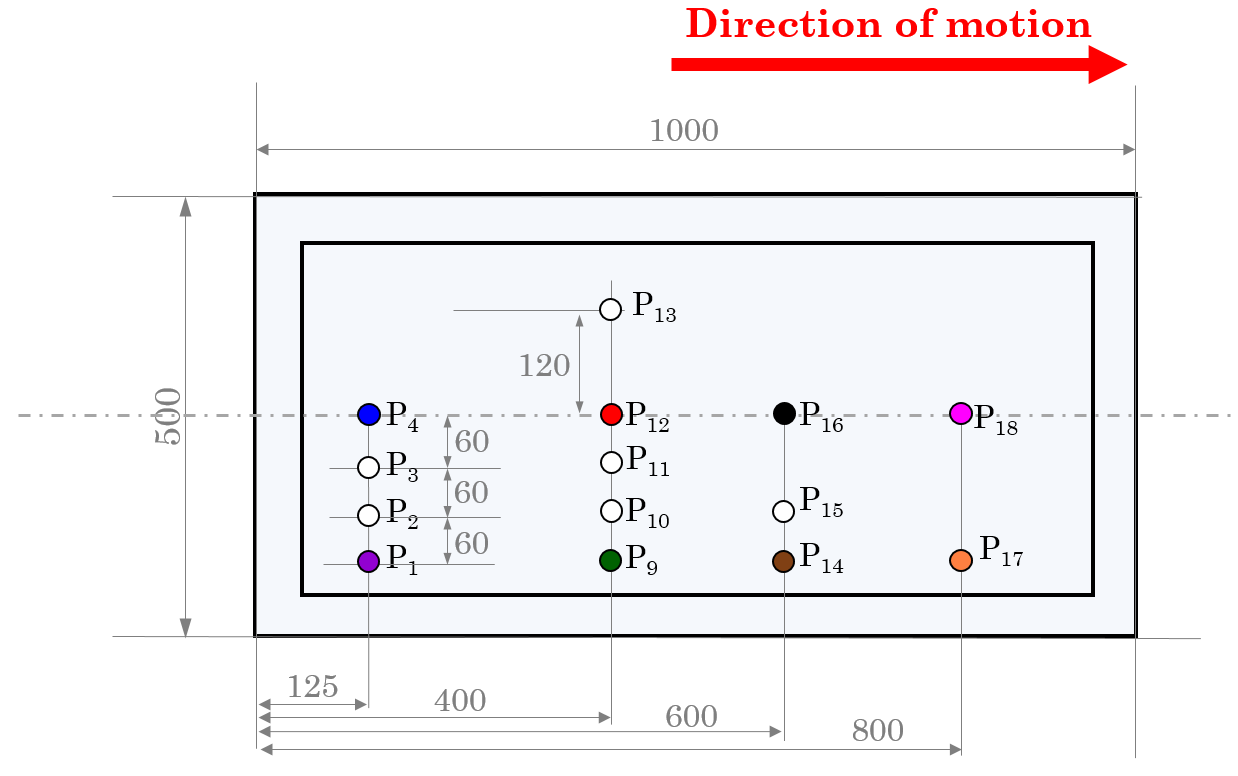}} \quad
\subfigure[Pressure probe positions (side view)]{\includegraphics[width=0.38\textwidth]{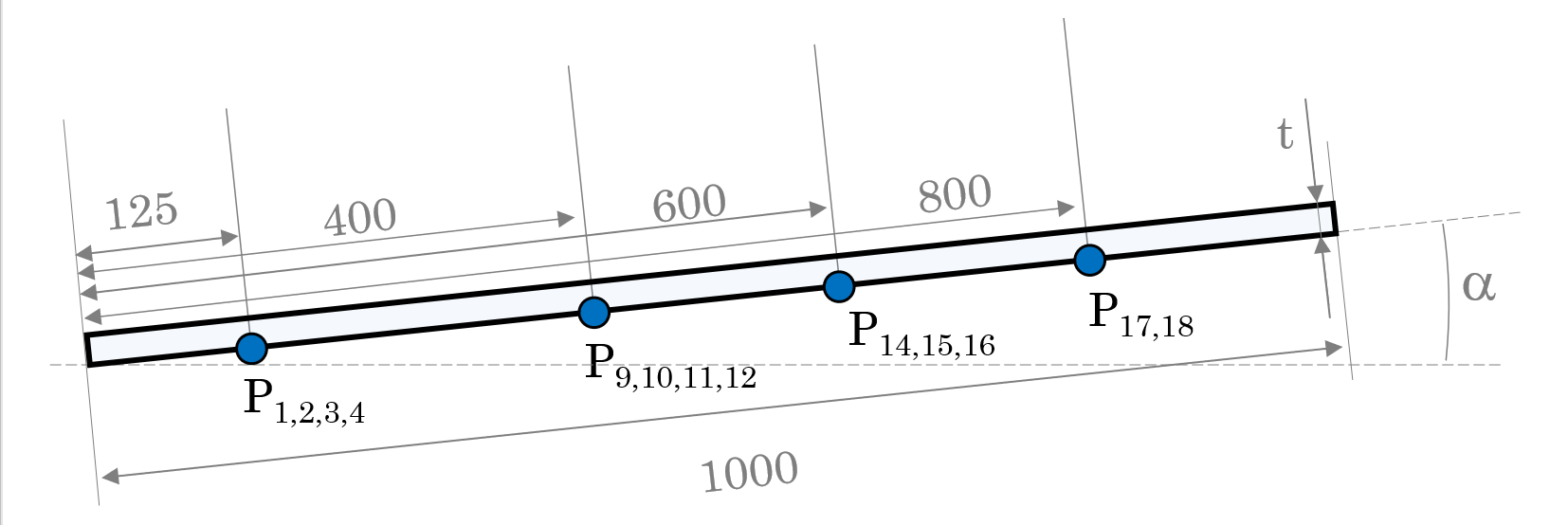}} \\
\subfigure[15\_10\_30 midline]{\includegraphics[width=0.45\textwidth]{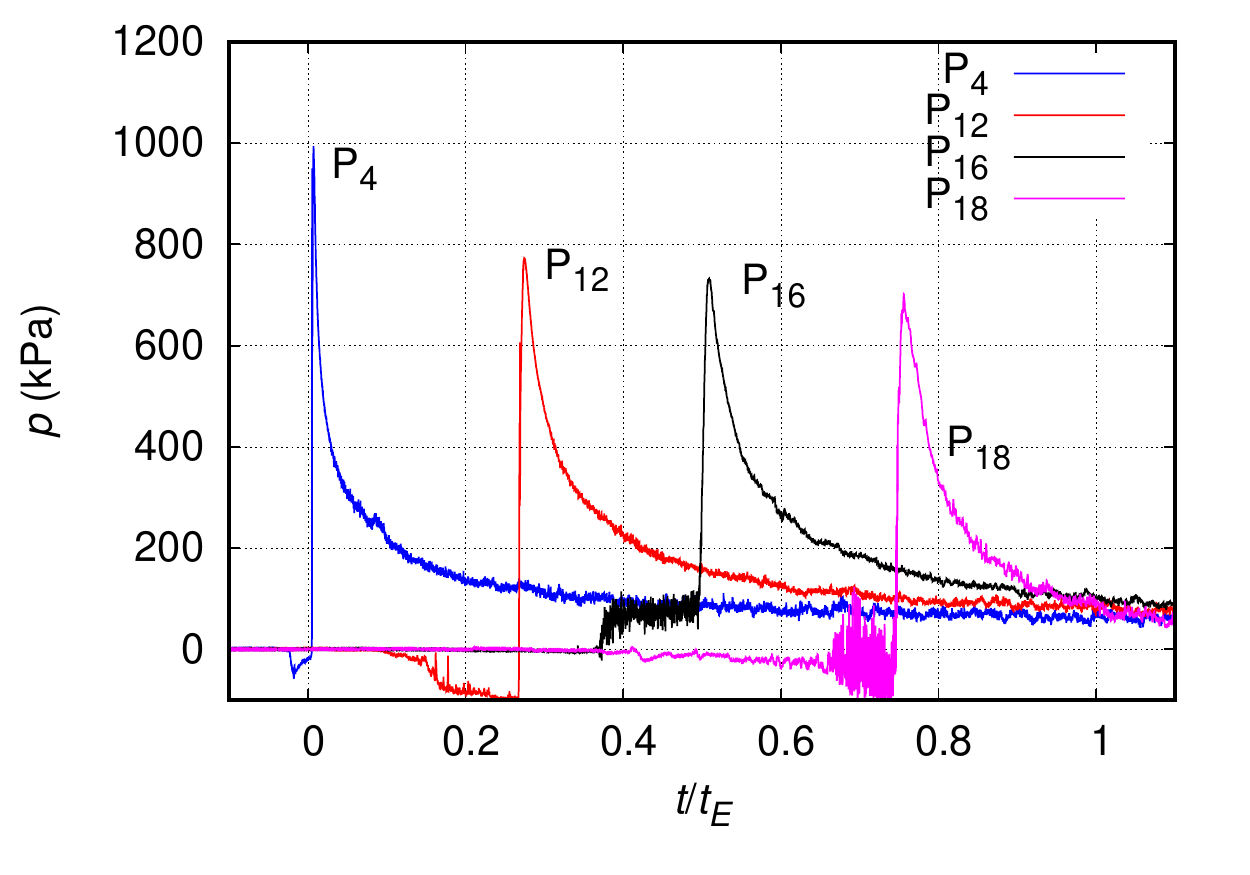}} \quad
\subfigure[15\_10\_30 side]{\includegraphics[width=0.45\textwidth]{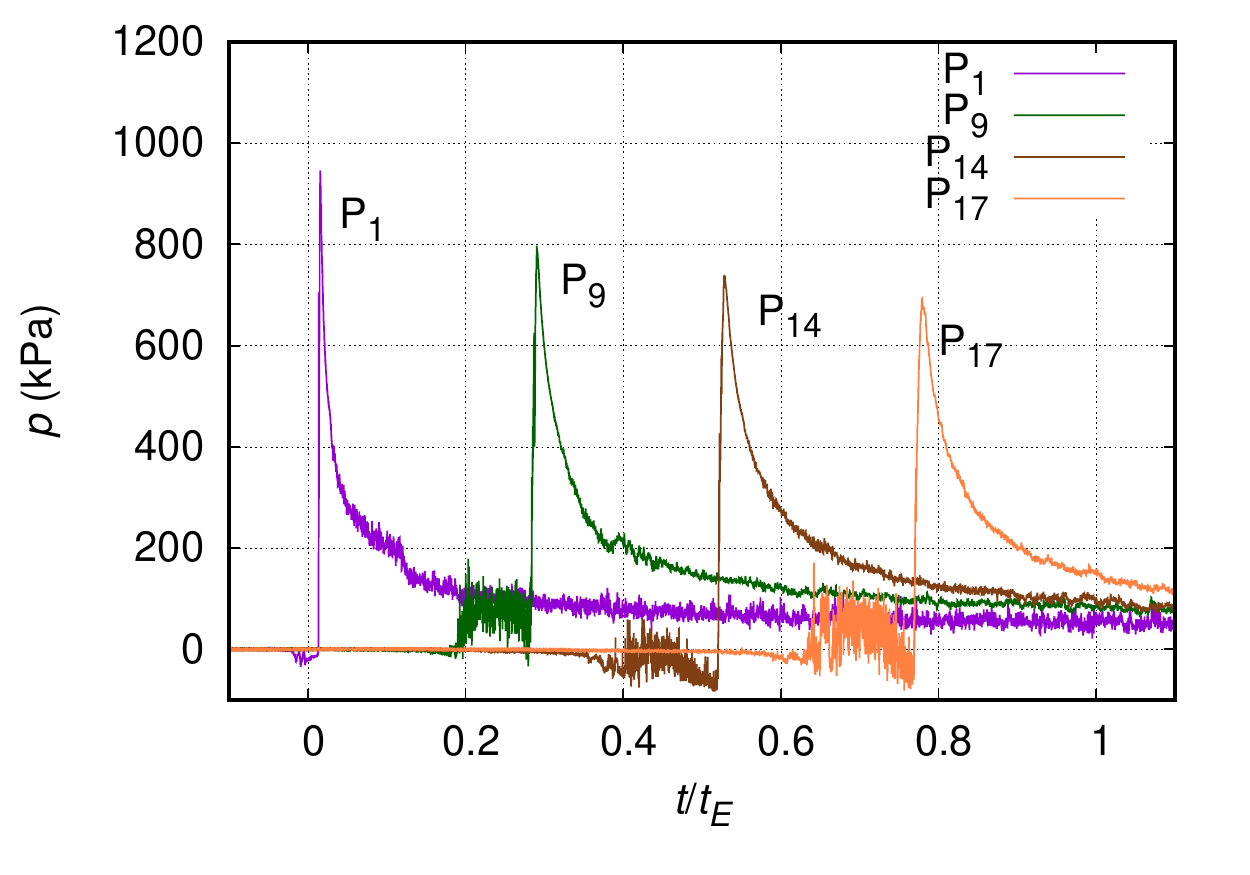}} \\
\subfigure[3\_10\_30 midline]{\includegraphics[width=0.45\textwidth]{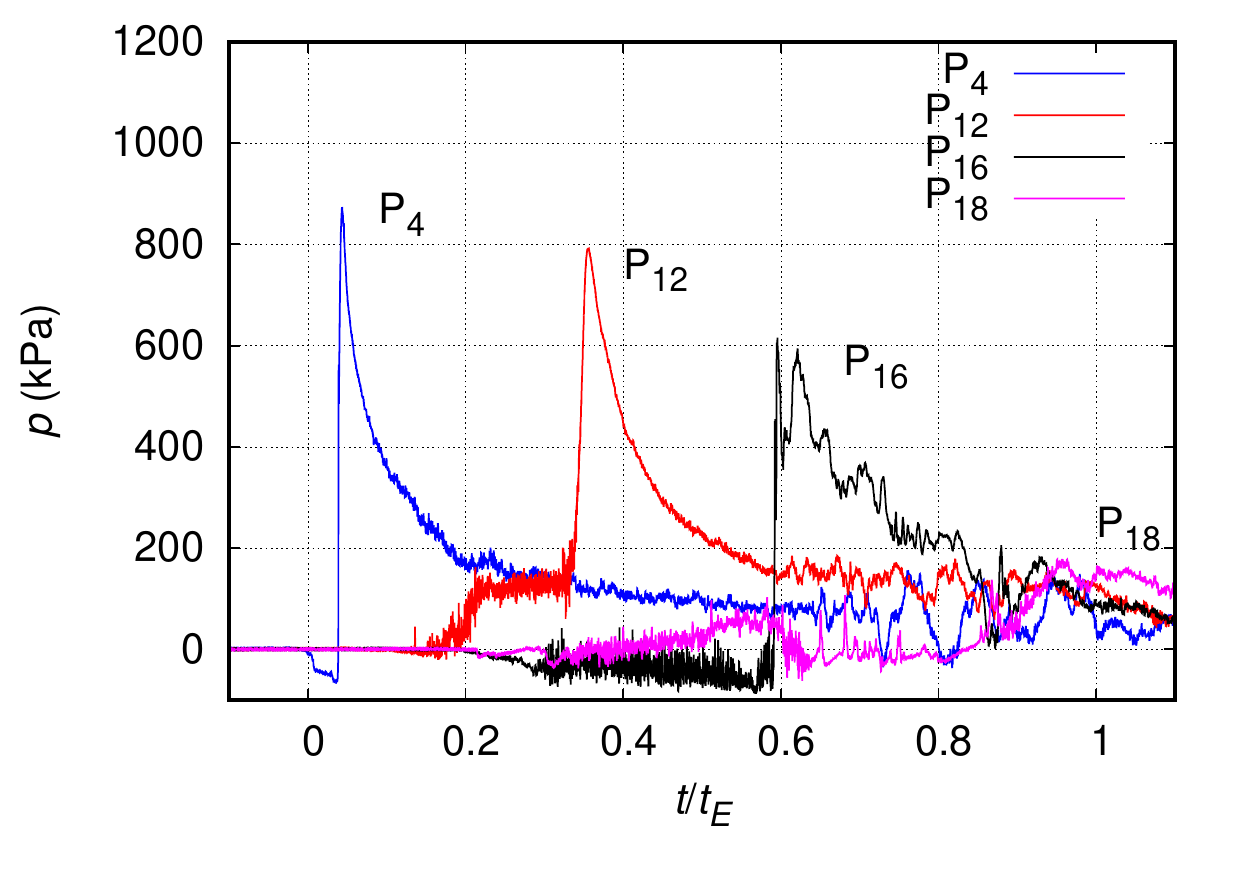}} \quad
\subfigure[3\_10\_30 side]{\includegraphics[width=0.45\textwidth]{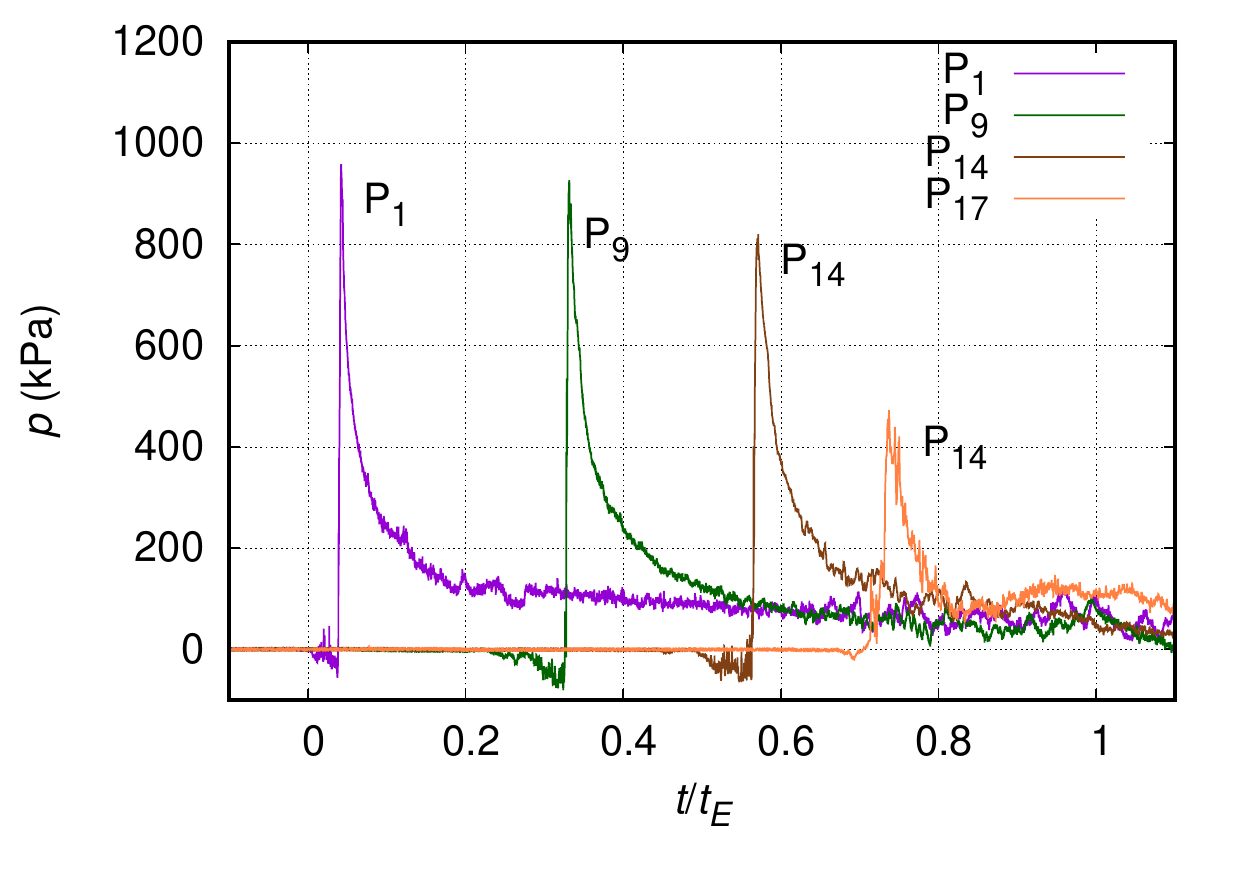}} \\
\subfigure[08\_10\_30 midline]{\includegraphics[width=0.45\textwidth]{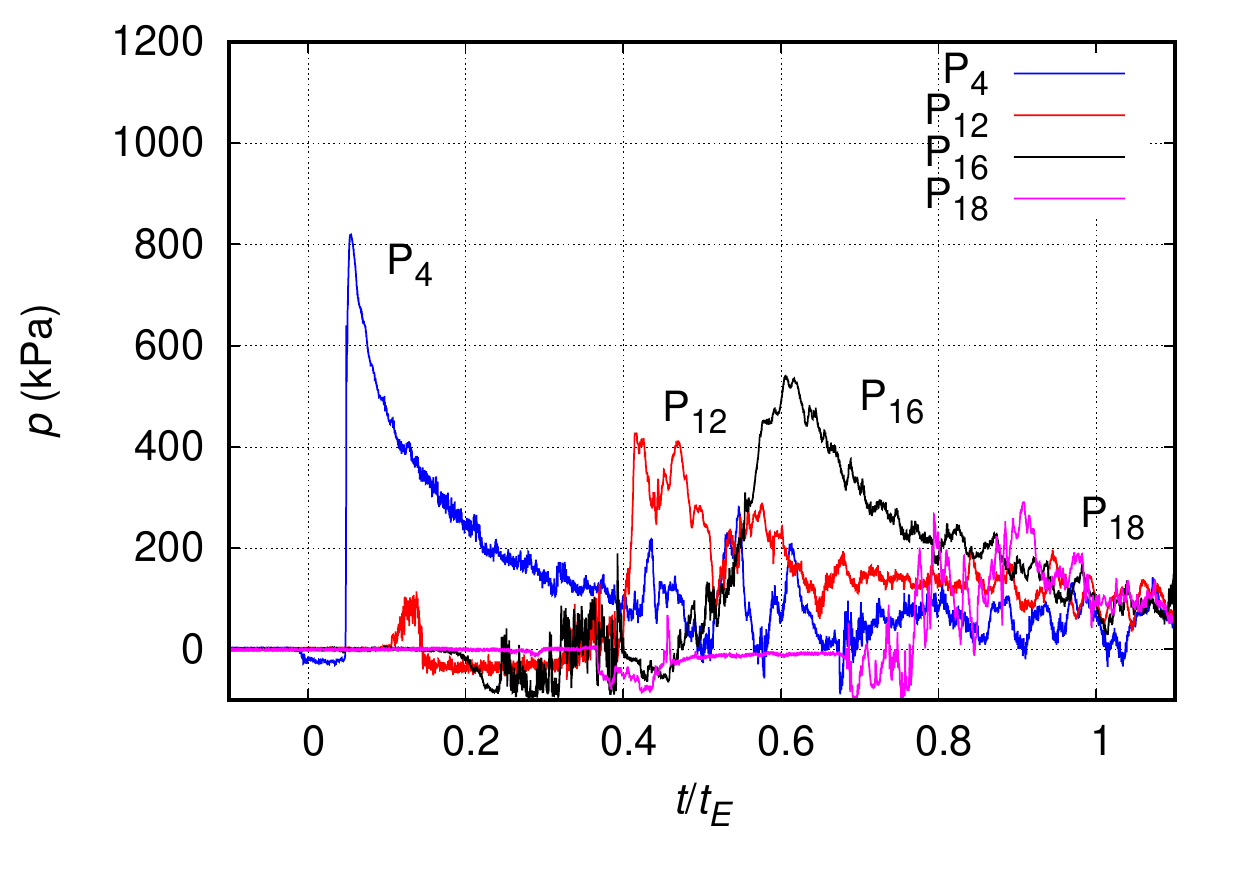}} \quad
\subfigure[08\_10\_30 side]{\includegraphics[width=0.45\textwidth]{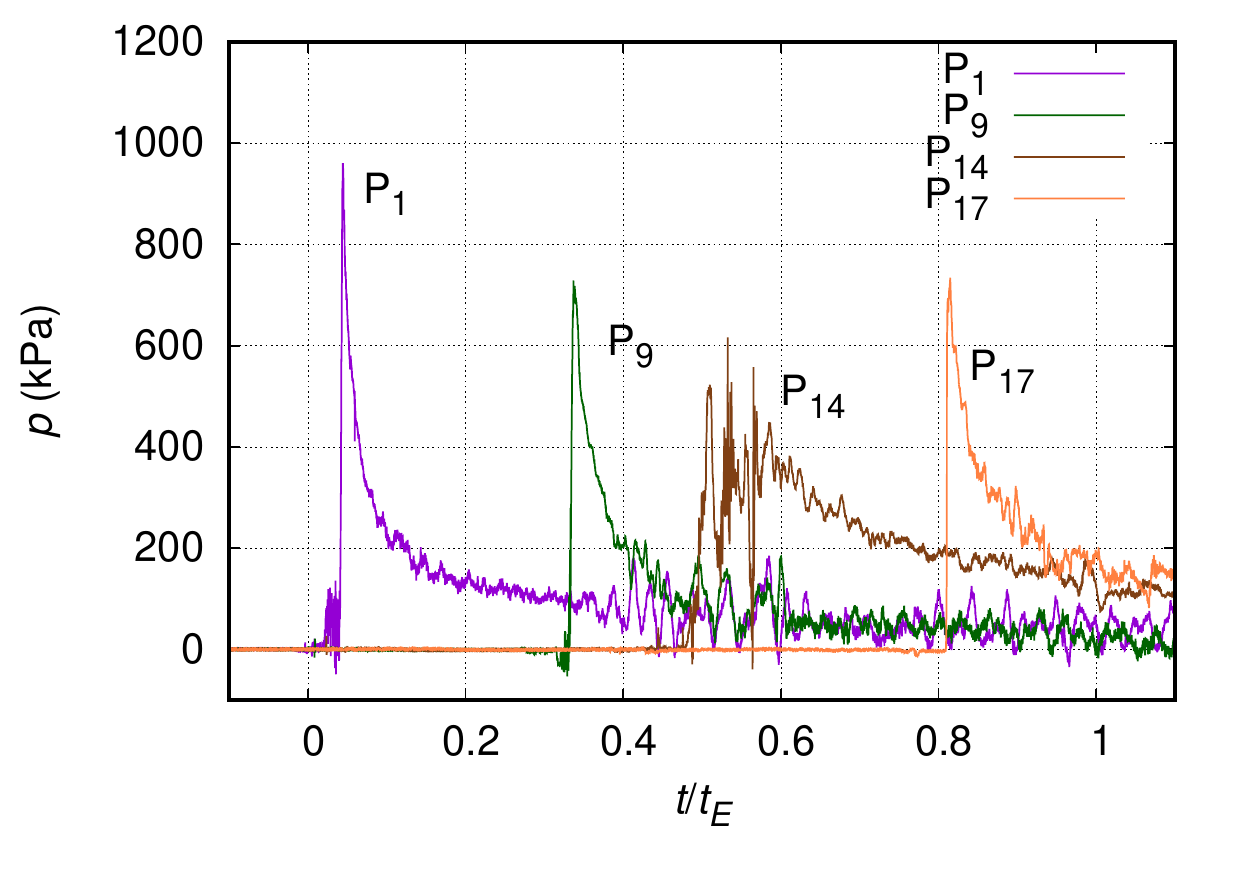}} \\
\caption{Time histories recorded by the pressure probes located along the midline and at the side for the tests XX\_10\_30, i.e at U=30~m/s, 
10$^{\circ}$~pitch angle for the plate 15~mm thick (top) 3~mm thick (centre) and 0.8~mm thick (bottom).}
\label{fig:X131_Pressure}
\end{figure}

In order to show the changes in the pressure distribution caused by the
structural deformations, in Figure \ref{fig:X131_Pressure} the time 
histories of pressure measured by the probes located along the midline
(left column) and at the side (right column) of the plate are provided for
the three different plate thickness. The comparison is established for the
test condition XX\_10\_30, i.e. U=30~m/s, $\alpha=10^{\circ}$ which is the one with the
lowest loading. For the reader convenience, the position of the pressure
probes on the plate is given in the figures on top. 
Some of the pressure time histories shown in Figure \ref{fig:X131_Pressure} display a short negative transient just before the sharp peak. This phenomenon was discussed in detail in \cite{iafrati2015high} with the support of the underwater videos, and it was found that such negative values are associated with the formation of vortices from the probe heads, which introduce an irregularity in the plate surface.
At such a high speed, the pressure in the core of the vortices drops down to the vapour pressure. As a consequence, cavitation above the pressure probe occurs concurrently with the passage of the thin spray. 
The irregularity in the plate surface is also responsible for the positive pressures observed just before the sharp rise at some probes, which is presumably due to the formation of a stagnation point, still associated with the same vortices \cite{iafrati2015high}.
The pressure time histories plotted in Figures \ref{fig:X131_Pressure}(d) and \ref{fig:X131_Pressure}(f) 
show that in the 3~mm plates the peaks at the side, i.e.
 P$_1$, P$_9$ and P$_{14}$, are slightly delayed compared to those observed 
for the thick plate, whereas no substantial changes are found in terms of 
the peak amplitude. A more significant reduction of the pressure peak is
observed instead for P$_{17}$.
The situation is different for the probes along the midline. In this case
the delay is more evident, particularly for P$_{12}$ and P$_{16}$ and,
more importantly, the peak is much lower for P$_{16}$ and completely
disappears for probe P$_{18}$, see Figures \ref{fig:X131_Pressure}(c) and (e).
This is of course a consequence of the structural deformation, which is more
pronounced in the middle of the plate than at the sides, where the thick
Aluminium frame and the bolt connection prevent the plate deformation.
The deformation of the plate grows as the plate penetrates into the water, 
and the total load on the plate increases. This is the reason why 
the changes are more pronounced for P$_{16}$ and P$_{18}$ rather than for P$_{12}$.
The effect of the plate deformation on the pressure measurements is of course more 
evident in the 0.8~mm plates, as shown in Figures
\ref{fig:X131_Pressure}(g) and (h). Even if a pressure peak, albeit of smaller amplitude, 
still occurs at the side, e.g. at P$_{9}$ for instance, it completely
disappears in the middle except for P$_{4}$, which is very close to the
boundary. This indicates that the classical spray observed for the thick
plates is disrupted by the structural deformation, and the resulting loading
becomes more irregular.

\begin{figure}[htbp]
\centering
\subfigure[Pressure probe positions (top view)]{\includegraphics[width=0.58\textwidth]{Figures/Pressure_schematic_Midline.png}} \quad
\subfigure[Pressure probe positions (side view)]{\includegraphics[width=0.38\textwidth]{Figures/Pressure_schematic_sideview.png}} \\
\subfigure[15\_04\_45 midline]{\includegraphics[width=0.45\textwidth]
{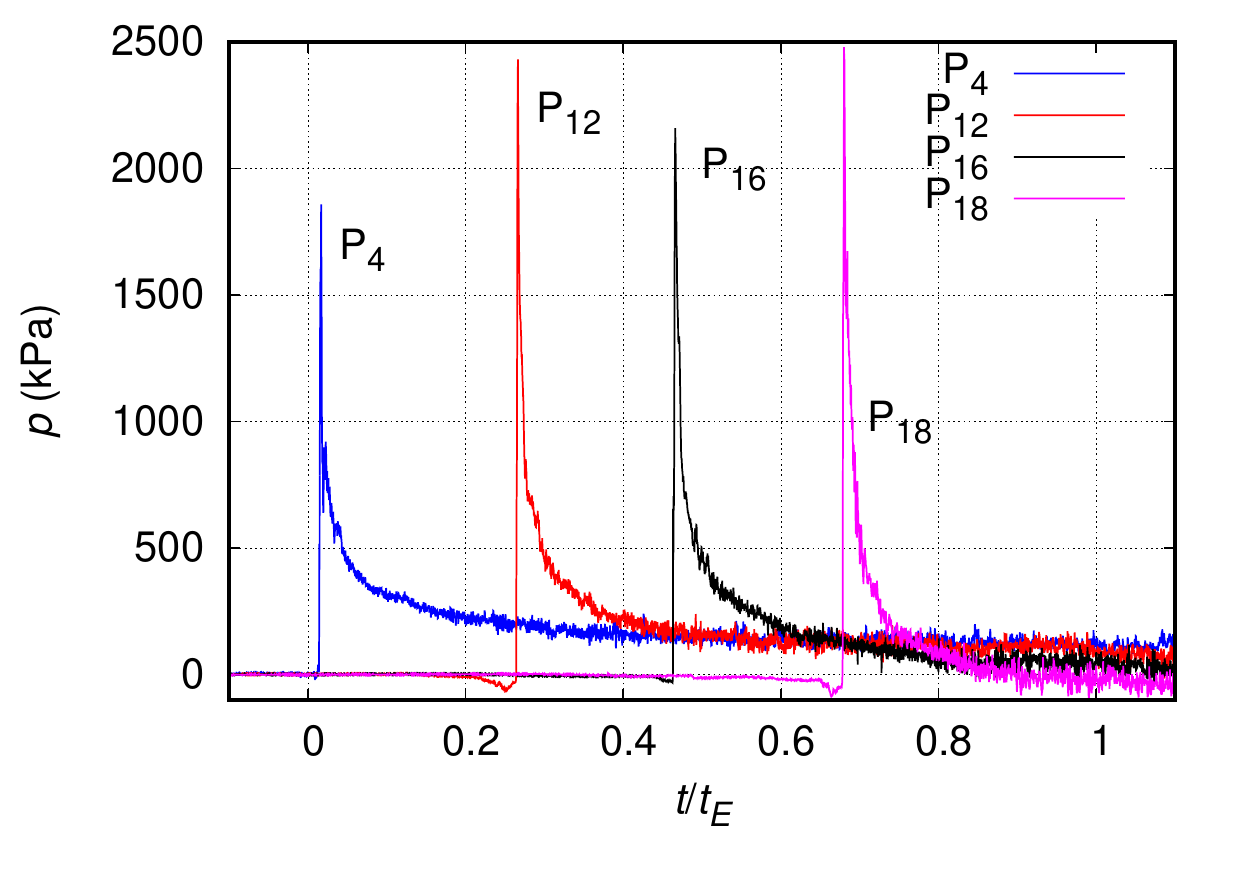}} \quad
\subfigure[15\_04\_45 side]{\includegraphics[width=0.45\textwidth]
{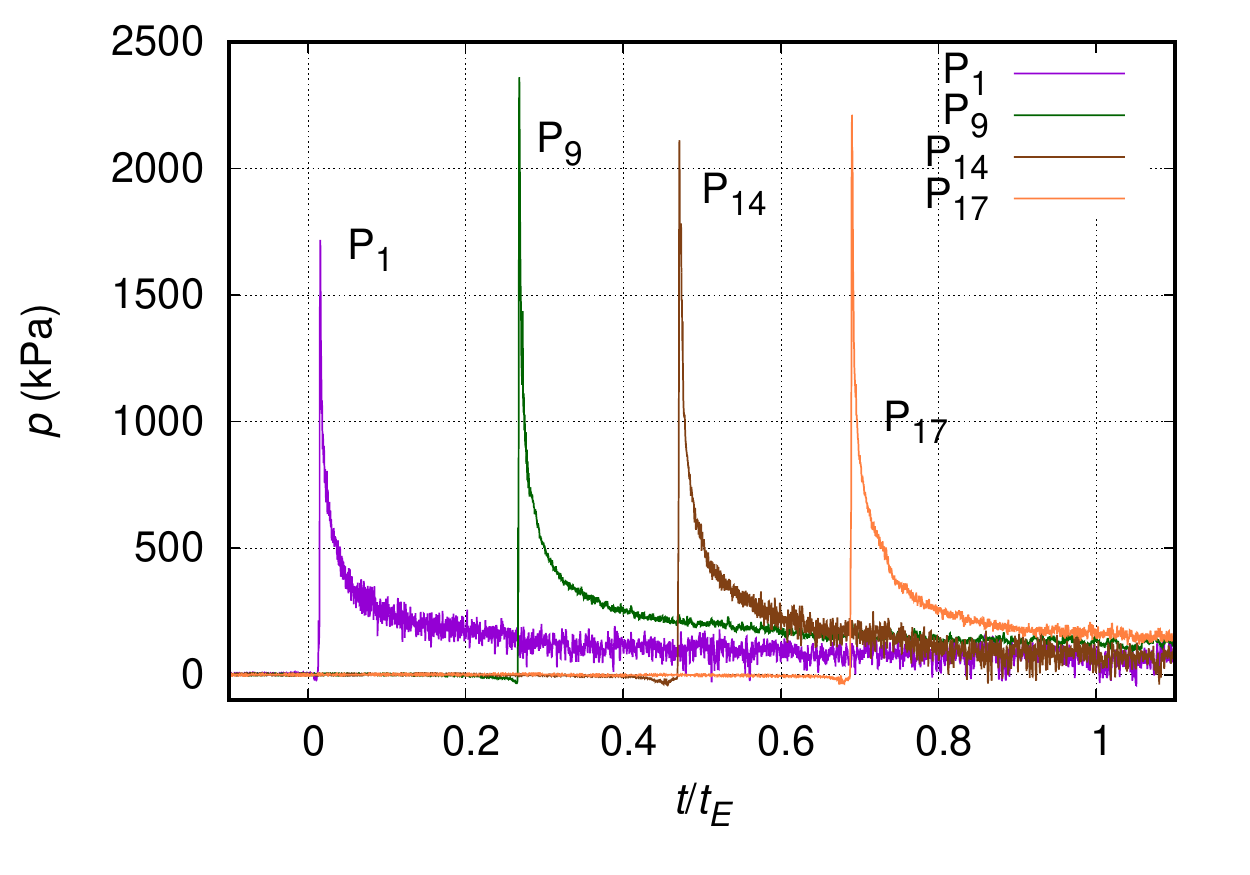}} \\
\subfigure[3\_04\_45 midline]{\includegraphics[width=0.45\textwidth]
{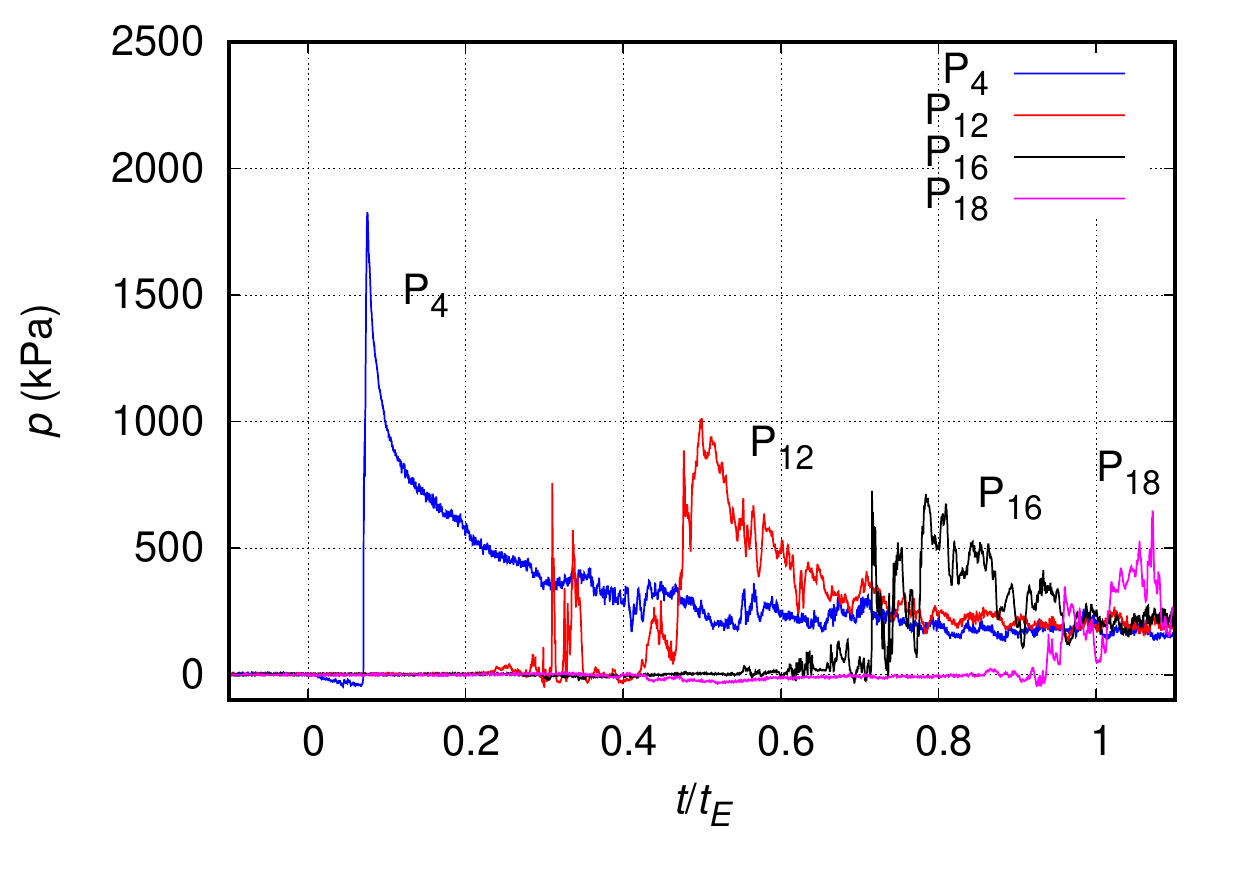}} \quad
\subfigure[3\_04\_45 side]{\includegraphics[width=0.45\textwidth]
{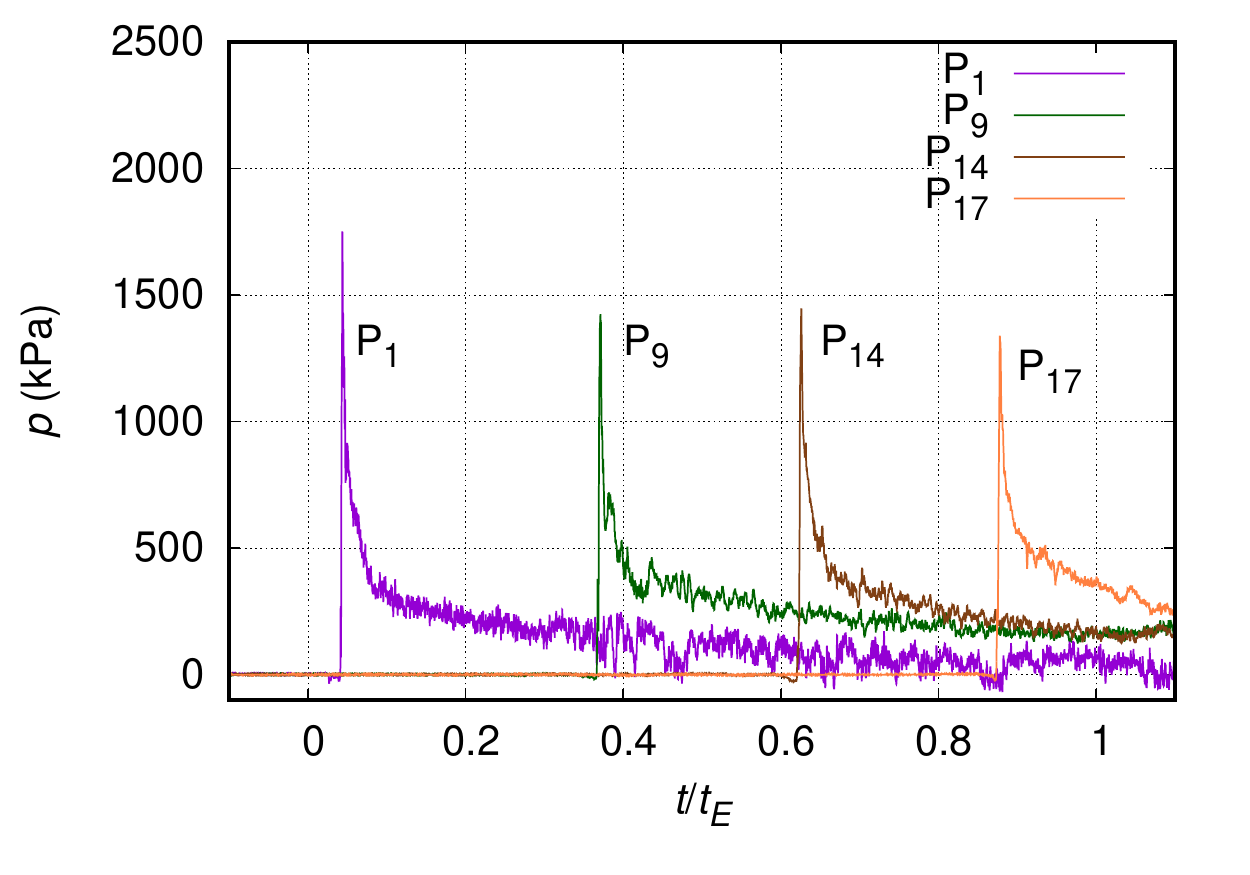}} \\
\caption{Time histories recorded by the pressure probes located along the midline and at the side for the tests XX\_04\_45, i.e at U=45~m/s,
4$^{\circ}$~pitch angle for the plate 15~mm thick (top) and 3~mm thick (bottom).}
\label{fig:X113_Pressure}
\end{figure}

Similar comments can be made for the cases XX\_04\_45, i.e. U=45~m/s, 
$\alpha=4^{\circ}$, see Figure \ref{fig:X113_Pressure}.
For such test condition, pressures have been measured for the 15~mm and 3~mm plates
only (see Table \ref{tab:AlTestConditions}). Owing to the higher velocity of such test condition, 
the pressure peaks measured for the thick plate are higher than those 
recorded in the XX\_10\_30  condition, even though the general behaviour of the 
signals is about the same. On the contrary, for the 3~mm-plate, the loss of 
correlation of the pressure signals is already evident at $P_{12}$, which 
means that the larger deformations caused by the
increased loading have a strong effect on the pressure distribution.
In the right column of Figure \ref{fig:X113_Pressure} the time histories of 
pressure measured at the side are shown. 
No significant differences are observed for the 15~mm plate between 
the pressures measured in the middle and at the side. For the 3~mm plate
the pressure at the side behaves similarly to that measured in the 15~mm-plate,
although for the 3~mm-plate the pressure peaks are much lower.
However, the pressure profiles in Figure \ref{fig:X113_Pressure}(f) 
are much more correlated than those in \ref{fig:X113_Pressure}(e), 
thus indicating that the strongest fluid-structure interaction occurs 
in the middle of the plate.
\subsubsection{Pressure changes in the transverse direction}
\label{pressuretransverse}
\begin{figure}[htbp]
\centering
\subfigure[Pressure probe positions]{\includegraphics[width=0.75\textwidth]
{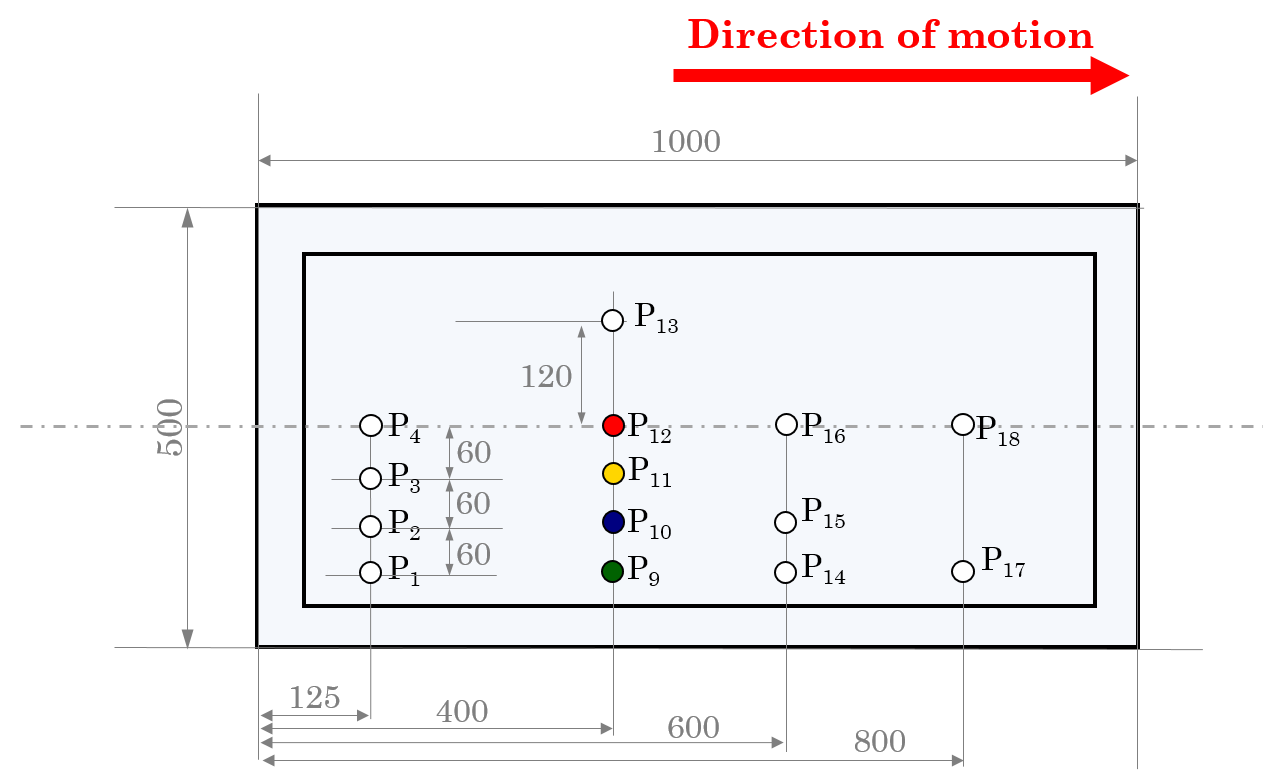}} \quad
\subfigure[15\_10\_30  (thickness 15~mm) ]{\includegraphics[width=0.48\textwidth]
{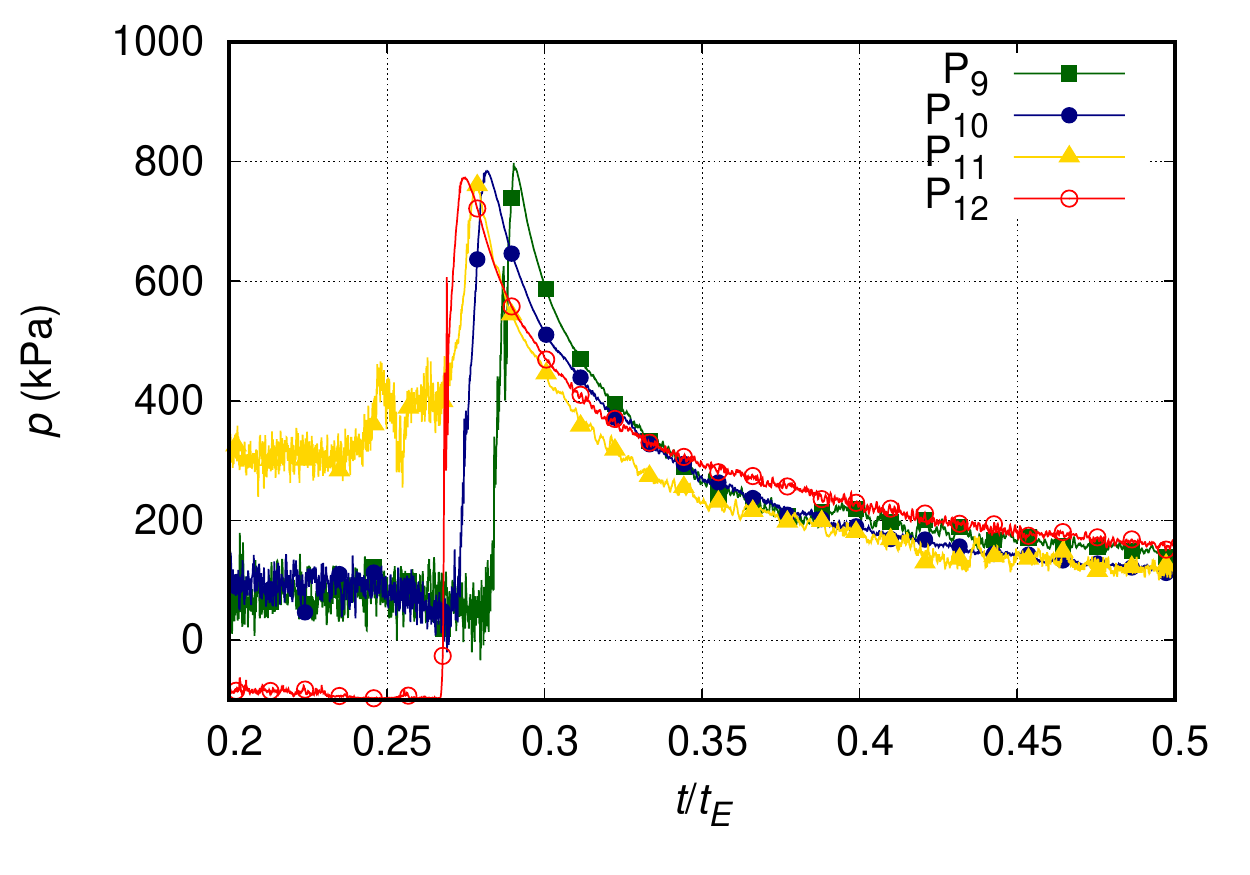}}  \\
\subfigure[3\_10\_30  (thickness 3~mm)]{\includegraphics[width=0.48\textwidth]{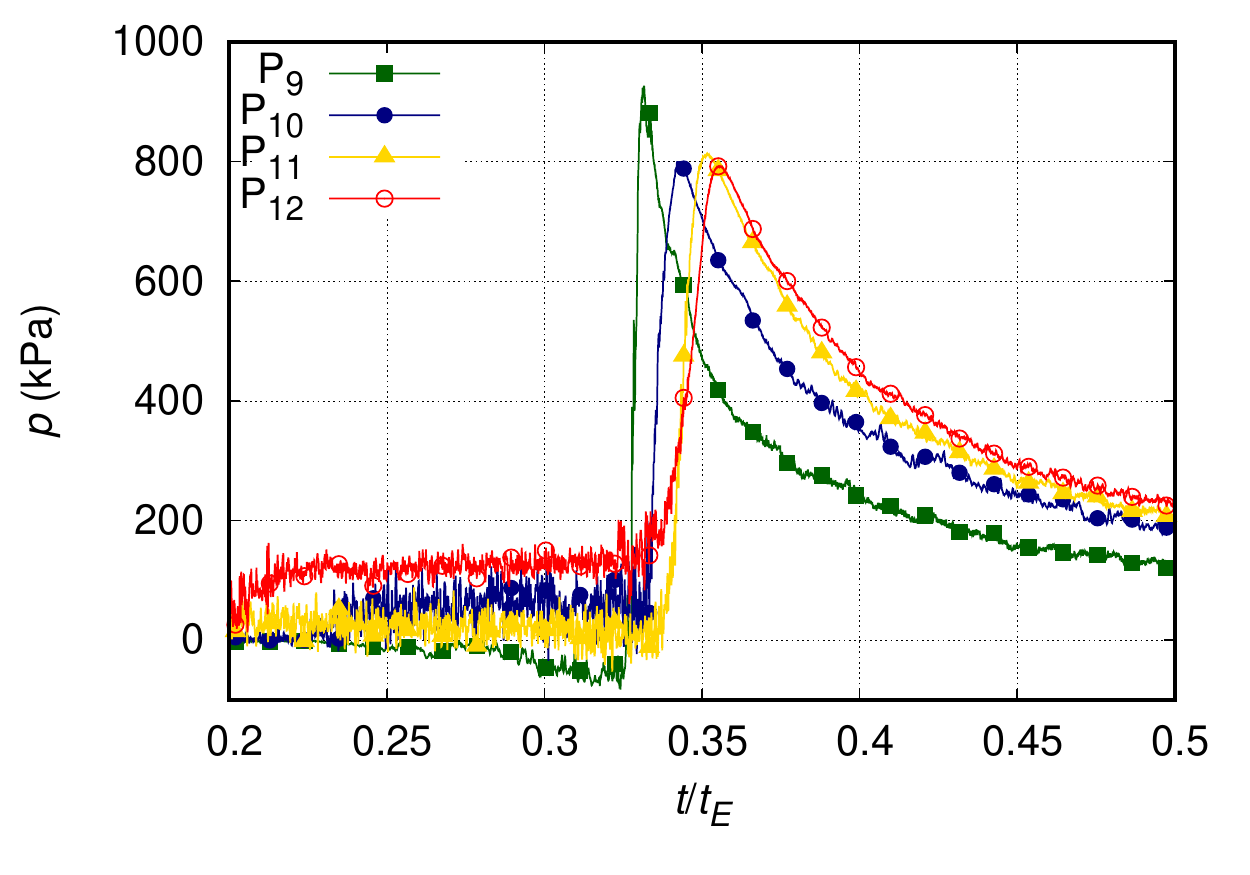}} \quad
\subfigure[08\_10\_30  (thickness 0.8~mm)]{\includegraphics[width=0.48\textwidth]{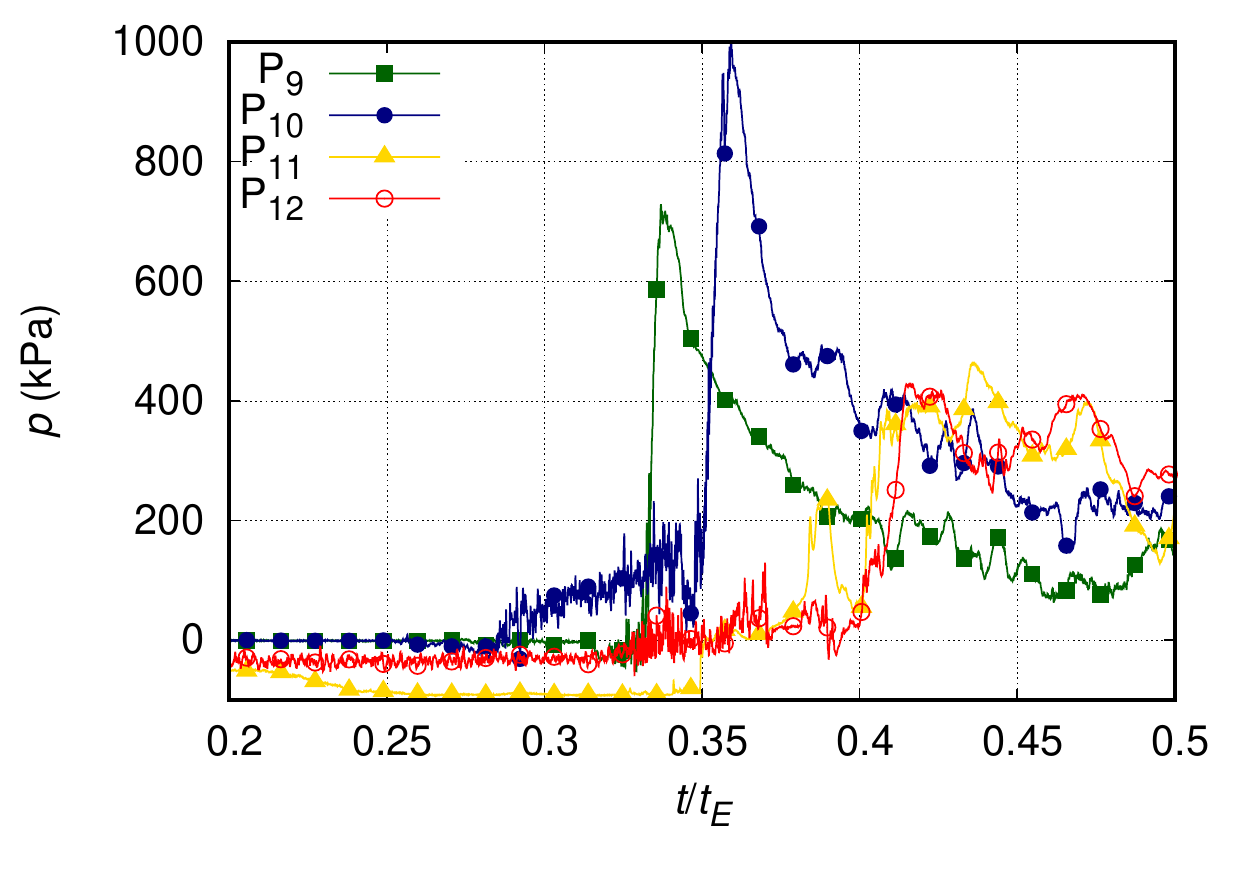}} \\
\caption{Time histories recorded by the pressure probes located at 0.4~m ahead of the trailing edge for the tests XX\_10\_30 , i.e at U=30 
m/s, 10$^{\circ}$~pitch angle for the 15~mm plates, the 3~mm plate and the 0.8~mm plate.} 
\label{fig:X131_Pressure_Spanwise}
\end{figure}

In Figure \ref{fig:X131_Pressure_Spanwise} the time histories of the 
pressure measured in the tests XX\_10\_30 , i.e. at 30~m/s, 10$^{\circ}$~pitch, by 
the probes located at 0.4~m from the trailing edge are shown.
It is possible to spot a clear different trend between the thick and 
the thin plates. In fact, in the thick plate the pressure 
peak first reaches the probe in the middle, i.e. P$_{12}$, and next the
other probes going from the midline to the side i.e. P$_{11}$, P$_{10}$ and 
P$_9$ in succession. 
Therefore, it is inferred that the spray root has a backward curvature, as
widely discussed in \cite{iafrati2016experimental}.
In the case of thin plates the pressure peak occurs at P$_{9}$ first and
then, in succession, at the probes P$_{10}$, P$_{11}$ and P$_{12}$, i.e. 
going from the side to the midline, thus denoting a forward curvature.
Furthermore, it can be noted that the occurrence of the first pressure 
peak in the thin plates is delayed compared to the thick plates.

\begin{figure}[htbp]
\centering
\subfigure[15\_10\_30 (thickness 15~mm)]{\includegraphics[width=0.31\textwidth]{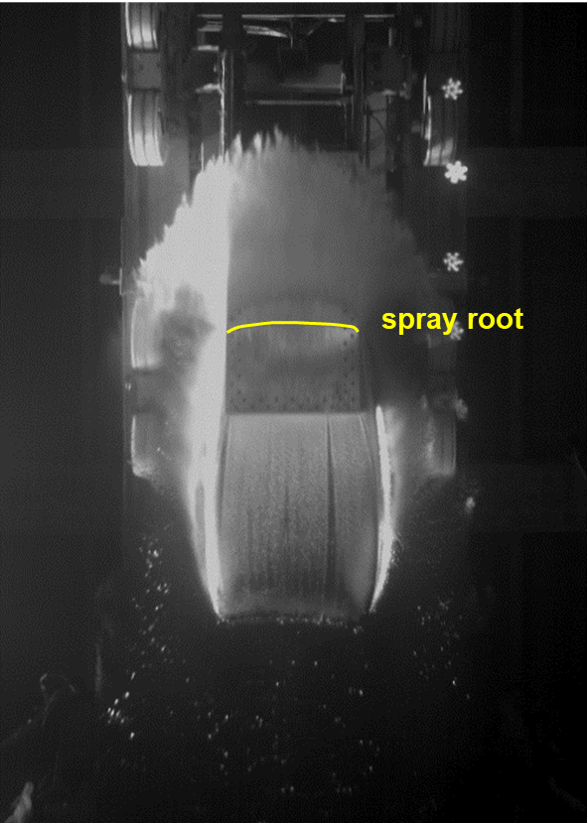}} \quad
\subfigure[3\_10\_30 (thickness 3~mm)]{\includegraphics[width=0.31\textwidth]{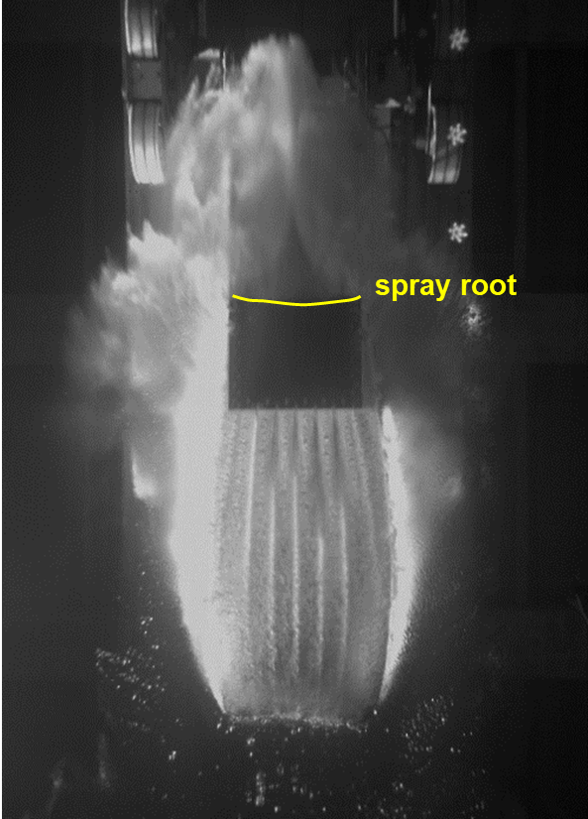}} \quad
\subfigure[08\_10\_30 (thickness 0.8~mm)]{\includegraphics[width=0.31\textwidth]{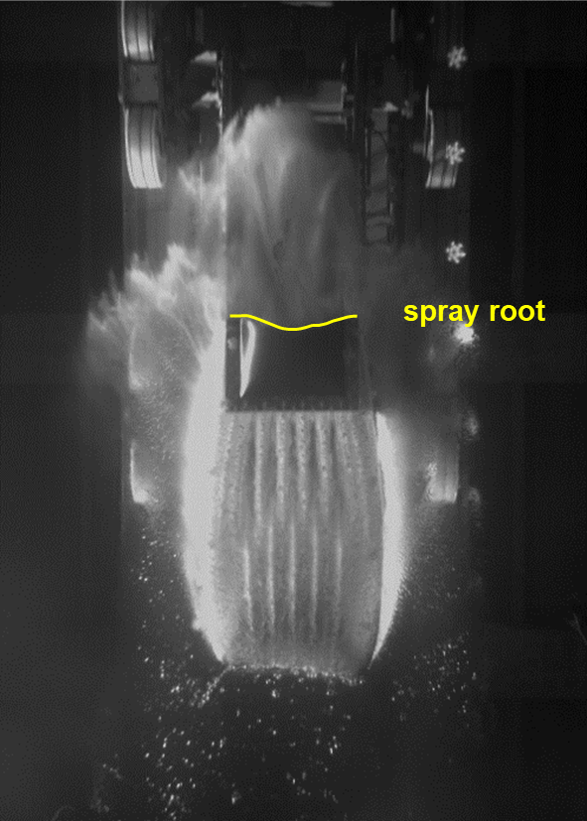}}
\caption{Underwater images at the instant in which the spray root is at the location of the 
probe P$_{12}$: test conditions XX\_10\_30, i.e. U = 30~m/s, $\alpha=10^{\circ}$ and thickness 15~mm (a), 3~mm (b) and 0.8~mm 
(c).}
\label{fig:JetRootShapeComparisonx131}
\end{figure}

The changes in the curvature of the spray root and the air entrapment 
in the central part of the plate during the impact phase 
can also be noticed in the underwater images, as displayed in
Figure \ref{fig:JetRootShapeComparisonx131} where the spray root line is
highlighted by the yellow line.
In this figure, in the case of the thick plate, the spray root line
exhibits a backward curvature. In the case of thin plates, a forward
curvature of the spray root is observed, which is more pronounced for the
0.8~mm plate. In the latter case, the shape of the spray root is strongly
affected by the structural deformation of the plate, which is the reason of
the sharp curvature in the profile near the inner frame.
The light reflections from the sides, particularly evident in 
Figure \ref{fig:JetRootShapeComparisonx131}(c), clearly highlight
the large out-of-plate deformation caused by the impact.

\begin{figure}[htbp]
\centering
\subfigure[Pressure probe positions]{\includegraphics[width=0.75\textwidth]{Figures/Pressure_schematic_spanwise.png}} \\
\subfigure[15\_04\_45 (thickness 15~mm)]{\includegraphics[width=0.48\textwidth]{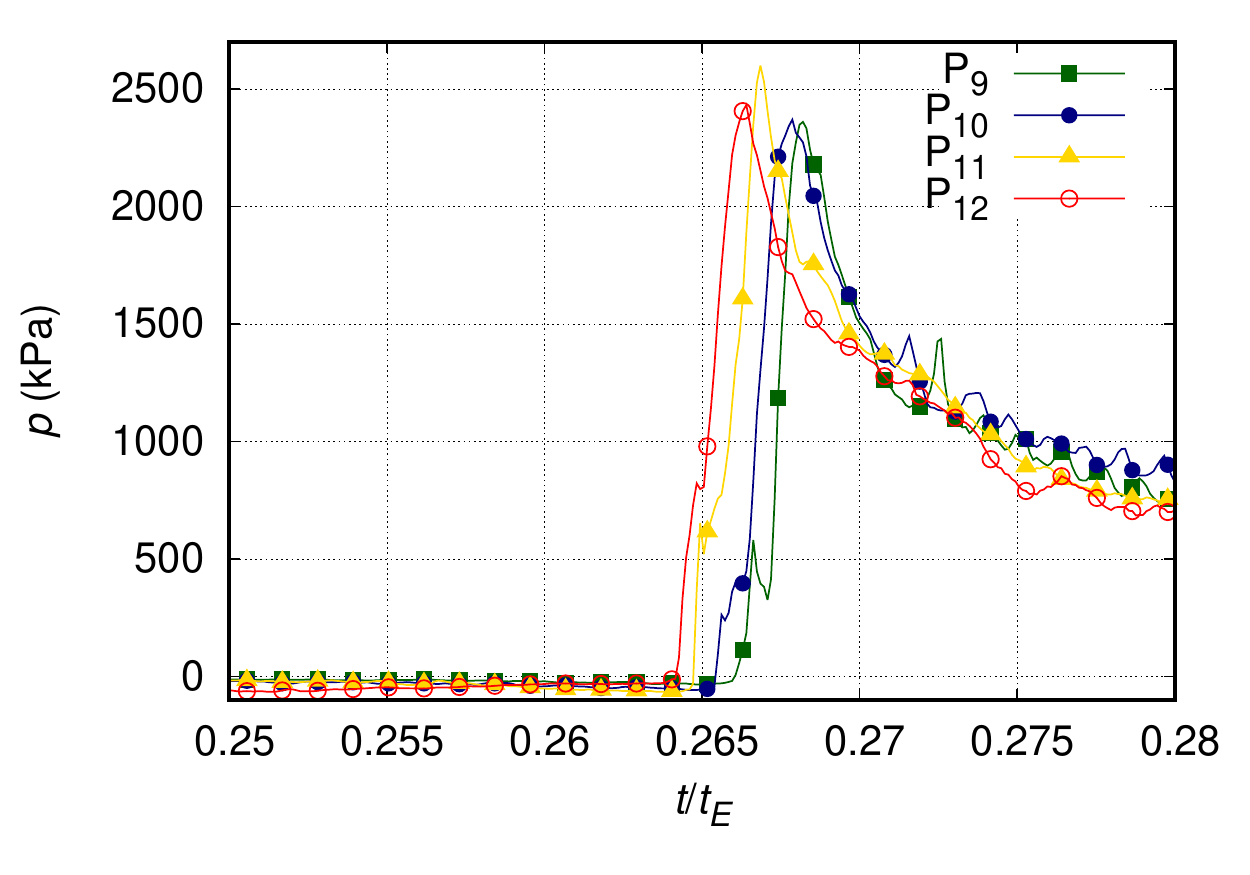}}  \quad
\subfigure[3\_04\_45 (thickness 3~mm)]{\includegraphics[width=0.48\textwidth]{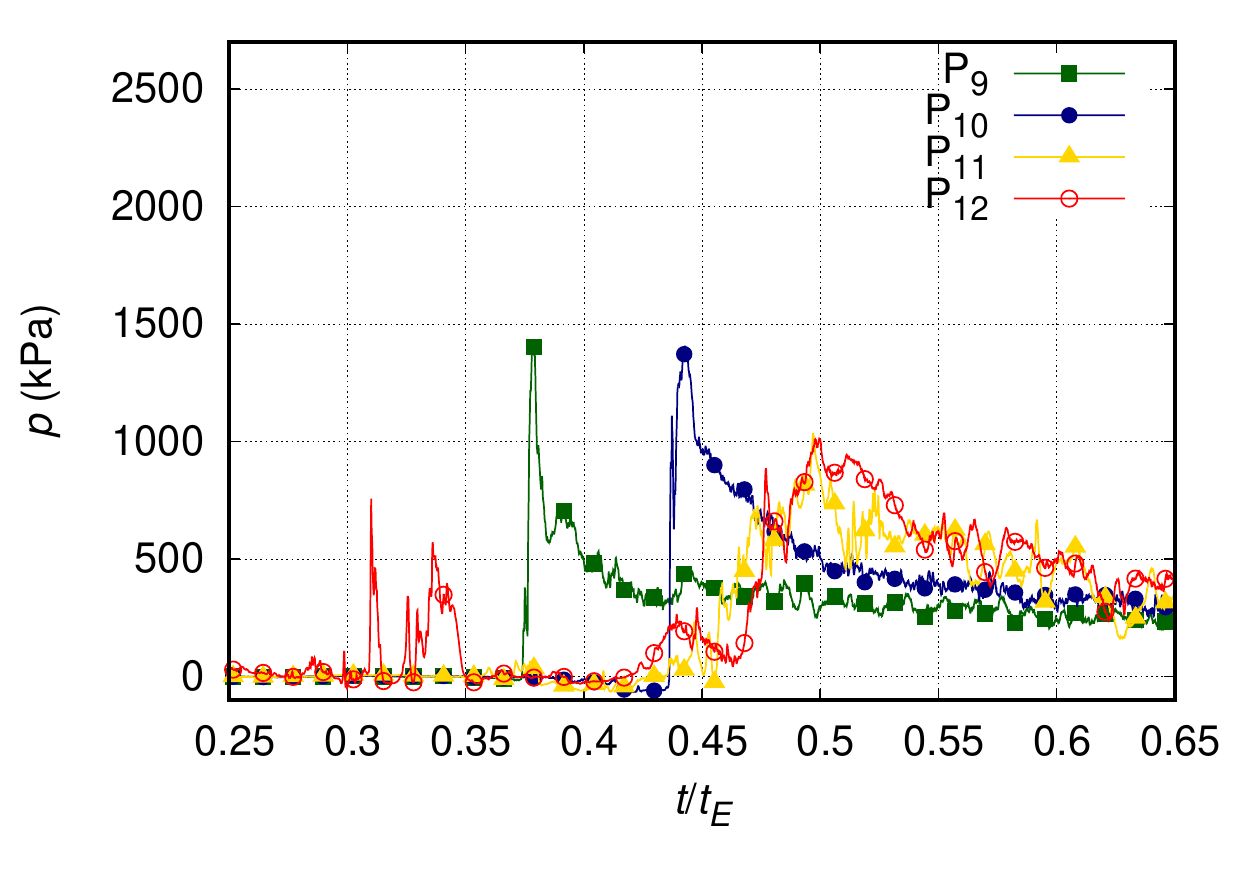}}
\caption{Time histories recorded by the pressure probes located at 0.4~m  ahead of the trailing edge of the plate for the test conditions XX\_04\_45, i.e at U=45~m/s, 4$^{\circ}$~pitch angle for the 15~mm plate and the 3~mm plate.}
\label{fig:X113_Pressure_Spanwise}
\end{figure}

The changes in the spray root curvature can also be noticed in the tests at 45~m/s, 
pitch~4$^{\circ}$, as shown in Figure \ref{fig:X113_Pressure_Spanwise}.
Also in this case, the sequence of the pressure peaks for the 3~mm plate 
is reversed with respect to that observed for the 15~mm plate. 
Moreover, for the 3~mm plate the pressure peaks exhibit an increased
spreading in time and their values are lower.

\begin{figure}[htbp]
\centering
\subfigure[15\_04\_45 (thickness 15~mm)]{\includegraphics[width=0.31\textwidth]{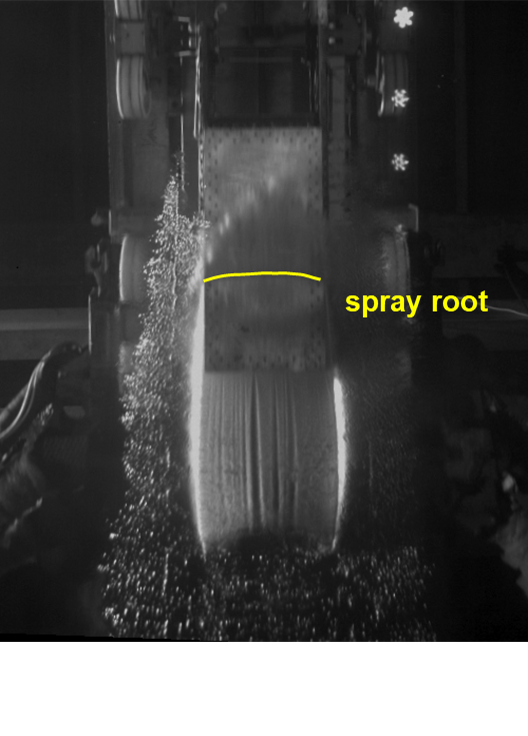}} \quad
\subfigure[3\_04\_45 (thickness 3~mm)]{\includegraphics[width=0.31\textwidth]{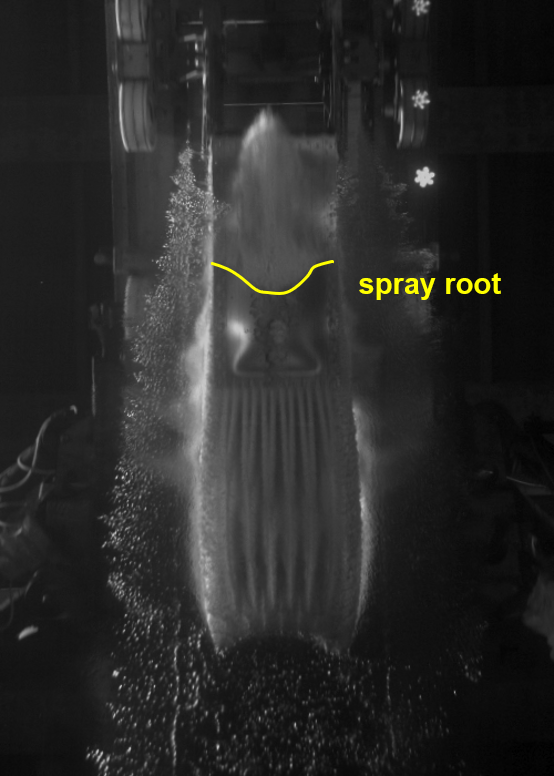}} \quad
\subfigure[08\_04\_45 (thickness 0.8~mm)]{\includegraphics[width=0.31\textwidth]{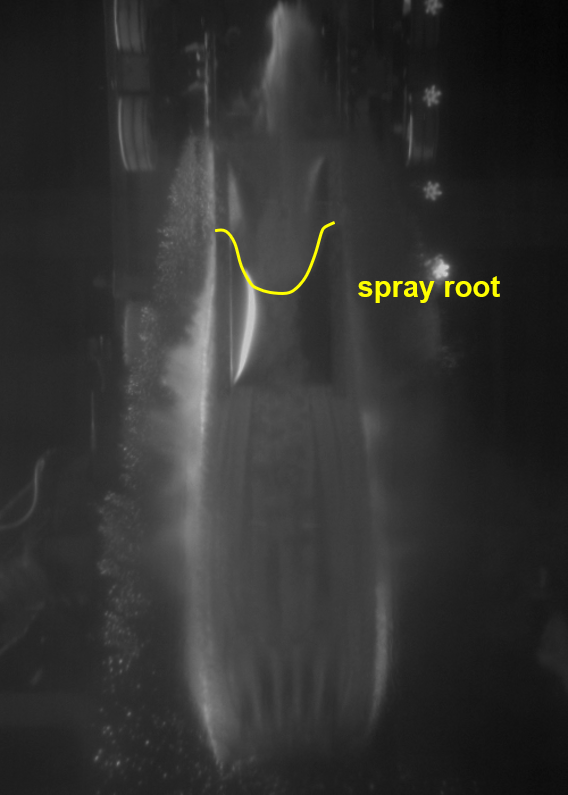}}
\caption{Underwater images at the instant in which the spray root is at the location of the 
probe P$_{12}$: test conditions XX\_04\_45, i.e. U = 45~m/s, $\alpha=4^{\circ}$ and thickness 15~mm (a), 3~mm (b) and 0.8~mm (c).}
\label{fig:JetRootShapeComparisonX113}
\end{figure}

Again, looking at the underwater images in Figure 
\ref{fig:JetRootShapeComparisonX113}, it is seen that the spray root
curvature in the 3~mm case is the opposite of the one observed for the thick
plate and it is much higher than that observed for the cases XX\_10\_30, due to
the higher loading and to the more significant fluid-structure interaction.
The effect is even more evident in the 0.8~mm case.
The large out-of-plane deformation is responsible for the spreading of the
pressure peaks in both space and time. The underwater images also 
highlight the formation of two jets at the sides converging towards the
midline.

\subsection{Combined analysis of pressure and strain measurements}

It is worth analysing simultaneously pressures and strains in order to 
better understand the relationship between the structural deformation and the
hydrodynamic loading, as well as their mutual interaction.

\begin{figure}[!p]
\centering
\subfigure[Pressure Probe and Strain Gauge 
Positions]{\includegraphics[width=0.6\textwidth]
{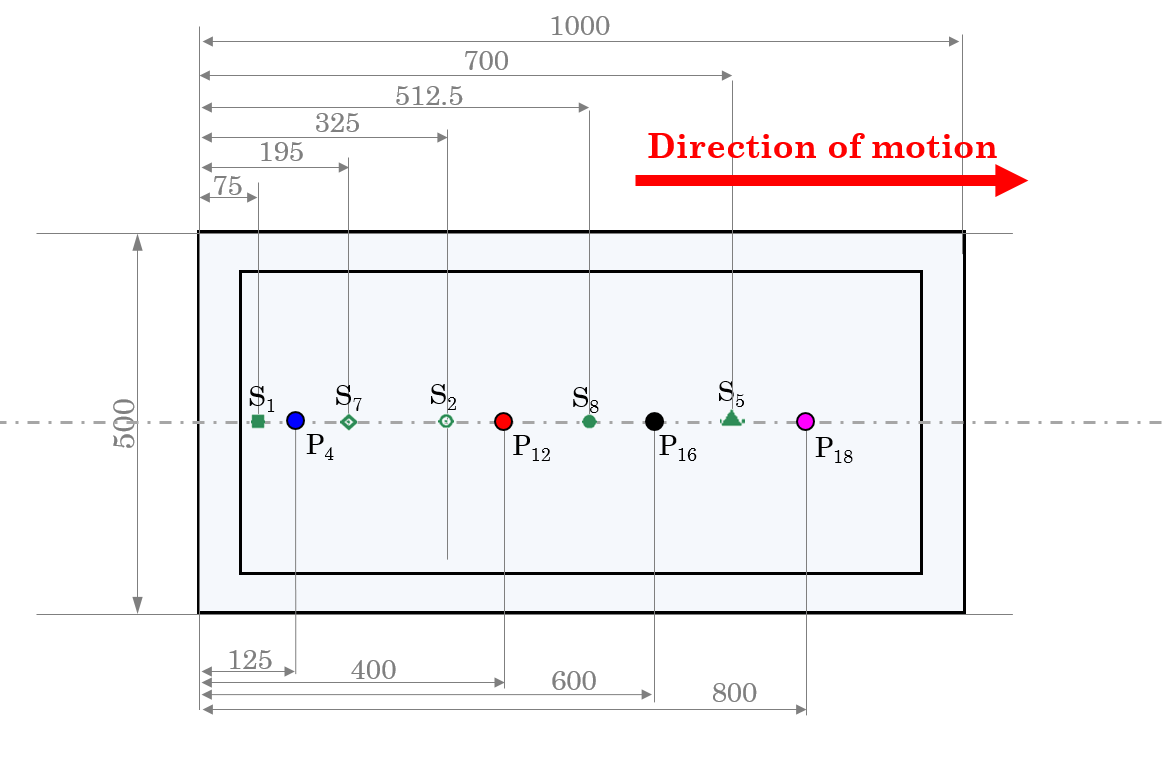}} \\
\subfigure[15\_10\_30 (thickness 15~mm)]{\includegraphics[width=0.7\textwidth]
{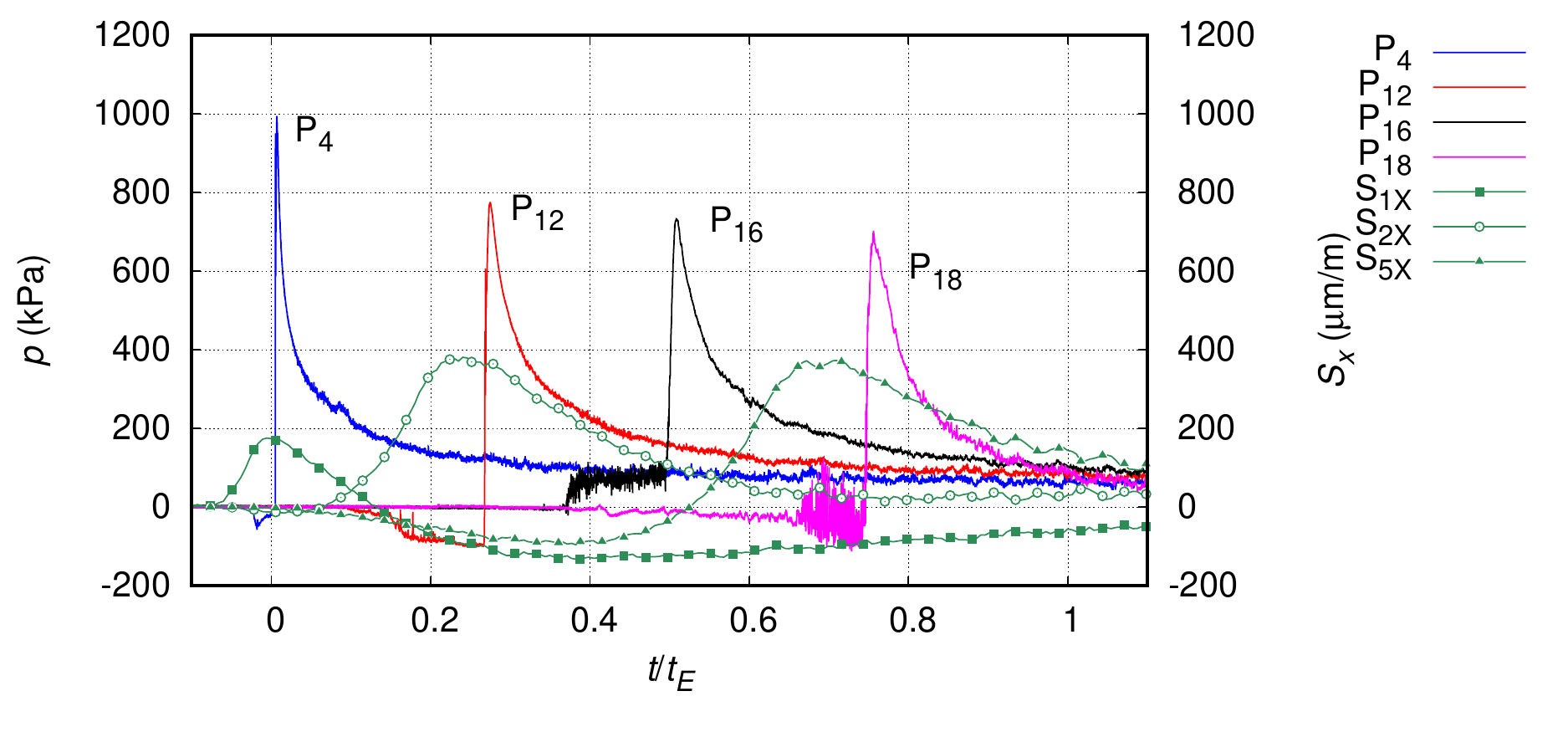}} \\
\subfigure[3\_10\_30 (thickness 3~mm)]{\includegraphics[width=0.7\textwidth]
{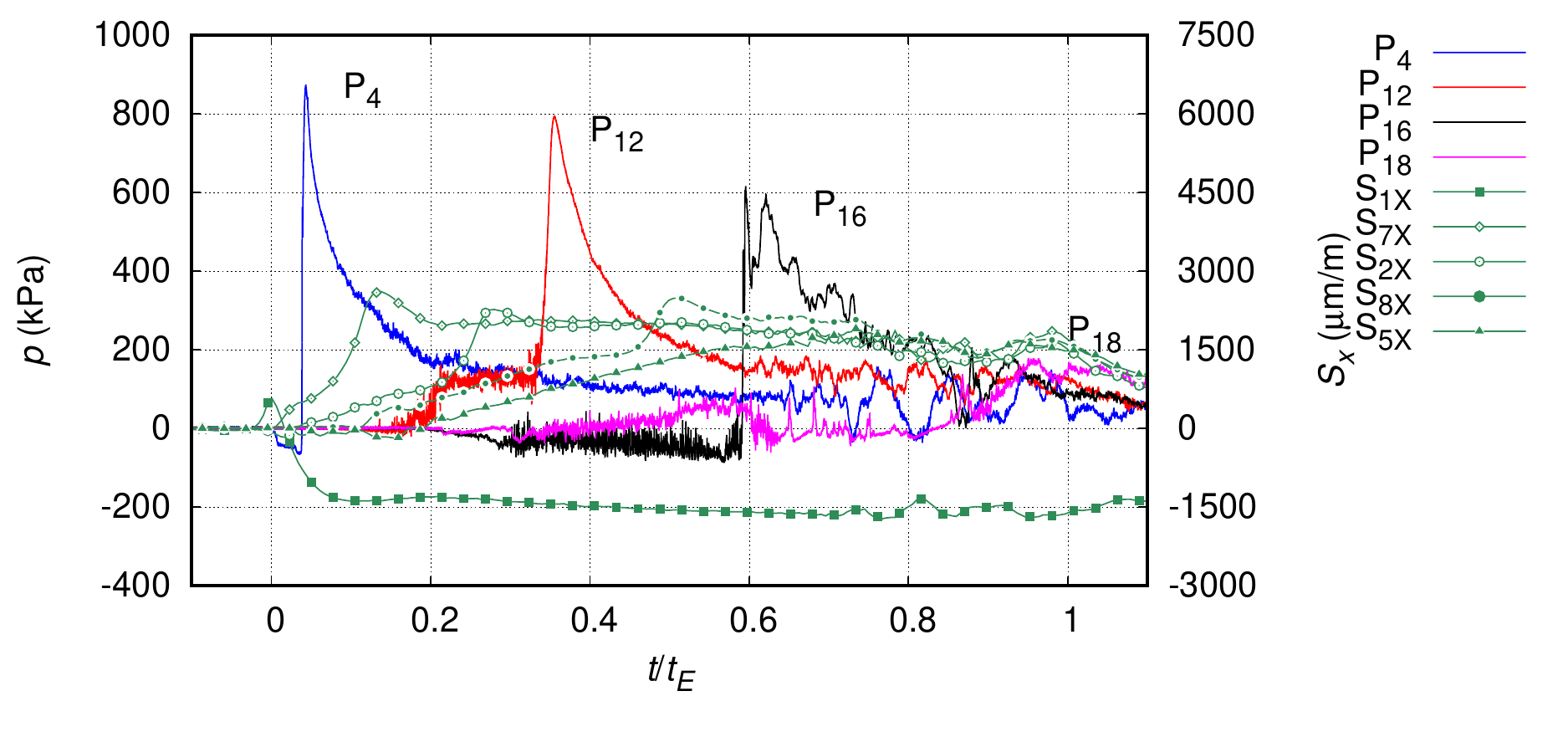}} \\
\subfigure[08\_10\_30 (thickness 0.8~mm)]{\includegraphics[width=0.7\textwidth]
{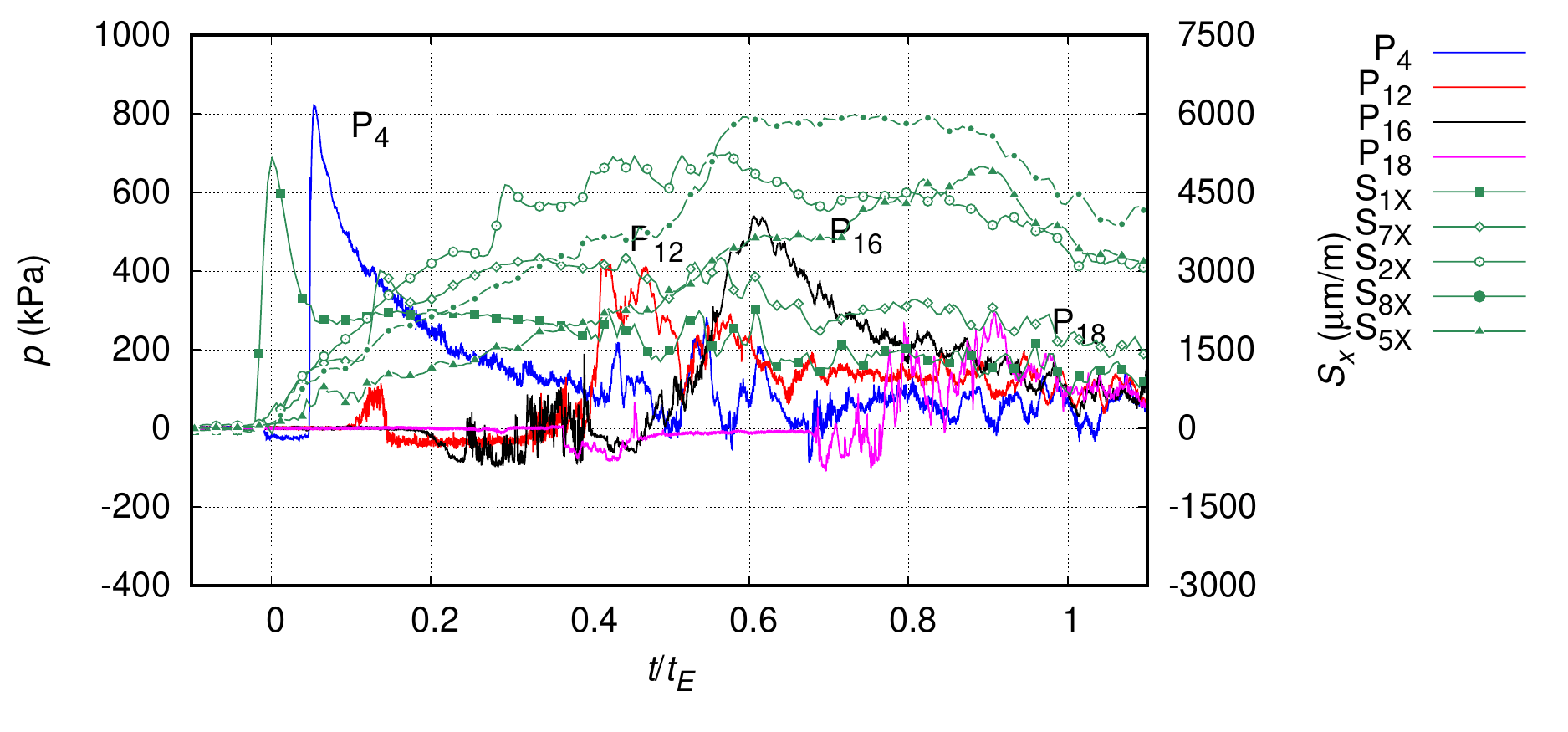}}
\caption{Time histories of the pressures and longitudinal strains measured along the midline for the tests XX\_10\_30, i.e U=30~m/s, 
10$^{\circ}$~pitch angle for 15~mm plate, the 3~mm plate and the 0.8~mm plate.}
\label{fig:X131_Pressure_Strains_CL}
\end{figure}

To this purpose, the pressures and the $x$-strains recorded along the midline
during the impact phase for the tests at 30~m/s, 10$^{\circ}$~pitch angle
and different thickness's are drawn in Figure 
\ref{fig:X131_Pressure_Strains_CL}.
In the case of the thick plate, there is a quite good correlation between
the pressures and the $x$-strains, with the strains attaining a maximum
concurrently with the passage of the pressure peak beneath.
After the passage of the peak, the strains gradually diminish and
approach zero as time elapses.
In the 3~mm plate, the $x$-strain still exhibits a maximum concurrent with
the passage of the pressure peak, but in this case the pulse width is narrower and less
pronounced. Furthermore, due to the reduced thickness of the plate, the structural deformation 
propagates much further ahead of the spray root and all the strains measured by the
gauges, except for S$_{1x}$, attain about the same strain level when reached by the pressure peak and 
remain quite flat afterwards, until the end of the impact phase. 
Owing to the disruption of the spray root and to the spreading of
the pressure peak, there is no evident maximum for S$_{5x}$, which also
approaches the strain values reached by the other gauges.
Of course the strain levels are much higher than in the 15~mm case.
In the 0.8~mm plate, a clear peak is seen for S$_{1x}$ and,
to a lesser extent, for S$_{7x}$. This is a consequence of the peaked
pressure distribution observed in the early stage of the impact, e.g. P$_4$.
Moving further, the pressure is spread over a wider area, which is the
reason why, rather than a clear peak, the strains exhibit an almost flat
region at high values and diminish afterwards. This is the case of
S$_{2x}$ for $t/t_E \in (0.3, 0.7)$, S$_{8x}$ for $t/t_E \in (0.6, 0.8)$
and S$_{5x}$ for $t/t_E \in (0.7, 0.9)$.
Differently from the 3~mm case, the strain levels vary substantially from
gauge to gauge. Moreover, the reduced thickness and the
complicated flow developing beneath the plate induce visible oscillations 
in the strains.

\begin{figure}[!p]
\centering
\subfigure[Pressure Probe and Strain Gauge 
Positions]{\includegraphics[width=0.6\textwidth]{Figures/SG_Pressure_Positions_Midline.png}} \\
\subfigure[15\_04\_45 (thickness 15~mm)]{\includegraphics[width=0.7\textwidth]{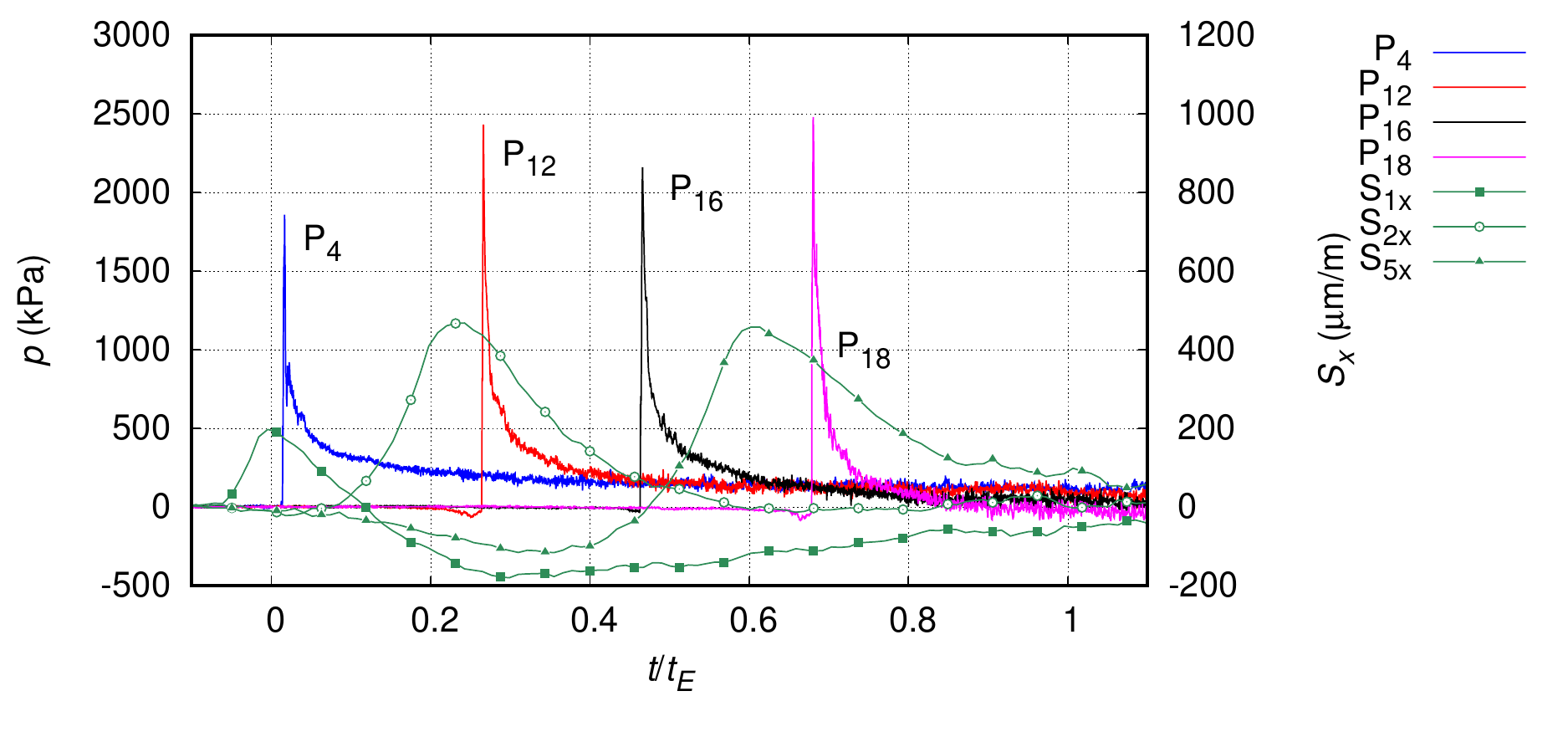}} \\
\subfigure[3\_04\_45 (thickness 3~mm)]{\includegraphics[width=0.7\textwidth]{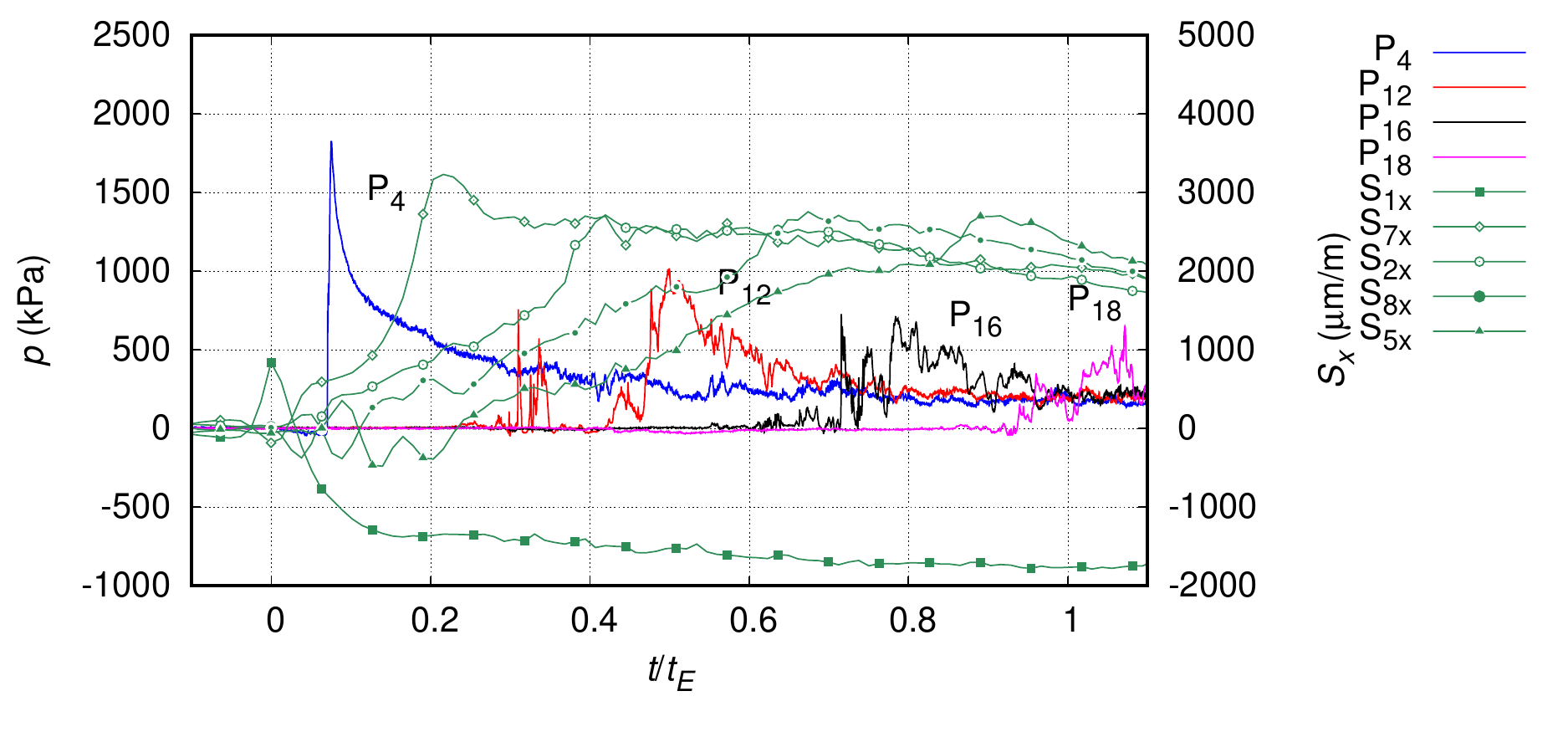}} 
\caption{Time histories of the pressures and longitudinal strains measured along the midline for the tests XX\_04\_45, i.e U=45~m/s,
 4$^{\circ}$~pitch angle for 15~mm plate and the 3~mm plate.}
\label{fig:X113_Pressure_Strains_CL}
\end{figure}

Similar conclusions can be drawn by looking at the data of the tests
performed at 45~m/s, 4$^{\circ}$~pitch angle for the 15~mm and 3~mm plates, 
shown in Figure \ref{fig:X113_Pressure_Strains_CL}.
At least for the 15~mm plate, owing to the smaller pitch angle, the 
pressure peak is much sharper than that observed in test condition 15\_10\_30, see
Figure \ref{fig:X131_Pressure_Strains_CL}(b), and there is a quite high
correlation between the pressure and the strain profiles.
Similarly to what found in the test condition 3\_10\_30, for the 3~mm plates the
strains grow gradually and the values of all gauges approach about the same level, 
even though this occurs at $t/t_E \simeq 0.8$ in the case 3\_04\_45 and at $t/t_E \simeq 0.6$ 
in the case 3\_10\_30, see Figure \ref{fig:X131_Pressure_Strains_CL}(b).

\subsection{Hydrodynamics loads}
\label{hyrodynamicloads}

As already noticed in \cite{iafrati2015high}, the tangential
load $F_x$ is not so significant during the impact phase. $F_x$
undergoes a sharp rise as soon as the spray root reaches the leading edge
of the plate, and the front face of the acquisition box drops down to the
still water level.

\begin{figure}[htbp]
\centering
\subfigure[XX\_10\_30 - $U = 30~m/s$ and $\alpha=10^{\circ}$]{\includegraphics[width=0.6\textwidth]{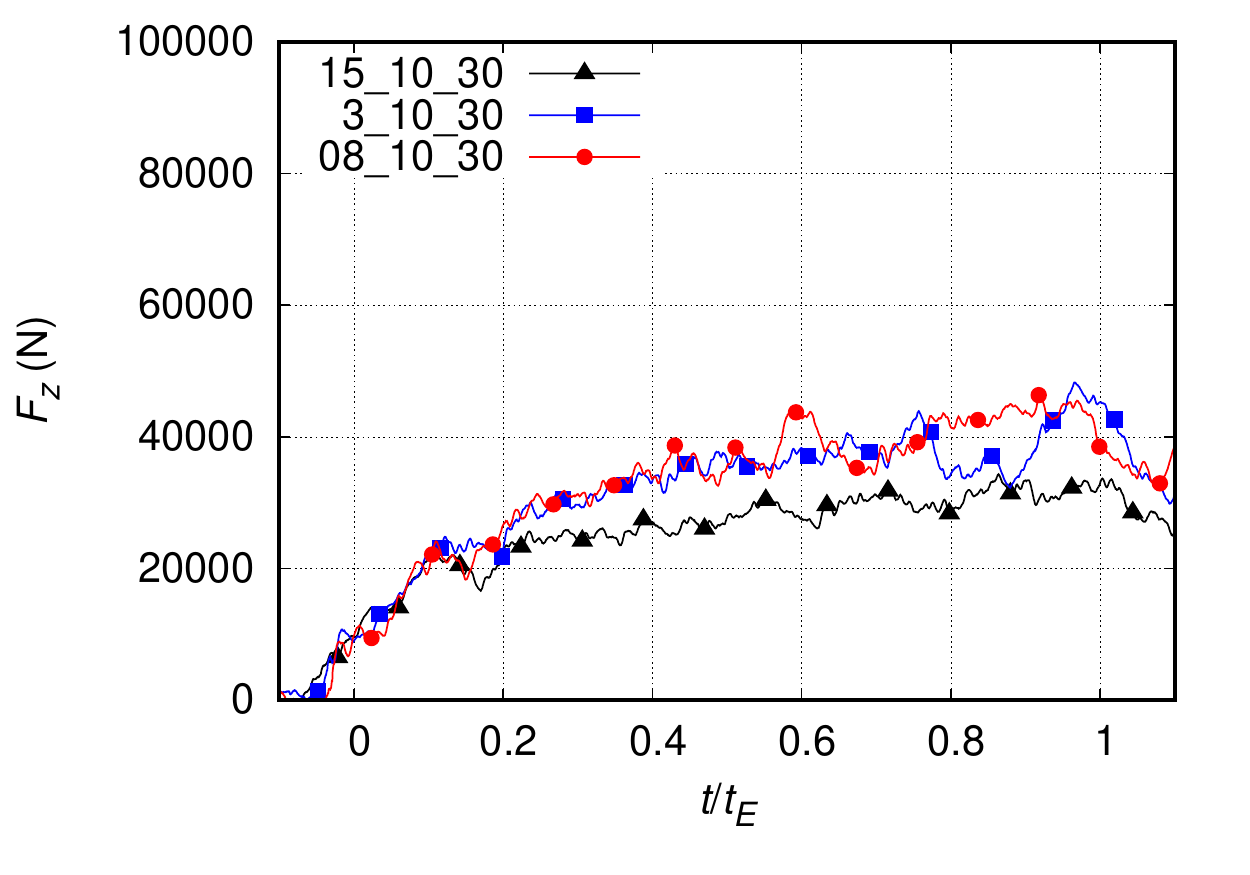}} 
\\
\subfigure[XX\_06\_40 - $U = 40~m/s$ and $\alpha=6^{\circ}$]{\includegraphics[width=0.6\textwidth]{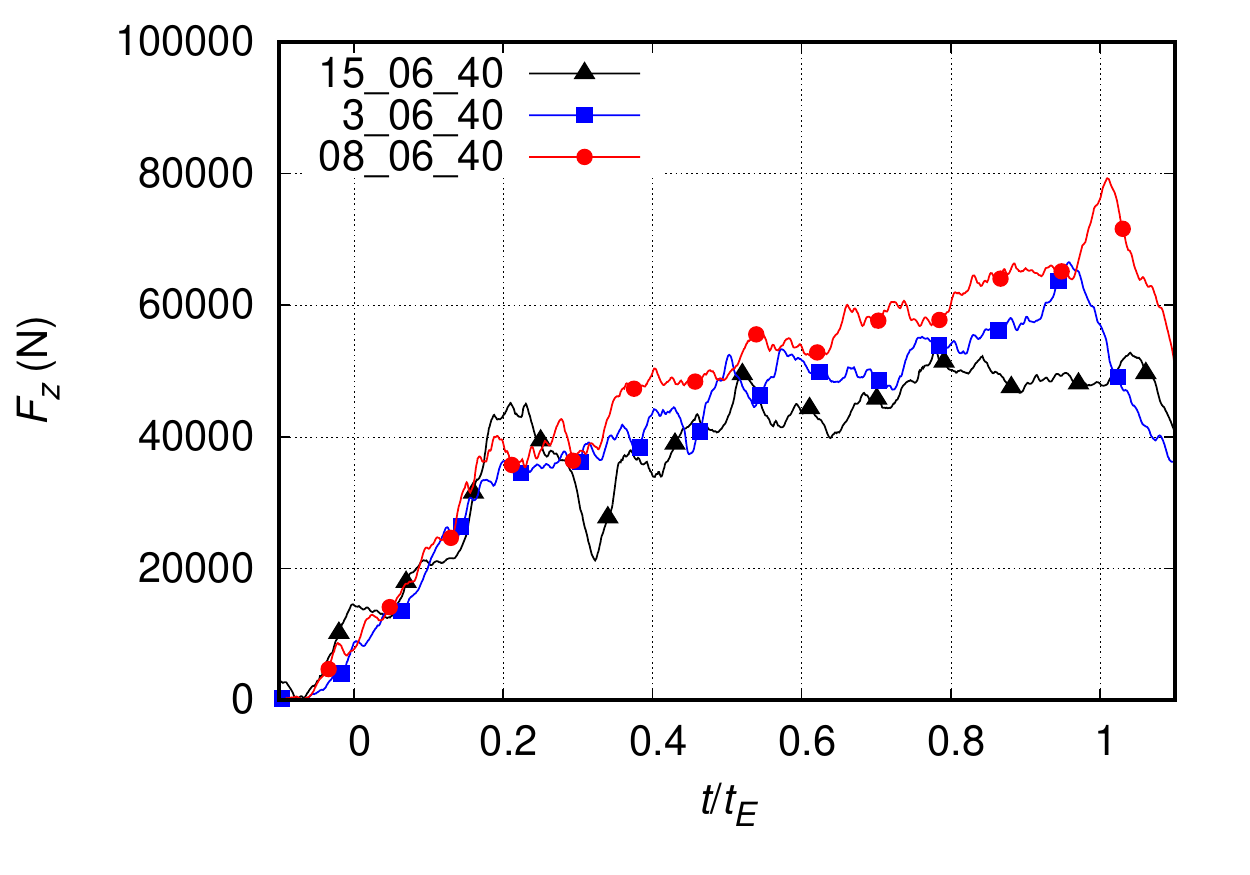}} 
\\
\subfigure[XX\_04\_45 - $U = 45~m/s$ and $\alpha=4^{\circ}$]{\includegraphics[width=0.6\textwidth]{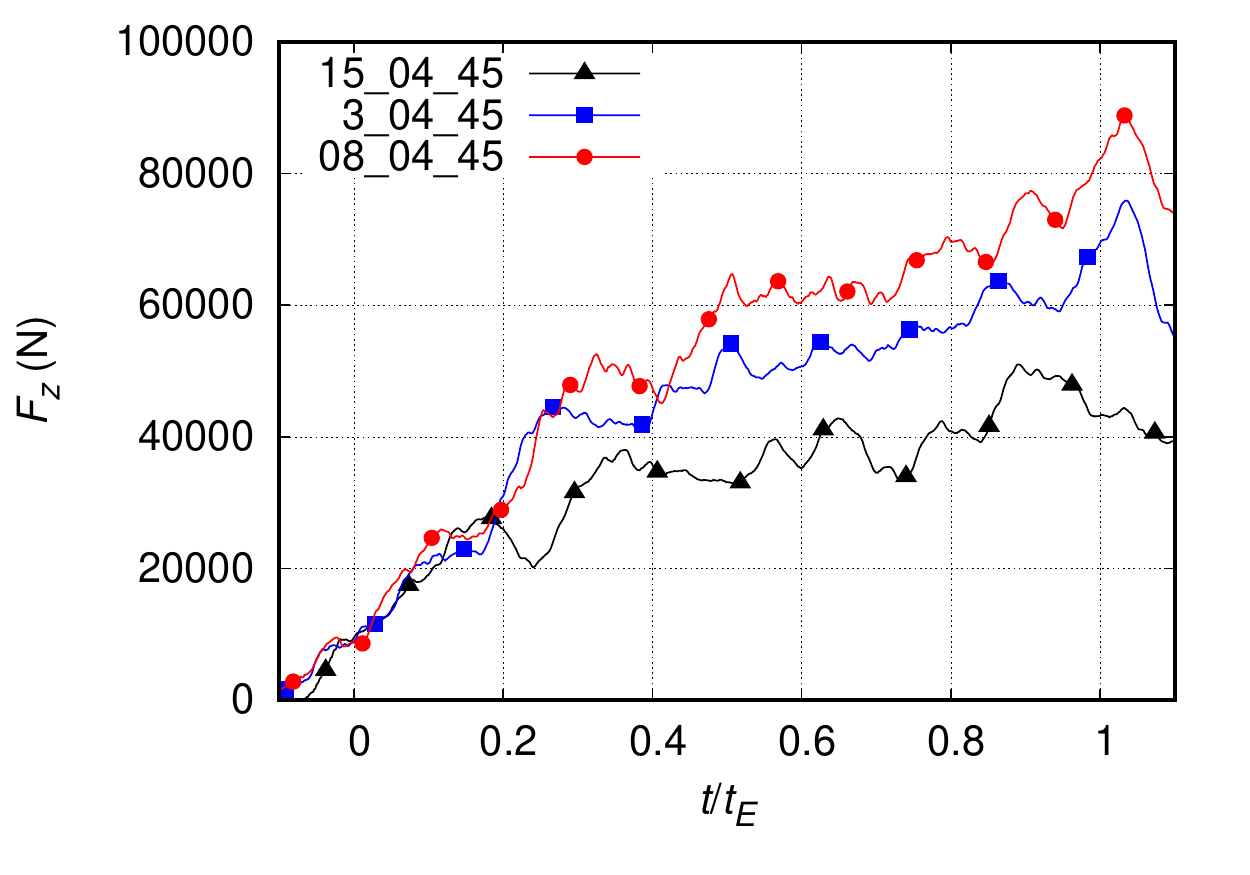}} 
\caption{Time histories of the normal force at different thickness's for the test conditions XX\_10\_30, XX\_06\_40 and
XX\_04\_45.}
\label{fig:ZLoads}
\end{figure}

Much more relevant is the normal force $F_z$\footnote{The sign convention for $F_z$ and $F_x$ is consistent with
the coordinate system shown in Figure \ref{fig:trolley_drawing}.}.
In Figure \ref{fig:ZLoads} comparisons between the time histories of the 
$z$-force measured in the tests XX\_10\_30 (i.e. $U = 30 $ 
m/s and $\alpha=10^{\circ}$), XX\_06\_40 ($U = 40 $~m/s and
$\alpha=6^{\circ}$) and XX\_04\_45 ($U = 45 $~m/s and 
$\alpha=4^{\circ}$) are shown. It is worth 
reminding that the rise of the $z$-force occurs slightly earlier than at
$t=0$, because the origin of time is chosen to be at the peak
of $S_{1x}$, which is located slightly ahead of the trailing edge, as explained in Section \ref{setup}.
In all cases there is an initial stage, lasting up to $t/t_E = 0.2$, 
during which the normal load is almost independent of the plate thickness.
This indicates that, even for the thinnest plates, the out-of-plane 
deformation in the early stage is not large enough to affect the loading.
However, in the following stage the curves start to diverge, with the
normal load increasing when reducing the plate thickness, further confirming that
the deformation of the plate affects the hydrodynamics
beneath the plate. Based on the above discussion on the behaviour of the
pressure, even though the intensity of the pressure peak drops
substantially as the structural deformation increases, presumably the integrated pressure along
the plate is higher. 
This result is consistent with what found numerically in \cite{siemann2017coupled},
 where it is shown that the loading increases with the structural deformation. It is also worth
remarking that the numerical results in \cite{siemann2017coupled}, obtained with
two different numerical approaches, show a very good agreement with the present experimental results for the 
15~mm plate, the 3~mm plate and the 0.8~mm plate at 6$^{\circ}$ pitch and at a horizontal speed of 40 m/s.
From a quantitative viewpoint when reducing the plate stiffness, the normal loads increases,
and the maximum values are from 30 to 50\% higher than those measured in the thick plate.

\begin{figure}[htbp]
	\centering
		\includegraphics[width=0.50\textwidth]{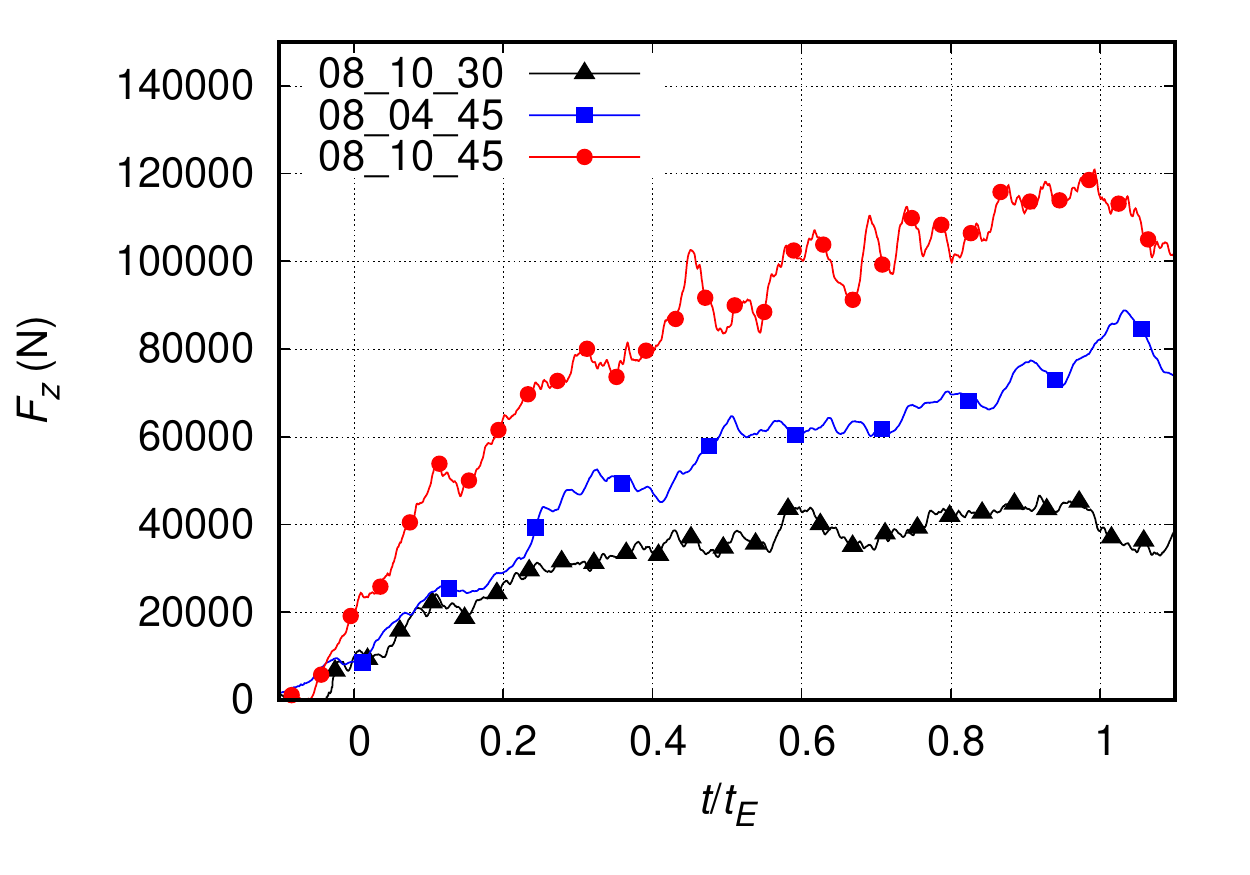}
	\caption{Time histories of the normal force for the 0.8~mm plate at different horizontal speeds and pitch angles.}
	\label{fig:ZForces_31XX_Comparison}
\end{figure}

\begin{figure}[htbp]
\centering
\subfigure[15\_XX\_YY - 15~mm]{\includegraphics[width=0.48\textwidth]{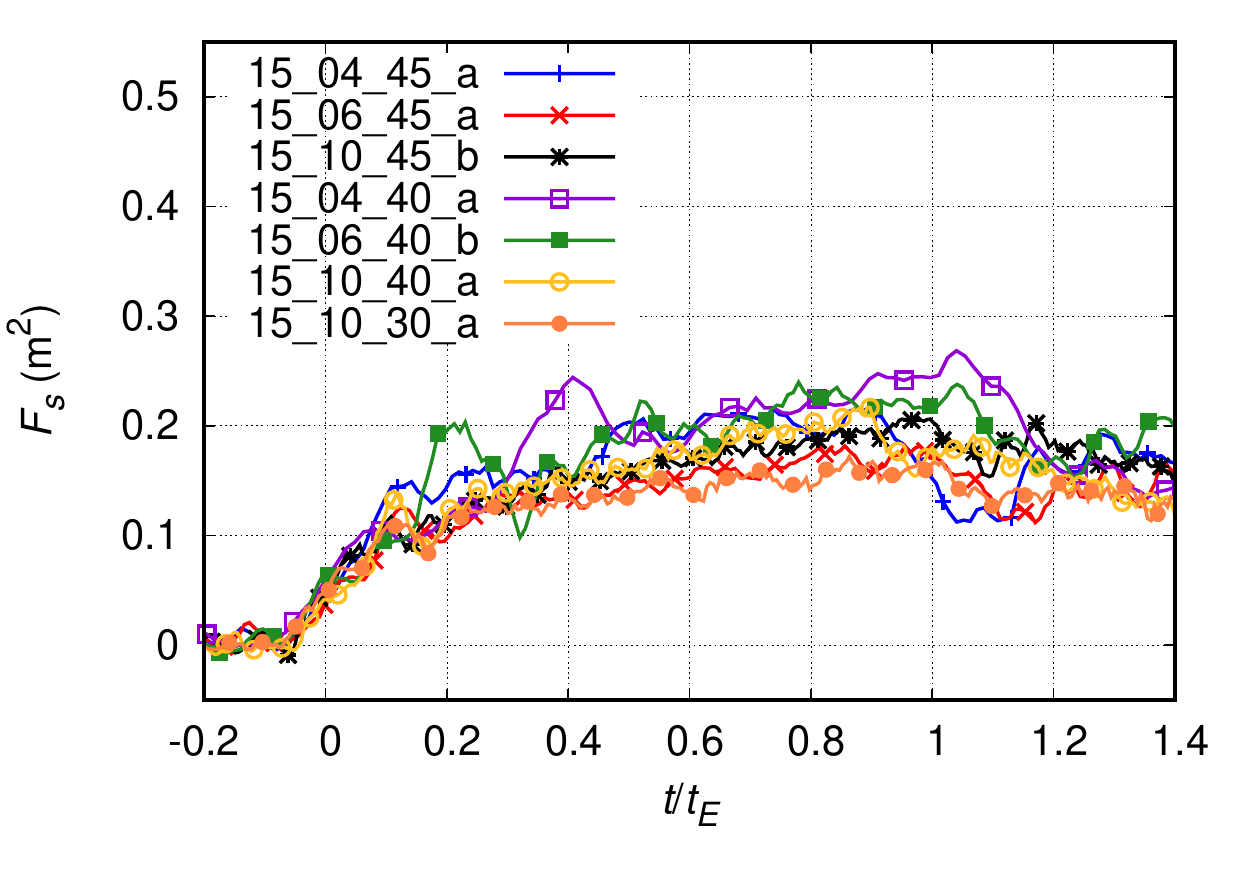}} \quad
\subfigure[3\_XX\_YY - 3~mm]{\includegraphics[width=0.48\textwidth]{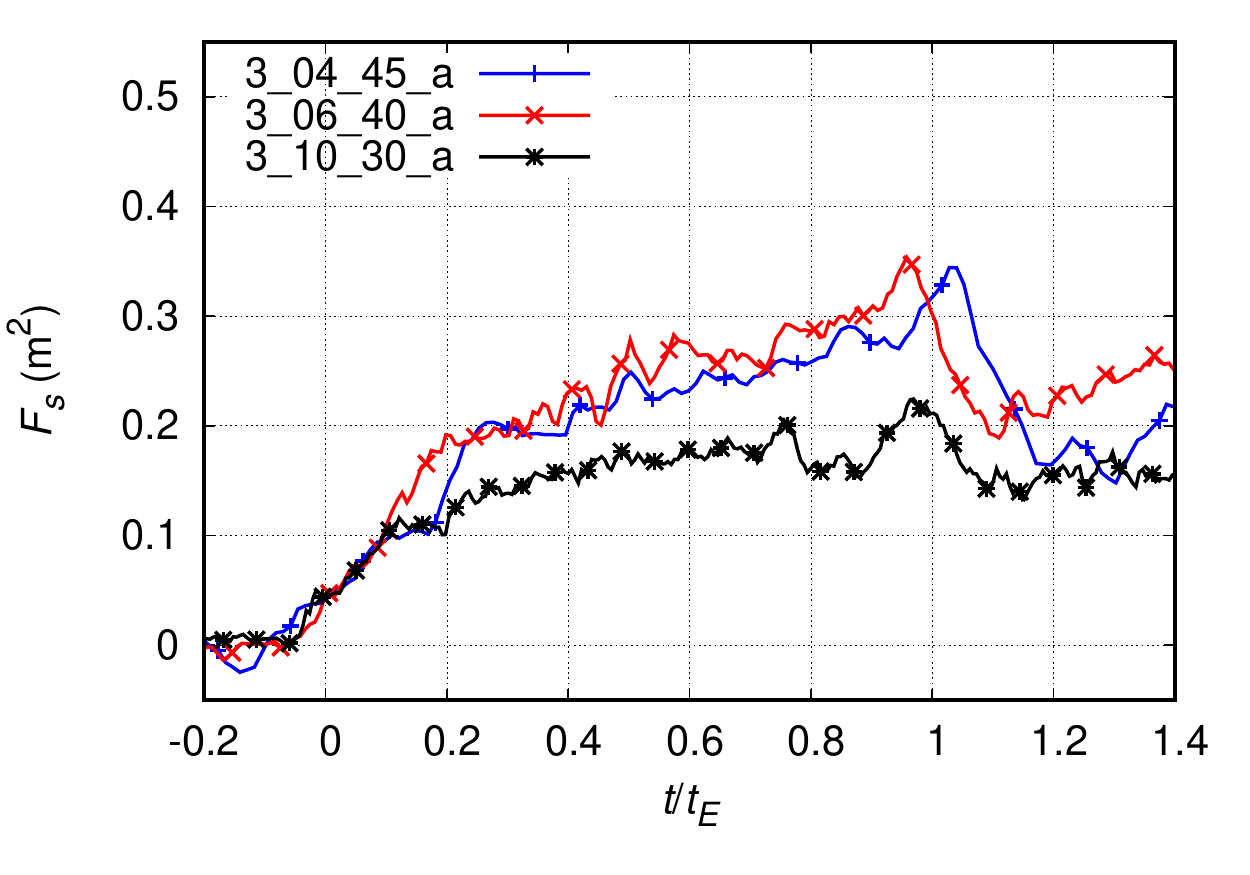}} \\
\subfigure[08\_XX\_YY - 0.8~mm]{\includegraphics[width=0.48\textwidth]{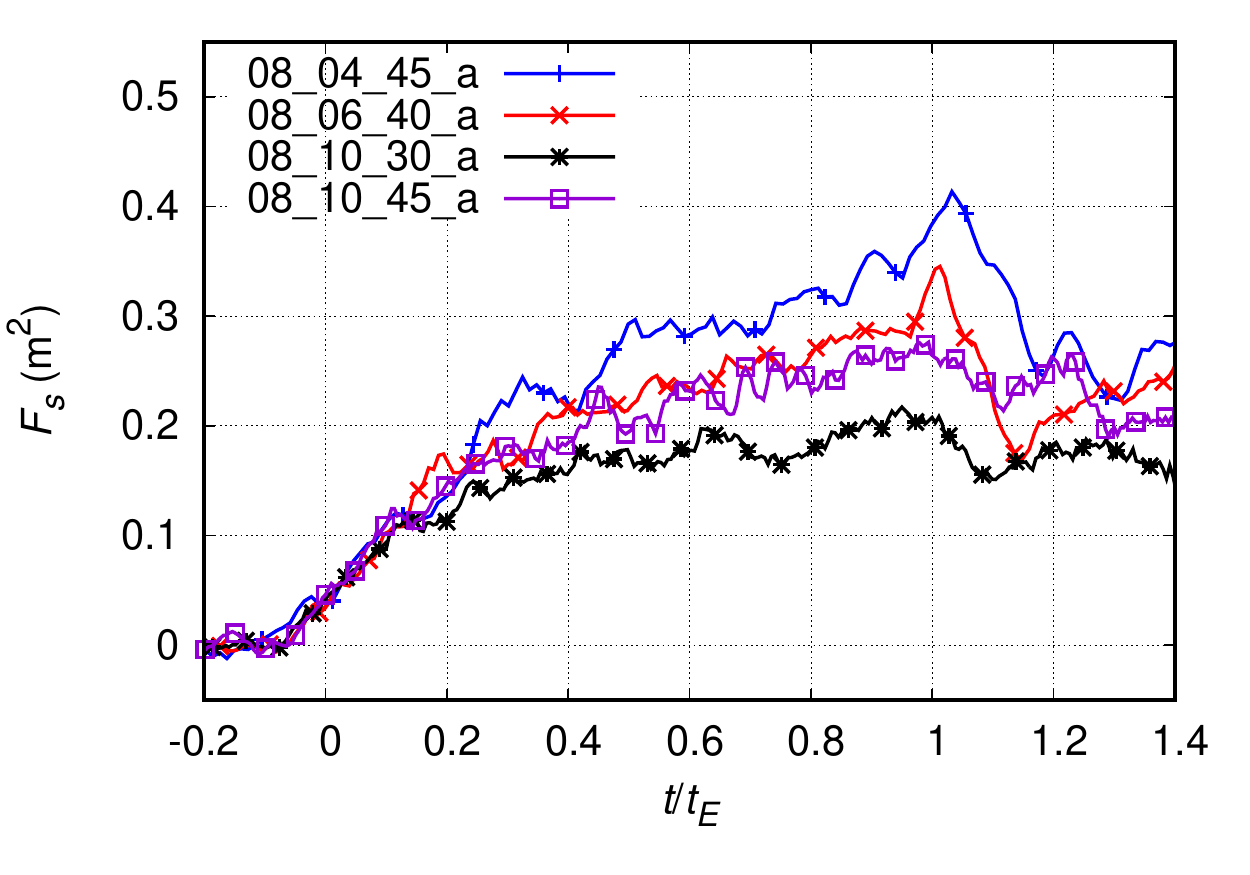}} 
\caption{Time histories of the scaled normal force for different test conditions.}
\label{fig:ZLoadsScaling}
\end{figure}

It is worth observing that the difference between the normal loading 
grows when increasing the horizontal 
speed and reducing the pitch angle. This result is valid for the thick plates, see
 \cite{iafrati2016experimental,iafrati2015high}, and it is also observed in the thin plates, as shown for example 
in Figure \ref{fig:ZForces_31XX_Comparison}, where the time histories 
of the normal load acting on the 0.8~mm plate at different pitch angles and horizontal speed are shown.
In \cite{iafrati2016experimental} the normal loading measured for the thick
plate were found to scale with the projection of the force exerted by the stagnation pressure $\rho (U^2+V^2)/2$,
assumed to be uniformly distributed all over the plate surface, onto the direction normal to the plate\footnote{In \cite{iafrati2016experimental} there was a mistake in the text and the data were scaled by $\rho (U^2+V^2)\sin(\alpha+\beta)$
and not by $U_n^2$ as indicated in the captions.}. The same scaling is adopted here, 
by drawing in Figure \ref{fig:ZLoadsScaling} the quantity $F_s = F_z/[\rho (U^2+V^2) \sin(\alpha+\beta)]$. 
Each plot shows $F_s$ at a different plate thickness, but at a different horizontal velocity
and pitch angle.
It can be seen that for the thick plates (15 mm) the time history of $F_s$ are 
fairly overlapped, with a limited scatter, thus confirming the validity of the scaling, 
which however is no longer valid for the thin plates. 
In fact, for the thin plates, there is an initial phase, up to $t/t_E =0.15 - 0.2$ 
in which the time histories of the scaled loads collapse. Then, as the plate penetrates water, the time histories 
start to diverge significantly, as a result of the plate deformation and of the fluid-structure interaction.
The scaled values grow when increasing the test speed and the pitch angle.
Since the increase in both parameters leads to an increase in the plate 
deformation, the results again indicate that the deformation amplifies the 
loading and, furthermore, that such amplification does not follow the scaling.
Clearly these effects are more significant 
for the thinnest plates (0.8~mm), where the deformations are larger.

The significant effect of the structural deformation on the loading
and the difficulties in the scaling have important consequences.
First of all, it is not always true that tests on rigid specimens provide
conservative estimates of the loading. Secondly, owing to the difficulties in
reproducing the correct scaling of the structure and the changes in the 
loads induced by the structural deformation, scaled model tests of deformable 
structures cannot be considered as completely reliable, and a verification at full-scale conditions,
 i.e. at the ditching conditions and on a structure  
representative of a real aircraft (especially with the same the panel thickness's), 
is advisable, whenever feasible.

\begin{figure}[htbp]                                                      
\centering
\subfigure[XX\_10\_30 - $U = 30~m/s$ and $\alpha=10^{\circ}$]{\includegraphics[width=0.6\textwidth]{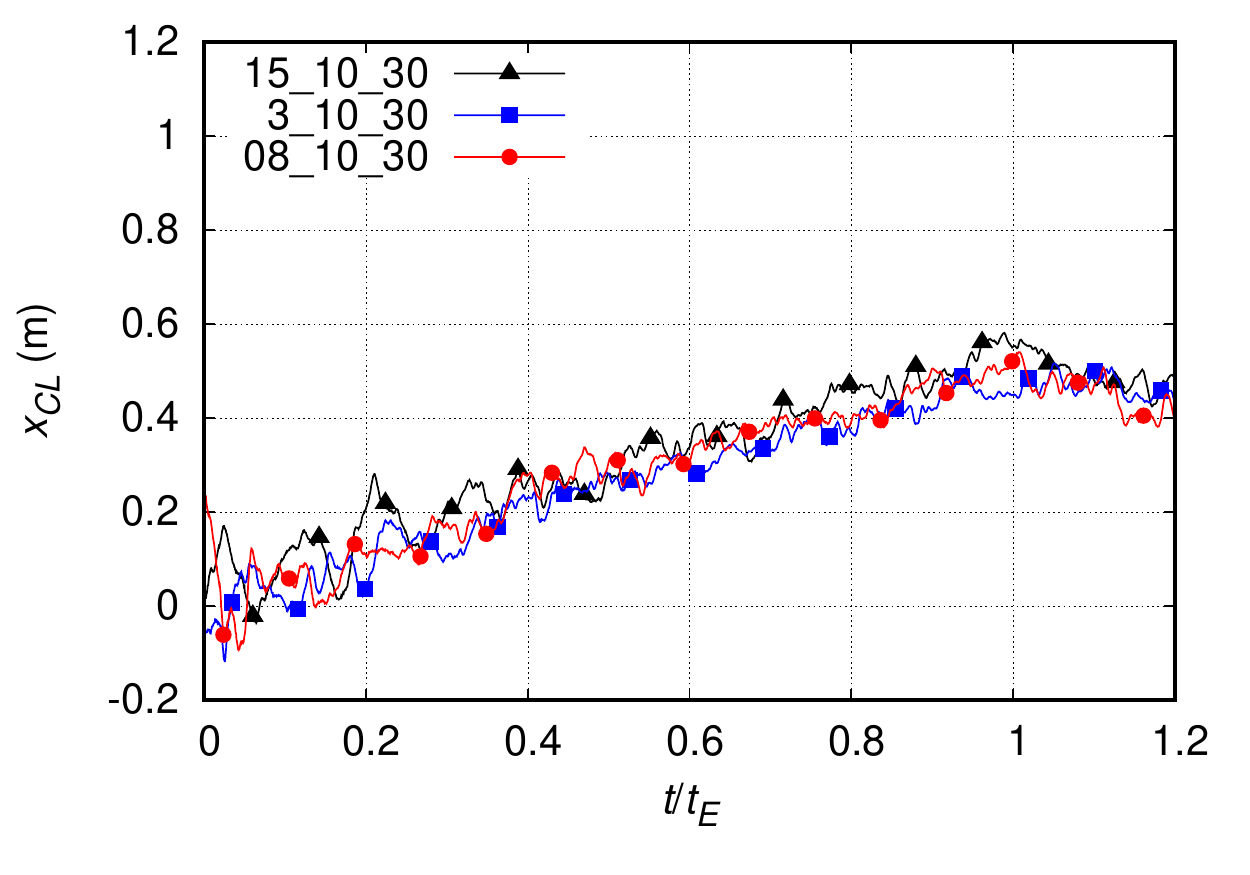}} 
\\
\subfigure[XX\_06\_40 - $U = 40~m/s$ and $\alpha=6^{\circ}$]{\includegraphics[width=0.6\textwidth]{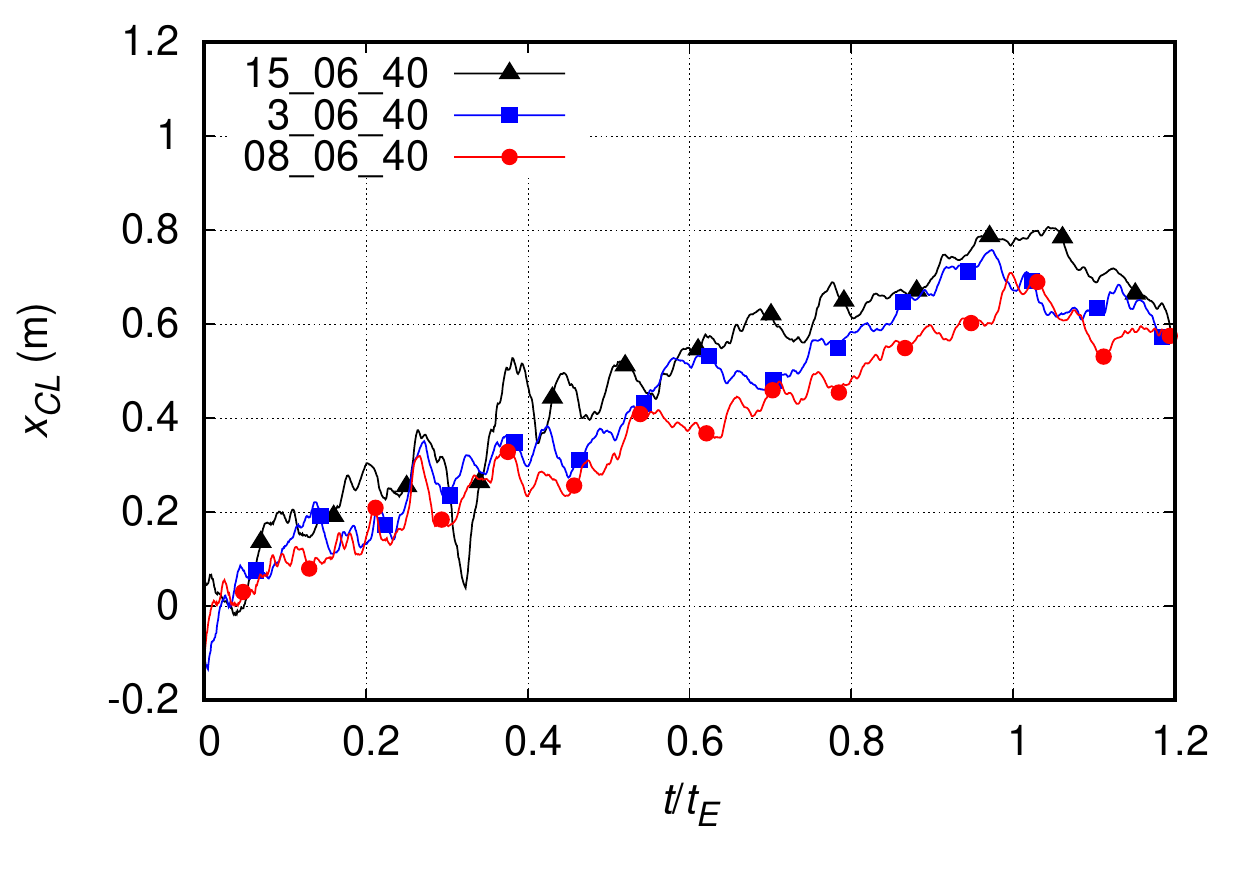}} 
\\
\subfigure[XX\_04\_45 - $U = 45~m/s$ and $\alpha=4^{\circ}$]{\includegraphics[width=0.6\textwidth]{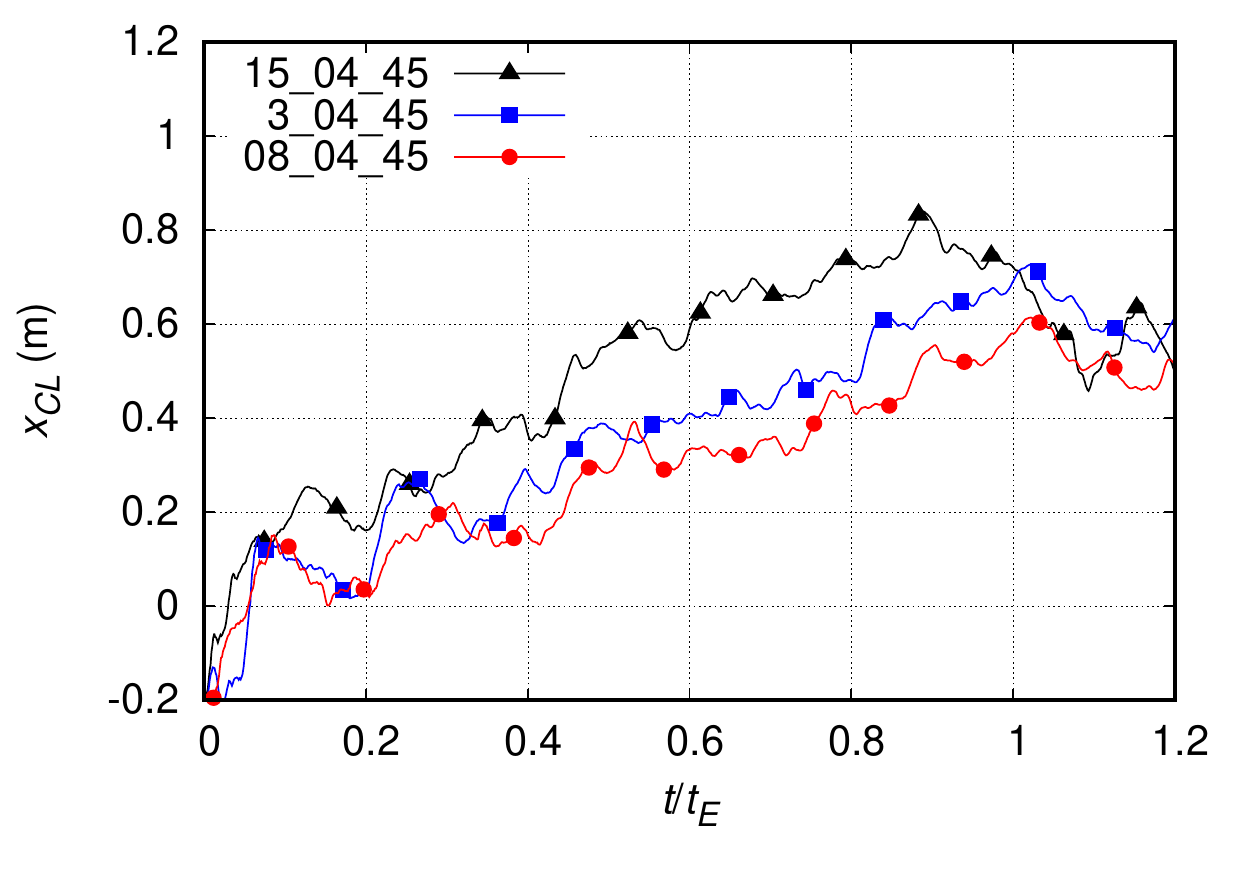}} 
\caption{Position of the centre of loads during the impact phase for different thickness's for the conditions XX\_10\_30, XX\_06\_40 and XX\_04\_45.}
\label{fig:Centre_of_Loads}
\end{figure}

By using the data of the forces provided by the rear and forward load cells, it is possible to retrieve an estimate of
the position of the centre of loads during the impact phase. This is shown in Figure \ref{fig:Centre_of_Loads}. 
The centre of loads is provided as its distance from the trailing edge $x_{CL}$ measured along the plate, as indicated in
Figure \ref{fig:trolley_drawing}, and it is given by:
	\[
	x_{CL}=x_0 + \frac{L \, F_{zF}}{F_{zR}+ F_{zF}}
\]
where $L$ is the distance between the two load cells (1.385 m), $x_0$ is the distance between the trailing edge and the rear load cell (0.02 m) and $F_{zF}$ and $F_{zR}$ are the forces measured by the forward and rear load cell respectively.%                                                                         
The position of the centre of loads at a certain time instant is of course related to the pressure distribution
over the plate at that instant. As the plate penetrates water, the centre of load moves at a certain speed towards the trailing edge, as a consequence of the
propagation of the spray root and of the corresponding pressure peak. 
At the end of the impact phase the pressure peak reaches the leading edge and leaves the plate surface, 
hence the centre of loads moves backwards.
For the cases XX\_10\_30 i.e. at U= 30~m/s and pitch angle~10$^{\circ}$, the structural deformations 
are not so relevant, thus there are no substantial 
changes in the position of the centre of loads. This means that even if the total loads 
tend to slightly increase as the thickness decreases, however the pressure 
distribution at each time instant is not particularly affected by the 
fluid-structure interaction. 
For the test conditions XX\_06\_40 and XX\_04\_45, instead, the centre of loads for the thin 
plate propagates more slowly towards the leading edge compared to the thick plates.
This is because, due to the deformation of the plate, the spray root propagation velocity is reduced, 
particularly in the middle of the plate, as shown in Figures \ref{fig:X113_Pressure} and \ref{fig:JetRootShapeComparisonX113}, 
and, correspondingly the associated pressure distribution is more spread compared to the case of the thick plates.
Owing to the increased loading, the reduction in the propagation velocity of the centre of loads is even more evident for the test condition XX\_04\_45 i.e. at 45~m/s and at 4$^{\circ}$ pitch angle, as shown in Figure \ref{fig:Centre_of_Loads}(c).
\section{Conclusions}
In this paper an analysis of the fluid-structure interaction phenomena occurring during the water impact of Aluminium plates of different thickness at high horizontal velocity has been carried out based on experimental measurements. Test conditions are, as much as possible, close to those of real aircraft ditching scenarios.

It has been found that the reduction of the thickness (and so of the stiffness) of the plate induces a higher level of strains, which go beyond the yielding limit in some cases. The structural behaviour changes substantially due to both the different stiffness and to the fluid-structure interaction, which leads to a quite different hydrodynamic loading. 

A careful analysis of the pressure and strains has revealed that the correlation between them is progressively reduced as the structural deformation increases with the reduction of the stiffness as well as with the penetration of the plate. 

By looking at the time histories of pressure, a general reduction of the pressure peaks is observed, which is accompanied by a spreading of the pressure distribution in both space and time. Particularly evident is the effect of the structural deformations on the propagation of the spray root. The pressure peak/spray root propagation speed is observed to be reduced for the thin plates compared to to thick plates. It is also found that the out-of-plane deformation leads to an inversion of the curvature of the spray root line compared to the case of the thick plates, as also attested by the underwater images.

An increase in the total loading with the increase in the structural deformation is observed. Such increase is noticed starting from about $t/t_E \simeq 0.2$ (i.e. from 1/5th of the impact phase) and progressively grows up to the end of the impact phase when it can be 30\% to 50\% higher than the corresponding loading in the thick plates. Differently from the thick plates, the scaling with the normal projection of the force exerted by the stagnation pressure is no longer valid in presence of large structural deformation, i.e. for $t/t_E > 0.2.$

Two practical conclusions can be drawn based on the above considerations. First of all, differently from what it is usually assumed and even without a strong hydroelastic coupling, tests on rigid structures do not always provide conservative estimates of the loads. Such a circumstance depends on the details of the structure and on the hydrodynamic loading, and it is difficult to judge in advance. Secondly, in presence of a strong fluid-structure interaction, such as that observed in these tests, the hydrodynamic loads are strongly dependent on the structural behaviour, and it is impossible to determine a proper scaling that works for geometry, velocity, loads and structure at the same time.

The above considerations pave the way for future and more focused investigations on the subject. First of all, a deeper comprehension of the strong fluid-structure interaction phenomena and, specifically, of the way in which the loading changes as a consequence of the structural deformation is deemed essential. Furthermore, such studies could also help to understand whether there is a way to scale the structure that can also account for the variation of the loads associated with the plate deformation. Finally, it would be interesting to analyse what happens in terms of the energy balance and the energy exchange between the fluid and the structure. Owing to the complexity of the problem and to the need of achieving a detailed characterization of the flow field, of the pressure distribution and of the resulting structural loading and deformation, such studies can only be addressed by high-fidelity, fully coupled, fluid-structural numerical models. To this purpose, the unique data coming from this rather extensive experimental campaign surely constitute a solid database for the verification and validation of such numerical tools.

\section*{Funding}

This project has been partly funded from the European Union's Horizon 2020 Research and Innovation Programme under Grant Agreement No. 724139 (H2020-SARAH: increased SAfety \& Robust certification for ditching of Aircrafts \& Helicopters).

\appendix

\section{Ratio of the wetting time to the first natural period of the plate}
\label{nat_frequencies}

As noticed in Faltinsen \cite{faltinsen1999water,faltinsen2005hydrodynamics}, the relevance of the 
fluid-structure interaction can be indicated by the ratio of the wetting time 
to the first wetted natural period $T_{11}$ (related to the mode 1-1) of the 
flat plate. 

The wetting time can be estimated on first approximation, as the duration of the impact phase $t_E$.
The first wetted natural period, instead, can be derived using a simplified approach, 
which is described hereafter.

In the following  $a$, $b$ and $h$ are the length, the width 
and the thickness of the plate respectively. Since the outer part of the plate is reinforced by the frame, it is not
allowed to deform, therefore it is assumed that the values of $a$ and $b$ are slightly reduced, being 0.9 and 0.4 m respectively, instead of 1 and 0.5.

As anticipated in Section \ref{setup}, the boundary condition on the edges of the plate can be considered a quasi-clamped boundary condition, in between a simply-supported condition at all sides (SS) and a clamped condition at all sides (C). This has been verified a posteriori by comparing the results of numerical modal FEM simulations and experimental modal tests, which were performed on the 3 mm plate. Given that, both these conditions are examined in this procedure. 

In the case of simply supported plate at all sides (SS) the first natural (dry) frequency can be estimated using the relationship (4.20) in Leissa \cite{leissa}:
\begin{equation}
	\label{f11_SS}
	f_{11}^{SS} = \frac{1}{2 \pi} \, \sqrt{ \frac{D}{ \rho^{\prime}_m} } \, \left[ \left( \frac{\pi}{a} \right)^2 + \left( \frac{\pi}{b} \right)^2 \right]
\end{equation}
In the previous relationships $D = \frac{E \, h^3}{12 \, (1-\nu^2)}$, $E$, $\nu$ and $\rho^{\prime}_m$ are the Young modulus, 
the Poisson's ratio and the mass of the material per unit area respectively. For the plate material, the Aluminium alloy AL2024 T3,
E=73.1~GPa, $\nu=0.33$ and $\rho^{\prime}_m=\rho_m \, h$, where $\rho_m$ is the density of the material i.e. 2780~kg/m$^3$.

In the case of clamped edges at all sides (C) the first natural frequency can be determined using the relationship
\begin{equation}
	\label{f11_C}
 \qquad f_{11}^{C} = \frac{1}{2 \pi} \, \sqrt{ \frac{D}{\rho^{\prime}_m} } \, \frac{K_{C}}{b^2} ,
\end{equation}
where $K_{C}$ is the value displayed in Table 4.28 (first row) in \cite{leissa}. $K_{C}$ depends on the ratio $a/b$, which is 2.25 in the present case. In \cite{leissa} it is shown that $K_{C}$ is 24.56 for $a/b=2$ and 23.76 for $a/b=2.5$. It is reasonable to assume that for $a/b=2.25$, $K_{C}$ is the average of the two values reported above, hence $K_{C}$=24.16.

To estimate the wetted natural frequency/period, it is assumed that the mass of the material per unit area is increased by
an amount $\rho^{\prime}_a$, which is the added mass per unit area of a flat plate initially floating on the free surface and moving suddenly downwards (heave).
On first approximation, the total added mass of the plate can be estimated as outlined in Newman \cite{newman}. In particular, in Table 4.3 (page 145) the value of the added mass of a flat plate of unit length and half-width $b$ that moves in a vertical direction in an unbounded domain is provided. In the case of plate impact, the added mass at heave, which we denote as $m_a$, is halved (see Newman \cite{newman}, Section 6.17), hence we obtain:
\begin{equation}
	m_a = \frac{1}{8} \, \pi \, \rho a b^2
\end{equation}
where $\rho$ is the water density. Assuming that the added mass is distributed uniformly over the plate surface, the value of $\rho^{\prime}_a$, the added mass per unit surface, is given by:
\begin{equation}
	\rho^{\prime}_a= \frac{m_a}{a \, b} = \frac{\pi}{8} \, \rho \, b
\end{equation}
The first natural wetted frequency can then be estimated using the relationships 
(\ref{f11_SS}) and (\ref{f11_C}), however replacing $\rho^{\prime}_m$ with $\rho^{\prime}_w \doteq \rho^{\prime}_m + \rho^{\prime}_a$: 
\begin{equation}
	f_{11 \, wet}^{SS} = \frac{1}{2 \pi} \, \sqrt{ \frac{D}{ \rho^{\prime}_w} } \, \left[ \left( \frac{\pi}{a} \right)^2 + \left( \frac{\pi}{b} \right)^2 \right] \qquad 	f_{11 \, wet}^{C} = \frac{1}{2 \pi} \, \sqrt{ \frac{D}{\rho^{\prime}_w} } \, \frac{24.16}{b^2} .
\end{equation}

For the quasi-clamped boundary condition, it can be assumed that the first natural wetted frequency is the average of these two values, i.e.
\begin{equation}
	 f_{11 \, wet}= 1/2 \left( f_{11 \, wet}^{SS} + f_{11 \, wet}^{C} \right)
\end{equation}

The ratio $t_E/T_{11}$ is shown in Table \ref{tab:TableRatio} for a combination
of three plate thicnkess's, namely 15~mm, 3~mm and 0.8~mm, and three pitch angles, 
namely $\alpha=4^{\circ}$, $\alpha=6^{\circ}$ and $\alpha=10^{\circ}$. It is worth observing that the values of $t_E$
in this case are not determined as described in Section \ref{setup}, 
but on first approximation as $t_E=L \, \sin{\alpha}/V$, which corresponds to the 
assumption that the flat plate does not deform significantly during the water entry.
\begin{table}[h!]
\centering
\begin{tabular}{|c|c|c|c|} \hline
	$t_E/T_{11}$ & $\alpha=4^{\circ}$ & $\alpha=6^{\circ}$ & $\alpha=10^{\circ}$ \\ \hline
	t=15~mm & 8.97 & 13.4 & 22.3 \\ \hline 
	t=3~mm & 0.88 & 1.31 & 2.18 \\ \hline 
	t=0.8~mm & 0.12 & 0.18 & 0.31 \\ \hline
\end{tabular}
\caption{Ratio of the wetting time to the first wetted natural period of the flat plates for the test conditions examined.}
\label{tab:TableRatio}
\end{table}
Faltinsen \cite{faltinsen2005hydrodynamics} reports that for a ratio $t_E/T_{11}$ 
higher than 1 a quasi-static approach can be applied to the transient problem
by assuming the structure as rigid, whereas for $t_E/T_{11} 
\leq 0.25$ a coupling between the hydrodynamic and the elastic 
problem is expected to occur. Based on the data shown in Table \ref{tab:TableRatio}, a significant fluid-structure interaction
is expected for the 0.8~mm plate 
at all pitch angles, but it could also be relevant for the 3~mm plate, 
especially at $\alpha=4^{\circ}$ and $\alpha=6^{\circ}$.

\section{Test repeatability}
\label{repeatability_section}

\begin{figure}[htbp]
\centering
\subfigure[S$_{1x}$]{\includegraphics[width=0.45\textwidth]{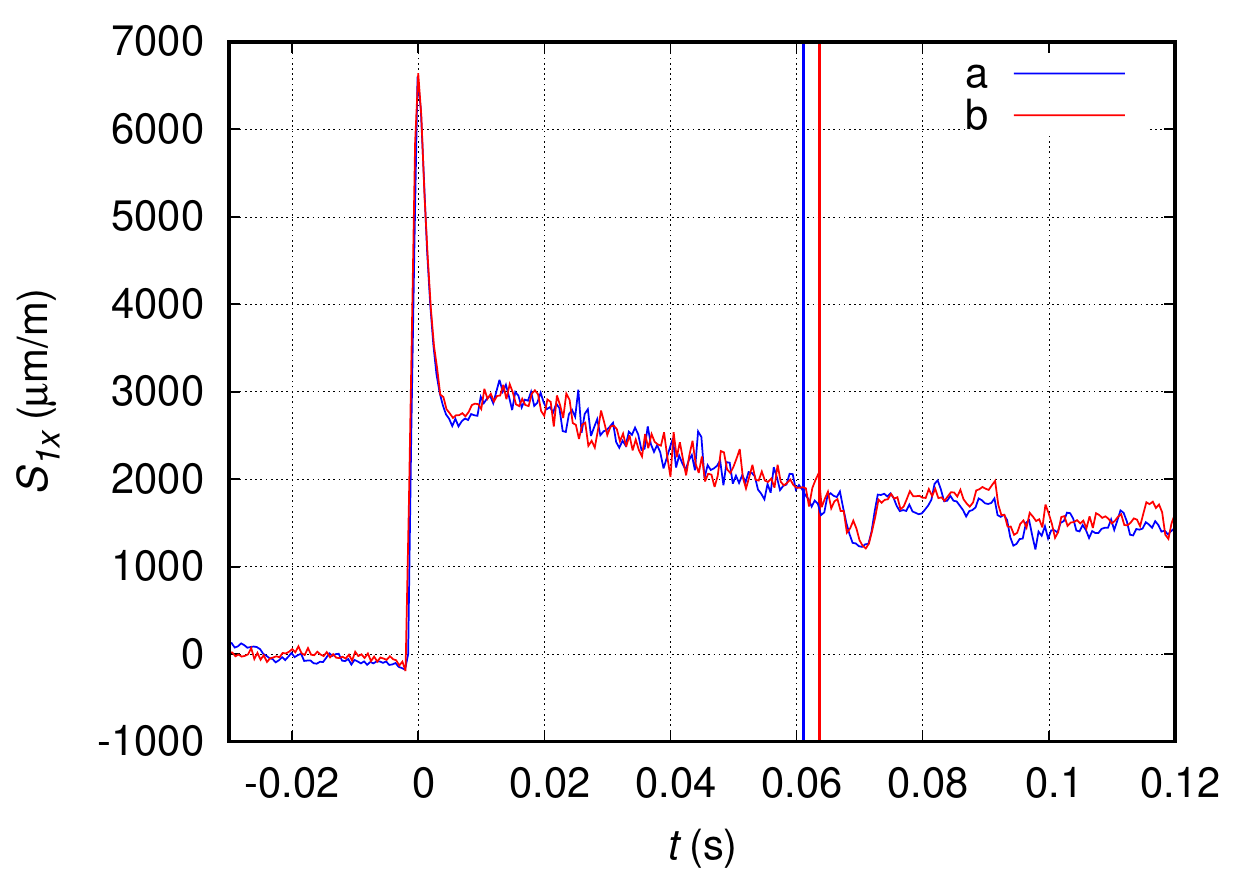}} \quad
\subfigure[S$_{2x}$]{\includegraphics[width=0.45\textwidth]{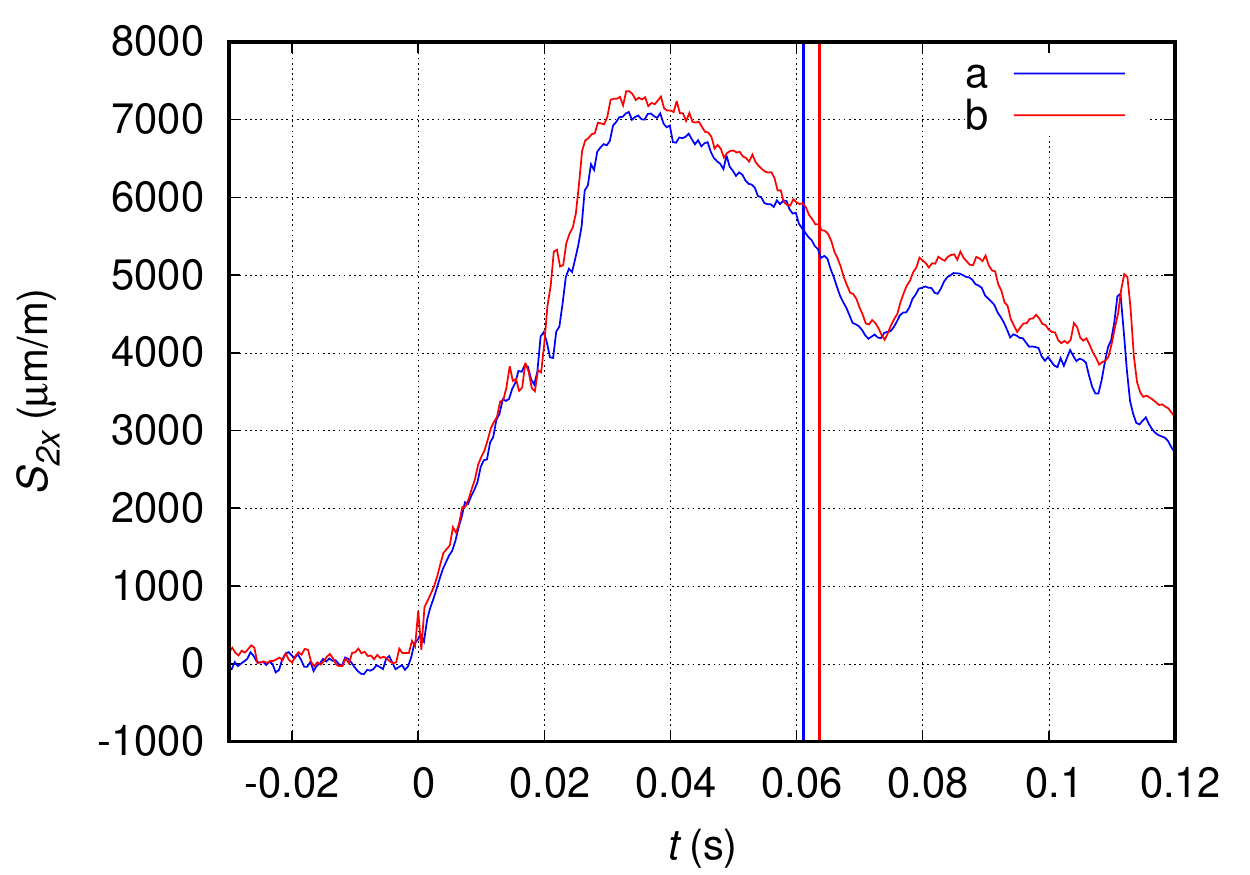}}
\\
\subfigure[S$_{3x}$]{\includegraphics[width=0.45\textwidth]{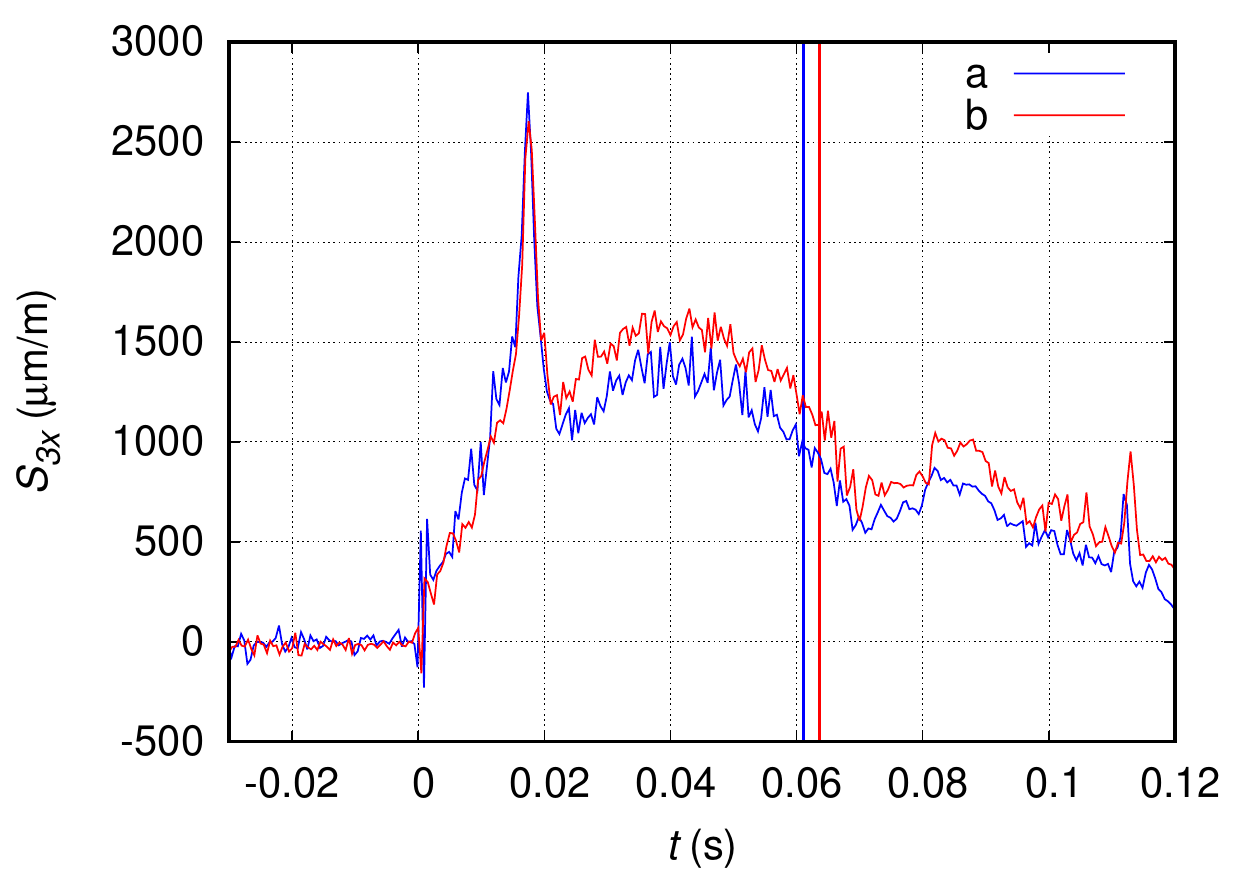}} \quad
\subfigure[S$_{5x}$]{\includegraphics[width=0.45\textwidth]{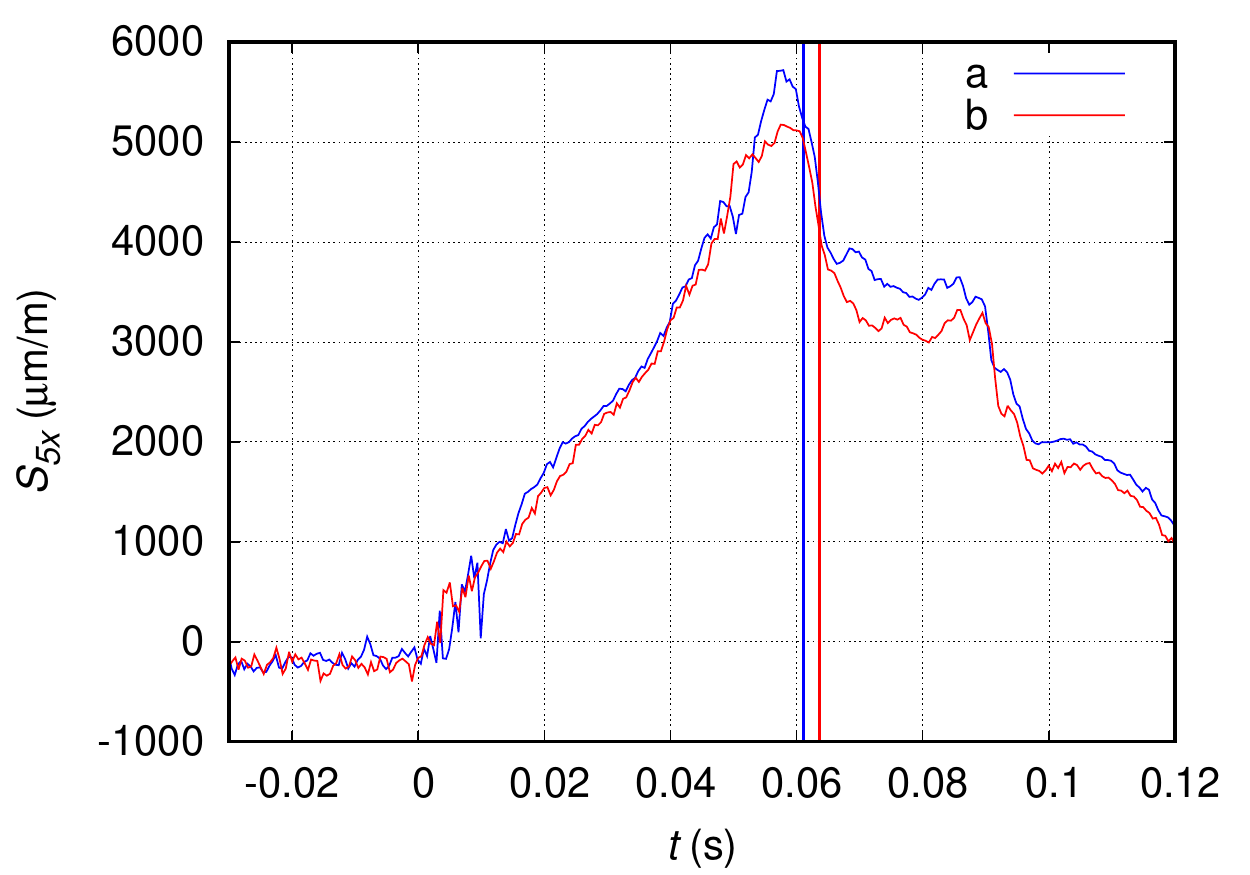}} 
\\
\subfigure[S$_{1y}$]{\includegraphics[width=0.45\textwidth]{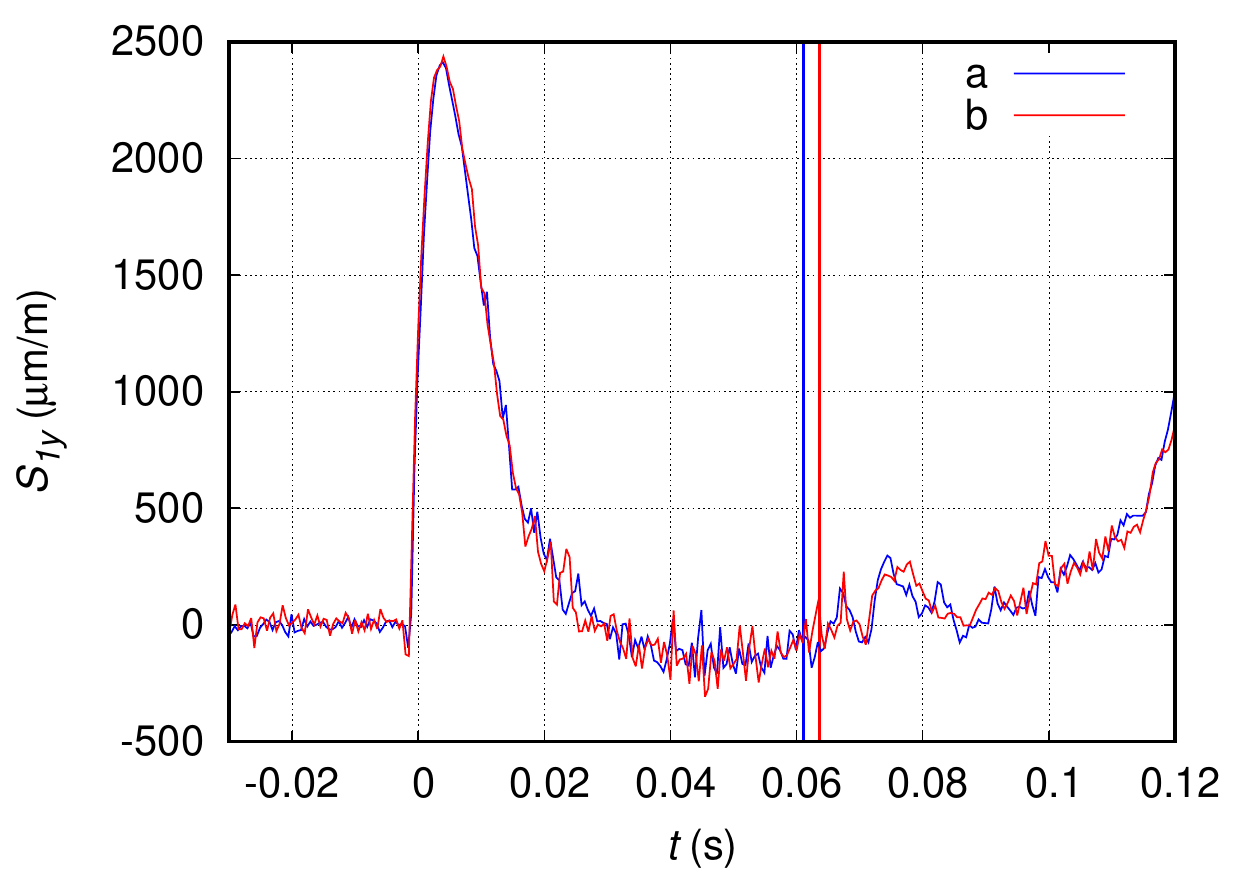}} \quad
\subfigure[S$_{2y}$]{\includegraphics[width=0.45\textwidth]{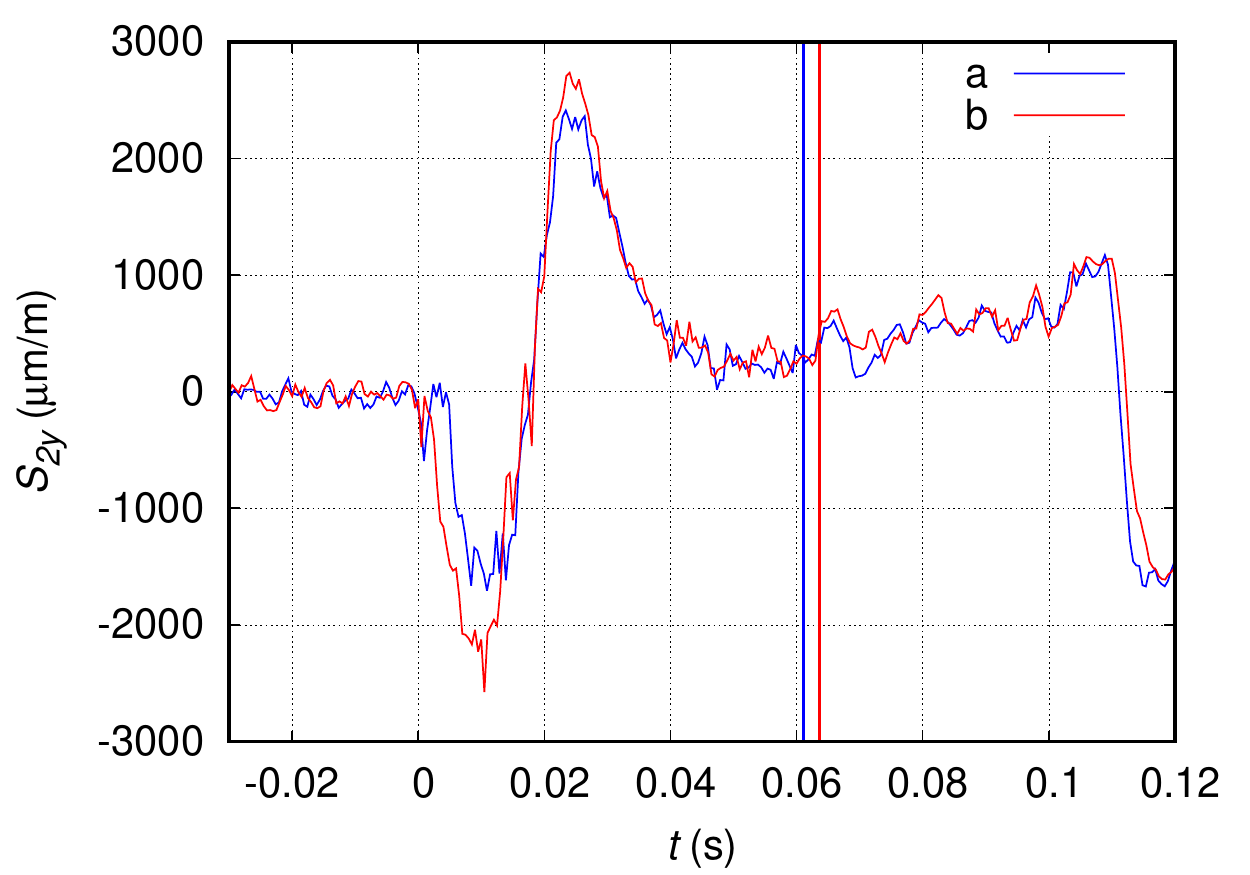}}
\\
\subfigure[S$_{3y}$]{\includegraphics[width=0.45\textwidth]{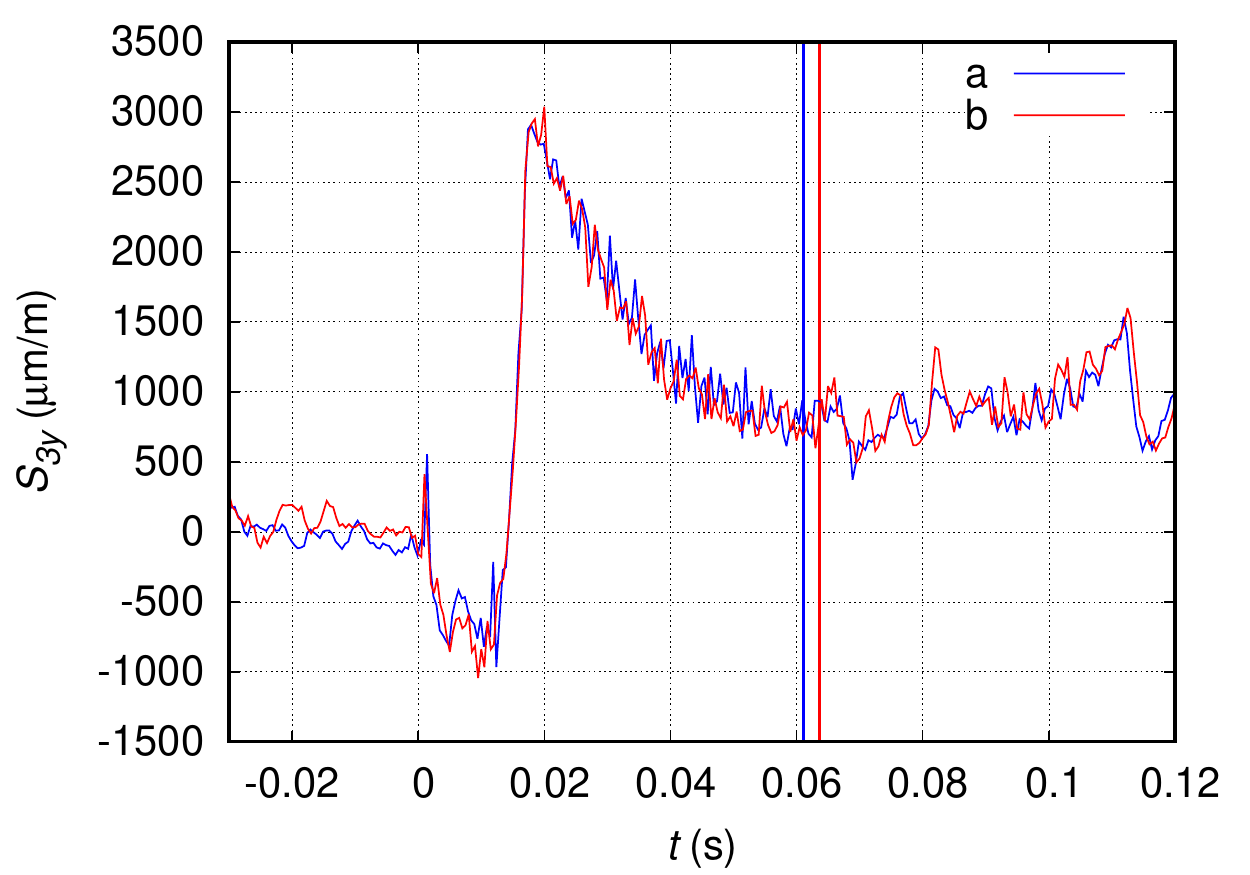}} \quad
\subfigure[S$_{5y}$]{\includegraphics[width=0.45\textwidth]{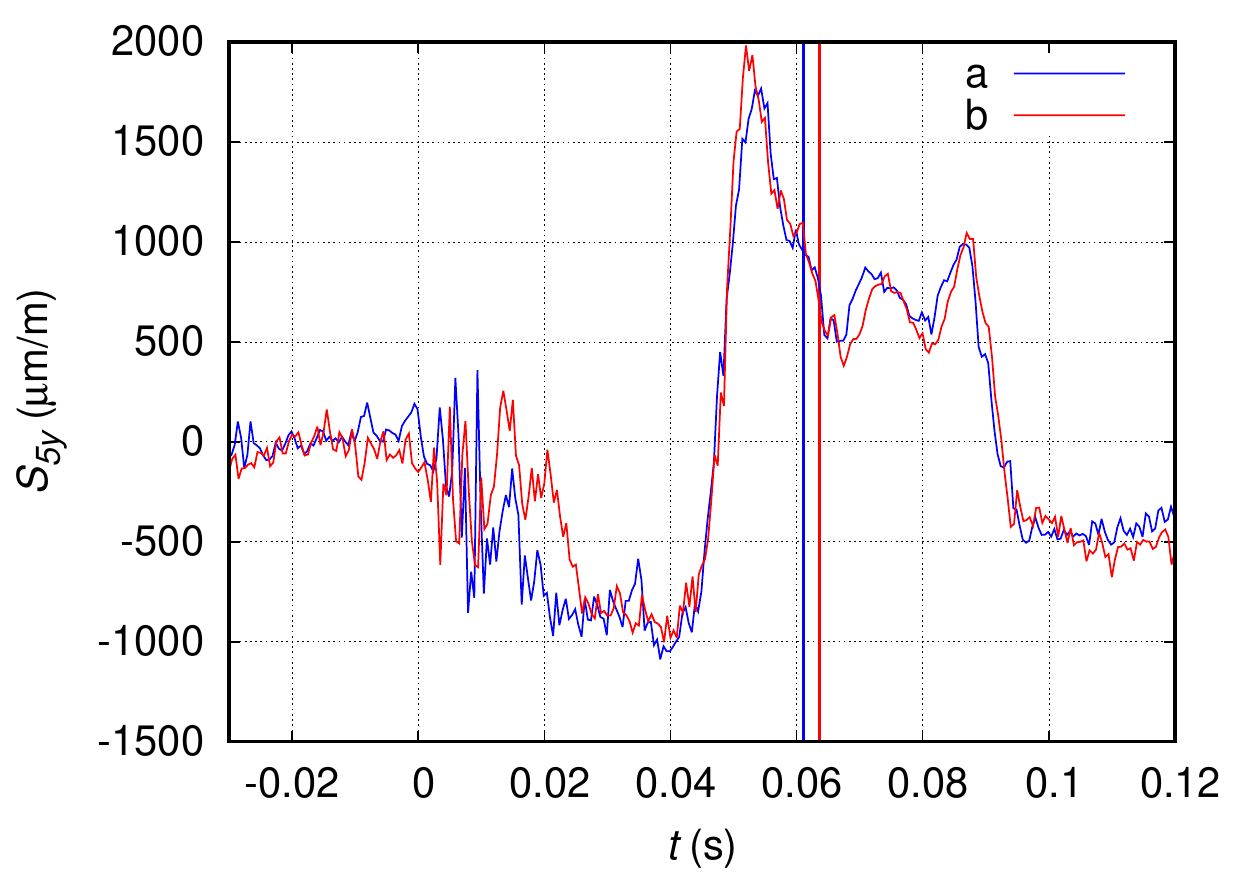}}
\caption{Comparison of the longitudinal and transverse strains recorded by the strain gauges 
$S_1$, $S_2$, $S_3$ and $S_5$ for the 0.8~mm plate at U=40~m/s, pitch angle~6$^{\circ}$ for the two 
repeats.}
\label{fig:SXY_3122_all}
\end{figure}

In order to assess the repeatability of the test conditions, 
the time histories of the strains $S_1$, $S_2$, $S_3$ and $S_5$ in the longitudinal
(S$_x$) and transverse (S$_y$) directions for the tests performed on
the 0.8~mm plate tested at 40~m/s and 6$^\circ$ are shown in Figure
\ref{fig:SXY_3122_all}.
In the figures, the time $t_E$ of each test is indicated with a vertical 
line of the same colour as that of the corresponding line plot.
A good repeatability is observed for the strain gauges located along the
midline, i.e. S$_1$, S$_2$ and S$_5$, as well as for the strain gauge
located at the side, i.e. S$_3$, for which a higher sensitivity to
the boundary conditions is expected.

\begin{figure}[htbp]
\centering
\subfigure[S$_{1x}$]{\includegraphics[width=0.45\textwidth]{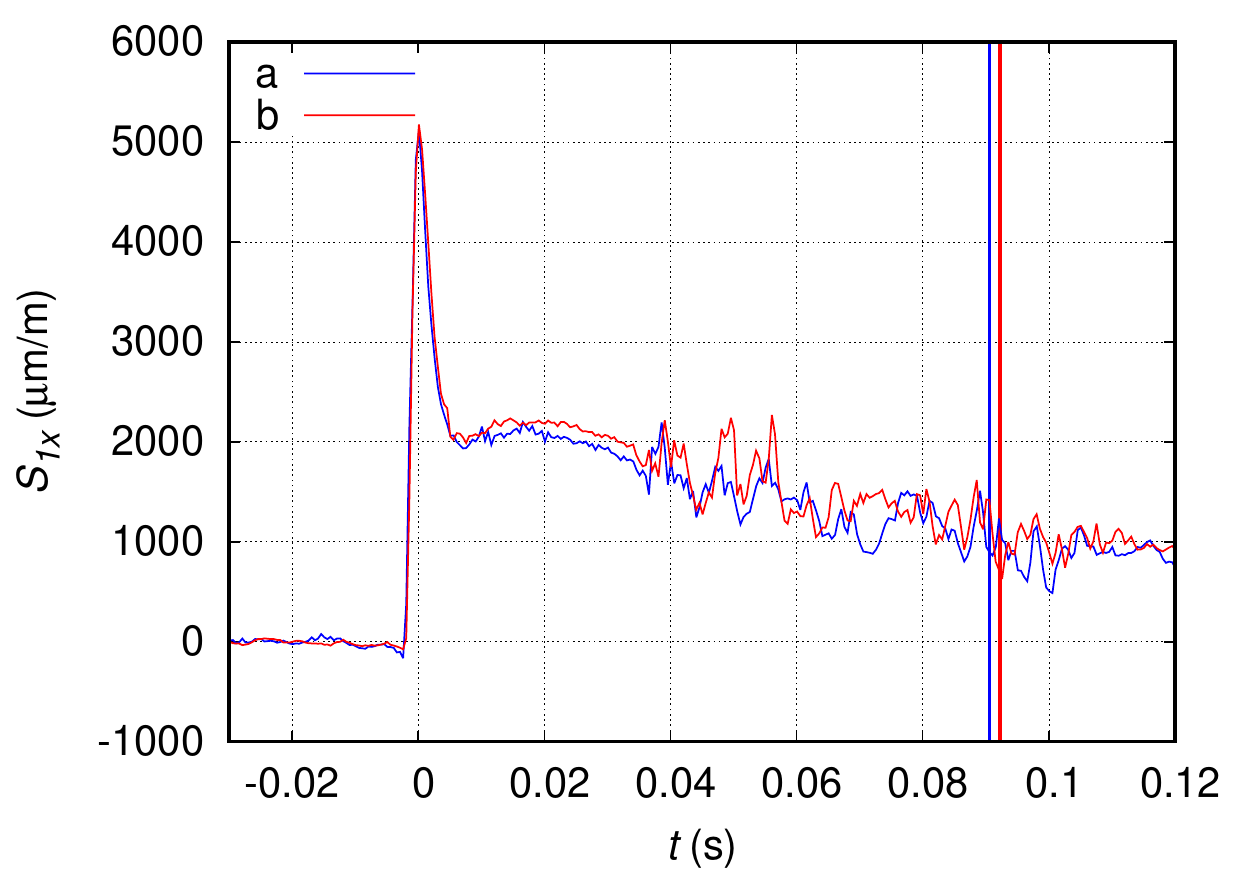}} \quad
\subfigure[S$_{2x}$]{\includegraphics[width=0.45\textwidth]{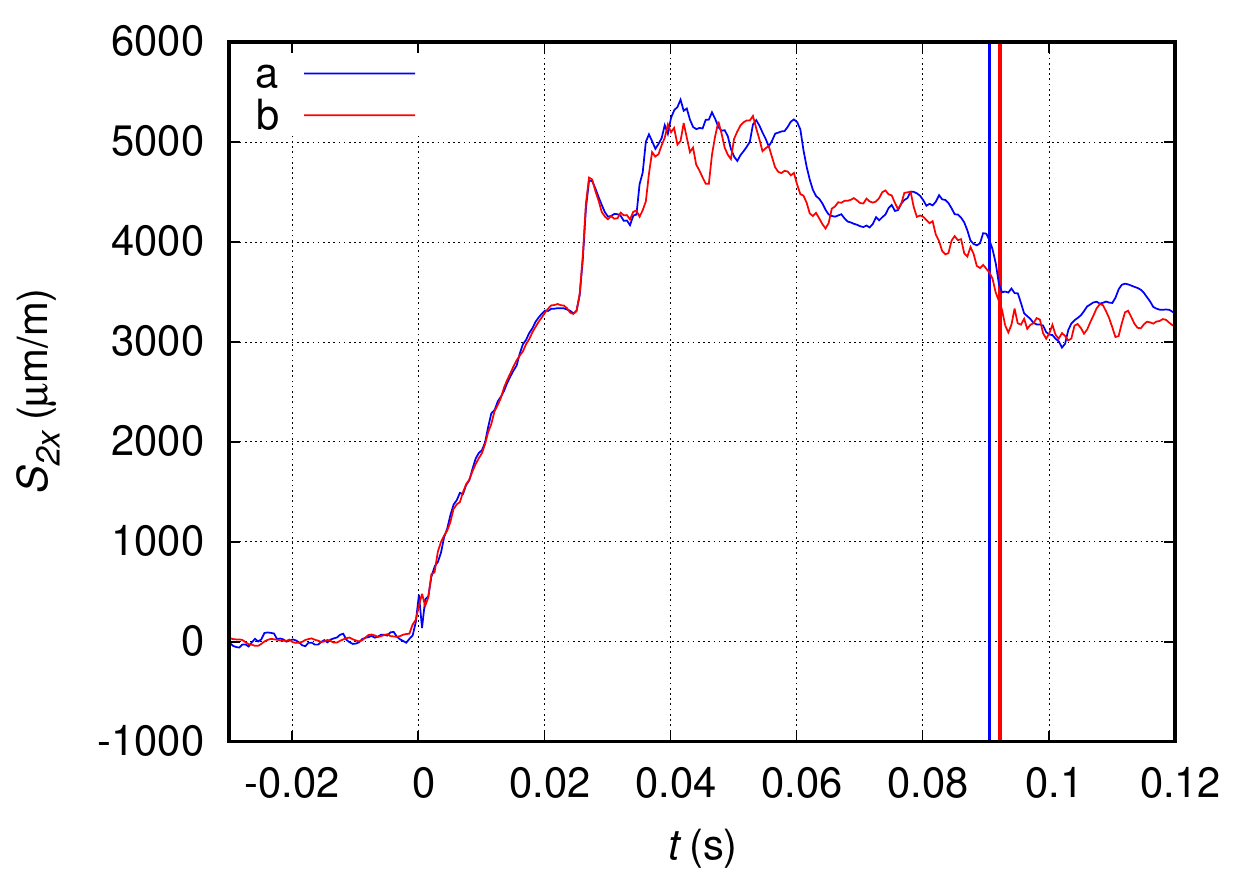}}
\\
\subfigure[S$_{3x}$]{\includegraphics[width=0.45\textwidth]{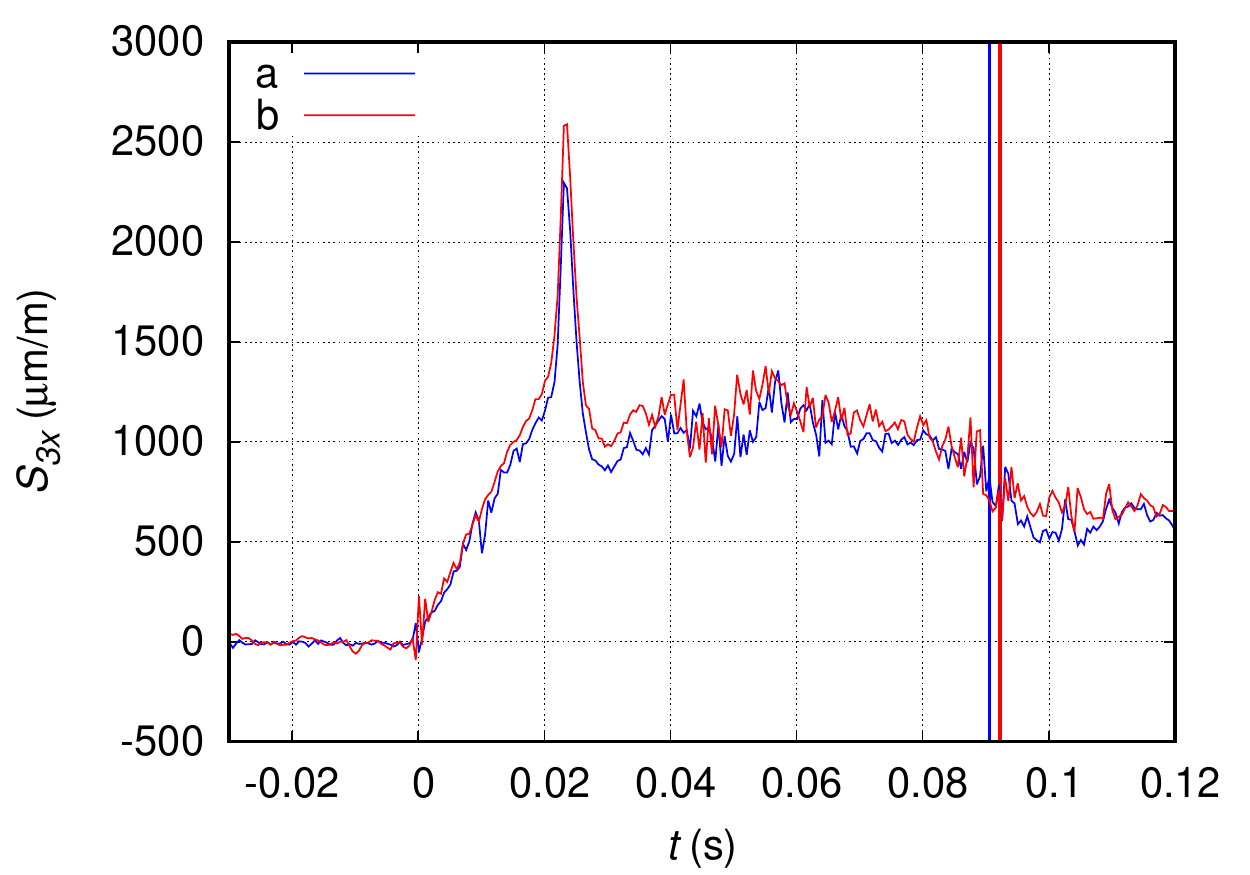}} \quad
\subfigure[S$_{5x}$]{\includegraphics[width=0.45\textwidth]{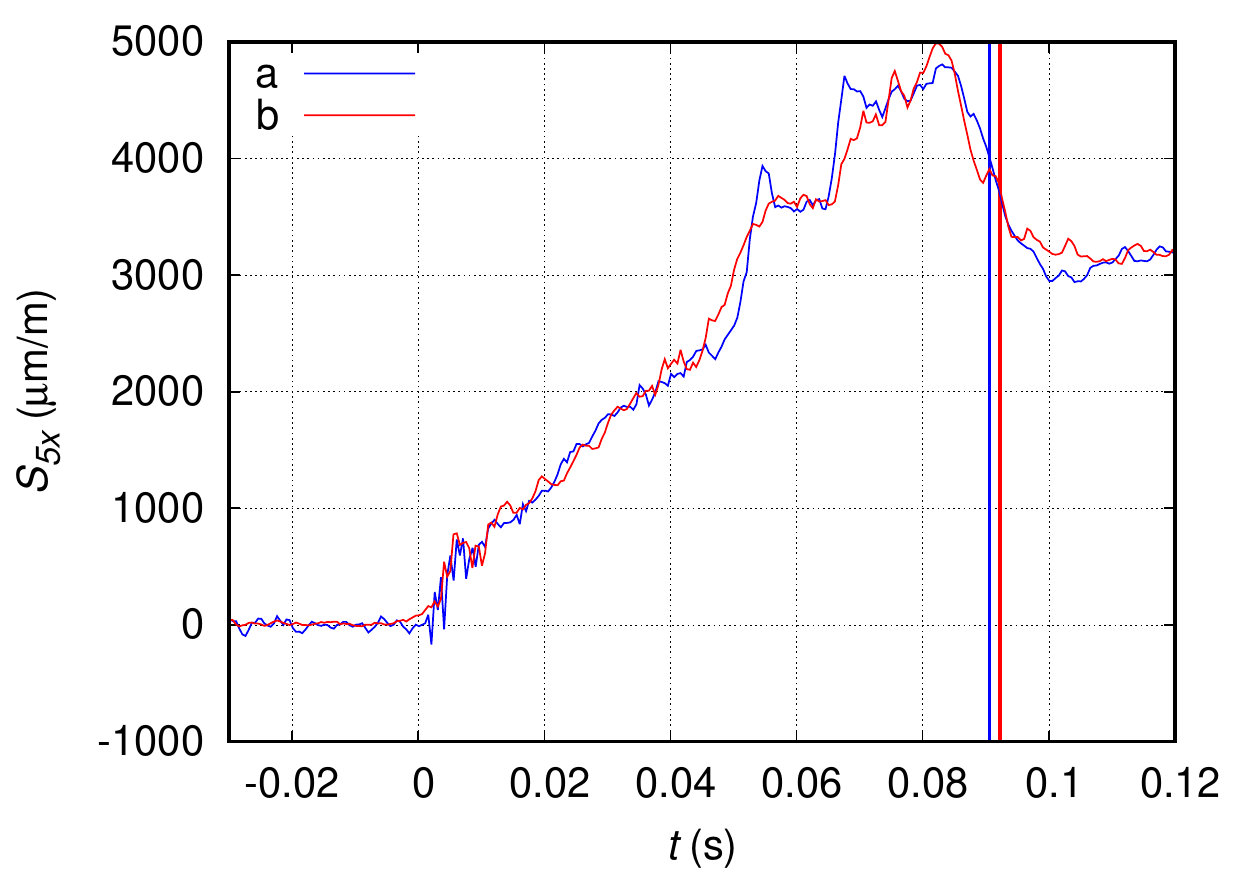}} 
\\
\subfigure[S$_{1y}$]{\includegraphics[width=0.45\textwidth]{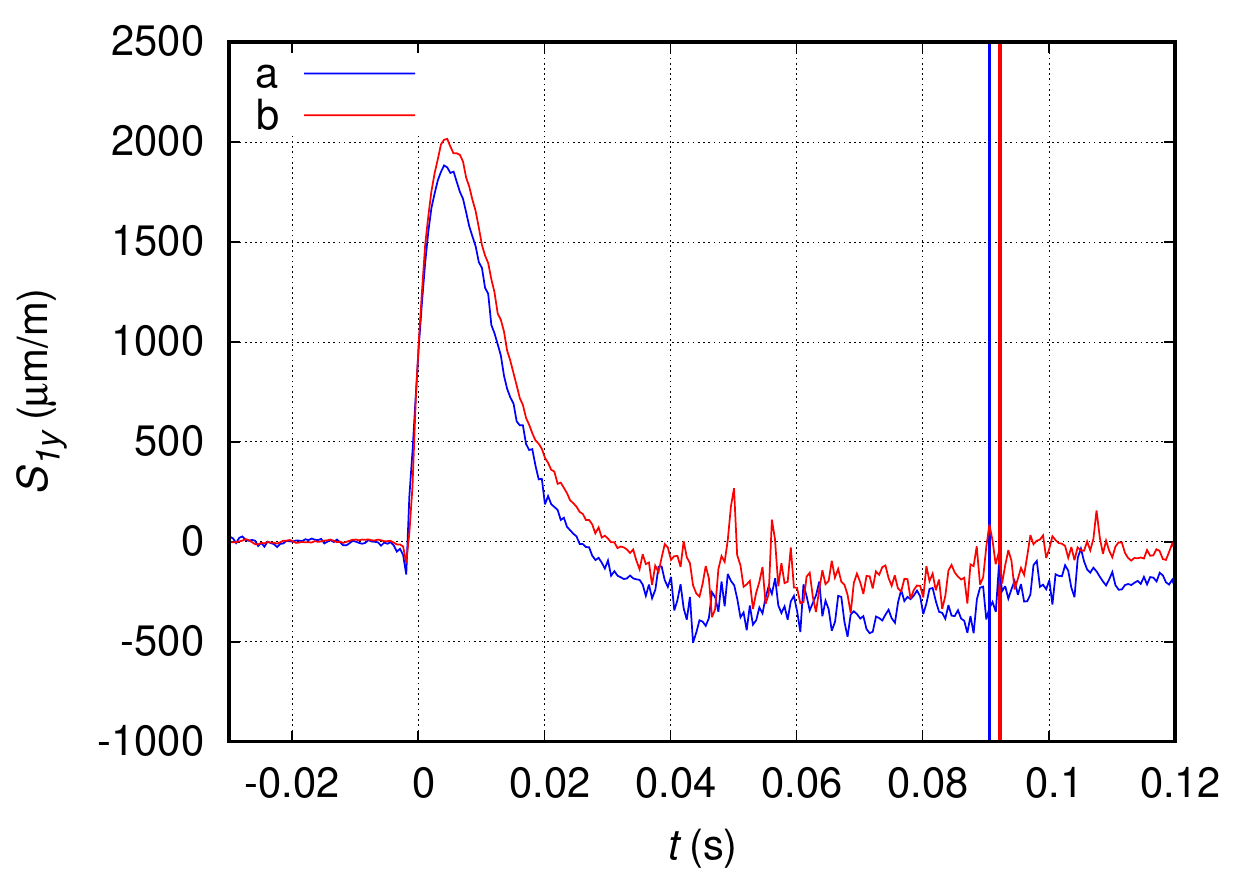}} \quad
\subfigure[S$_{2y}$]{\includegraphics[width=0.45\textwidth]{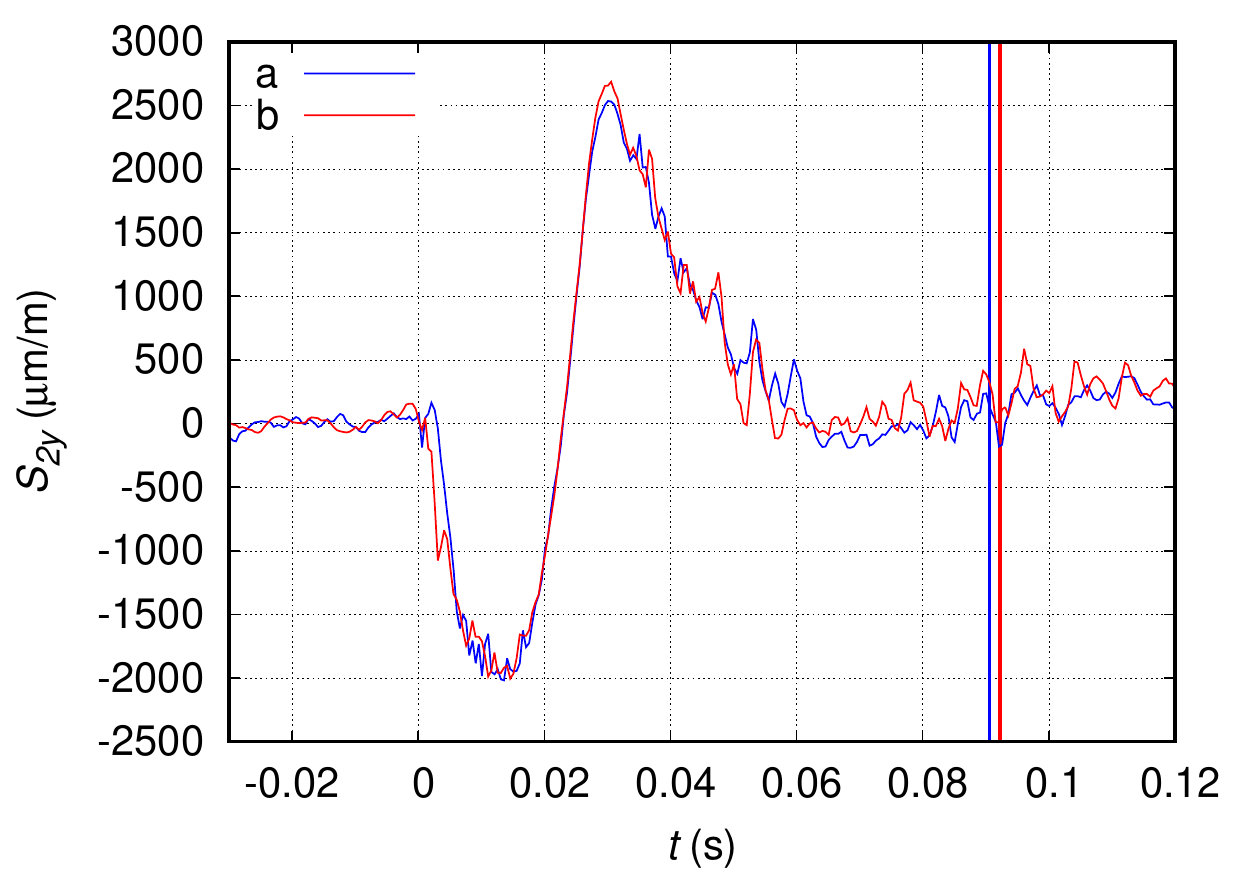}}
\\
\subfigure[S$_{3y}$]{\includegraphics[width=0.45\textwidth]{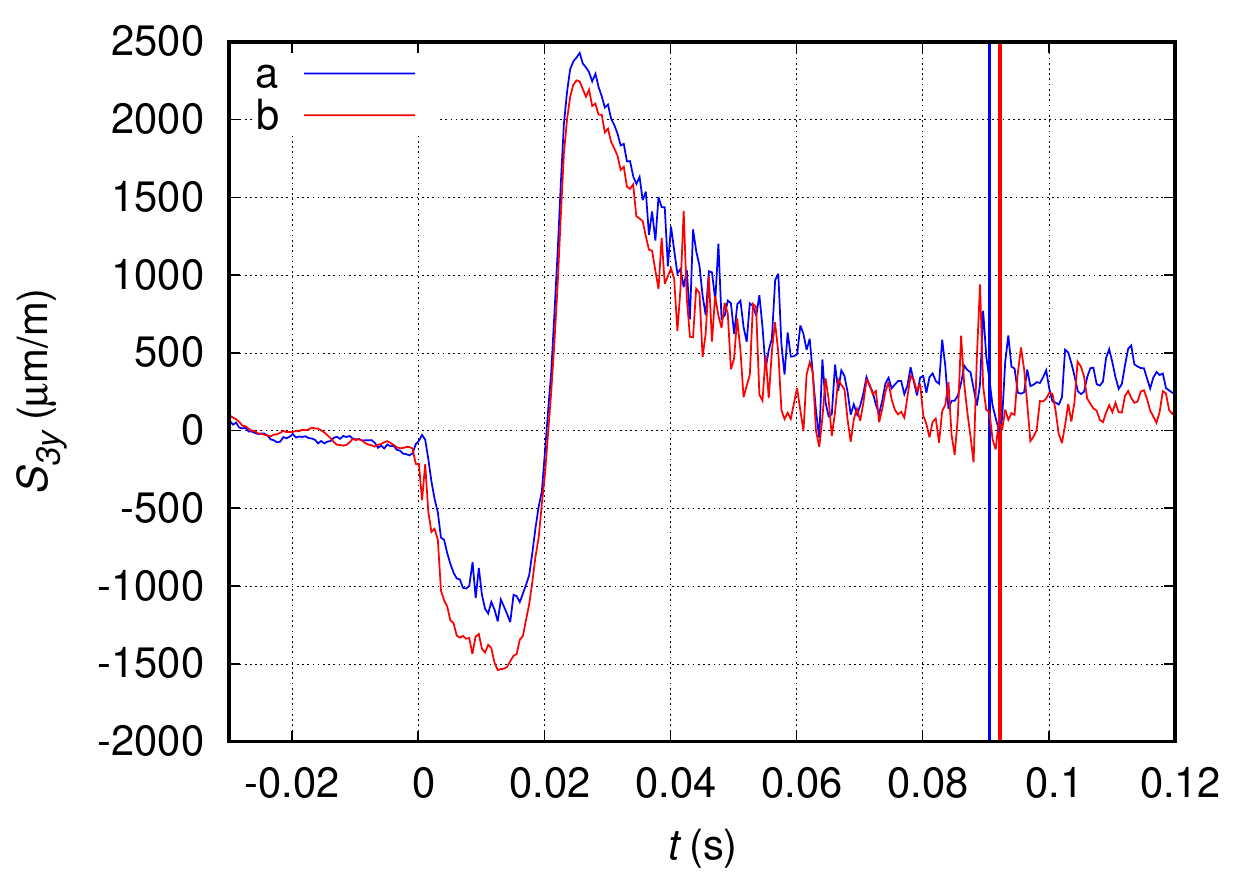}} \quad
\subfigure[S$_{5y}$]{\includegraphics[width=0.45\textwidth]{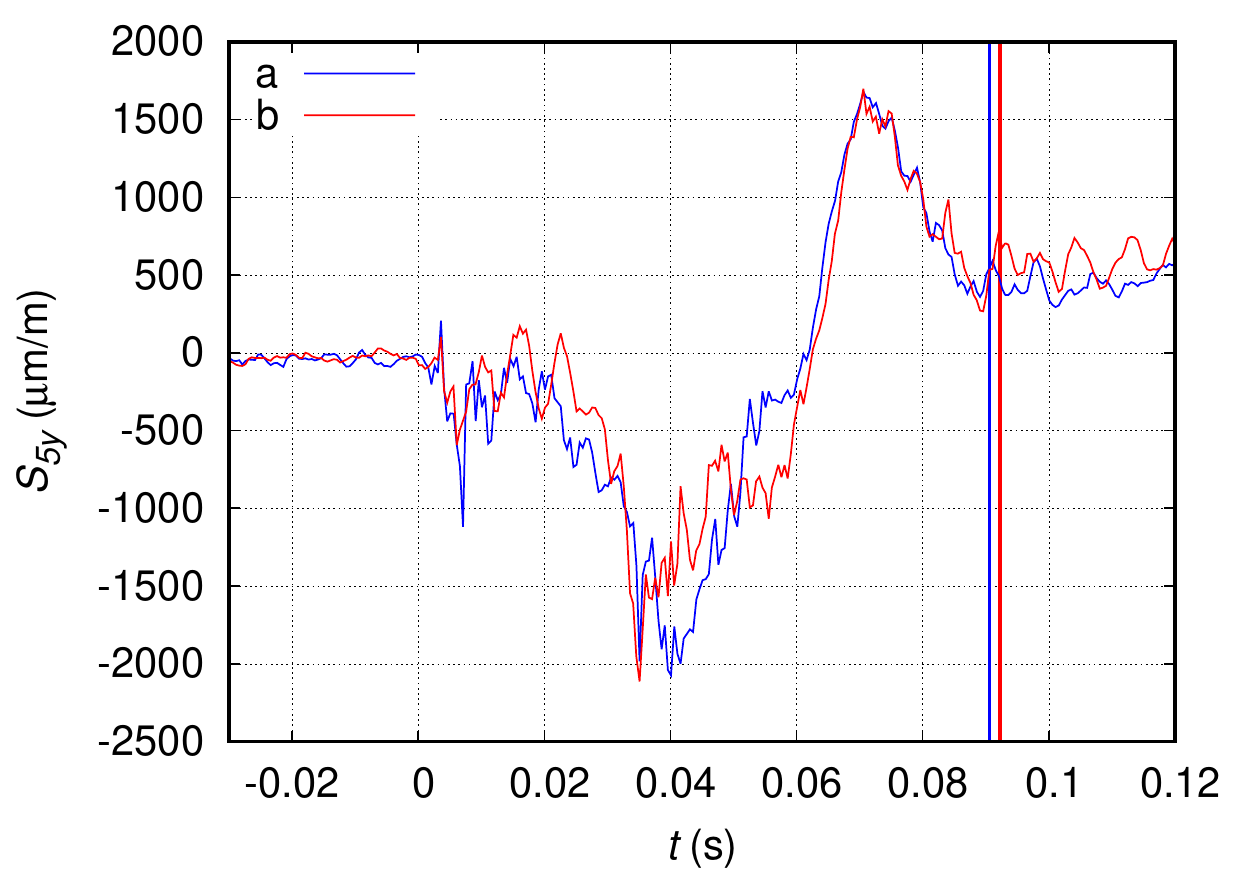}}
\caption{Comparison of the longitudinal and transverse strains recorded by the strain gauges 
$S_1$, $S_2$, $S_3$ and $S_5$ for the 0.8~mm plate at U=30~m/s, pitch angle~10$^{\circ}$ for the two 
repeats.}
\label{fig:SXY_3131_all}
\end{figure}

A similar comparison is established in Figure 
\ref{fig:SXY_3131_all} for the test at U=30~m/s, thickness 0.8~mm and 
10$^{\circ}$~pitch.
In this case, pressure probes were installed on one plate only.
The good agreement that is observed enables to say that there are no significant 
changes in terms of the structural behaviour induced neither by the presence of the
holes drilled to install the pressure probes, nor by the
pressure probes themselves.

\begin{figure}[htbp]
\centering   
\subfigure[$P_{4}$]{\includegraphics[width=0.45\textwidth]{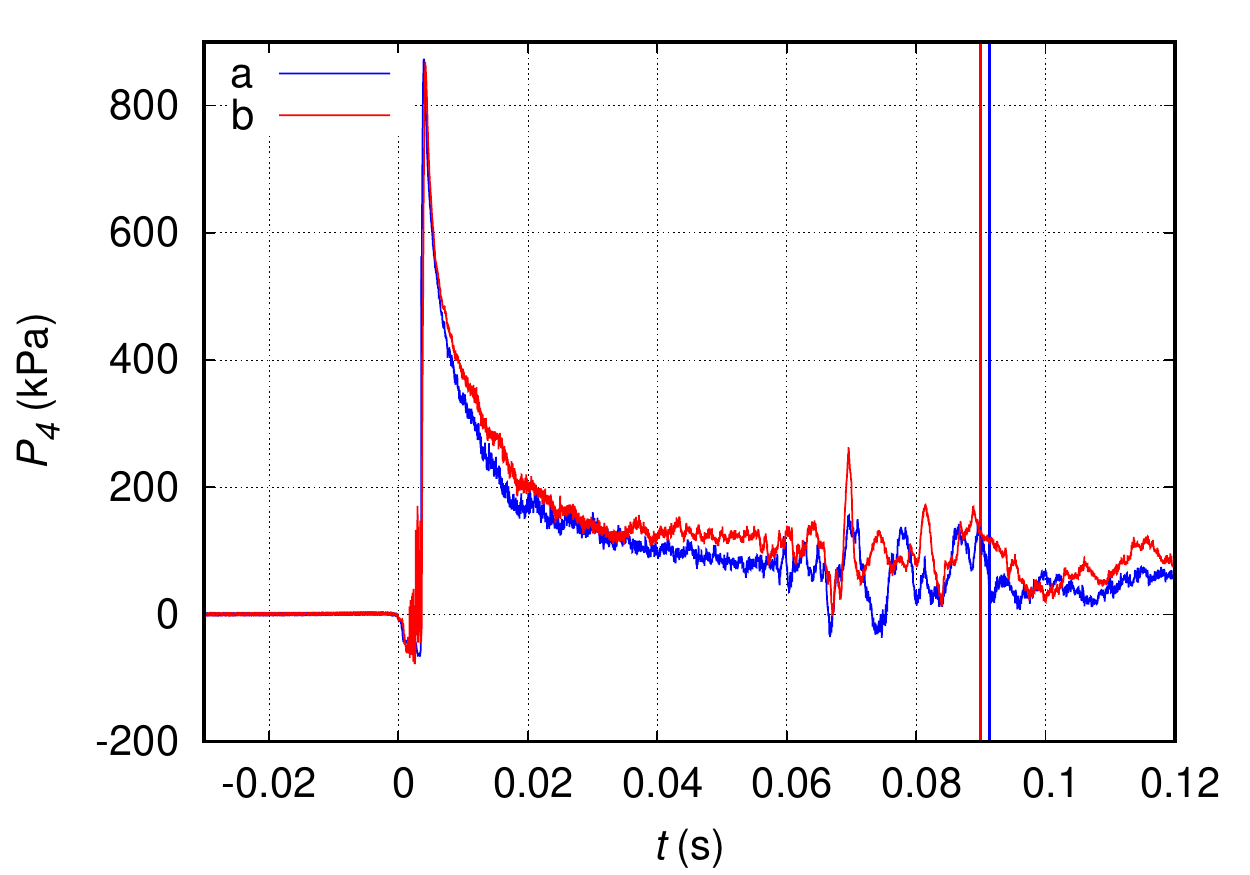}} \quad
\subfigure[$P_{12}$]{\includegraphics[width=0.45\textwidth]{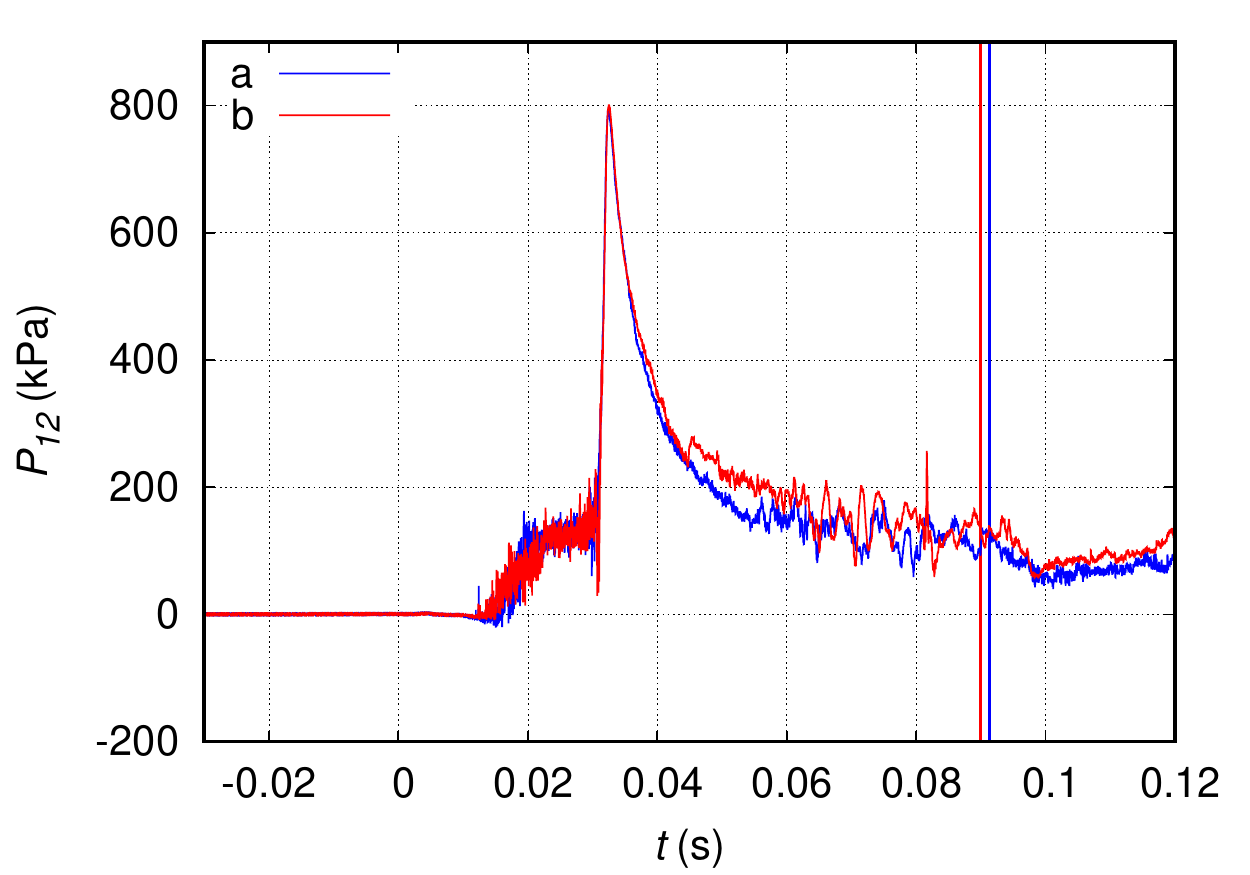}}
\\
\subfigure[$P_{16}$]{\includegraphics[width=0.45\textwidth]{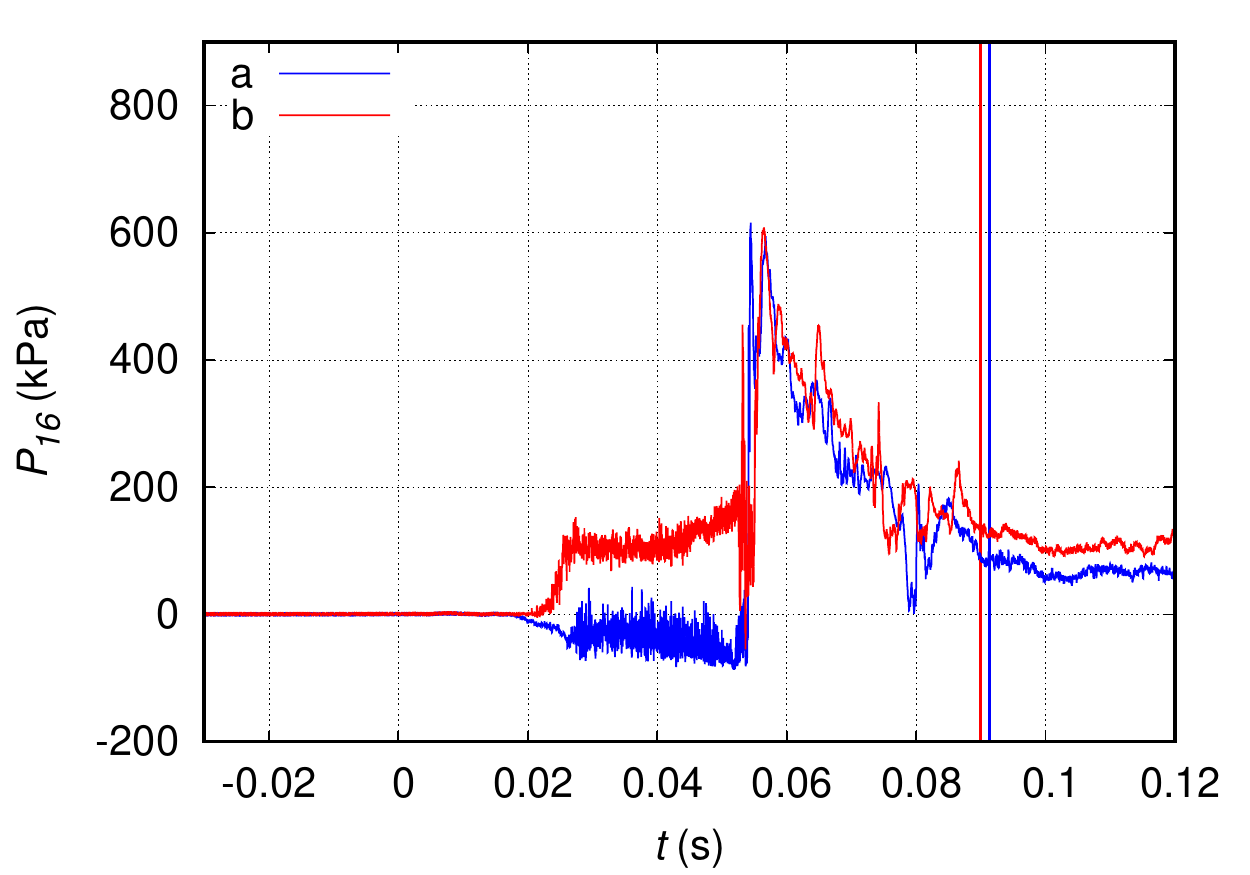}} \quad
\subfigure[$P_{18}$]{\includegraphics[width=0.45\textwidth]{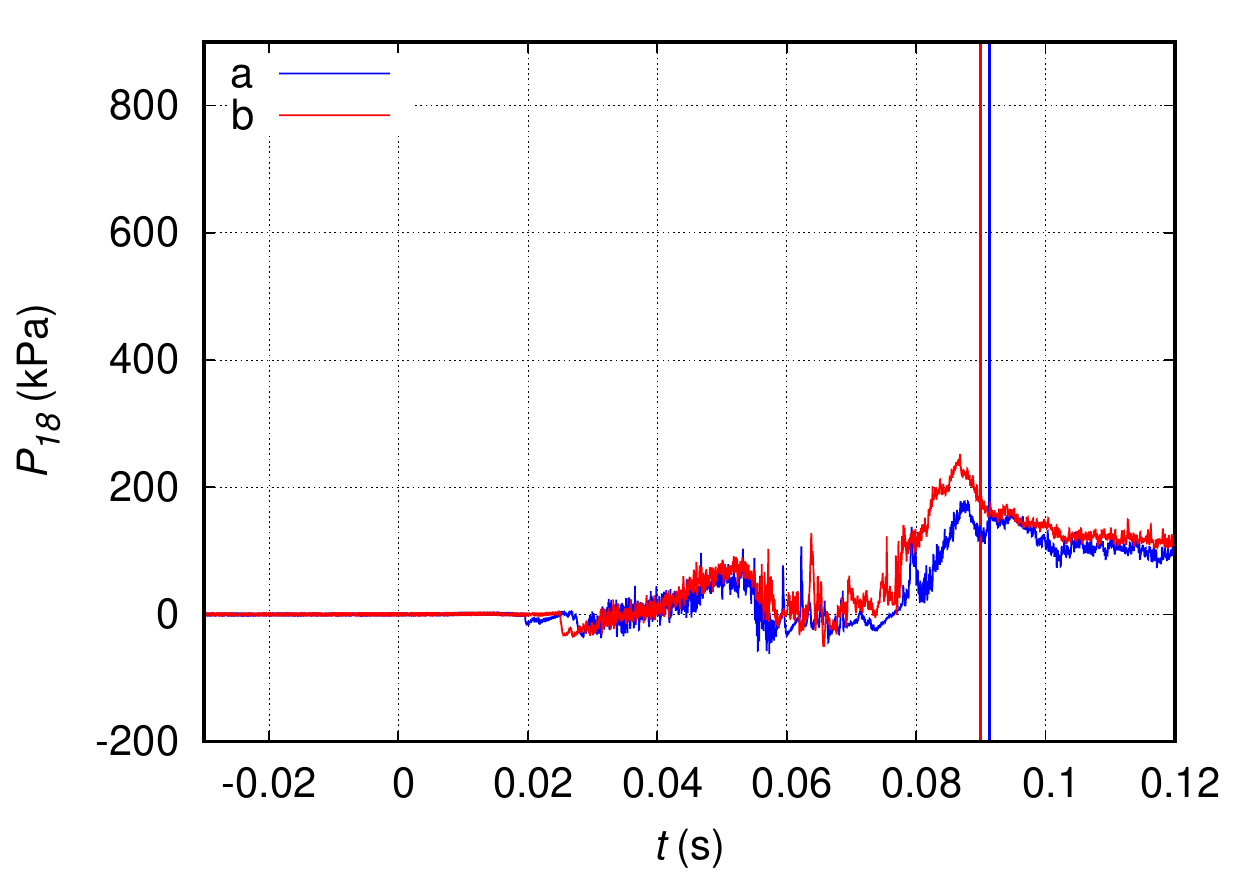}}
\\
\subfigure[$P_{10}$]{\includegraphics[width=0.45\textwidth]{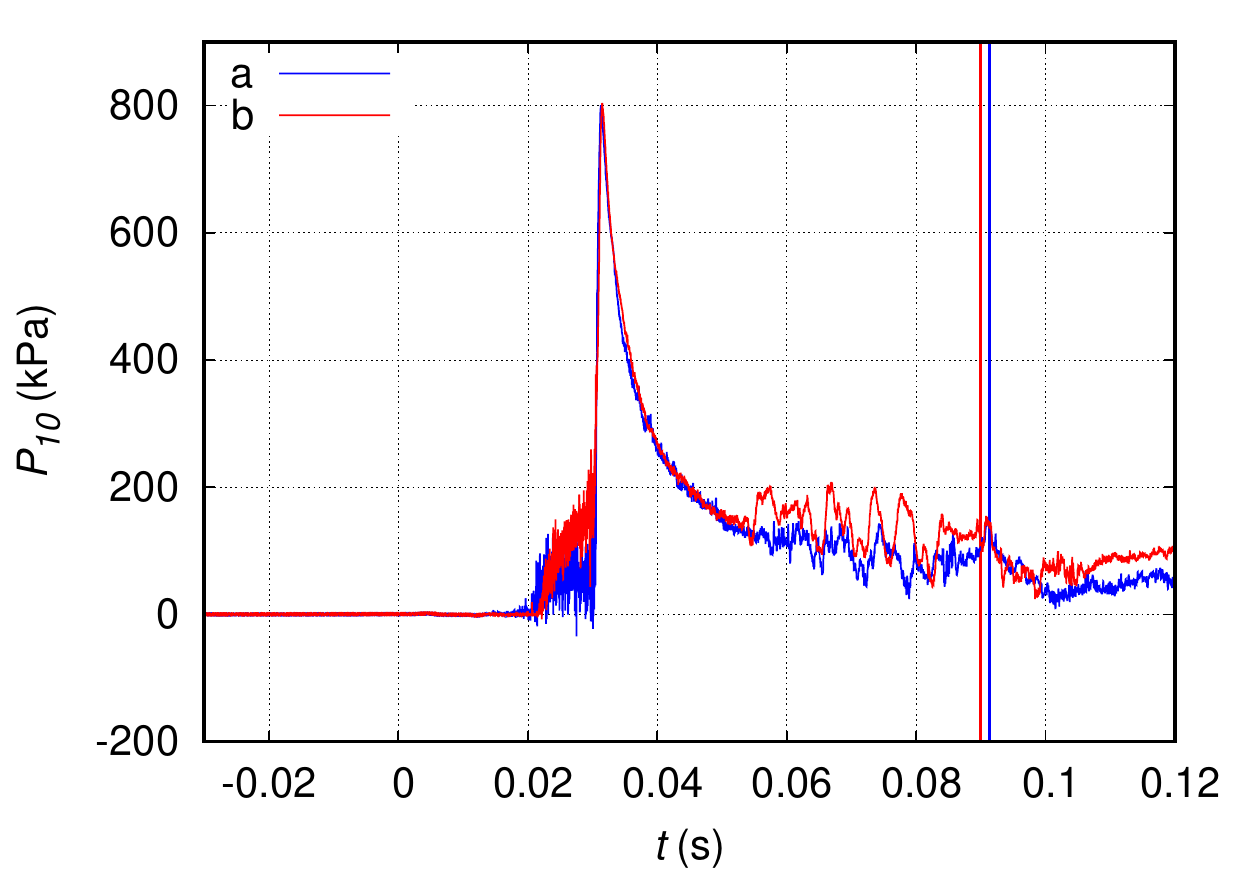}} \quad
\subfigure[$P_{13}$]{\includegraphics[width=0.45\textwidth]{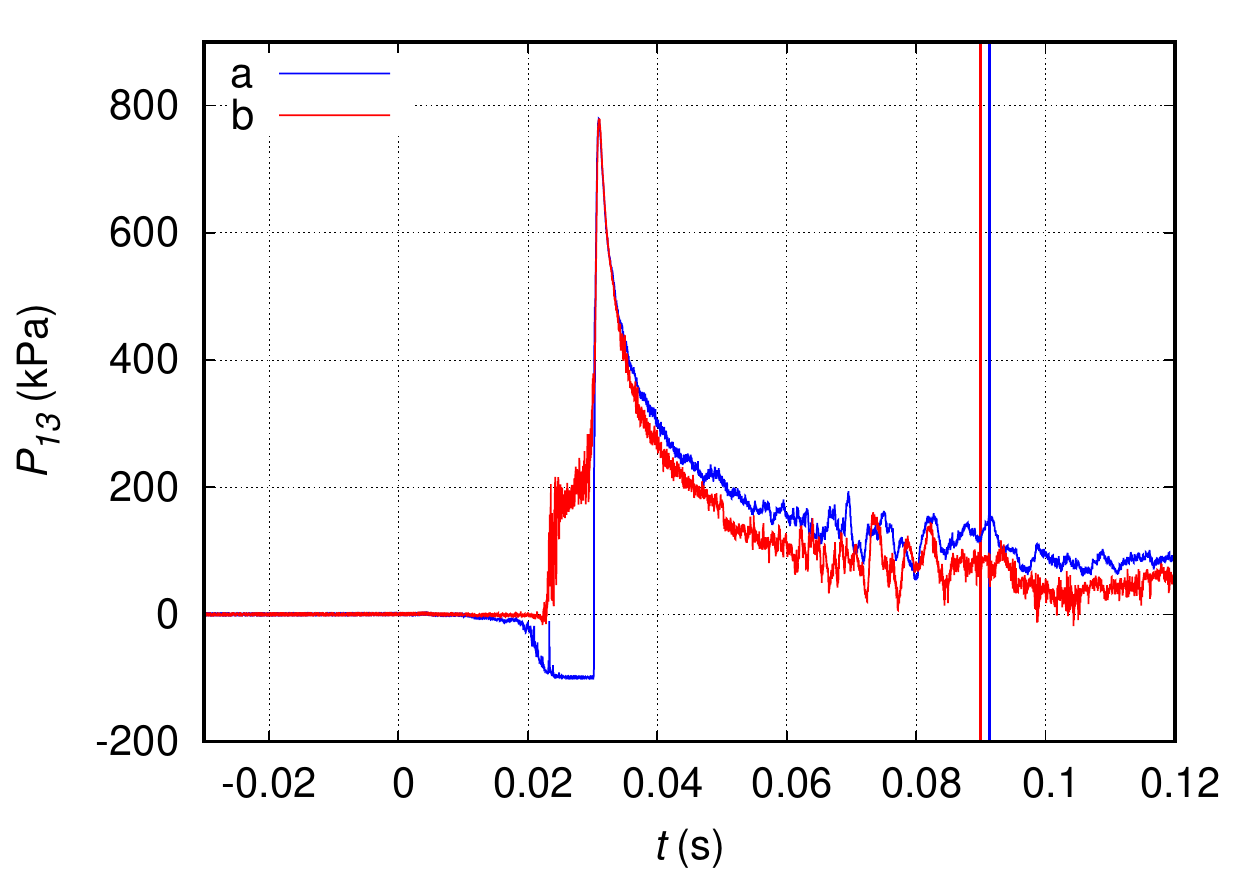}}
\caption{Comparison of pressure time histories for the test condition 3\_10\_30, i.e. at U=30~m/s, thickness 3~mm and pitch angle~10$^{\circ}$
for the two repeats.}
\label{fig:P_2131_some}
\end{figure}

As for the pressure measurements, the repeatability is assessed for the
tests performed at 30~m/s, 10$^{\circ}$~pitch, for the 3~mm plate (Figure 
\ref{fig:P_2131_some}). Comparisons are shown for the probes located along
the midline (P$_4$, P$_{12}$, P$_{16}$ and P$_{18}$) and at the sides
($P_{10}$ and P$_{13}$). Being P$_{10}$ and P$_{13}$ located in symmetric 
positions, their comparison also provides an indication
of the symmetry of the loading and, to some extent, of the structural
deformation as well. 

\begin{figure}[htbp]
\centering
\subfigure[$F_x$]{\includegraphics[width=0.45\textwidth]{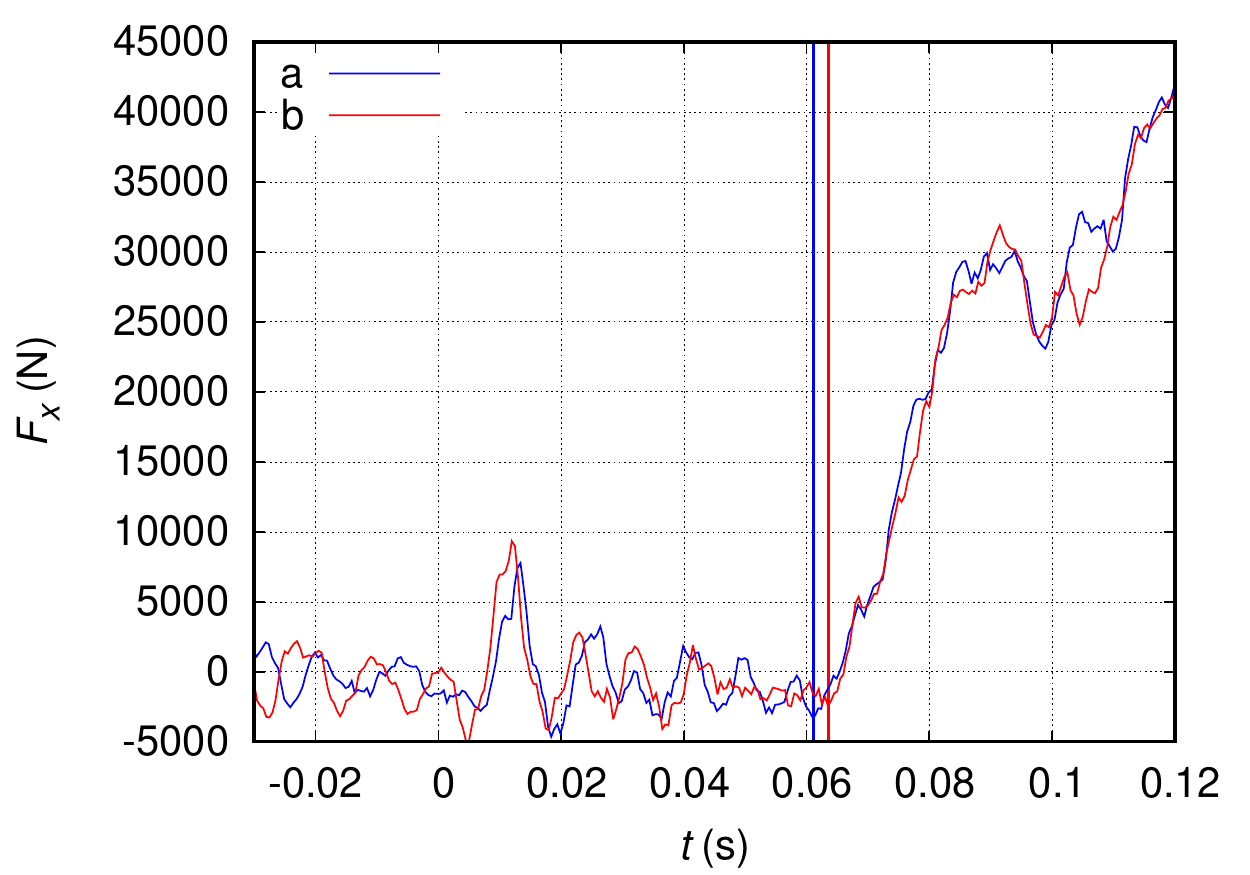}} \quad
\subfigure[$F_z$]{\includegraphics[width=0.45\textwidth]{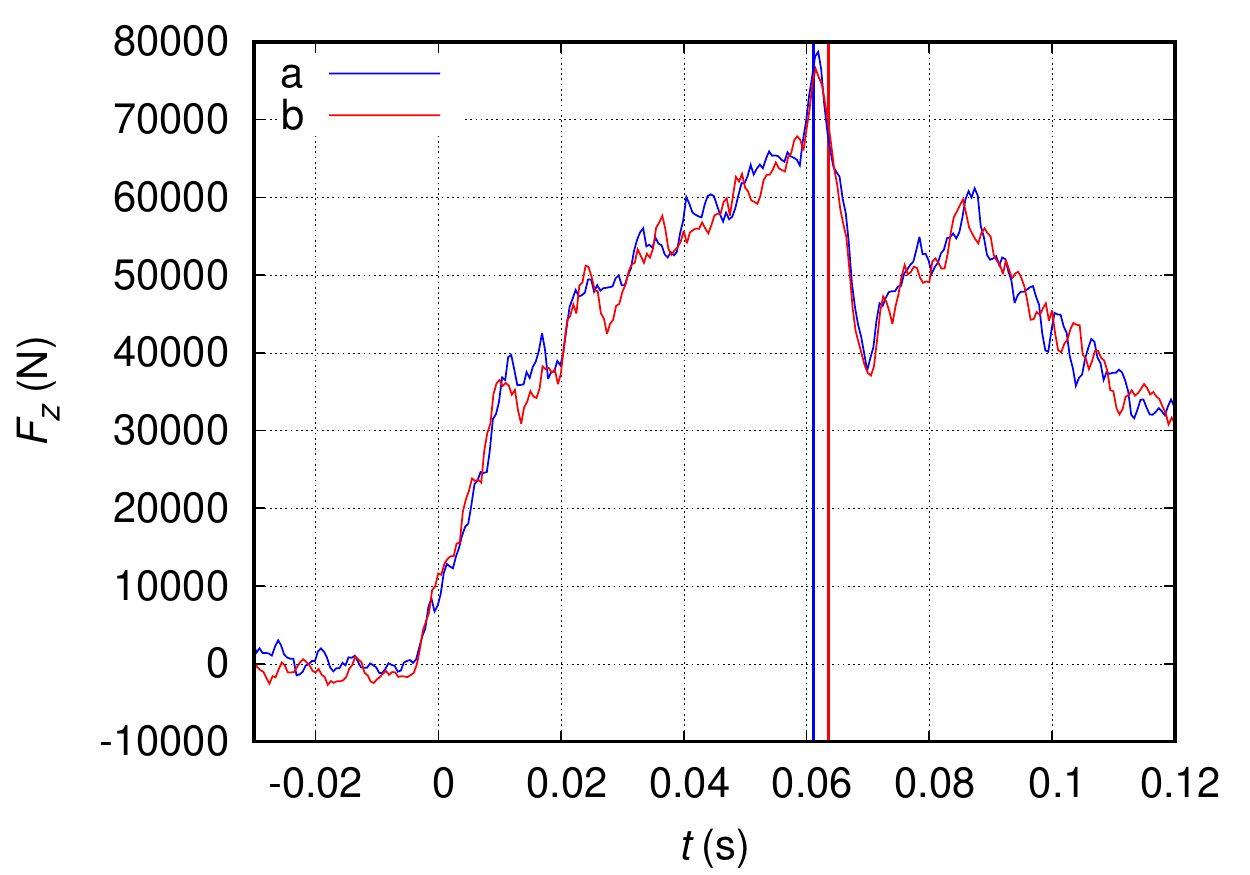}}
\caption{Comparison of the total forces for the
test condition 08\_06\_40, i.e. at U=40~m/s, thickness 0.8~mm and pitch angle~6$^{\circ}$ for the two repeats.}
\label{fig:Forces_3122_all}
\end{figure}

Finally, a comparison between the total load measurements the in $x$ and $z$ direction is shown in Figure
\ref{fig:Forces_3122_all} for the tests of the 0.8~mm plates at 40~m/s, 
pitch angle~6$^{\circ}$.
Even in the test conditions shown in Figure \ref{fig:Forces_3122_all}, in 
which the horizontal velocity and the pitch angle are quite high and the 
thickness is low, a good repeatability is observed. 
Figure \ref{fig:Forces_3122_all} also 
shows that $t_E$ corresponds to the time at which $F_x$ starts to rise and
$F_z$ takes the maximum.

\bibliography{paper_biblio}

\end{document}